\documentclass[iop,apj,numberedappendix,appendixfloats,linenumbers]{emulateapj} 

\usepackage[export]{adjustbox}

\usepackage{amssymb}
\usepackage[usenames, dvipsnames]{color}
\usepackage{makecell}
\usepackage{color}
\usepackage{natbib}
\usepackage{verbatim}
\usepackage{times}
\usepackage{etoolbox}
\usepackage{amsmath}

\usepackage[citecolor=blue,hypertexnames=false]{hyperref}

\hypersetup{
    colorlinks=true,       
    linkcolor=blue,                 
    filecolor=magenta,      
    urlcolor=blue
}

\graphicspath {{Images/PDF_Files/}{Images/}{Images/PNG_Files/}}

\newcommand{\hi}{H~{\sc i}}
\newcommand{\hii}{H~{\sc ii}}
\newcommand{\htwo}{H$_2$}
\newcommand{\gray}{$\gamma$-ray}
\newcommand{\grays}{$\gamma$-rays}
\newcommand{\fermilat}{{\it Fermi}--LAT}

\newcommand{\p}{$\pm$ }

\submitted{Accepted for publication in The Astrophysical Journal}
\shorttitle{\textit{Fermi}-LAT Observations Towards the Outer Halo of M31}
\shortauthors{Karwin, Murgia, Campbell, and Moskalenko}

\usepackage[pagewise]{lineno}

\begin{document}

\renewcommand{\linenumberfont}{\tiny}

\title{\textit{Fermi}-LAT Observations of $\gamma$-Ray Emission Towards the Outer Halo of M31}

\author{Christopher M. Karwin\altaffilmark{$\mathrm{\dagger}$}, Simona Murgia\altaffilmark{$\mathrm{\ddagger}$}, and Sheldon Campbell\altaffilmark{$\mathrm{\mathparagraph}$}}
\affil{\textit{Department of Physics and Astronomy, University of California, Irvine, CA 92697, USA}}

\author{Igor V. Moskalenko\altaffilmark{$\mathrm{*}$}}
\affil{\textit{Hansen Experimental Physics Laboratory and Kavli Institute for Particle Astrophysics and Cosmology, \\ Stanford University, Stanford, CA 94305, USA}}

\altaffiltext{$\mathrm{\dagger}$}{ckarwin@uci.edu}
\altaffiltext{$\mathrm{\ddagger}$}{smurgia@uci.edu}
\altaffiltext{$\mathrm{\mathparagraph}$}{sheldoc@uci.edu}
\altaffiltext{$\mathrm{*}$}{imos@stanford.edu}
\begin{abstract}

The Andromeda Galaxy is the closest spiral galaxy to us and has been the subject of numerous studies. It harbors a massive dark matter (DM) halo which may span up to $\sim$600 kpc across and comprises $\sim$90\% of the galaxy's total mass. This halo size translates into a large diameter of 42$^\circ$ on the sky for an M31--Milky Way (MW) distance of 785 kpc, but its presumably low surface brightness makes it challenging to detect with $\gamma$-ray telescopes. Using 7.6 years of \textit{Fermi} Large Area Telescope (\fermilat) observations, we make a detailed study of the $\gamma$-ray emission between 1--100 GeV towards M31's outer halo, with a total field radius of $60^\circ$ centered at M31, and perform an in-depth analysis of the systematic uncertainties related to the observations. We use the cosmic ray (CR) propagation code GALPROP to construct specialized interstellar emission models (IEMs) to characterize the foreground $\gamma$-ray emission from the MW, including a self-consistent determination of the isotropic component. We find evidence for an extended excess that appears to be distinct from the conventional MW foreground, having a total radial extension upwards of $\sim$120--200 kpc from the center of M31. We discuss plausible interpretations of the excess emission but emphasize that uncertainties in the MW foreground, and in particular, modeling of the \hi-related components, have not been fully explored and may impact the results.

\end{abstract}

\section{Introduction}

\setcounter{footnote}{0}

The Andromeda Galaxy, also known as M31, is very similar to the MW. It has a spiral structure and is comprised of multiple components including a central super-massive black hole, bulge, galactic disk (the disk of stars, gas, and dust), stellar halo, and circumgalactic medium, all of which have been studied  extensively~\citep{roberts1893selection,slipher1913radial,pease1918rotation,hubble1929spiral,babcock1939rotation,mayall1951comparison,arp1964spiral,Rubin:1970zza,roberts1975rotation,henderson1979model,beck1982distribution,brinks1984high,blitz1999high,ibata2001giant,deHeij:2002ne,Ferguson:2002yi,Braun:2003ey,galleti20042mass,zucker2004new,ibata2005accretion,barmby2006dusty,GildePaz:2006bw,Ibata:2007xz,Li:2007ud,faria2007probing,huxor2008globular,richardson2008nature,braun2009wide,McConnachie:2009up,Garcia:2009hu,Saglia:2009tp,2010A&A...511A..89C,peacock2010m31,Hammer:2010ug,Mackey:2010ix,Li:2010kf,2012ApJ...745..121L,McConnachie:2012vd,Lewis:2012dj,Bate:2013jha,veljanoski2014outer,huxor2014outer,Ade:2014bjw,bernard2015nature,lehner2015evidence,bernard2015nature,mcmonigal2015major,Conn:2016nnu,kerp2016survey}. Furthermore, the Andromeda Galaxy, like all galaxies, is thought to reside within a massive DM halo \citep{Rubin:1970zza,roberts1975rotation,faber1979masses,Bullock:1999he,carignan2006extended,seigar2008revised,Banerjee:2008kt,tamm2012stellar,Velliscig:2015ffa}. The DM halo of M31 is predicted to extend to roughly 300 kpc from its center and have a mass on the order of  $10^{12} M_\odot$, which amounts to approximately $90\%$ of the galaxy's total mass~\citep{Klypin:2001xu,seigar2008revised,2010A&A...511A..89C,tamm2012stellar,Fardal:2013asa,Shull:2014uia,lehner2015evidence}. For cold DM, the halo is also predicted to contain a large amount of substructure~\citep{Braun:1998ik,blitz1999high,deHeij:2002ne,Braun:2003ey,Diemand:2006ik,Kuhlen:2007ku,Springel:2008cc,Zemp:2008gw,Moline:2016pbm}, a subset of which hosts M31's population of satellite dwarf galaxies \citep{McConnachie:2012vd,martin2013pandas,collins2013kinematic,Ibata:2013rh,pawlowski2013dwarf,Conn:2013iu}. The combined M31 system, together with a similar system in the MW, are the primary components of the Local Group. The distance from the MW to M31 is approximately 785 kpc~\citep{Stanek:1998cu,McConnachie:2004dv,conn2012bayesian}, making it relatively nearby. Consequently, M31 appears extended on the sky. Because of this accessibility, M31 offers a prime target for studying galaxies; and indeed, a wealth of information has been gained from observations in all  wavelengths of the electromagnetic spectrum, e.g., see the references provided at the beginning of the introduction.

The \textit{Fermi} Large Area Telescope (\fermilat) is the first instrument to significantly detect M31 in \grays\ \citep{Fermi-LAT:2010kib,ogelman2010discovery}. Prior to \fermilat\ other pioneering experiments  set limits on a tentative signal~\citep{fichtel1974high,pollock1981search,sreekumar1994study,Hartman:1999fc}, with the first space-based \gray\ observatories dating back to 1962~\citep{kraushaar1962search,kraushaar1972high}. Note that M31 has not been significantly detected by any ground-based \gray\ telescopes, which are typically sensitive to energies above $\sim$100 GeV~\citep{Abeysekara:2014ffg,Funk:2015ena,Bird:2015npa,tinivella2016review}. 

The initial M31 analysis performed by the \fermilat\ Collaboration modeled M31 both as a point source and an extended source, finding marginal preference for extension at the confidence level of 1.8$\sigma$~\citep{Fermi-LAT:2010kib}. In order to search for extension, a uniform intensity elliptical template is employed, where the parameters of the ellipse are estimated from the IRIS 100 $\mu$m observation of M31~\citep{MivilleDeschenes:2004ci}. This emission traces a convolution of the interstellar gas and recent massive star formation activity~\citep{Yun:2001jx,Reddy:2003xn,Fermi-LAT:2010kib} and can be used as a template for modeling the \gray\ emission.

Since the initial detection further studies have been conducted~\citep{Dugger:2010ys,Li:2013qya,Pshirkov:2015hda,Pshirkov:2016qhu,Ackermann:2017nya}. A significant detection of extended \gray\ emission with a total extension of 0.9$^\circ$ was reported by~\citet{Pshirkov:2016qhu}, where the morphology of the detected signal consists of two bubbles symmetrically located perpendicular to the M31 disk, akin to the MW Fermi bubbles. Most recently the \emph{Fermi}-LAT Collaboration has published their updated analysis of M31~\citep{Ackermann:2017nya}. This study detects M31 with a significance of nearly $10\sigma$, and evidence for extension is found at the confidence level of $4\sigma$. Of the models tested, the best-fit morphology consists of a uniform-brightness circular disk with a radius of 0.4$^\circ$ centered at M31. The \gray\ signal is not found to be correlated with regions rich in gas or star formation activity, as was first pointed out by~\citet{Pshirkov:2016qhu}. 

In this work we make a detailed study of the $\gamma$-ray emission observed towards the outer halo of M31, including the construction of specialized interstellar emission models to characterize the foreground emission from the MW, and an in-depth evaluation of the systematic uncertainties related to the observations. Our ultimate goal is to test for a $\gamma$-ray signal exhibiting spherical symmetry with respect to the center of M31, since there are numerous physical motivations for such a signal. 

In general, disk galaxies like M31 may be surrounded by extended CR halos~\citep{Feldmann:2012rx,Pshirkov:2015hda}. Depending on the strength of the magnetic fields in the outer galaxy, the CR halo may extend as far as a few hundred kpc from the galactic disk. However, the actual extent remains highly uncertain. The density of CRs in the outer halo is predicted to be up to 10\% of that found in the disk~\citep{Feldmann:2012rx}. Disk galaxies like M31 are also surrounded by a circumgalactic medium, which is loosely defined as a halo of gas (primarily ionized hydrogen) in different phases which may extend as far as the galaxy's virial radius~\citep{Gupta:2012rh,Feldmann:2012rx,lehner2015evidence,Pshirkov:2015hda,howk2017project}. In addition, the stellar halo of M31 is observed to have an extension $\gtrsim$50 kpc~\citep{Ibata:2007xz,McConnachie:2009up,Mackey:2010ix}. CR interactions with the radiation field of the stellar halo and/or the circumgalactic gas could generate $\gamma$-ray emission.

\begin{figure*}[tbh!]
\centering
\includegraphics[width=0.49\textwidth]{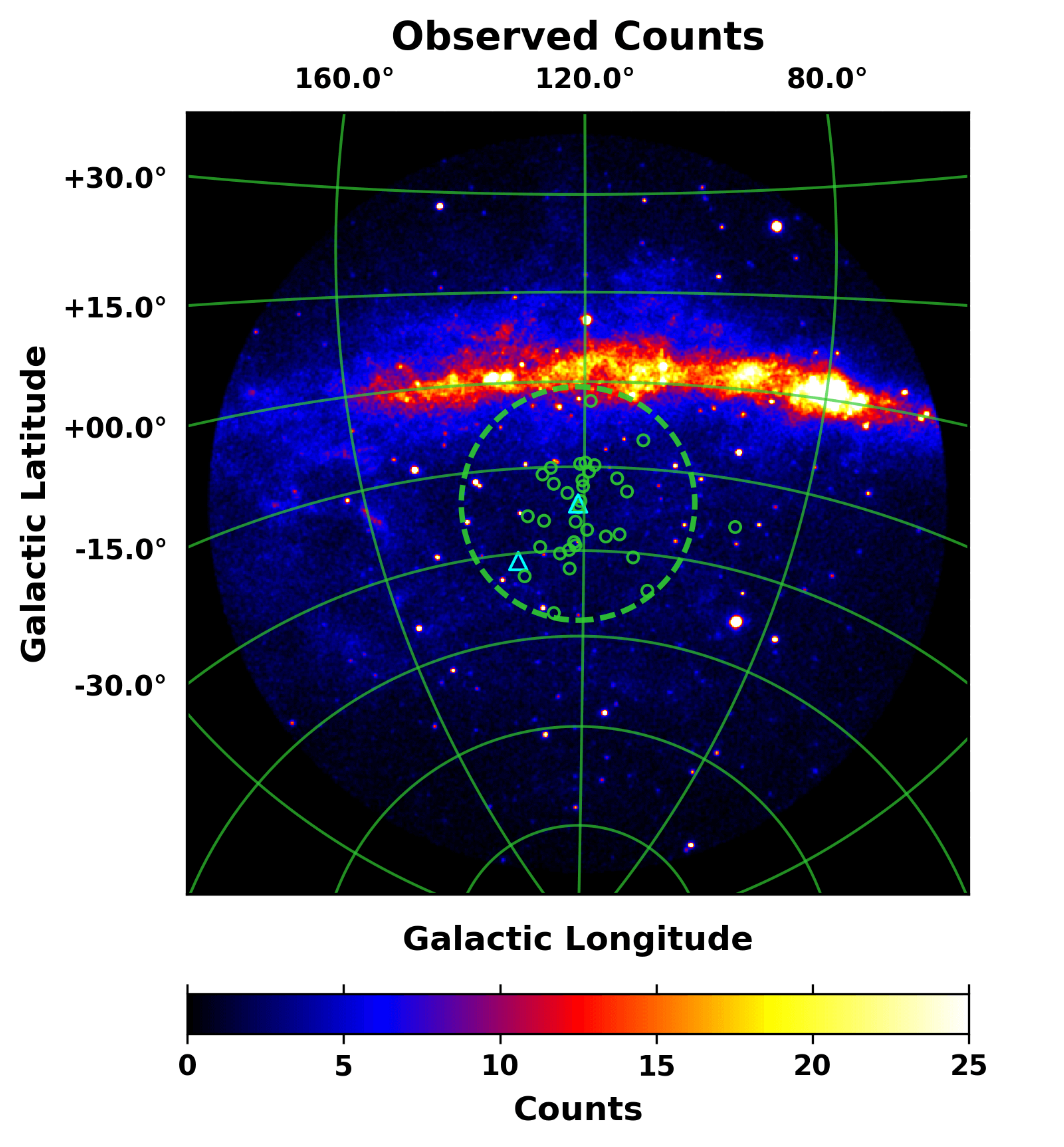}
\includegraphics[width=0.49\textwidth]{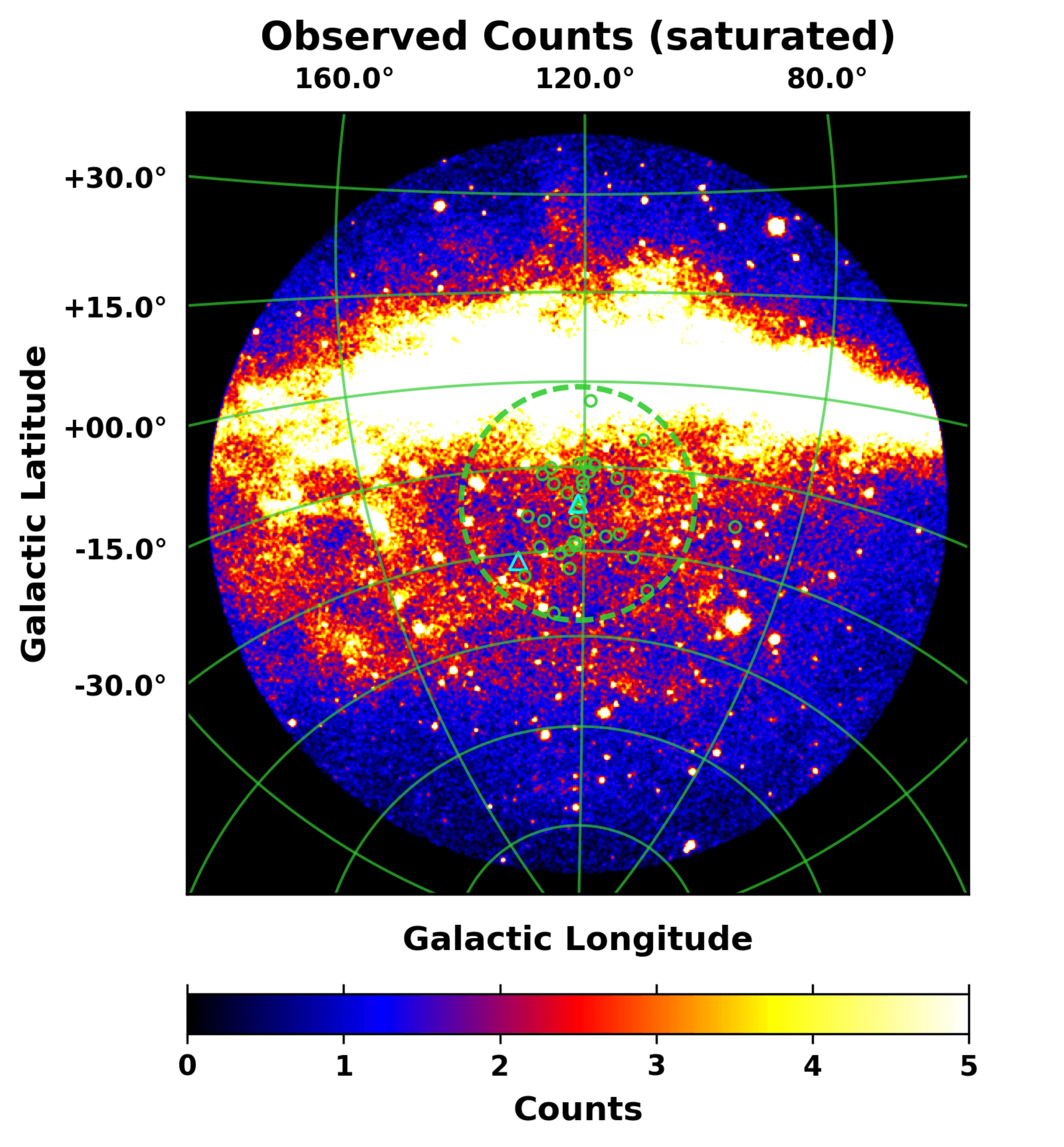}
\caption{Observed counts (left) and saturated counts (right) for a $60^\circ$ radius centered at M31, and an energy range of 1--100 GeV. The green dashed circle ($21^\circ$ in radius) corresponds to a 300 kpc projected radius centered at M31, for an M31--MW distance of 785 kpc, i.e.\ the canonical virial radius of M31. Also shown is M31's population of dwarf galaxies. M31 and M33 are shown with cyan triangles, and the other dwarfs are shown with $1^\circ$ green circles, each centered at the optical center of the respective galaxy. The sizes of the circles are a bit arbitrary, although they roughly correspond to the point spread function (PSF, 68\% containment angle) of \textit{Fermi}-LAT, which at 1 GeV is $\sim$$1^\circ$. Most of the MW dwarfs are not detected by \textit{Fermi}-LAT, and so we do not necessarily expect the individual M31 dwarfs to be detected. The primary purpose of the overlay is to provide a qualitative representation of the extent of M31's outer halo, and to show its relationship to the MW disk. Note that $\sim$3 dwarfs (which are thought to be gravitationally bound to M31) reach as for as $\sim$300 kpc, with one dwarf (And XXVIII) reaching as far as $\sim$360 kpc, as seen in the figure.} 
\label{fig:observed_counts}
\end{figure*}

Some hints of the extent and distribution of the M31 halo may be gained from observations of the distributions of well-studied objects, clearly tied to the M31 system. In Section~\ref{sec:gas_related_emission} we compare the distribution of the observed $\gamma$-ray emission in the M31 field to such features as M31's population of globular clusters~\citep{galleti20042mass,huxor2008globular,peacock2010m31,Mackey:2010ix,veljanoski2014outer,huxor2014outer} and M31's population of satellite dwarf galaxies~\citep{McConnachie:2012vd,martin2013pandas,collins2013kinematic}. We note that \fermilat\ does not detect most of the MW dwarfs~\citep{Ackermann:2015zua}, and likewise we do not necessarily expect to detect most of the individual M31 dwarfs. The dwarfs are included here primarily as a qualitative gauge of the extent of M31's DM halo, and more generally, in support of formulating the most comprehensive picture possible of the M31 region. We also compare the observed $\gamma$-ray emission to the M31 cloud~\citep{blitz1999high,kerp2016survey}, which is a highly extended lopsided gas cloud centered in projection on M31. It remains uncertain whether the M31 cloud resides in M31 or the MW, although most recently~\citet{kerp2016survey} have argued that M31's disk is physically connected to the M31 cloud.

Lastly, we note that due to its mass and proximity, the detection sensitivity of M31 to DM searches with \grays\ is competitive with the MW dwarf spheroidal galaxies, particularly if the signal is sufficiently boosted by substructures~\citep{Falvard:2002ny,Fornengo:2004kj,Mack:2008wu,Dugger:2010ys,Conrad:2015bsa,Gaskins:2016cha}. Moreover, M31 is predicted to be the brightest extragalactic source of DM annihilation~\citep{lisanti2018search,lisanti2018mapping}. At a distance of  $\sim$785 kpc from the MW \citep{Stanek:1998cu,McConnachie:2004dv,conn2012bayesian} and with a virial radius of  a few hundred kpc~\citep{Klypin:2001xu,seigar2008revised,2010A&A...511A..89C,tamm2012stellar,Fardal:2013asa,Shull:2014uia,lehner2015evidence}, the diameter of M31's DM halo covers  $\gtrsim$42$^{\circ}$ across the sky. However, there is a high level of uncertainty regarding the exact nature of the halo geometry, extent, and substructure content~\citep{Kamionkowski:1997xg,Braun:1998ik,blitz1999high,deHeij:2002ne,Braun:2003ey,Helmi:2003pp,Bailin:2004wu,Allgood:2005eu,Bett:2006zy,Hayashi:2006es,Kuhlen:2007ku,Banerjee:2008kt,Zemp:2008gw,Saha:2009dt,Law:2009yq,Banerjee:2011rr,Velliscig:2015ffa,Bernal:2016guq,garrison2017not}.

Our analysis proceeds as follows. In Section~\ref{sec:Data and Models} we describe our data selection and modeling of the interstellar emission. In Section~\ref{sec:FM31 Baseline} we present the baseline analysis of the M31 field and perform a template fit, including the addition of M31-related components to the model.  In Section~\ref{sec:smooth_residual_emission} we compare the radial intensity profile and emission spectrum of the M31-related components to corresponding predictions for DM annihilation towards the outer halo of M31, including contributions from both the M31 halo and the MW halo in the line of sight. In Section~\ref{sec:gas_related_emission} we compare the structured $\gamma$-ray emission in the M31 field to a number of complementary M31-related observations. Section \ref{sec:fianl} provides an extended summary of the analysis and results. Supplemental information is provided in Appendices. In Appendix~\ref{sec:IEM_Summary} we briefly describe the models for diffuse Galactic foreground emission. In Appendix~\ref{sec:different_IEMs} we consider some additional systematics pertaining to the observations. Appendix~\ref{sec:DM} provides the details of calculations of the DM profiles discussed in the paper.

\section{Data and Models} \label{sec:Data and Models}

\subsection{Data} \label{sec:Data Selection}

The \textit{Fermi Gamma-ray Space Telescope} was launched on June 11, 2008. The main instrument on board \textit{Fermi} is the Large Area Telescope. It consists of an array of 16 tracker modules, 16 calorimeter modules, and a segmented anti-coincidence detector. \fermilat\ is sensitive to $\gamma$-rays in the energy range from approximately 20 MeV to above 300 GeV. A full description of the telescope, including  performance specifications, can be found in~\citet{Atwood:2009ez}, \citet{Abdo:2009gy}, and \citet{Ackermann:2012kna}.  

Our region of interest (ROI) is a region with a radius of $60^\circ$ centered at the position of M31, $(l,b) = (121.17^{\circ}, -21.57^{\circ})$. We employ front and back converting events corresponding to the P8R2\_CLEAN\_V6 selection. The events have energies in the range 1--100 GeV and have been collected from 2008-08-04 to 2016-03-16 (7.6 years). The data are divided into 20 bins equally spaced in logarithmic energy, with $0.2^\circ\times0.2^\circ$ pixel size.
The analysis is carried out with the \fermilat\ ScienceTools  (version v10r0p5)\footnote{Available at \url{http://fermi.gsfc.nasa.gov/ssc/data/analysis}}. In  particular, the binned maximum likelihood fits are performed with the {\it gtlike} package.

Figure~\ref{fig:observed_counts} shows the total observed counts between 1--100 GeV for the full ROI. Two different count ranges are displayed. The map on the left shows the full range. The bright emission along 0$^\circ$ latitude corresponds to the plane of the MW. The map on the right shows the saturated counts map, emphasizing the lower counts at higher latitudes.  Overlaid is a green dashed circle ($21^\circ$ in radius) corresponding to a 300 kpc projected radius centered at M31, for an M31-MW distance of 785 kpc, i.e.~the canonical virial radius of M31. Also shown is M31's population of dwarf galaxies. The primary purpose of the overlay is to provide a qualitative representation of the extent of M31's outer halo, and to show its relationship to the MW disk. Note that we divide the full ROI into subregions, and our primary field of interest is a $28^\circ \times 28^\circ$ square region centered at M31, which we refer to as field M31 (FM31), as further discussed below.

\subsection{Foreground Model and Isotropic Emission} \label{sec:iem}

The foreground emission from the MW and the isotropic component (the latter includes unresolved extragalactic diffuse $\gamma$-ray emission, residual instrumental background, and possibly contributions from other Galactic components which have a roughly isotropic distribution) are the dominant  contributions in \grays\ towards the M31 region. We use the CR propagation code GALPROP\footnote{Available at \url{https://galprop.stanford.edu}}(v56) to construct specialized interstellar emission models (IEMs) to characterize the MW foreground emission, including a self-consistent determination of the isotropic component. These foreground models are physically motivated and \emph{are not} subject to the same caveats\footnote{The list of caveats on the \fermilat\ diffuse model is available at \url{https://fermi.gsfc.nasa.gov/ssc/data/analysis/LAT_caveats.html} \label{caveats}} for extended source analysis as the default IEM provided by the \textit{Fermi}-LAT Collaboration for point source analysis (hereafter FSSC IEM)~\citep{Acero:2016qlg}. Here we provide a brief description of the GALPROP model \citep{Moskalenko:1997gh,Moskalenko:1998gw,Strong:1998pw,Strong:1998fr,2006ApJ...642..902P,Strong:2007nh,Vladimirov:2010aq,Johannesson:2016rlh,porter2017high,Johannesson:2018bit,PhysRevC.98.034611}, and more details are given in Appendix~\ref{sec:IEM_Summary}.

The GALPROP model calculates self-consistently spectra and abundances of Galactic CR species and associated diffuse emissions (radio, X-rays, \gray{s}) in 2D and 3D. The CR injection and propagation parameters are derived from local CR measurements. The Galactic propagation includes all stable and long-lived particles and isotopes ($e^\pm$, $\bar{p}$, H-Ni) and all relevant processes in the interstellar medium. The radial distribution of the CR source density is parametrized as  
\begin{equation} \label{eq:1}
\rho(r) = \left(\frac{r + r_1}{r_\odot + r_1}\right)^{a}\times \exp \left(-b \times \frac{r - r_\odot}{r_\odot + r_1}\right),
\end{equation}
where $r$ is the Galactocentric radius, $r_\odot= 8.5$ kpc, and the parameter $r_1$ regulates the CR density at $r=0$. The injection spectra of CR species are described by the rigidity (R) dependent function
\begin{equation}  \label{eq:2}
q(R) \propto (R/R_0)^{-\gamma_0}\prod_{i=0}^2\bigg[1 + (R/R_i)^\frac{\gamma_i - \gamma_{i+1}}{s_i}\bigg]^{s_i},
\end{equation}
where $\gamma_i (i =0, 1, 2, 3)$ are the spectral indices, $R_i (i = 0, 1, 2)$ are the break rigidities, $s_i$ are the smoothing parameters ($s_i=\mp0.15$ for $|\gamma_i |\lessgtr |\gamma_{i+1} |$), and the numerical values of all parameters are given in Table~\ref{tab:GALPROP_parameters}. Some parameters are not in use, so for $p$ and He, we have only $\gamma_{i=0, 1, 2}$ and $R_{i=0, 1}$.

\begin{deluxetable}{lcc}[tbh!]
\tablecolumns{5}
\tablewidth{0mm}
\tablecaption{GALPROP Model Parameters\label{tab:GALPROP_parameters}}
\tablehead{
\colhead{Parameter} &
\colhead{M31 IEM} &
\colhead{IG IEM}}
\startdata
\tablenotemark{a} $z$ [kpc] &4 &6 \\
\tablenotemark{a} $r$ [kpc] & 20 &30  \\
\tablenotemark{b} $a$  &1.5 &1.64 \\
\tablenotemark{b} $b$  & 3.5  &4.01 \\
\tablenotemark{b} $r_1$  & 0.0  &0.55 \\
\tablenotemark{c} $D_0$ [10$^{28}$ cm$^2$ s$^{-1}$]               & 4.3&7.87\\
\tablenotemark{c} $\delta$               &0.395    & 0.33\\
\tablenotemark{c} $\eta$                & 0.91      &1.0\\
\tablenotemark{c} Alfv\'en speed, $v_{\rm A}$ [km s$^{-1}$] &28.6 &34.8\\
\tablenotemark{d} $v_{\rm conv,0}$  [km s$^{-1}$]           &12.4       &\nodata  \\
\tablenotemark{d} $dv_{\rm conv}/dz$  [km s$^{-1}$ kpc$^{-1}$ ]         &10.2     &\nodata\\
\tablenotemark{e} $R_{p,0}$ [GV]       &7 &11.6 \\
\tablenotemark{e} $R_{p,1}$ [GV]       &360  &\nodata\\
\tablenotemark{e} $\gamma_{p,0}$ & 1.69     &1.90\\
\tablenotemark{e} $\gamma_{p,1}$ & 2.44     &2.39\\
\tablenotemark{e} $\gamma_{p,2}$ & 2.295    &\nodata\\
\tablenotemark{e} $R_{\rm He,0}$ [GV] &7 &\nodata\\
\tablenotemark{e} $R_{\rm He,1}$ [GV] &330 &\nodata\\
\tablenotemark{e} $\gamma_{\rm He,0}$ &1.71&\nodata\\
\tablenotemark{e} $\gamma_{\rm He,1}$ &2.38&\nodata\\
\tablenotemark{e} $\gamma_{\rm He,2}$ &2.21&\nodata\\
\tablenotemark{e} $R_{e,0}$ [GV] &0.19 &\nodata \\
\tablenotemark{e} $R_{e,1}$ [GV] &6 &2.18 \\
\tablenotemark{e} $R_{e,2}$ [GV]     &95  & 2171.7  \\
\tablenotemark{e} $\gamma_{e,0}$ &2.57&\nodata\\
\tablenotemark{e} $\gamma_{e,1}$ &1.40&1.6\\
\tablenotemark{e} $\gamma_{e,2}$ &2.80    &2.43\\
\tablenotemark{e} $\gamma_{e,3}$ &2.40& 4.0 \\
\tablenotemark{f} $J_{p}$ [$\mathrm{10^{-9} \ cm^{-2} \  s^{-1} \ sr^{-1} \ MeV^{-1}}$] &4.63&4.0\\
\tablenotemark{f} $J_{e}$ [$\mathrm{10^{-11} \ cm^{-2} \ s^{-1} \ sr^{-1} \ MeV^{-1}}$] &1.44&0.011  \\ 
\tablenotemark{g} A5 [kpc] &8--10&8--10\\
\tablenotemark{g} A6 [kpc] &10--11.5&10--50\\
\tablenotemark{g} A7 [kpc] &11.5--16.5&\nodata\\
\tablenotemark{g} A8 [kpc] &16.5--50&\nodata\\
\tablenotemark{h} IC Formalism&Anisotropic&Isotropic

\enddata
\tablecomments{For reference, we also give corresponding values for the (``Yusifov'') IEMs used in~\citet{TheFermi-LAT:2015kwa} for the analysis of the inner Galaxy (IG).}
\tablenotetext{a}{Halo geometry: $z$ is the height above the Galactic plane, and $r$ is the radius.}
\tablenotetext{b}{CR source density. The parameters correspond to Eq.~(\ref{eq:1}).}
\tablenotetext{c}{Diffusion: $D(R)$ $\propto \beta^\eta R^\delta$. $D(R)$ is normalized to $D_0$ at 4.5 GV.}
\tablenotetext{d}{Convection: $v_{\rm conv}(z)=v_{\rm conv,0}+(dv_{\rm conv}/dz)z$.}
\tablenotetext{e}{Injection spectra: The spectral shape of the injection spectrum is the same for all CR nuclei except for protons. The parameters correspond to Eq.~(\ref{eq:2}).}
\tablenotetext{f}{The proton and electron flux are normalized at the Solar location at a kinetic energy of 100 GeV. Note that for the IG IEM the electron normalization is at a kinetic energy of 25 GeV.}
\tablenotetext{g}{Boundaries for the annuli which define the IEM. Only A5 (local annulus) and beyond contribute to the foreground emission for FM31.}
\tablenotetext{h}{Formalism for the inverse Compton (IC) component.}
\end{deluxetable}

\begin{figure}[tbh!]
\centering

\includegraphics[width=0.48\textwidth]{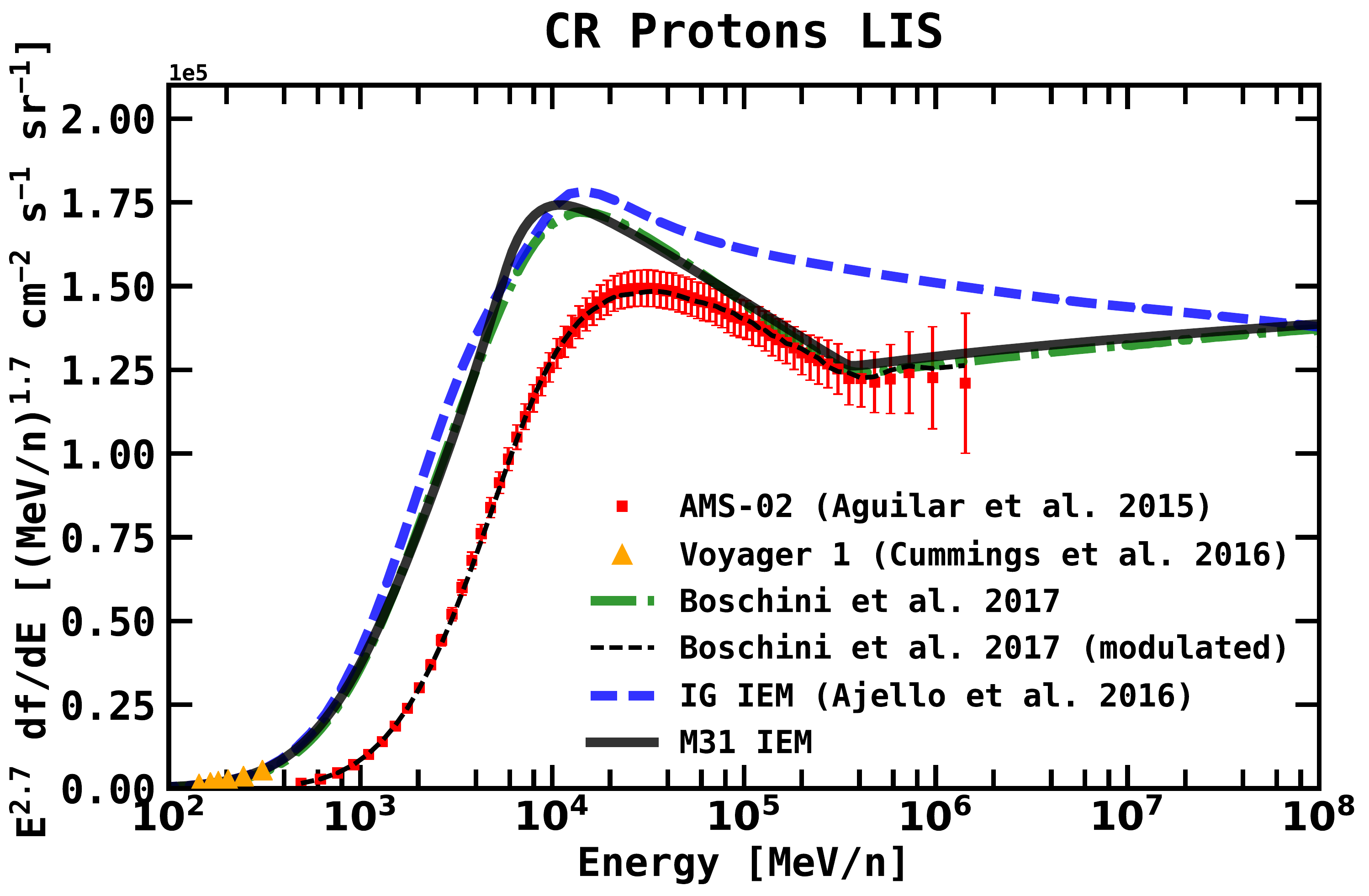}
\includegraphics[width=0.48\textwidth]{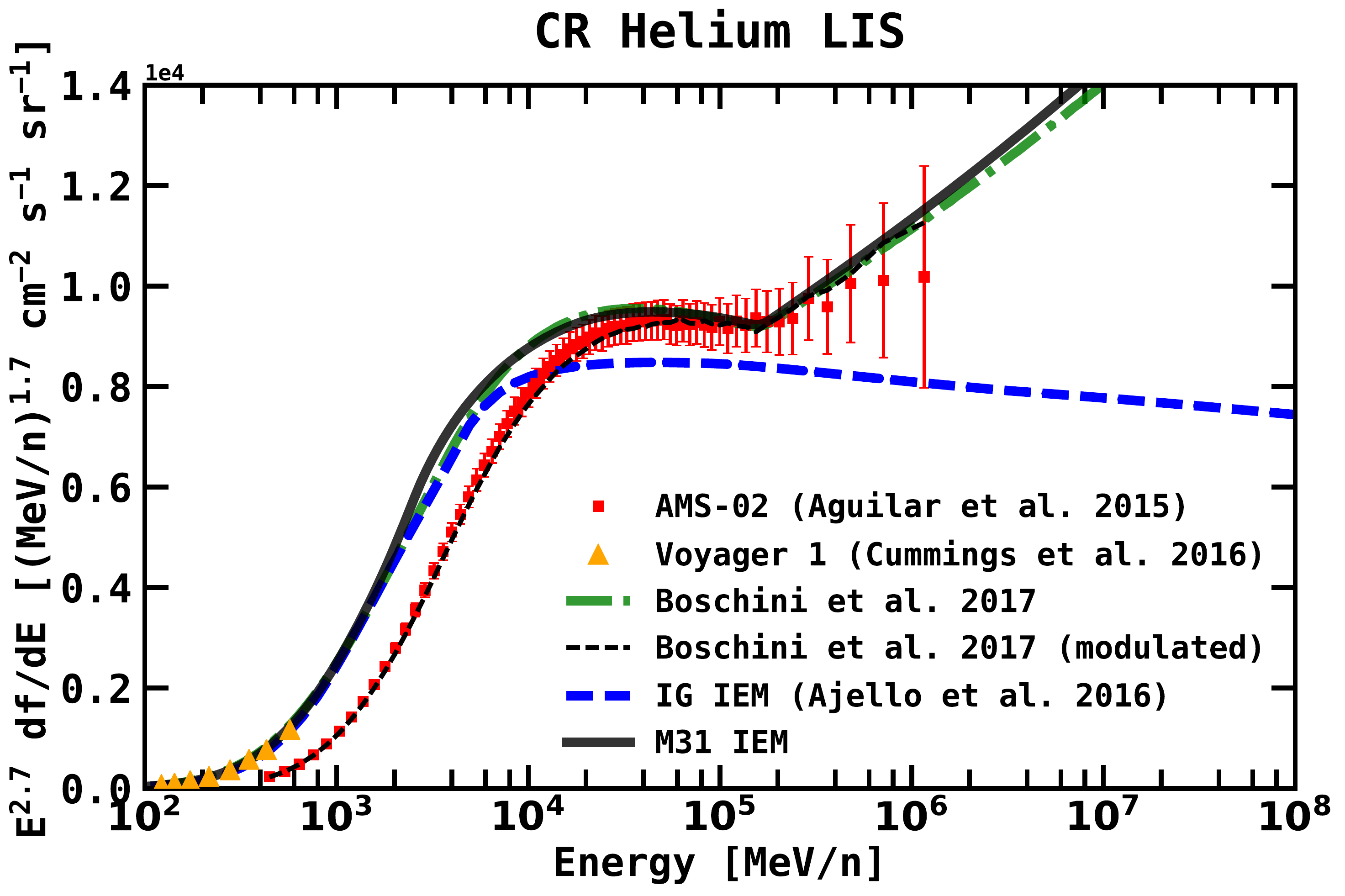}
\includegraphics[width=0.48\textwidth]{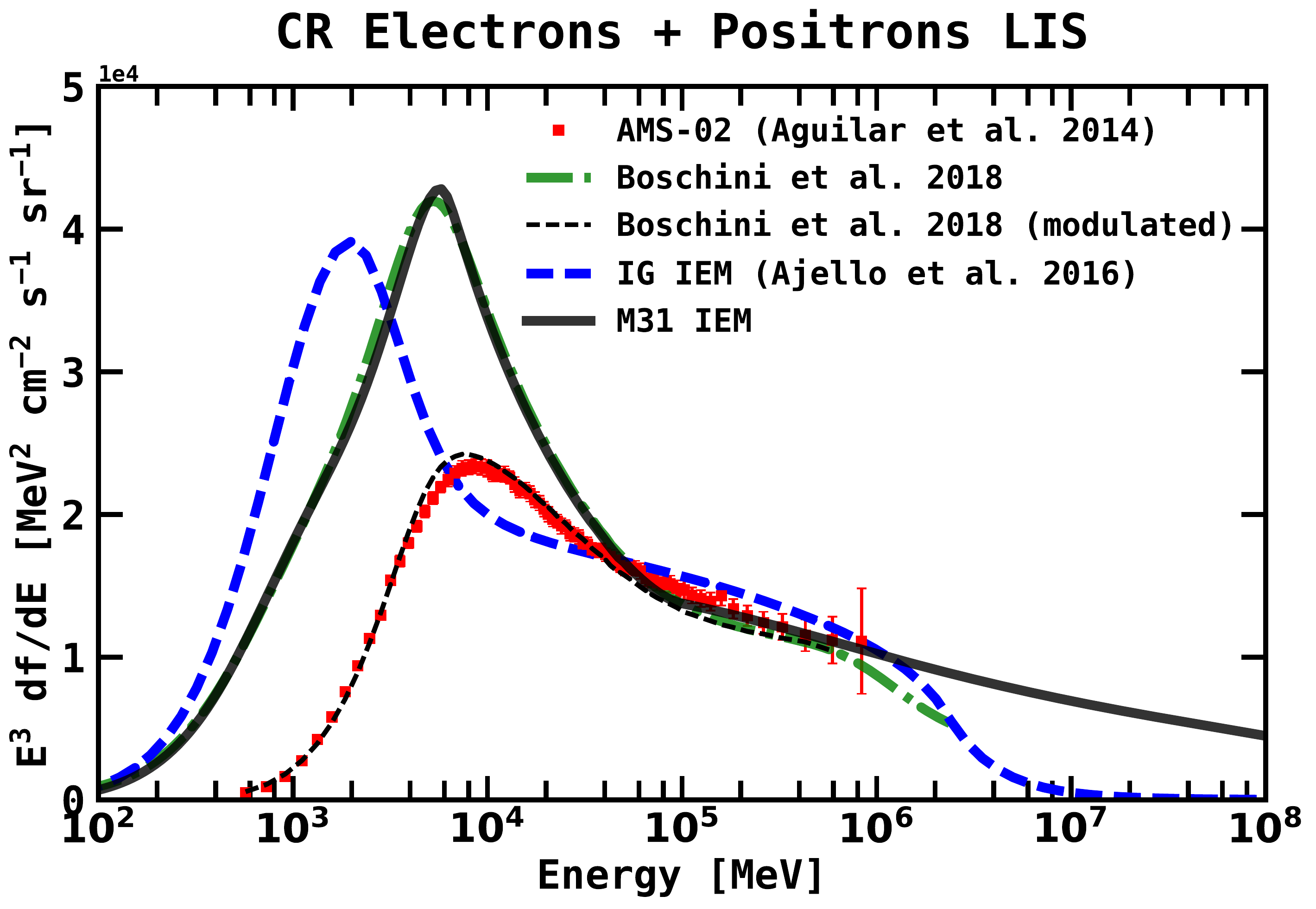}

\caption{The local interstellar spectra (LIS) for CR protons (top), He (middle), and all electrons ($e^- +e^+$) (bottom). The latest AMS-02 measurements from~\citet{PhysRevLett.113.221102,PhysRevLett.114.171103,PhysRevLett.115.211101} are shown with red squares. The green dashed line shows the results from~\citet{Boschini:2017fxq,Boschini:2018zdv}, which employ GALPROP and HelMod together in an iterative manner to derive the LIS. We adopt their derived GALPROP CR parameters, and the LIS for our IEM (M31 IEM: solid black line) are roughly the same. The thin dotted black line shows the LIS modulated with HelMod \citep{Boschini:2017fxq,Boschini:2018zdv}. Yellow triangles show the Voyager 1 $p$ and He data in the local interstellar medium \citep{2016ApJ...831...18C}. Voyager 1 electron data are below 100 MeV and, therefore, are not shown. In addition we show the LIS for the (``Yusifov'') IEM in~\citet{TheFermi-LAT:2015kwa}, which we use as a reference model in our study of the systematics for the M31 field (see Appendix~\ref{sec:IG_IEMs}).}
\label{fig:CR_LIS}
\end{figure}

Heliospheric propagation is calculated using the dedicated code HelMod\footnote{Available at \url{http://www.helmod.org/}}. HelMod is a 2D Monte Carlo code for heliospheric propagation of CRs, which describes the solar modulation in a physically motivated way. It was demonstrated that the calculated CR spectra are in a good agreement with measurements including measurements outside of the ecliptic plane at different levels of solar activity and the polarity of the magnetic field. The result of the combined iterative application of the GALPROP and HelMod codes is a series of local interstellar spectra (LIS) for CR $e^-$, $e^+$, $p$, He, C, and O nuclei \citep{Boschini:2017fxq,Boschini:2018zdv,2018ApJ...858...61B} that effectively disentangle two tremendous tasks such as Galactic and heliospheric propagation.

For our analysis we used a GALPROP-based combined diffusion-convection-reacceleration model with a uniform spatial diffusion coefficient and a single power law index over the entire rigidity range as described in detail in \citet{Boschini:2017fxq}. Since the distribution of supernova remnants (SNRs), conventional CR sources, is not well determined due to the observational bias and the limited lifetime of their shells, other tracers are often employed. In our calculations we use the distribution of pulsars \citep{yusifov2004revisiting} that are the final state of evolution of massive stars and can be observed for millions of years. The same distribution was used in the analysis of the \gray{} emission from the Inner Galaxy (IG) \citep{TheFermi-LAT:2015kwa}. 

We adopt the best-fit GALPROP parameters from~\citet{Boschini:2017fxq,Boschini:2018zdv}, which are summarized in Table~\ref{tab:GALPROP_parameters}. The spectral shape of the injection spectrum is the same for all CR nuclei except for protons. The corresponding CR spectra are plotted in Figure~\ref{fig:CR_LIS}. Also plotted in Figure~\ref{fig:CR_LIS} are the latest AMS-02 measurements from~\citet{PhysRevLett.113.221102,PhysRevLett.114.171103,PhysRevLett.115.211101} and Voyager 1 $p$ and He data in the local interstellar medium \citep{2016ApJ...831...18C}. The modulated LIS are taken from \citet{Boschini:2017fxq,Boschini:2018zdv} and correspond to the time frame of the published AMS-02 data. In addition, we plot the LIS for the (``Yusifov'') IEMs used in~\citet{TheFermi-LAT:2015kwa} for the analysis of the inner Galaxy (IG), which we use as a reference model in our study of the systematics for the M31 field (see Appendix~\ref{sec:IG_IEMs}). Overall, the LIS for the M31 model are in good agreement with the AMS-02 data. 

We note that there is a small discrepancy in the modulated all-electron  ($e^- + e^+$) spectrum between $\sim$4--10 GeV that, however, does not affect our results. Electrons in this energy range do not contribute much to the observed diffuse emission. The upscattered photon energy is $\epsilon_1\sim\epsilon_0\gamma^2$, where $\epsilon_0$ and $\gamma$ are the energy of the background photon and the Lorentz-factor of the CR electron, correspondingly. For our range of interest $\epsilon_1$$\sim$5 GeV, we need CR electrons of $\sim$35 GeV for $\epsilon_0\sim$1 eV optical photons and even higher for IR and CMB, while the number density of optical photons in the ISM is very small. Additionally, we perform several systematic tests throughout this work, including fits with three different IEMs (M31, IG, and FSSC IEMs), as well as a fit in a tuning region surrounding FM31 on the south.

\begin{figure*}[tbh!]
\centering
\includegraphics[width=1\textwidth]{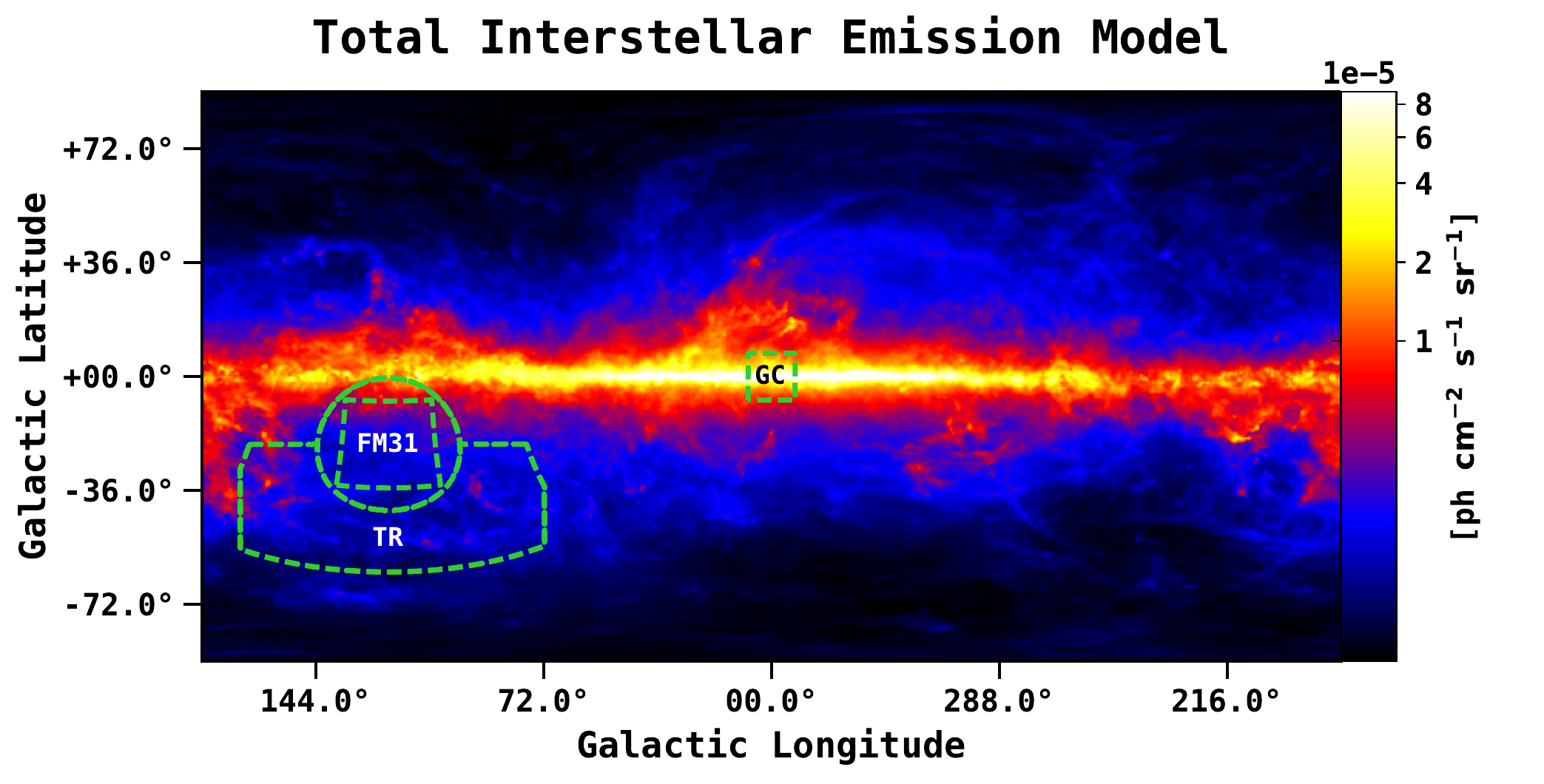}
\caption{The total interstellar emission model (IEM) for the MW integrated in the energy range 1--100 GeV. The color corresponds to the intensity, and is shown in logarithmic scale. The intensity level is for the initial GALPROP output, before tuning to the $\gamma$-ray data. The map is shown in a Plate Carr\'{e}e projection, and the pixel size is 0.25 deg/pix. The model has contributions from $\pi^0$-decay, (anisotropic) IC emission, and Bremsstrahlung. Overlaid is the region of interest (ROI) used in this analysis. From the observed counts (Figure~\ref{fig:observed_counts}) we cut an $84^\circ\times84^\circ$ ROI, which is centered at M31. The green dashed circle is the 300 kpc boundary corresponding to M31's canonical virial radius (of $\sim$$21^\circ$), as also shown in Figure~\ref{fig:observed_counts}. We label the field within the virial radius as field M31 (FM31), and the region outside (and south of latitudes of $-21.57^\circ$) we label as the tuning region (TR). Longitude cuts are made on the ROI at $l=168^\circ \ \mathrm{and} \ l=72^\circ$, as discussed in the text. For reference we also show the Galactic center region (GC), which corresponds to a $15^\circ\times15^\circ$ square centered at the GC.}
\label{fig:galactic_diffuse_schematic}
\end{figure*}

\begin{figure}[tbh!]
\centering
\includegraphics[width=0.5\textwidth]{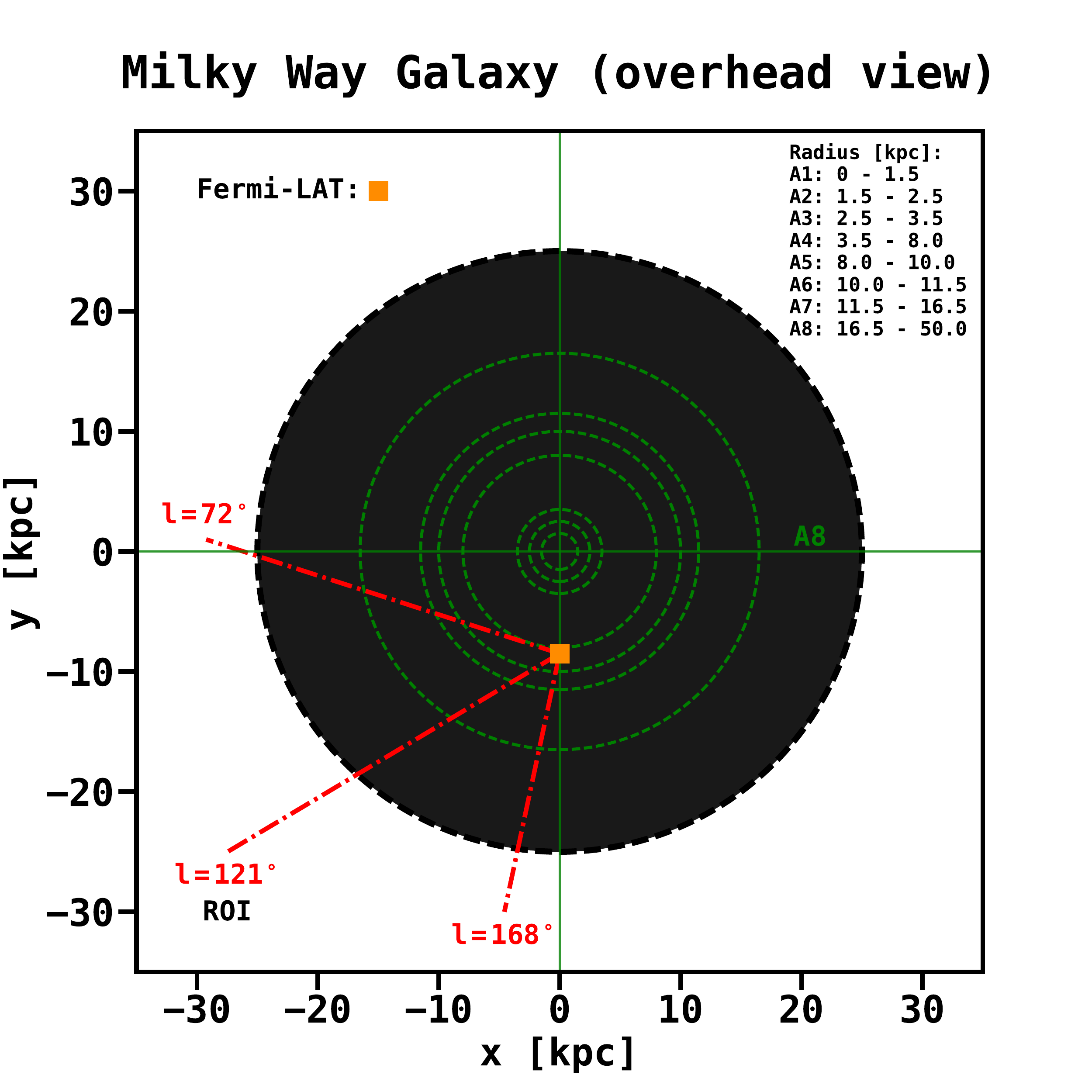}
\caption{Schematic of the eight concentric circles which define  the annuli (A1--A8) in the IEM, as described in the text. The  ranges in Galactocentric radii are reported in the legend. Note that the full extension of A8 is not shown. Only A5--A8 contribute to the Galactic foreground emission for the field used in this analysis.}
\label{fig:measurement_schematic}
\end{figure}

\begin{figure*}[tbh!]
\centering
\includegraphics[width=0.49\textwidth]{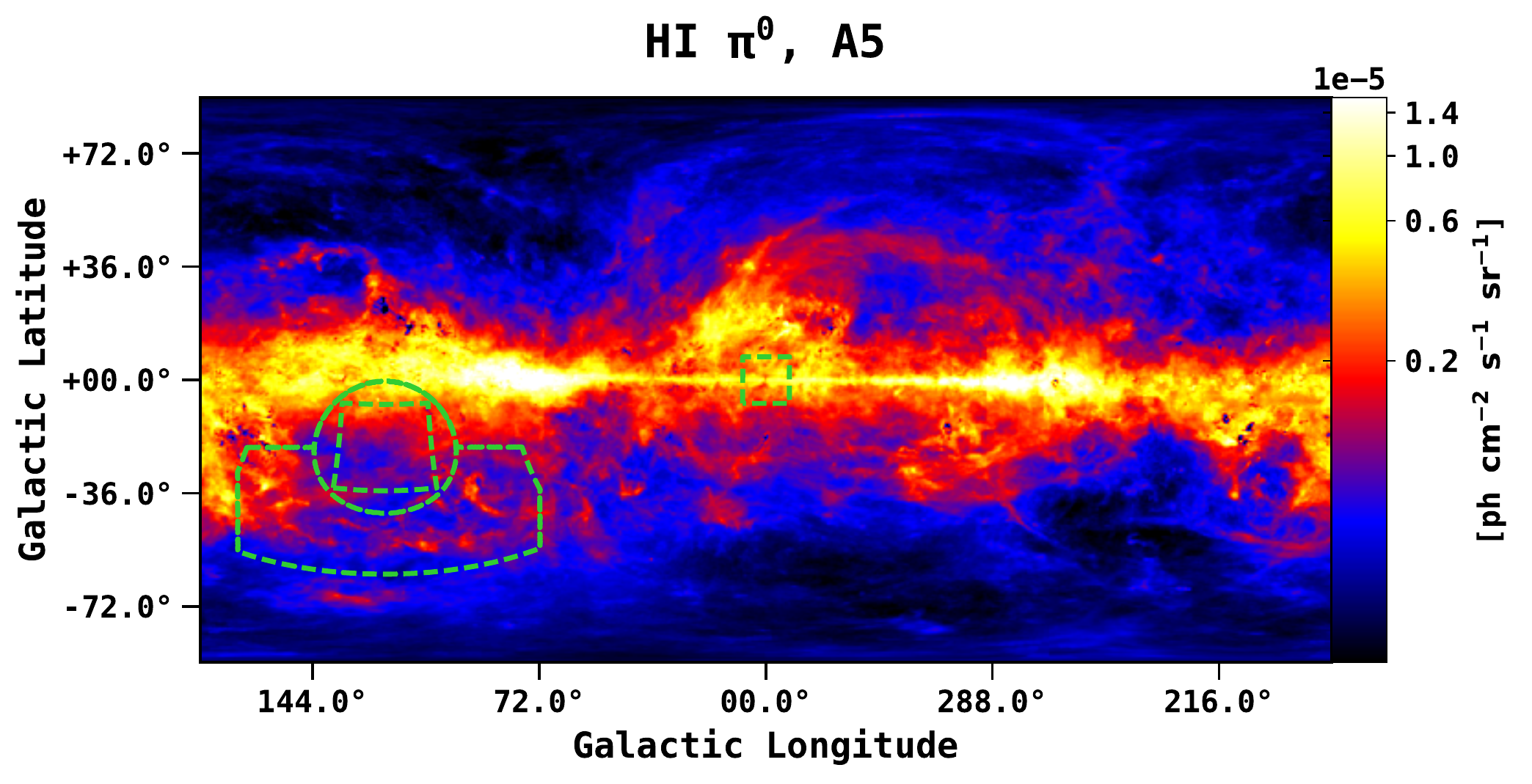}
\includegraphics[width=0.49\textwidth]{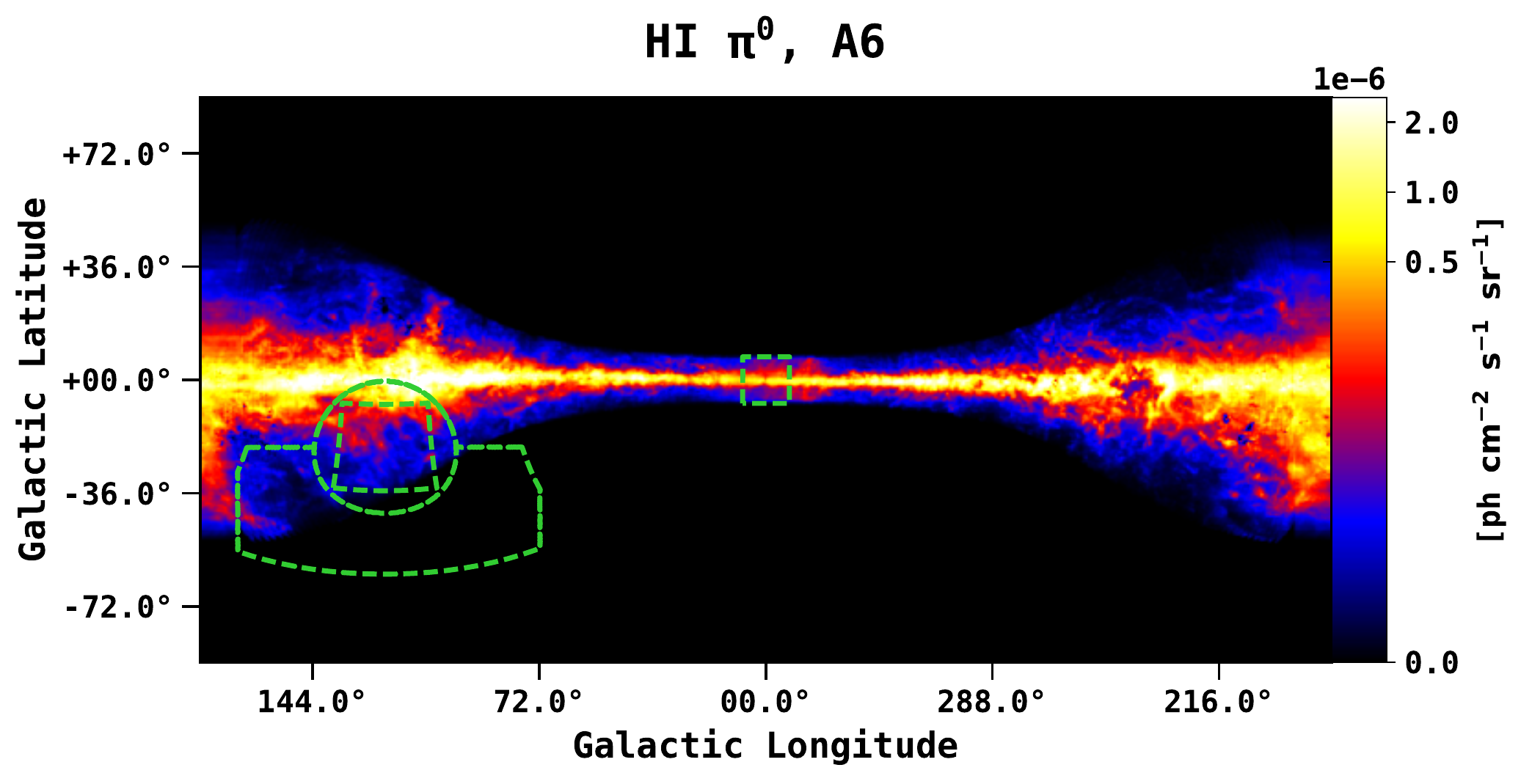}
\includegraphics[width=0.49\textwidth]{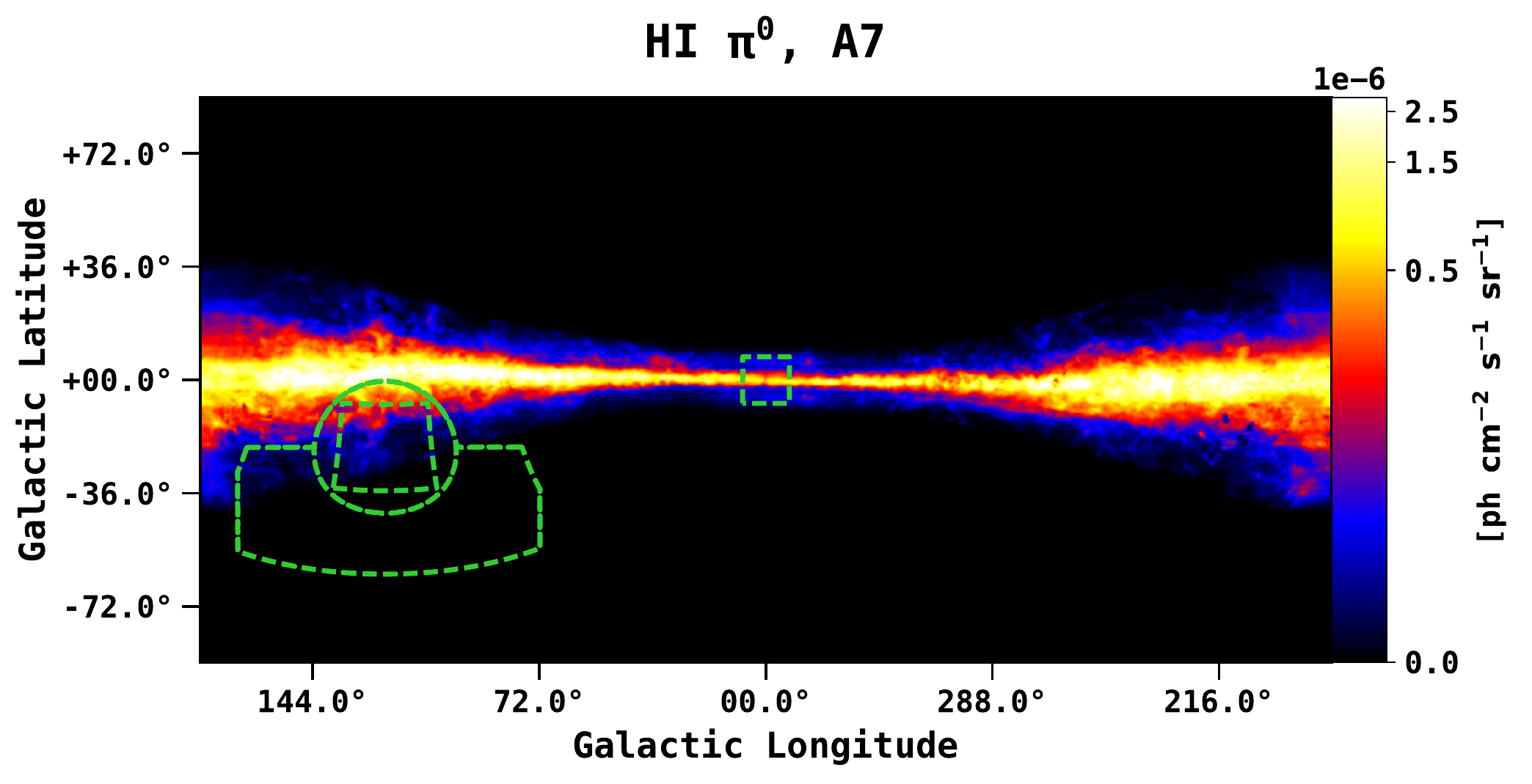}
\includegraphics[width=0.49\textwidth]{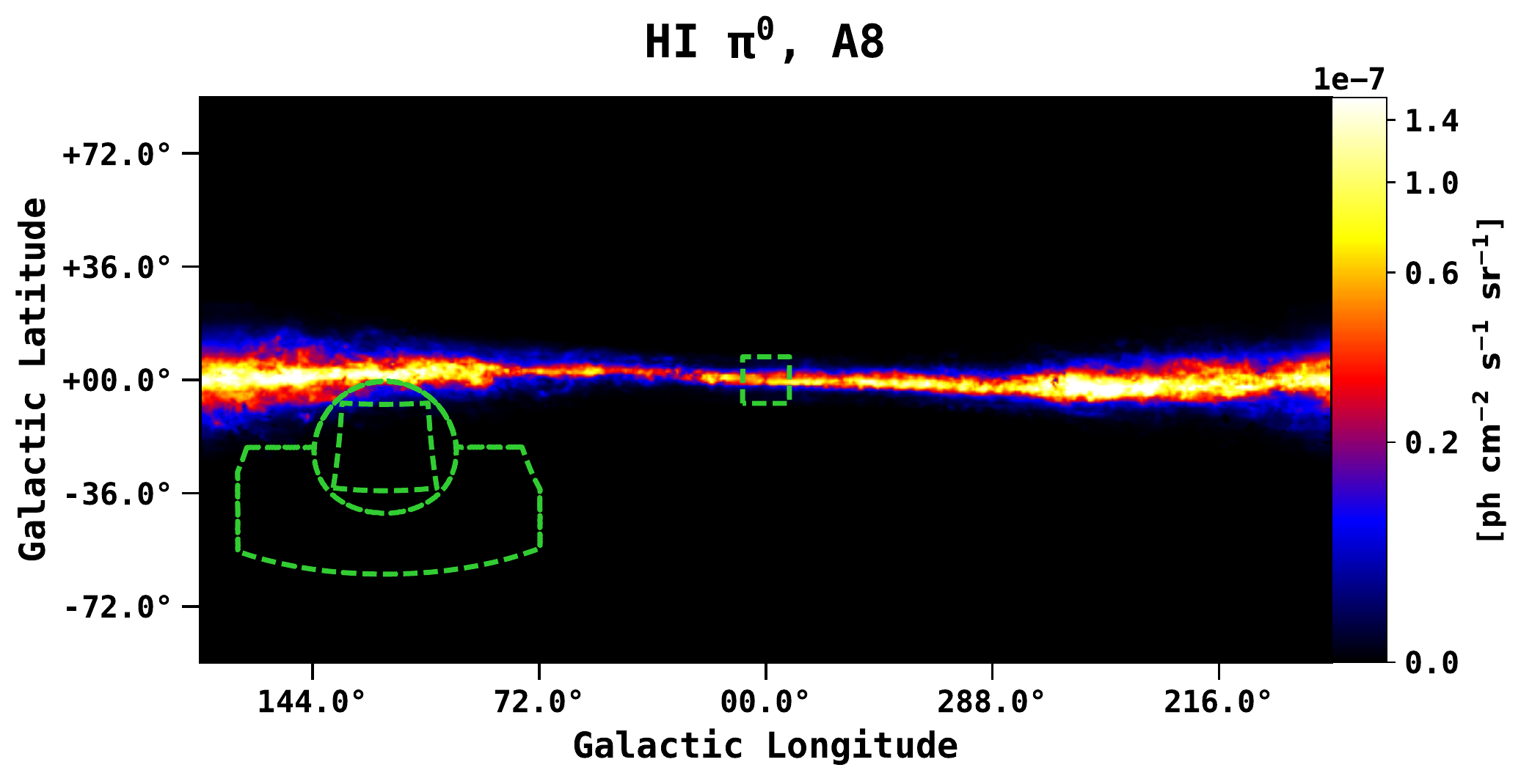}
\includegraphics[width=0.49\textwidth]{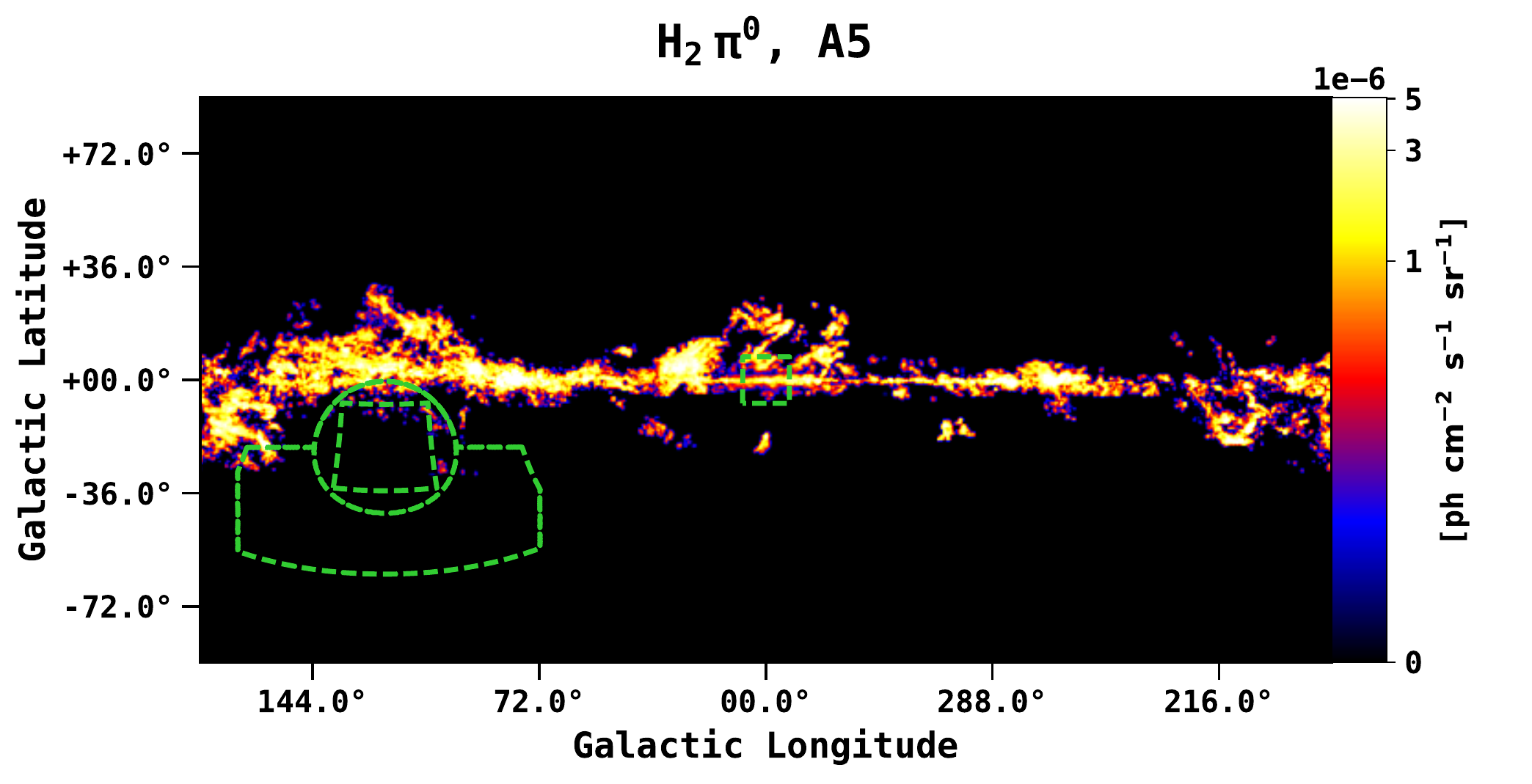}
\includegraphics[width=0.49\textwidth]{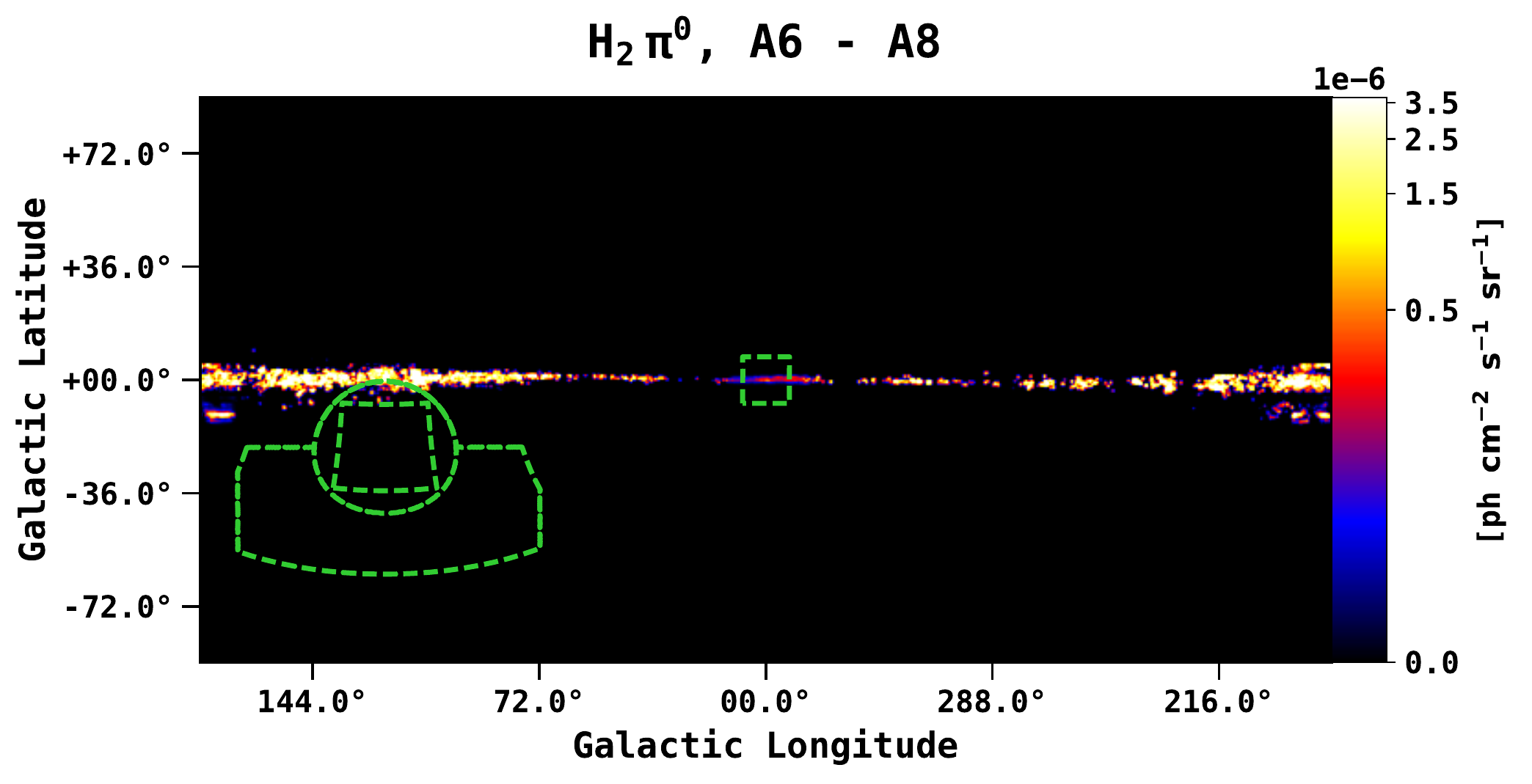}
\includegraphics[width=0.49\textwidth]{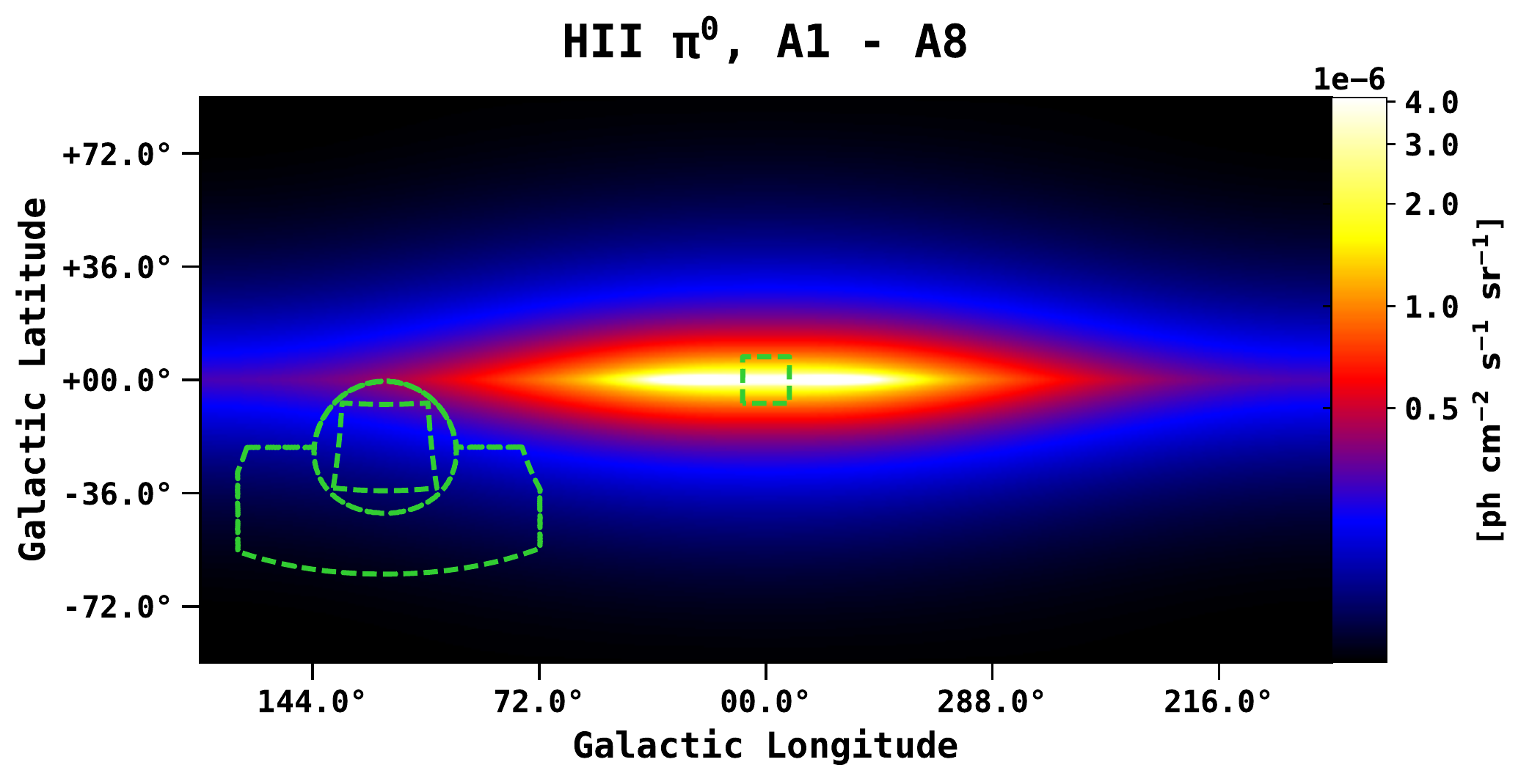}
\includegraphics[width=0.49\textwidth]{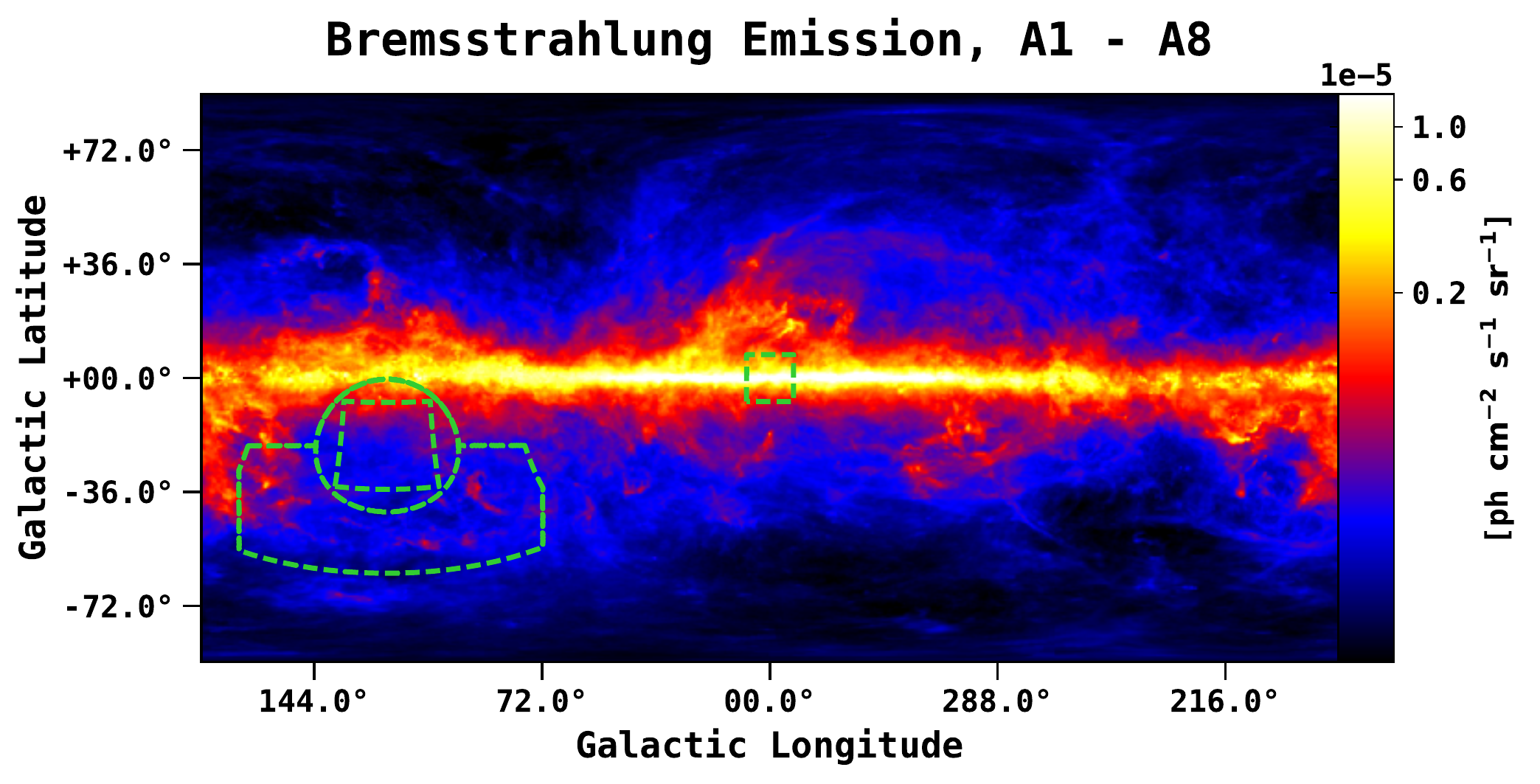}
\caption{Gas-related components of the IEM ($\pi^0$-decay related to \hi, \hii, and \htwo, and Bremsstrahlung emission) integrated in the energy range 1--100 GeV. The components correspond to different annuli, as indicated above each plot. The color corresponds to the intensity, and is shown in logarithmic scale. The intensity level is for the initial GALPROP outputs, before tuning to the $\gamma$-ray data. The maps are shown in a Plate Carr\'{e}e projection, and the pixel size is 0.25 deg/pix. Overlaid is the ROI used in this analysis, as well as the GC region (see Figure~\ref{fig:galactic_diffuse_schematic}).}
\label{fig:maps_1}
\end{figure*}

\begin{figure}[tbh!]
\centering
\includegraphics[width=0.49\textwidth]{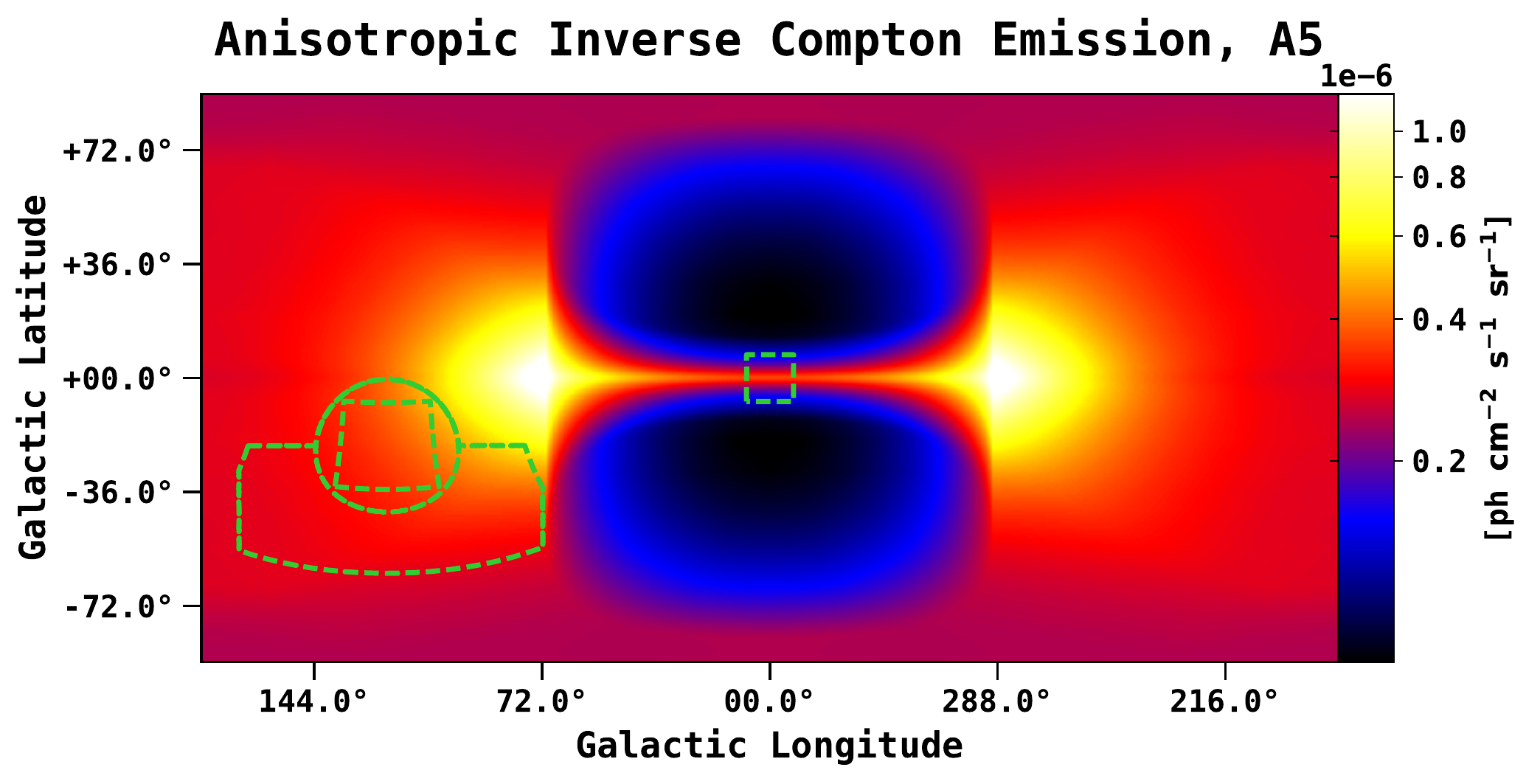}
\includegraphics[width=0.49\textwidth]{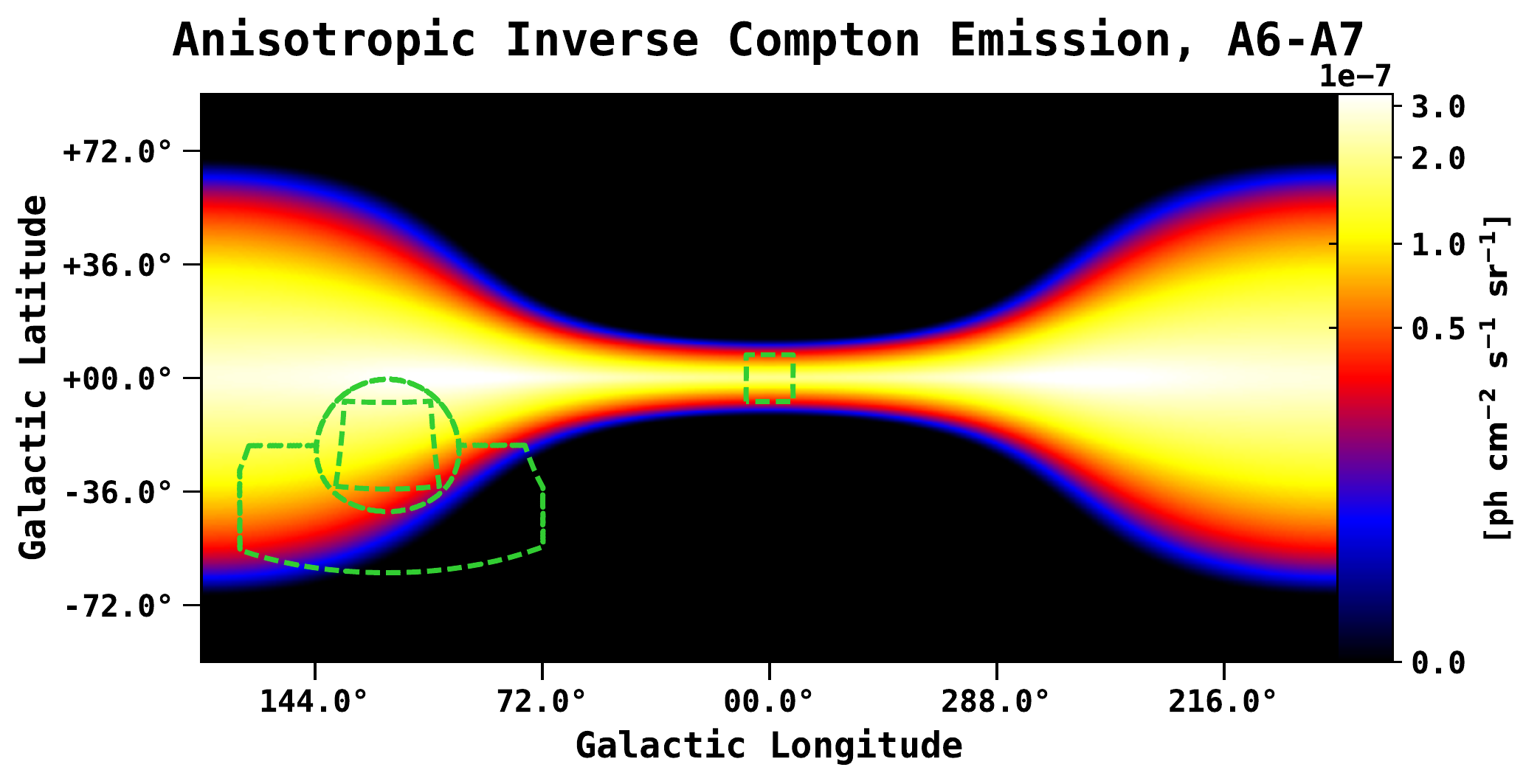}
\includegraphics[width=0.49\textwidth]{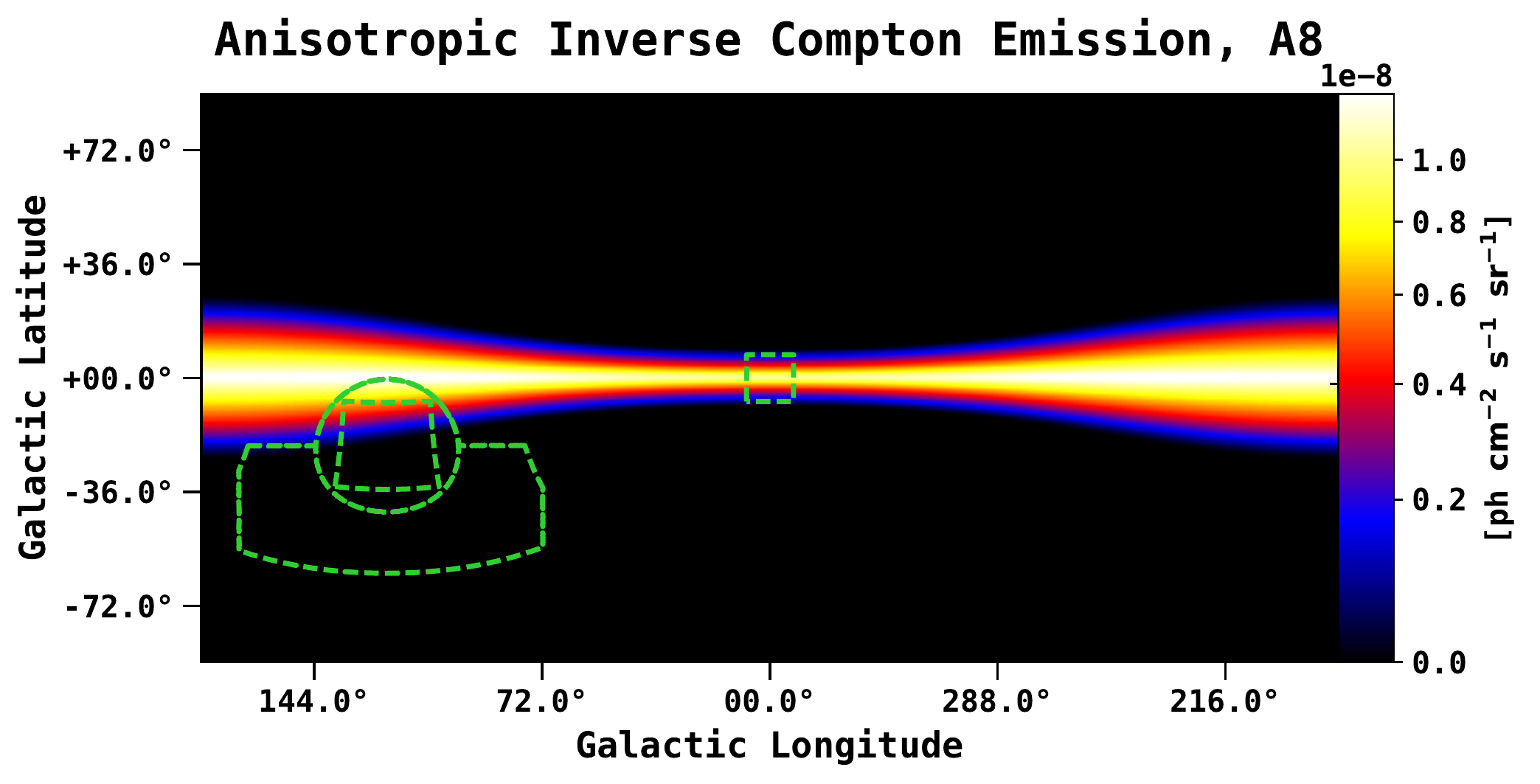}
\caption{Anisotropic Inverse Compton (AIC) components of the interstellar emission model for the MW in the energy range 1--100 GeV. The color corresponds to the intensity, and is shown in logarithmic scale. The intensity level is for the initial GALPROP outputs, before tuning to the $\gamma$-ray data. The map is shown in a Plate Carr\'{e}e projection, and the pixel size is 0.25 deg/pix. The IC A6 and A7 components are highly degenerate, and so we combine them into a single map A6$+$A7. Overlaid is the ROI used in this analysis, as well as the GC region (see Figure~\ref{fig:galactic_diffuse_schematic}). Note that we use the anisotropic IC maps as our default component. Unless otherwise stated, all reference to the IC component implies the anisotropic formalism.}
\label{fig:maps_2} 
\end{figure}

\begin{figure}[tbh!]
\centering
\includegraphics[width=0.49\textwidth]{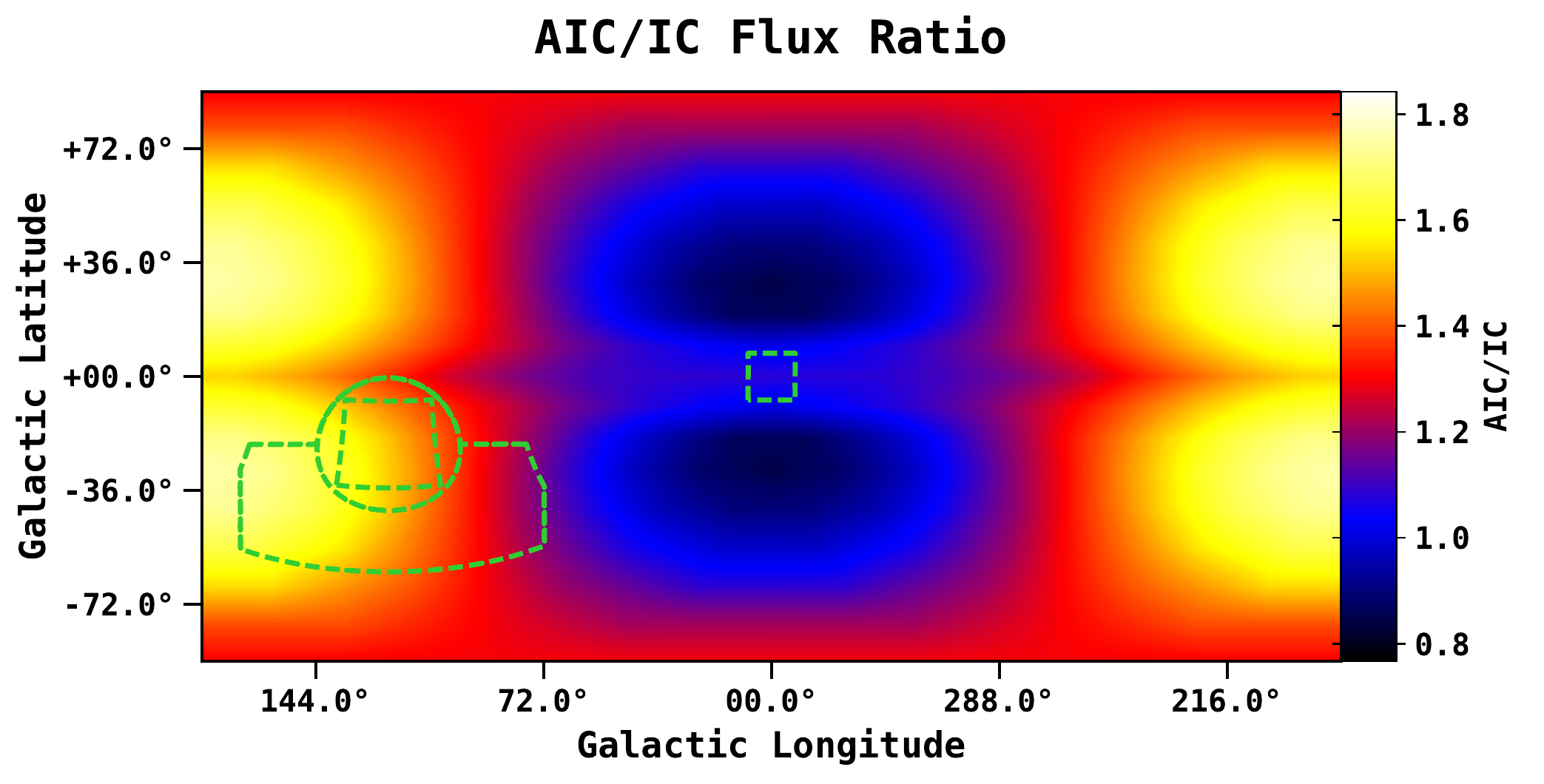}
\includegraphics[width=0.40\textwidth]{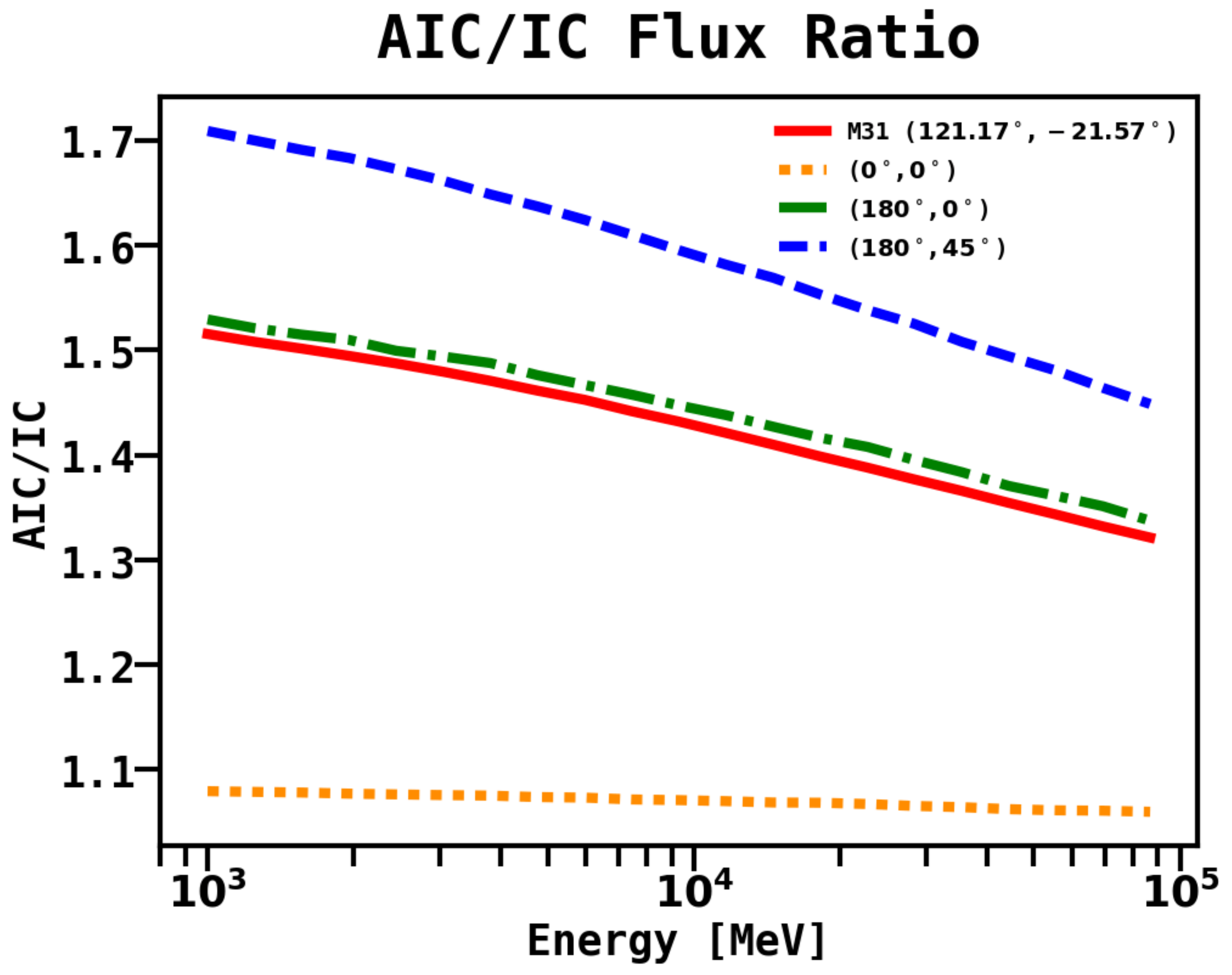}
\caption{The IEM employs the anisotropic IC sky maps, as discussed in the text. For comparison we show the differential flux ratio (AIC/IC) between the anisotropic (AIC) and isotropic (IC) inverse Compton components (all-sky). The top figure shows the spatial variation of the ratio at 1 GeV. The bottom figure shows the energy dependence of the ratio for 4 different spatial points, including M31. The ratio is close to unity towards the GC, increases with Galactic longitude and latitude, and reaches maximum at mid-latitudes towards the outer Galaxy. Note that we use the anisotropic IC maps as our default component. Unless otherwise stated, all reference to the IC component implies the anisotropic formalism.}
\label{fig:AIC_ratio}
\end{figure}

Figure~\ref {fig:galactic_diffuse_schematic} shows the total IEM in the energy range 1--100 GeV. The model includes $\pi^0$-decay, inverse Compton (IC), and Bremsstrahlung components. Overlaid is the ROI used in this analysis. From the observed counts (Figure~\ref{fig:observed_counts}) we cut an $84^\circ\times84^\circ$ ROI, which is centered at M31. The green dashed circle is the 300 kpc boundary corresponding to M31's canonical virial radius (of $\sim$$21^\circ$), as also shown in Figure~\ref{fig:observed_counts}. 

We label the field within the virial radius as FM31, and the region outside (and below latitudes of  $-21.57^\circ$) we label as the tuning region (TR). Longitude cuts are made on the ROI at $l=168^\circ \ \mathrm{and} \ l=72^\circ$. The former cut is made to stay away from the outer Galaxy, where the gas distribution becomes more uncertain, due to the method used for placing the gas at Galactocentric radii, i.e.\ Doppler shifted 21-cm emission. The latter cut is made to prevent the observations from including additional model component (i.e.~A4, as described below), which would further complicate the analysis. 

The \gray\ maps generated by GALPROP correspond to ranges in Galactocentric radii, and their boundaries are shown in Figure~\ref{fig:measurement_schematic} (A1--A8), which also depicts an overhead view of the annuli. The line of sight for the ROI, as seen from the location of the Solar system, is indicated with dash-dot red lines. Maps for the individual processes are shown in Figures~\ref{fig:maps_1} and~\ref{fig:maps_2}.

\begin{figure}[tbh!]
\centering
\includegraphics[width=0.49\textwidth]{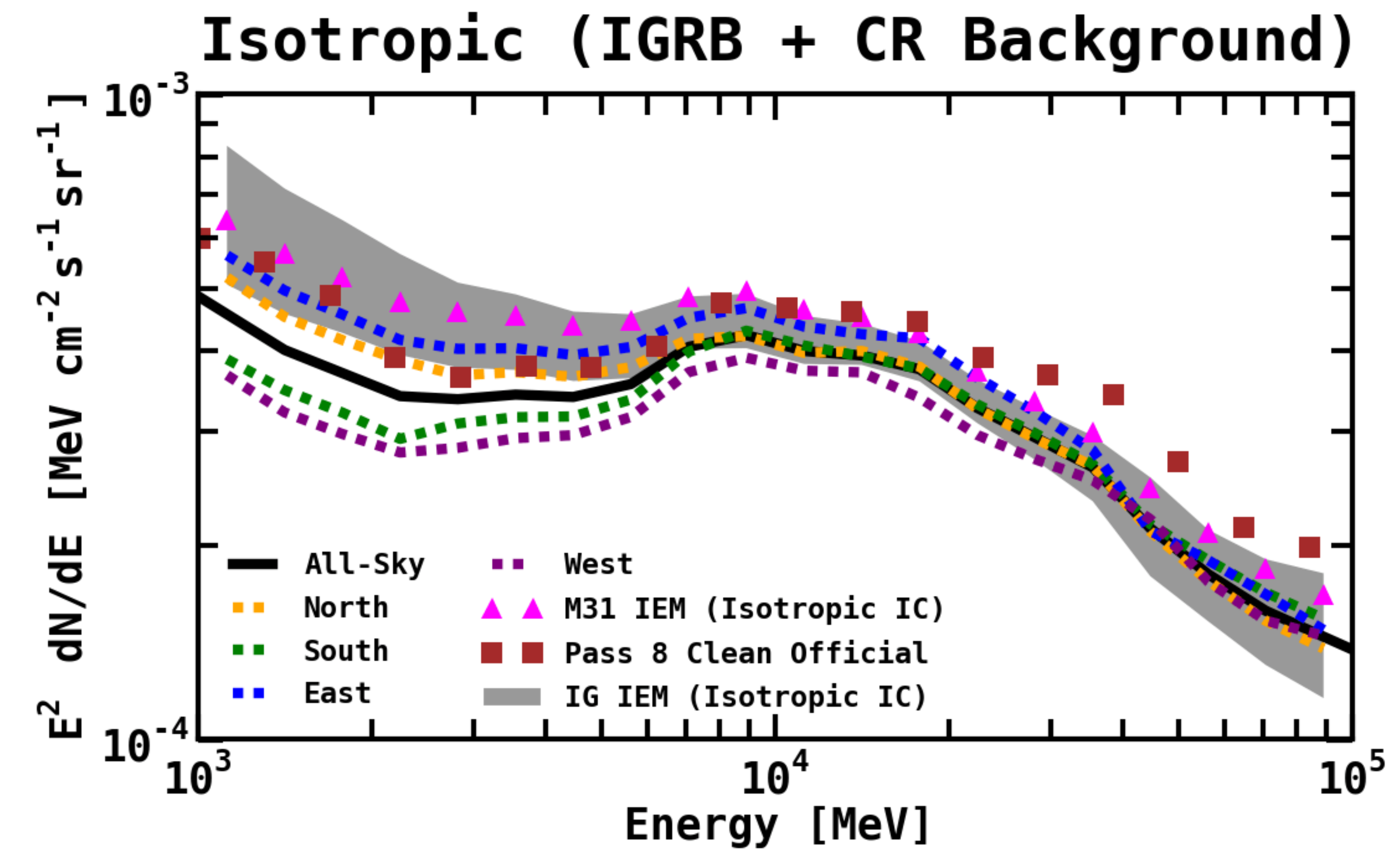}
\caption{The spectrum of the isotropic component has a dependence on the IEM and the ROI used for the calculation, as well as the data set. For the M31 IEM (which uses the AIC sky maps) we calculate the \textbf{all-sky} (solid black line) isotropic component in the following region: $|b| \geq 30^\circ, \ 45^\circ \leq  l \leq 315^\circ$. We also calculate the isotropic component in the different sky regions: \textbf{north}: $b\geq 30^\circ, \ 45^\circ \leq  l \leq 315^\circ$ (orange dashed line); \textbf{south}:  $b\leq -30^\circ,\ 45^\circ \leq  l \leq 315^\circ$ (green dashed line); \textbf{east}: $|b|\geq 30^\circ, \ 180^\circ \leq  l \leq 315^\circ$ (blue dashed line); and \textbf{west}: $|b|\geq 30^\circ, \ 45^\circ \leq  l \leq 180^\circ$ (purple dashed line). See Table~\ref{tab:norm_isotropic} for the corresponding best-fit normalizations. Magenta triangles show the all-sky isotropic component for the M31 IEM derived using the isotropic IC formalism. The brown squares show the official FSSC isotropic spectrum (iso\_P8R2\_CLEAN\_V6\_v06). The gray band is our calculated isotropic systematic uncertainty for the IG IEM, which uses the isotropic IC formalism (see Appendix~\ref{sec:IG_IEMs}).}
\label{fig:Isotropic_Sytematics}
\end{figure}

The \hi\ maps GALPROP employs are based on LAB\footnote{The Leiden/Argentine/Bonn Milky Way \hi\ survey} + GASS\footnote{GALEX Arecibo SDSS Survey; GALEX = the Galaxy Evolution Explorer,  SDSS = Sloan Digital Sky Survey} data, which for our ROI corresponds to LAB data only~\citep{kalberla2005leiden}. We note that there is a newer EBHIS\footnote{The Effelsberg-Bonn \hi\ Survey} survey that covers the whole northern sky, but for our purposes the LAB survey suffices. Besides, the development of the new \hi\ maps for GALPROP based on the EBHIS survey would require a dedicated study. The \hi-related $\gamma$-ray emission depends on the \hi\ column density, which depends on the spin temperature of the gas. We assume a uniform spin temperature of 150 K. The gas is placed at Galactocentric radii based on the Doppler-shifted velocity and Galactic rotation models. FM31 has a significant emission associated with \hi\ gas. The emission is dominated by A5, with further contribution from A6--A7. 

On the other hand, there is very little contribution from \htwo, which is concentrated primarily along the Galactic disk. The emission in FM31 only comes from A5. The 2.6 mm line of the $^{12}\rm{CO}$ molecular $J = 1 \rightarrow 0$ transition is used as a tracer of \htwo, assuming a proportionality between the integrated line intensity of CO, $W(\rm{CO})$, and the column density of \htwo, $N(\rm{H_2})$, given by the factor $X_{\rm{CO}}$. We use the $X_{\rm{CO}}$ values from~\citet{TheFermi-LAT:2015kwa}, which are tabulated at different Galactocentric radii with power law interpolation. In particular, the values relevant for this analysis are $1.4\times10^{20}$, $7.2\times10^{19}$, and $7.0\times10^{20}$ (cm$^{-2}$ K$^{-1}$  km$^{-1}$ s ), for radii 7.5, 8.7, and 11.0 (kpc), respectively. 

The foreground emission from \hii\ is subdominant. Modeling of this component is based on pulsar dispersion measurements. We use the model from~\citet{gaensler2008vertical}. 

The distribution of He in the interstellar gas is assumed to follow that of hydrogen, with a He/H ratio of 0.11 by number. Heavier elements in the gas are neglected. 

Our model also accounts for the dark neutral medium (DNM), or dark gas, which is a component of the interstellar medium that is not well traced by 21-cm emission or CO emission, as described in~\citet{grenier2005unveiling}, \citet{Ackermann:2012pya}, and~\citet{Acero:2016qlg}. For any particular region the DNM comprises unknown fractions of cold dense \hi\ and CO-free or CO-quiet \htwo. Details for the determination of the DNM component are described in~\citet{Ackermann:2012pya}. 

In summary, a template for the DNM is constructed by creating a map of ``excess'' dust column density $\rm{E(B-V)_{res}}$.  A gas-to-dust ratio is obtained for both \hi\ and CO using a linear fit of the N(\hi) map and W(CO) map to the $\rm{E(B-V)}$ reddening map of~\citet{schlegel1998maps}. In general, the method is all-sky, and a constant gas-to-dust ratio is assumed throughout the Galaxy. Subtracting the correlated parts from the total dust results in the residual dust emission, $\rm{E(B-V)_{res}}$, which is then associated with the DNM.  In the current study the DNM is incorporated into the \hi\ templates; see \citet{Ackermann:2012pya} for details.

The IC component arises from up-scattered low-energy photons of the Galactic interstellar radiation field (ISRF) by CR electrons and positrons. The ISRF (optical, infrared, and cosmic microwave background) is the result of the emission by stars, and scattering, absorption, and re-emission of absorbed starlight by dust in the interstellar medium. The ISRF is highly anisotropic since it is dominated by the radiation from the Galactic plane. An observer in the Galactic plane thus sees mostly head-on scatterings even if the distribution of the CR electrons is isotropic. This is especially evident when considering inverse Compton scattering by electrons in the halo, i.e.\ the diffuse emission at high Galactic latitudes. 

We employ the anisotropic formalism of the IC component \citep{Moskalenko:1998gw}. From the GALPROP code we use the standard ISRF model file (standard.dat) and standard scaling factors of 1.0 for optical, infrared, and microwave components. In Figure~\ref{fig:AIC_ratio} we show the differential flux ratio (AIC/IC) between the anisotropic (AIC) and isotropic (IC) inverse Compton components (all-sky). The top figure shows the spatial variation of the ratio at 1 GeV. The ratio is close to unity towards the GC, increases with Galactic longitude and latitude, and reaches maximum at mid-latitudes towards the outer Galaxy. The bottom figure shows the energy dependence of the ratio for 4 different spatial points, including M31. Note that unless otherwise stated, \emph{all reference to the IC component implies the anisotropic formalism.} Also, the $\gamma$-ray skymaps for IC A6 and A7 are highly degenerate, and so we combine them into a single map A6$+$A7.

The IC component anti-correlates with the isotropic component. The isotropic component includes unresolved extragalactic diffuse emission, residual instrumental background, and possibly contributions from other Galactic components which have a roughly isotropic distribution. The spectrum of the isotropic component depends on the IEM and the ROI used for the calculation. The spectrum also depends on the data set, since the residual instrumental background differs between data sets. We calculate the isotropic component self-consistently with the M31 IEM, and the spectrum is shown in Figure~\ref{fig:Isotropic_Sytematics}. Table~\ref{tab:norm_isotropic} gives the corresponding best-fit normalizations for the diffuse components.

The main calculation is performed over the full sky excluding regions around the Galactic plane and the Inner Galaxy: $|b| \geq 30^\circ, \ 45^\circ \leq  l \leq 315^\circ$. We note that even though it is not actually an all-sky fit, we refer to it as ``all-sky'' for simplicity hereafter. The fit includes 3FGL sources fixed, sun and moon templates fixed, \citet{wolleben2007new} component (Loop I two-component spatial template), all-sky $\pi^0$-decay and (anisotropic) IC normalization scaled, and all-sky Bremsstrahlung fixed. Besides, we calculate the isotropic component in the different sky regions: north, south, east, and west, as detailed in Figure~\ref{fig:Isotropic_Sytematics}. Also shown are the isotropic components resulting from the M31 IEM using the isotropic IC formalism, the FSSC IEM, and the IG IEM (which uses the isotropic IC formalism). At lower energies the intensities of the spectra calculated in the south and west (both regions associated with the M31 system) are lower than that of the spectra calculated in the north and east. Correspondingly, the IC normalizations are higher for the south and west.  Interestingly, independently of the IEM used in the fit, the isotropic spectrum features a bump at $\sim$10 GeV.

\begin{deluxetable}{lllll}[tbh!]
\tablecolumns{5}
\tablewidth{0mm}
\tablecaption{Normalizations for Calculations of the Isotropic Component\label{tab:norm_isotropic}}
\tablehead{
\colhead{Region} &
\colhead{$\pi^0$} &
\colhead{AIC}}
\startdata
All-sky &1.319 \p 0.005 &1.55 \p 0.04 \\
North&1.430 \p 0.010 &1.14 \p 0.05 \\
South &1.284 \p 0.006 &1.86 \p 0.05 \\
East&1.397 \p 0.009 &1.07 \p 0.05\\
West&1.287 \p 0.006 &1.88 \p 0.05 
\enddata
\tablecomments{See Figure~\ref{fig:Isotropic_Sytematics} for definition of the regions.}
\end{deluxetable}

\begin{figure}[tbh!]
\centering
\includegraphics[width=0.45\textwidth]{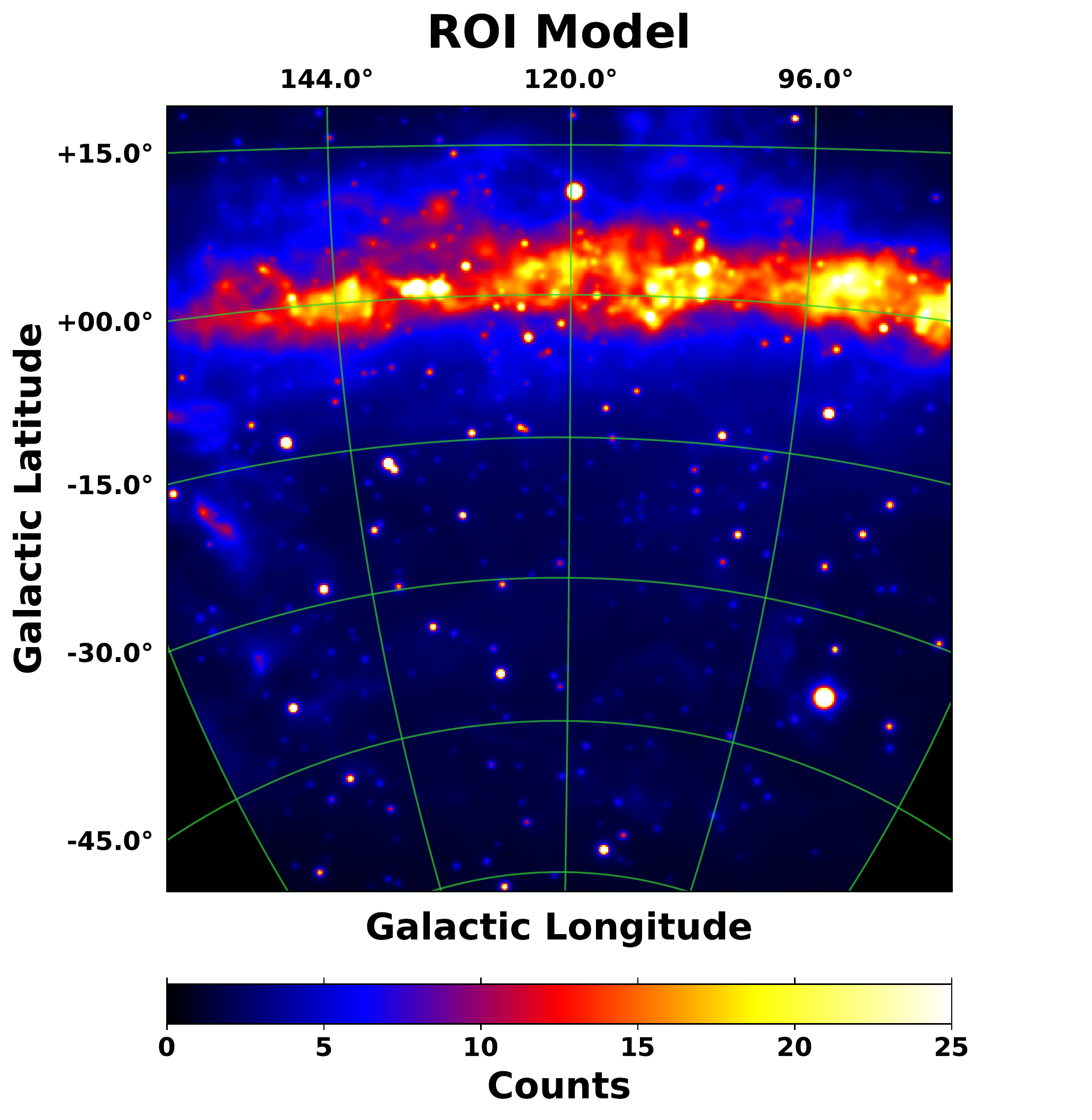}
\includegraphics[width=0.45\textwidth]{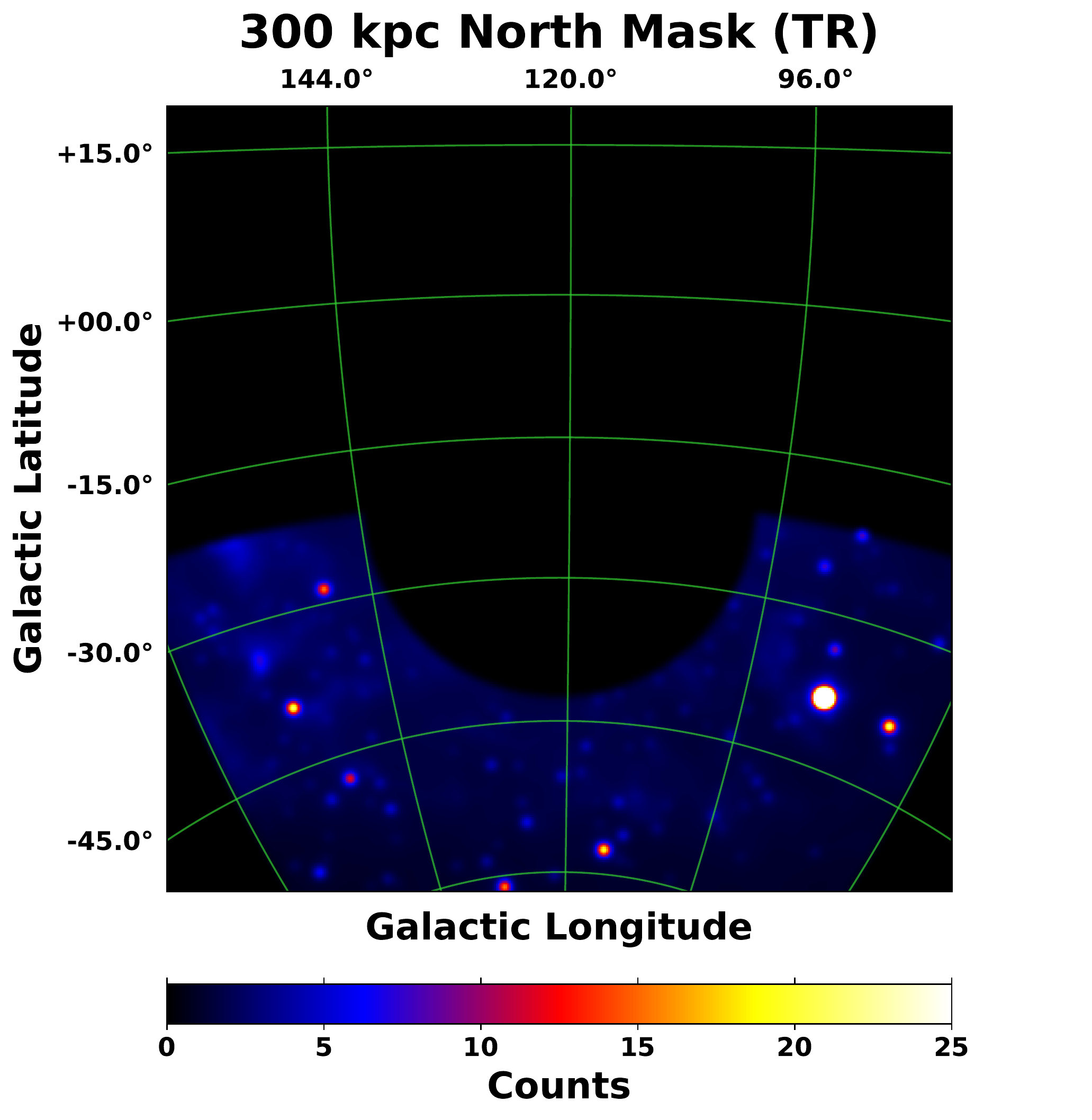}
\caption{Total model counts for the full ROI. For the tuning region (TR) we mask within the 300 kpc circle and latitudes above $-21.57^\circ$, as discussed in the text.} 
\label{fig:TR}
\end{figure}

In general, the model contains inherent systematic uncertainties due to a number of different factors, including the correlations between the different model components, uncertainties related to the determination of the DNM, and the presence of any un-modeled spatial variation in the spin temperature, CR density, and/or ISRF density. These issues will be addressed throughout this analysis.

\subsection{Tuning the IEM} \label{sec:tuning}

Figure~\ref{fig:TR} shows the total model counts for the full ROI. The bottom panel shows the TR, for which we mask the 300 kpc circle around M31 and latitudes north of $-21.57^\circ$. The primary purpose of the TR is to fit the normalization of the isotropic component. The isotropic component by definition is an all-sky average, but it may have some local spatial variations, since the instrumental background may also vary over the sky. The TR is also used to set the initial normalizations of the IC components, since they are anti-correlated with the isotropic component. 

The fit is performed by uniformly scaling each diffuse component as well as all 3FGL sources in the region. Note that the model includes all sources within $70^\circ$ of the ROI center, but only the sources in the TR are scaled in the fit. As a test, we also perform the fit by keeping fixed the 3FGL sources in the TR, and we find that the best-fit normalizations of the diffuse components are not very sensitive to the scaling of the point sources. Likewise, it is not necessary to scale the point sources outside of the TR, which are included in order to account for the spillover of the instrumental PSF. The fit uses the spectral shape of the isotropic spectrum derived from the all-sky analysis. The \hii\ component is fixed to its GALPROP prediction, since it it subdominant compared to the other components. The Bremsstrahlung component possesses a normalization of 1.0 \p 0.6, consistent with the GALPROP prediction. In our further fits in the FM31 region these components remain fixed to their all-sky GALPROP predictions. 

\begin{figure}[tb!]
\centering
\includegraphics[width=0.49\textwidth]{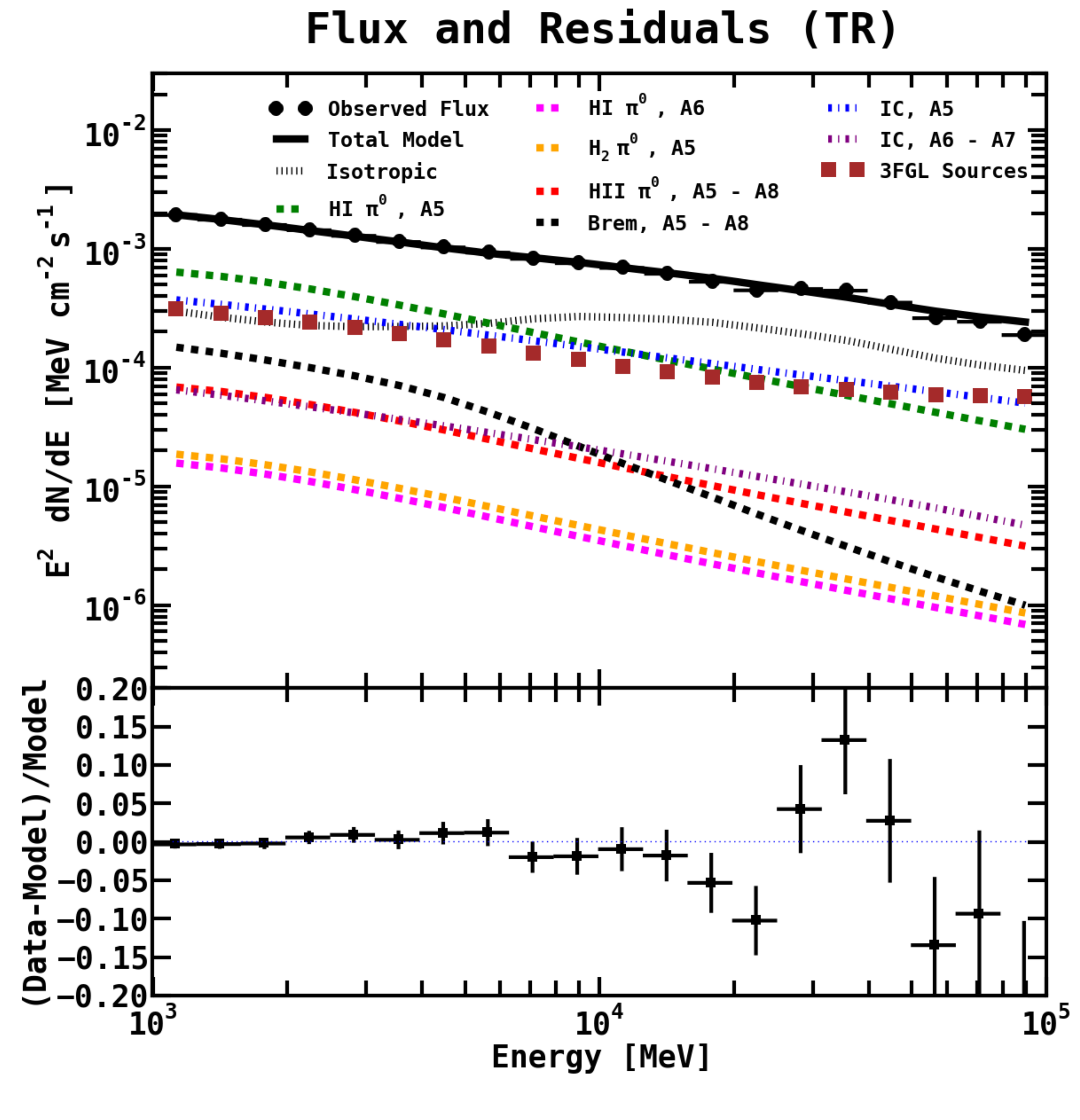}
\caption{Flux (upper panel) and fractional count residuals (lower panel) for the fit in the TR. The \hii\ component is fixed to its GALPROP prediction. The normalizations of all other diffuse components are freely scaled, as well as all 3FGL sources in the region. The residuals show fairly good agreement over the entire energy range.}
\label{fig:flux_and_residuals_TR}
\end{figure}

Figure~\ref{fig:flux_and_residuals_TR} shows the best-fit spectra and fractional count residuals resulting from the fit in the TR. The corresponding best-fit normalizations and integrated flux are reported in Table~\ref{tab:norm_TR}. The isotropic component possesses a normalization of 1.06 \p 0.04, consistent with the all-sky average. The \hi\ $\pi^0$ A6 component shows a fairly high normalization with respect to the model prediction, which is likely related to the fact that it only contributes near the edge of the region. 

The fractional residuals are fairly flat over the entire energy range, but somewhat worsen at higher energies, although they remain consistent with statistical fluctuations. We note that there does appear to be a subtle systematic bias in the fractional residuals, where the data are being over-modeled between $\sim$6--20 GeV and $\sim$50--100 GeV, with excess emission between $\sim$20--50 GeV. This may be due to the spectral shape of the 3FGL sources in the region that is not properly accounted. For the sources we use their spectral parameterizations rather than the binned data points, which may or may not be a good representation of the true spectra at high energies where the statistical fluctuations are significant.

\begin{figure}
\centering
\includegraphics[width=0.4\textwidth]{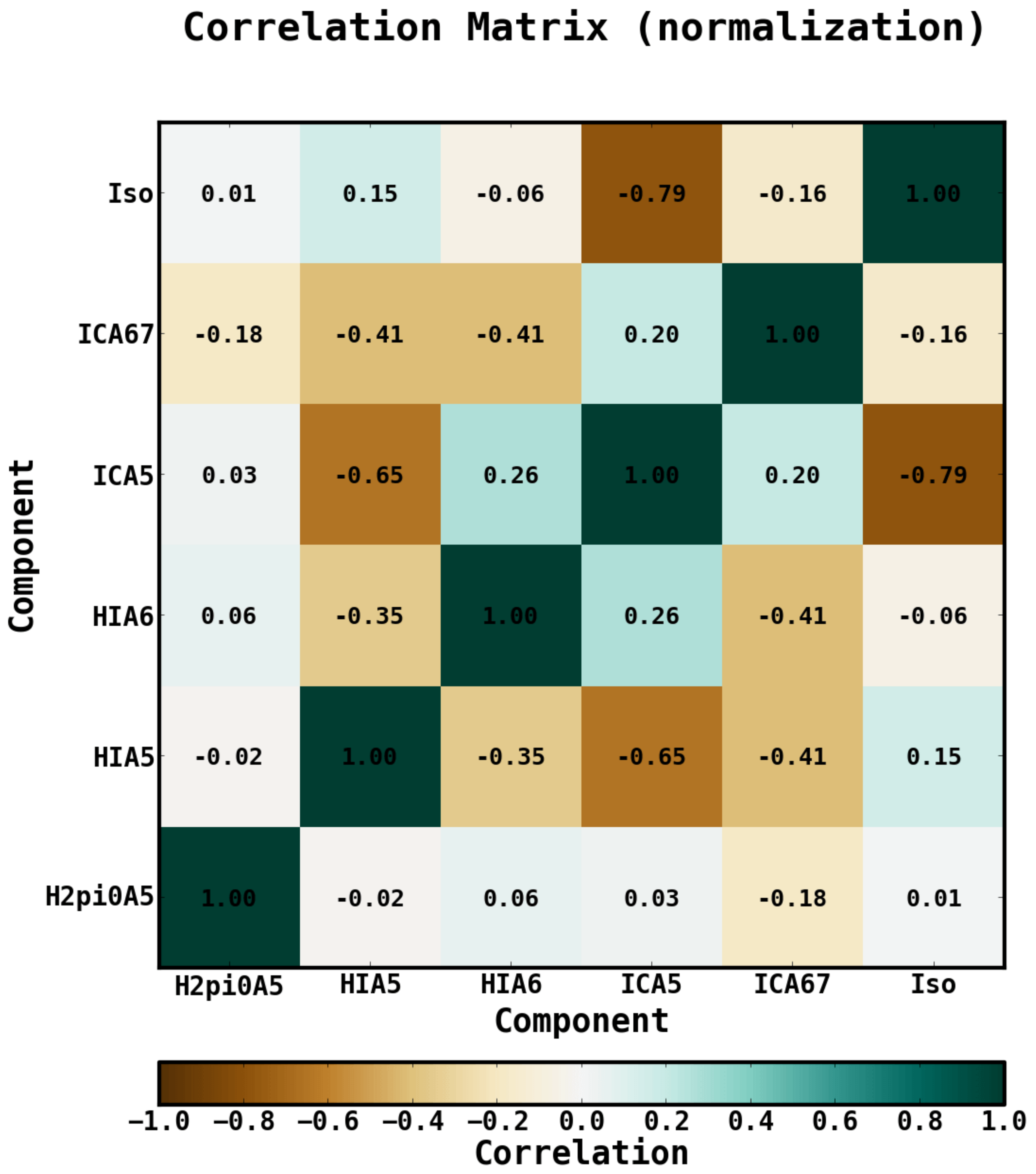}
\caption{Correlation matrix for the fit in the TR. For brevity IC A6 -- A7 is labeled as ICA67, and the isotropic component is labeled as Iso.}
\label{fig:TR_correlation}
\end{figure}

Figure~\ref{fig:TR_correlation} shows the correlation matrix\footnote{The correlation ($C$) of two parameters $A$ and $B$ is defined in terms of the covariance ($cov$) and the standard deviation ($\sigma$): $C = cov_{AB}\times (\sigma_A \sigma_B)^{-1}.$} for the fit. The isotropic component is anti-correlated with the IC components. The IC components are also anti-correlated with the \hi\ A5 component. The \htwo{} component shows very little correlation with the other components, but its contribution is very minimal in the TR. 

\begin{figure*}
\centering
\includegraphics[width=0.33\textwidth]{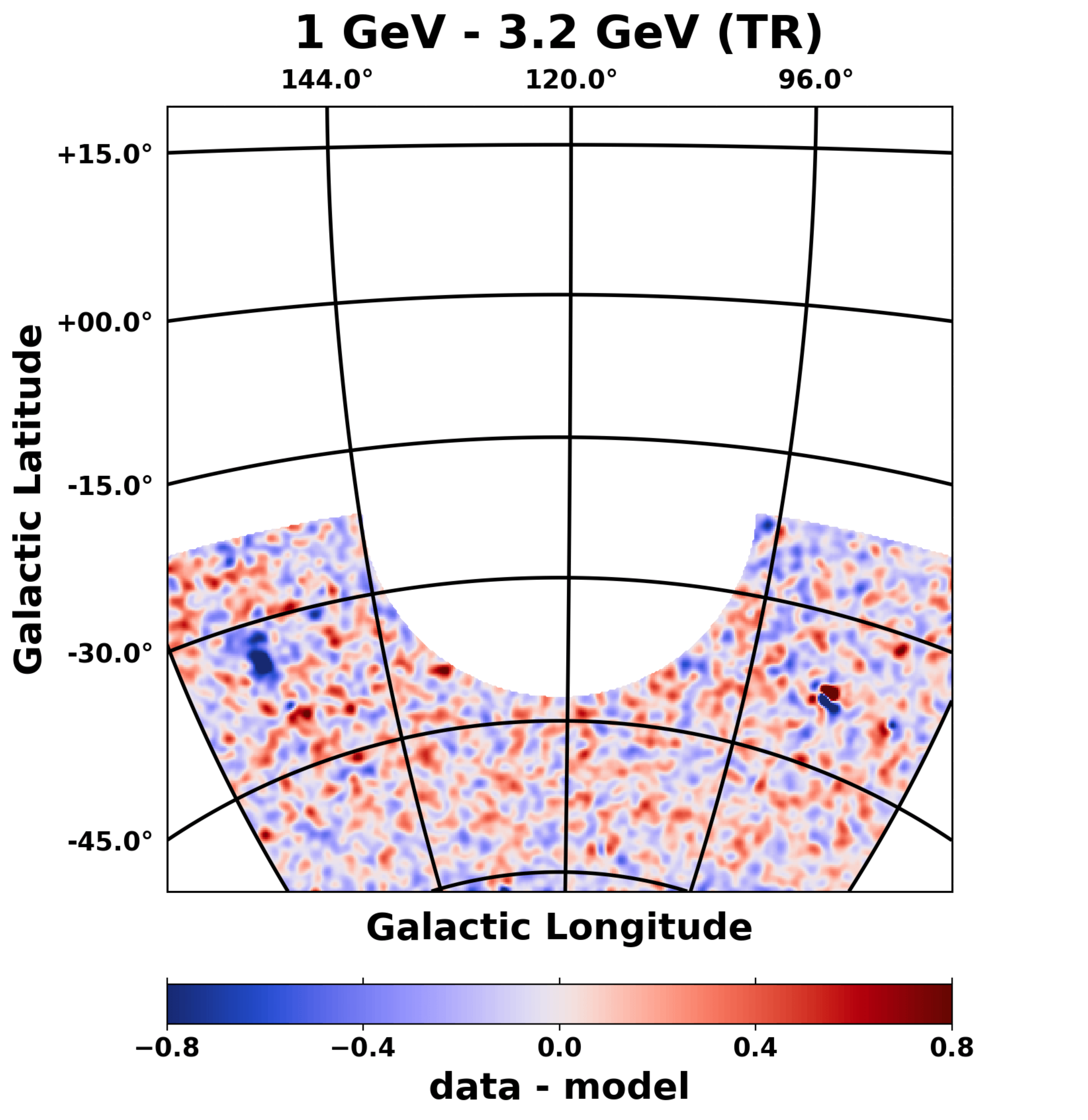}
\includegraphics[width=0.33\textwidth]{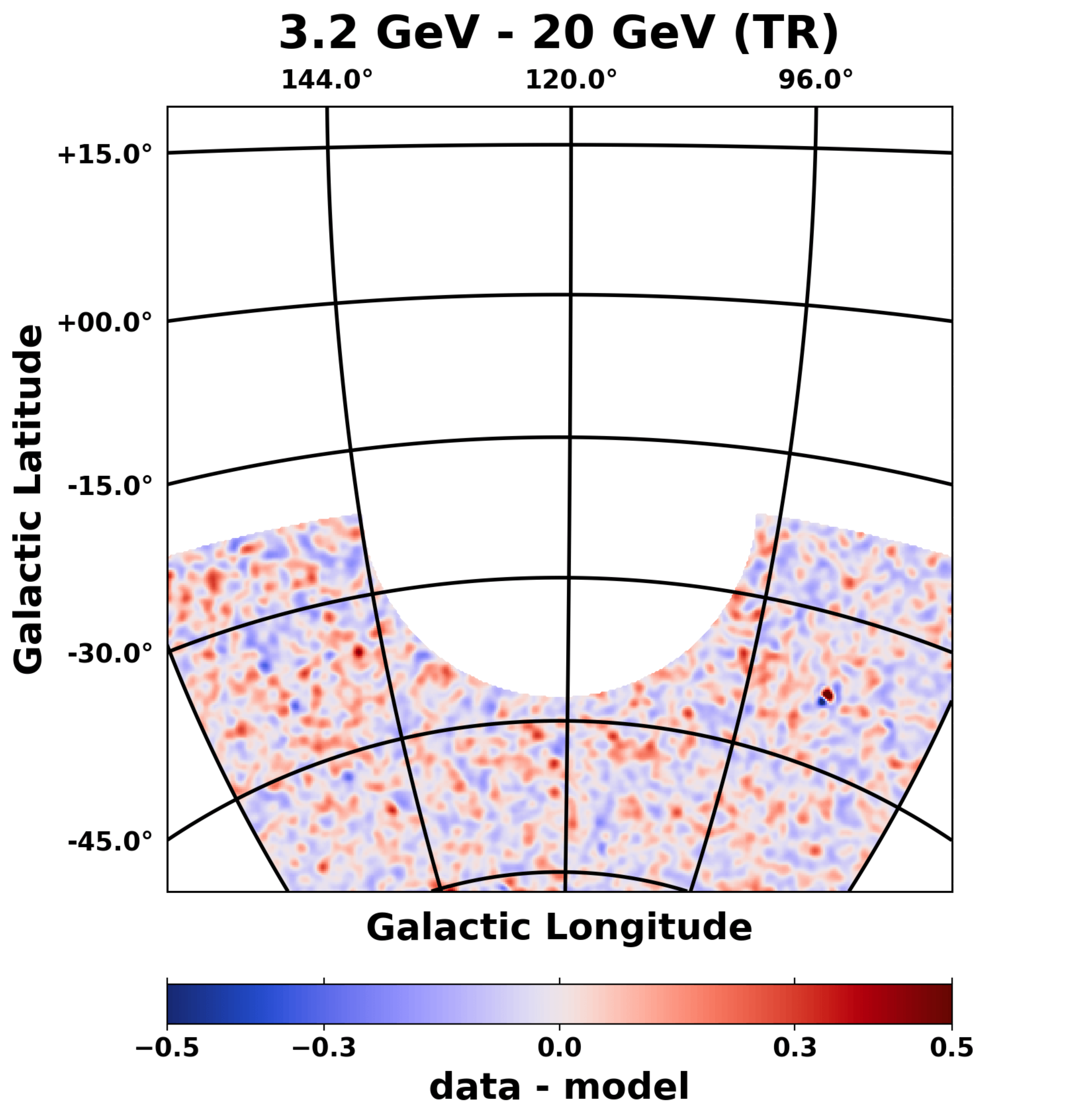}
\includegraphics[width=0.33\textwidth]{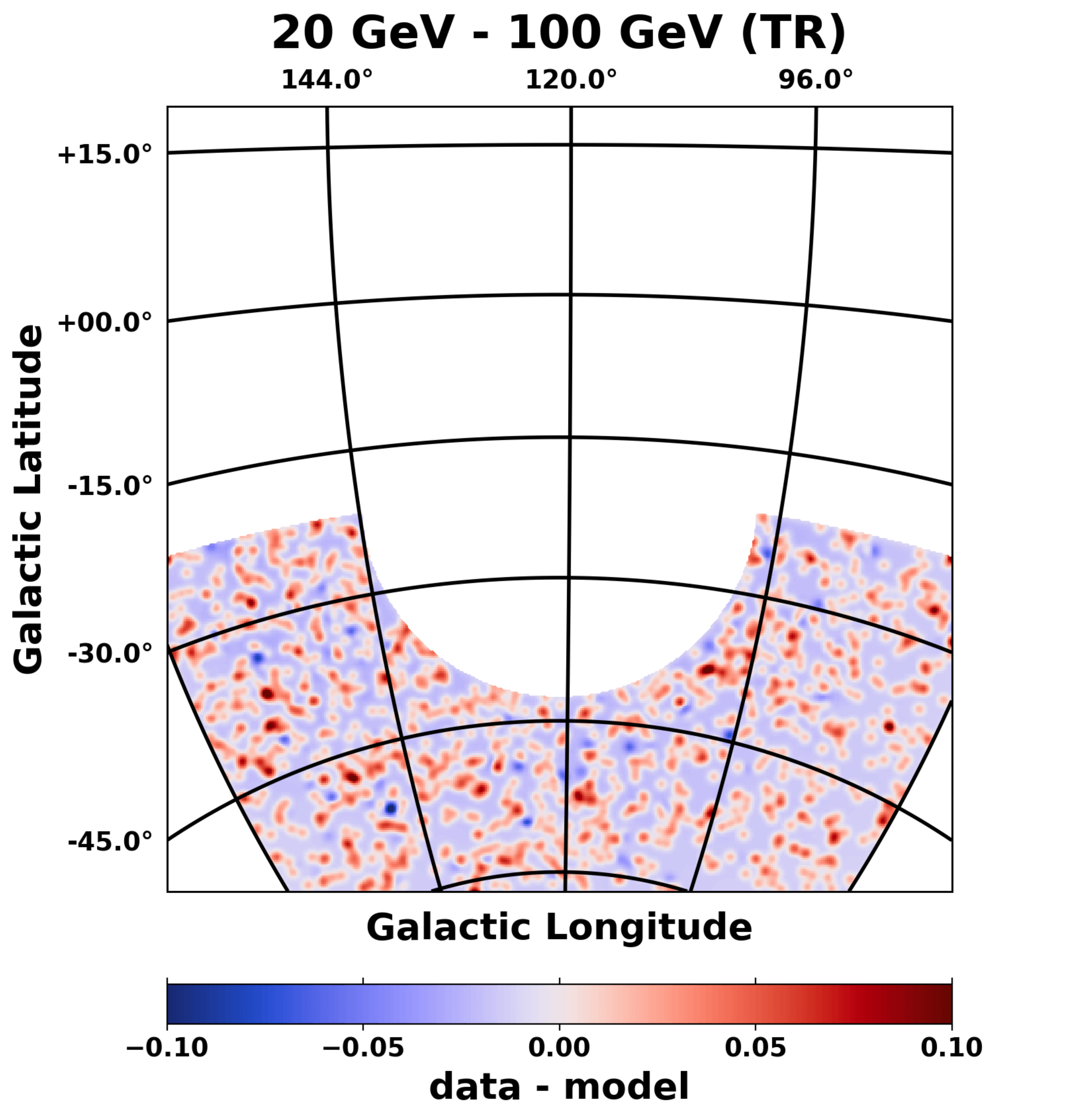}

\caption{Spatial count residuals (data $-$ model) resulting from the fit in the TR for three different energy bands, as indicated above each plot. The energy bins are chosen to coincide with an excess which is later observed in the fractional energy residuals for the fit in FM31, as discussed in the text. The color scale corresponds to counts/pixel, and the pixel size is $0.2^\circ \times 0.2^\circ$. The images are smoothed using a $1^\circ$ Gaussian kernel. This value corresponds to the PSF (68\% containment angle) of \textit{Fermi}-LAT, which at 1 GeV is $\sim$$1^\circ$.}
\label{fig:spatial_residuals_TR}
\end{figure*}

Figure~\ref{fig:spatial_residuals_TR} shows the spatial count residuals for three different energy bins, as indicated above each plot. The bins are chosen to coincide with positive residual emission which is observed in FM31, as discussed in Section~\ref{sec:FM31 Baseline}. Residuals are shown using a colormap from the colorcet package~\citep{kovesi2015good}. 

Two notable features can be observed in the residuals. Near $(l,b)$$\approx$$(156^\circ, -35^\circ)$ a deep hole can be seen in the first energy bin. Comparing to the \hi\ column density maps (see Figure~\ref{fig:maps_1}), this over-modeling is likely related to a feature in the gas. Note that the hole also contains a BL LAC (3FGL J0258.0+2030). The second notable feature is located near $(l,b)$$\approx$$(84^\circ, -40^\circ)$. This is a flat spectral radio quasar (3FGL J2254.0+1608). As a test, these trouble-regions were masked and it is found that they do not significantly impact the normalizations of the diffuse components. Otherwise the residual maps in all three energy bins are pretty smooth, exhibiting no obvious features.

\section{Analysis of the M31 Field} \label{sec:FM31 Baseline}
 
\subsection{Baseline Fit and Point Source Finding Procedure} \label{sec:psalgorithm}

The data set employed in this work is approximately  two times larger than the one used to derive the 3FGL. Therefore, in conjunction with the baseline fit, we search for additional point sources in FM31 to account for any un-modeled point-like structure that may otherwise contribute to the residual emission. The procedure we employ is  similar to the one developed in~\citet{TheFermi-LAT:2015kwa}. The point sources are initially modeled with the 3FGL. A maximum likelihood fit is performed by freeing the normalization of the 3FGL sources, as well as the \hi- and \htwo-related components. The top of FM31 also has contribution from IC A8, and its normalization is freed in the fit.  The normalizations of the isotropic and IC components (A5 and A6 -- A7) remain fixed to their best-fit values obtained in the TR. The \hii\ and Bremsstrahlung components are fixed to their GALPROP predictions. Note that the Bremsstrahlung component possesses a normalization of 1.0 \p 0.6 in the TR, consistent with the GALPROP prediction.

\begin{deluxetable}{lcccccc}
\tablecolumns{7}
\tablewidth{0mm}
\tablecaption{Baseline Values for the IEM Components in the TR \label{tab:norm_TR}}
\tablehead{
\colhead{Component} &
\colhead{Normalization} &
\colhead{Flux  ($\times 10^{-9})$}&
\colhead{Intensity ($\times 10^{-8})$}\\
& 
&
\colhead{(ph cm$^{-2}$ s$^{-1}$)} &
\colhead{(ph cm$^{-2}$ s$^{-1}$ sr$^{-1}$)}
}
\startdata
\hi\ $\pi^0$, A5 &1.10 \p 0.03&439.4 \p11.0 &153.1 \p 3.8 \\
\hi\ $\pi^0$, A6 &5.0 \p 1.3  &10.6 \p 2.8 &3.7 \p 1.0 \\
\htwo\ $\pi^0$, A5 &2.1 \p 0.1&12.6 \p 0.7 &4.4 \p 0.3 \\
Bremsstrahlung & 1.0 \p 0.6&100.4 \p 58.3&35.0 \p 20.3\\
IC, A5 &2.3 \p 0.1   &274.7 \p 14.0&95.7 \p 4.9 \\
IC, A6 -- A7&3.5 \p 0.4  &45.7 \p 4.8 &15.9 \p 1.7\\
Isotropic  &1.06 \p 0.04  &248.1 \p 10.4 &86.4 \p 3.6 
\enddata
\tablecomments{The normalizations of the diffuse components are freely scaled, as well as all 3FGL sources in the region. The fit uses the all-sky isotropic spectrum. Intensities are calculated by using the total area of the TR, which is 0.287 sr. Note that the reported errors are 1$\rm{\sigma}$ statistical only (and likewise for all tables).}
\end{deluxetable}

\begin{deluxetable}{lccccc}[tbhp!]
\tablecolumns{6}
\tablewidth{0mm}
\tablecaption{New point sources for FM31 \label{tab:PS_FM31}}

\tablehead{
\colhead{Name} &
\colhead{TS} &
\colhead{$l$} &
\colhead{$b$} &
\colhead{Index} &
\colhead{Flux\,($\times 10^{-10})$ }\\
& 
&
\colhead{(deg)} &
\colhead{(deg)} &
\colhead{$\alpha$}&
\colhead{(ph cm$^{-2}$ s$^{-1}$)}
}
\startdata
FM31\_1 & 34 & 124.58 & $-32.60$ & $2.61$ \p 0.34 & 2.9 \p 0.7\\
FM31\_2 & 31 & 122.66 & $-29.25$ & $2.78$ \p 0.33 & 2.8 \p 0.7 \\
FM31\_3 & 31 & 117.71 & $-26.83$ & $2.33$ \p 0.27 & 2.5 \p 0.6 \\
FM31\_4 & 29 & 131.86 & $-27.70$ & $2.14$ \p 0.24 & 1.9 \p 0.5 \\
FM31\_5 & 24 & 127.49 & $-9.62$ & $3.81$ \p 0.67 & 3.9 \p 0.9 \\
FM31\_6 & 23 & 129.91 & $-10.13$ & $3.09$ \p 0.39 & 3.4 \p 0.9 \\
FM31\_7 & 18 & 128.32 & $-10.58$ & $2.25$ \p 0.31 & 2.3 \p 0.8 \\
FM31\_8 & 18 & 111.53 & $-22.79$ & $3.32$ \p 0.55 & 2.7 \p 0.8 \\
FM31\_9 & 17 & 118.05 & $-31.02$ & $2.41$ \p 0.34 & 1.7 \p 0.6 \\
FM31\_10 & 17 & 119.73 & $-25.66$ & $4.26$ \p 1.26 & 2.1 \p 0.6 \\
FM31\_11 & 16 & 110.44 & $-25.71$ & $2.90$ \p 0.47 & 2.1 \p 0.7 \\
FM31\_12 & 15 & 108.73 & $-29.55$ & $2.17$ \p 0.36 & 1.5 \p 0.6\\
FM31\_13 & 14 & 126.34 & $-11.63$ & $3.12$ \p 0.57 & 2.4 \p 0.8 \\
FM31\_14 & 14 & 118.27 & $-9.50$ & $3.97$ \p 0.96 & 2.7 \p 0.9 \\
FM31\_15 & 13 & 110.61 & $-33.64$ & $3.90$ \p 0.95 & 1.8 \p 0.6 \\
FM31\_16 & 13 & 120.13 & $-30.65$ & $2.81$ \p 0.55 & 1.7 \p 0.6 \\
FM31\_17 & 12 & 133.80 & $-8.37$ & $2.29$ \p 0.44 & 1.7 \p 0.8\\
FM31\_18 & 11 & 126.84 & $-20.78$ & $2.23$ \p 0.37 & 1.3 \p 0.5 \\
FM31\_19 & 11 & 106.53 & $-28.95$ & $4.85$ \p 1.60 & 1.7 \p 0.6 \\
FM31\_20 & 11 & 116.65 & $-25.21$ & $5.39$ \p 1.48 & 1.6 \p 0.6 \\
FM31\_21 & 10 & 127.83 & $-27.92$ & $2.48$ \p 0.45 & 1.3  \p 0.5
\enddata
\tablecomments{The sources are fit with a power law spectral model $dN/dE \propto E^{-\alpha}$. The table gives the best-fit index, as well as the total flux, integrated between 1 GeV--100 GeV.}
\end{deluxetable}


\begin{figure}[tbh!]
\centering
\includegraphics[width=0.49\textwidth]{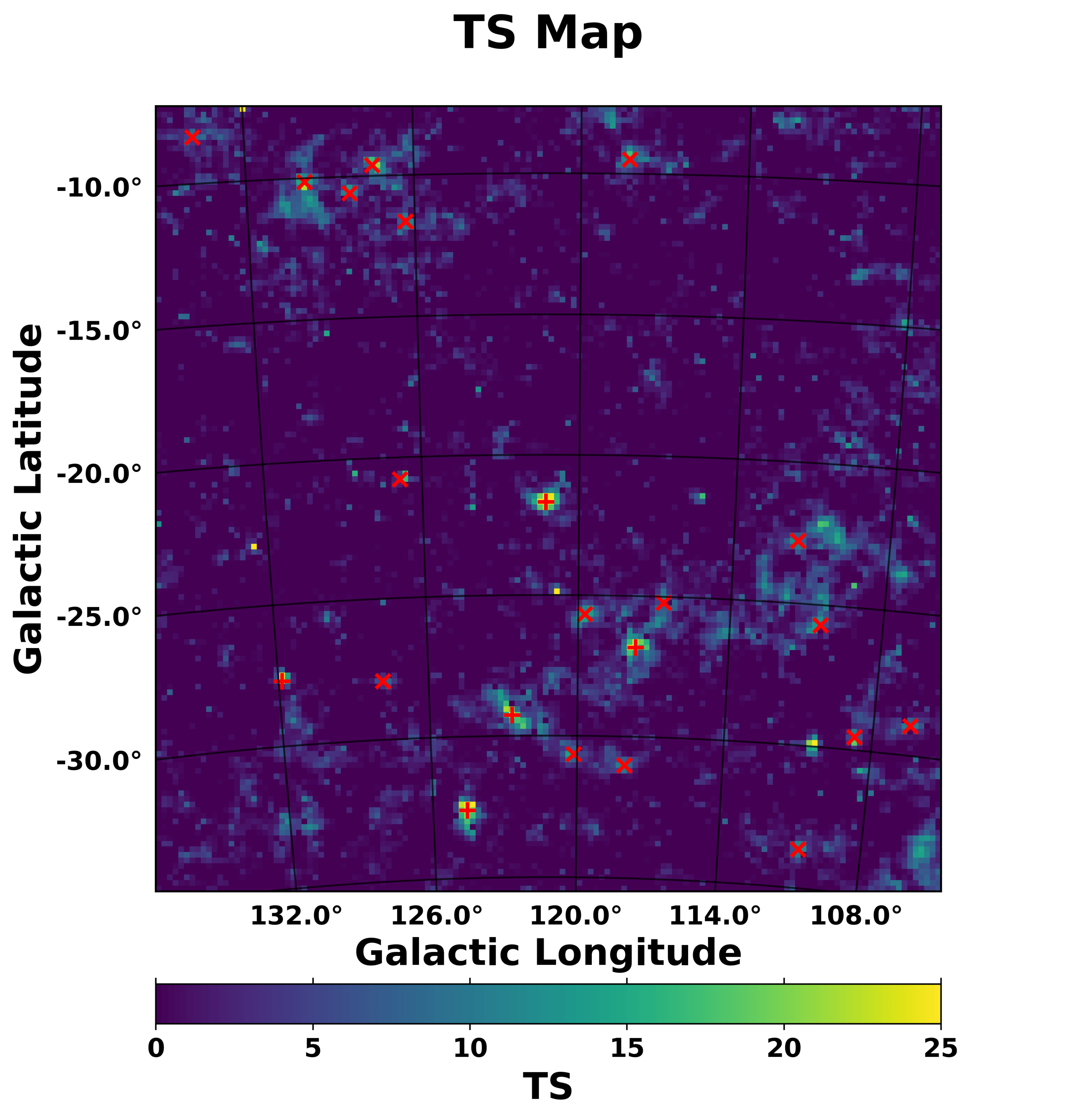}
\caption{The TS map is calculated after the baseline fit in FM31 (tuned). Overlaid are the additional point sources that we found using our point source finding procedure. Red crosses represent new sources with TS$\geq$25 and red slanted crosses represent new sources with 9$\leq$TS$<$25.}
\label{fig:TS_map}
\end{figure}

A wavelet transform is applied to the residual map to find additional point source candidates. We employ PGWave~\citep{1997ApJ...483..350D}, included in the \fermilat\ ScienceTools, which finds the positions of the point source candidates according to a user-specified signal-to-noise criterion (we use 3$\sigma$) based on the assumption of a locally flat background. Since PGWave does not provide spectral information, we model the spectrum of each point source candidate with a power law function and determine the initial values of the parameters via a maximum likelihood fit in the field, while all other components are held constant. 

The determination of the spectrum is further refined by performing additional maximum likelihood fits concurrently with the other components in the region, i.e.\ 3FGL point sources, \hi\ A5--A7, and \htwo{} A5. All point sources within a 30$^\circ$ radius of the field center are included in the model; however, only sources within a 20$^\circ$ radius are fit. The extra padding is included to account for the instrumental PSF. Owing to the large number of point sources involved, the fit is performed iteratively starting with the point sources (and point source candidates) with largest  significance of detection. All point source candidates with a test statistic (TS)\footnote{For a more complete explanation of the TS resulting from a likelihood fit see~\citet{1996ApJ...461..396M} and \url{https://fermi.gsfc.nasa.gov/ssc/data/analysis/documentation/Cicerone/Cicerone\_Likelihood/}} TS$\geq$9 are added to the model. Parameters for the additional point sources are summarized in Table~\ref{tab:PS_FM31}. 

Figure~\ref{fig:TS_map} shows the TS map calculated after the initial fit in FM31, before finding additional point sources. To reduce computational time, all components are held fixed to their best-fit values obtained in the initial fit. The TS map is calculated using the \textit{gttsmap} function included in the ScienceTools. Note that we do not include an M31 template for the calculation. Overlaid on the map are the additional point sources that we found using our point source finding procedure. In total we found 4 sources with TS$\geq$25 (besides the M31 source), and 17 sources with 9$\leq$TS$<$25. A point source is found corresponding to the M31 disk, but this source is removed for the baseline fit, and no M31 component is included (likewise the M31 source is not listed in Table~\ref{tab:PS_FM31}). Many of the new sources are correlated with large-scale structures which are also visible in the residual maps, and they are likely spurious sources which are actually features in the diffuse emission.

\begin{figure}[tb!]
\centering
\includegraphics[width=0.49\textwidth]{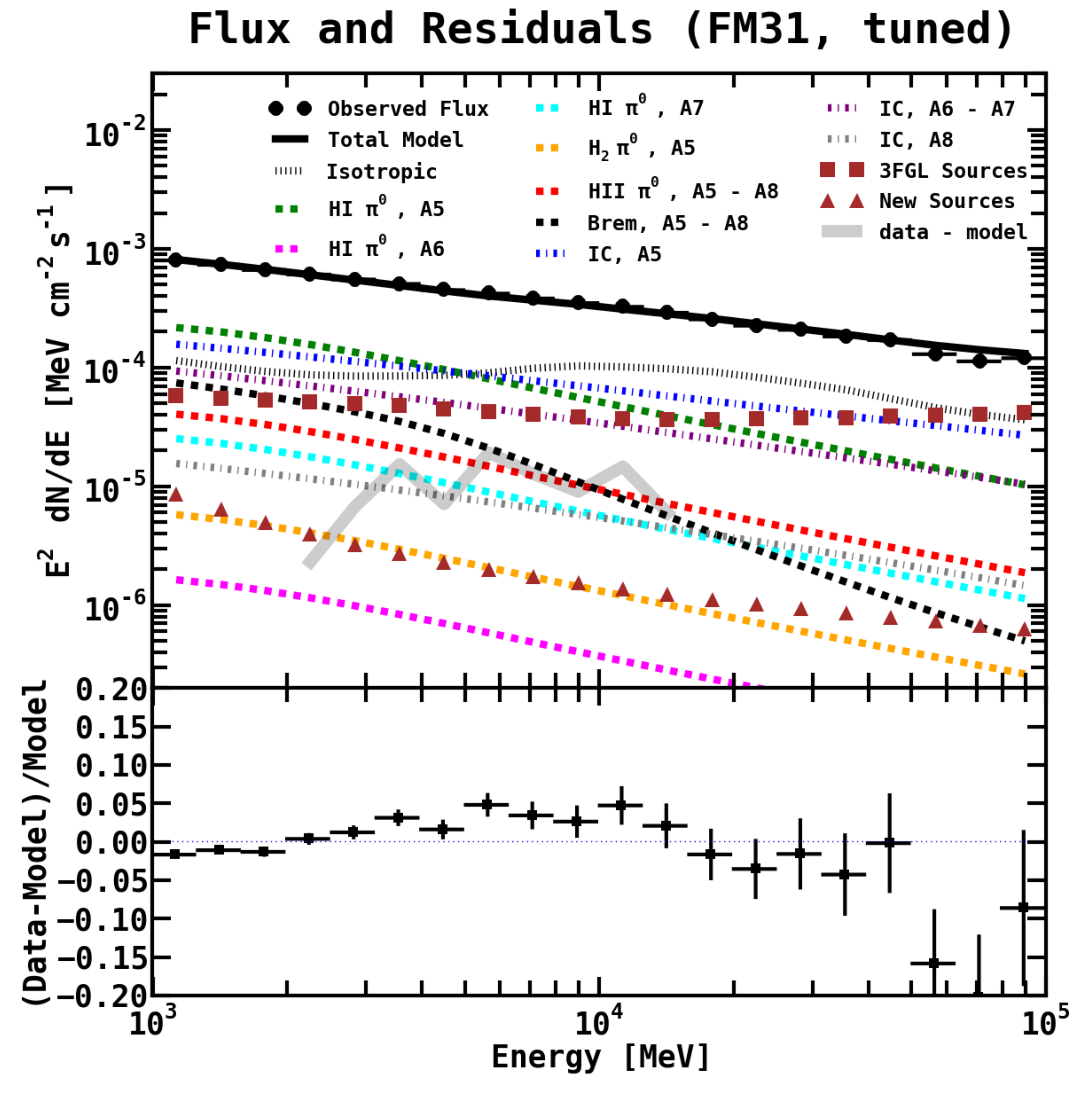}
\caption{Flux (upper panel) and fractional count residuals (lower panel) for the fit in FM31 (tuned). The \hii\ and Bremsstrahlung components are fixed to their GALPROP predictions. The normalizations of the IC (A5 and A6 -- A7) and isotropic components are held fixed to the values obtained in the tuning region. The normalizations of the \hi- and \htwo-related components are fit to the $\gamma$-ray data in FM31, as well as 3FGL sources within $20^\circ$ of M31, and additional point sources which we find using our point source finding procedure. Note that the top of FM31 has contribution from IC A8, and its normalization is also freed in the fit. The fractional residuals show an excess between $\sim$3--20 GeV reaching a level of $\sim$4\% (error bars show 1$\sigma$ statistical error). Above and below this range the data are being over-modeled as the fit tries to balance the excess with the negative residuals. This is in contrast to the fit in the TR, which shows fairly good agreement over the entire energy range. For reference, the residuals (data -- model) are also plotted in the upper panel (faint gray band).}
\label{fig:flux_and_residuals_FM31_Tuned}
\end{figure}

Figure~\ref{fig:flux_and_residuals_FM31_Tuned} shows the final results for the flux and count residuals for the baseline fit in FM31, including additional point sources, with the normalizations of the isotropic and IC components fixed to their best-fit values obtained in the TR. The corresponding best-fit normalizations and integrated flux are reported in Table~\ref{tab:norm_FM31_Tuned}. Note that the reported errors are 1$\sigma$ statistical error only. 

Below $\sim$5 GeV the emission is dominated by \hi\ A5, IC A5, and the isotropic component, in order of highest to lowest. A cross-over then occurs, and above $\sim$5 GeV the order is reversed. The 3FGL sources also become more dominant at higher energies. The cumulative spectrum of the additional point sources is consistent with that of the 3FGL sources, although the flux is roughly an order of magnitude less.

The fractional residuals show an excess between $\sim$3--20 GeV at the level of $\sim$4\%, and the data is being somewhat over-modeled above and below this range. The over-modeling is expected as the fit tries to balance the excess with the negative residuals. This is in contrast to the TR which shows fairly good agreement over the entire energy range. The normalizations of \hi\ A5 and A6 are low with respect to the GALPROP predictions, and likewise with respect to the values obtained in the TR and the all-sky fit. The normalization of \hi\ A7 is high with respect to the GALPROP prediction. The normalization of \htwo{} is also high, but its contribution is minimal in FM31.

\begin{figure*}[tbh!]
\centering
\includegraphics[width=0.33\textwidth]{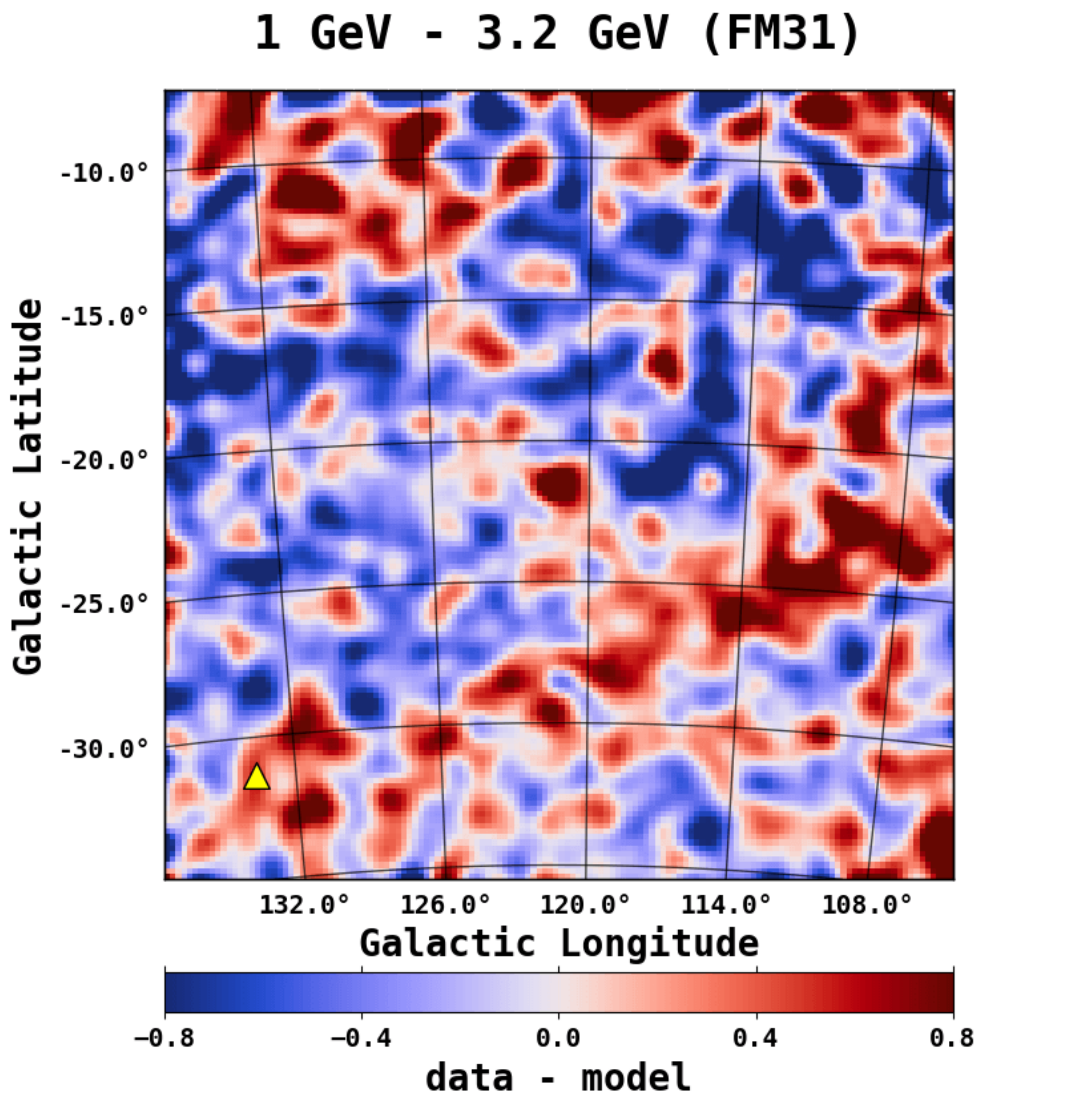}
\includegraphics[width=0.33\textwidth]{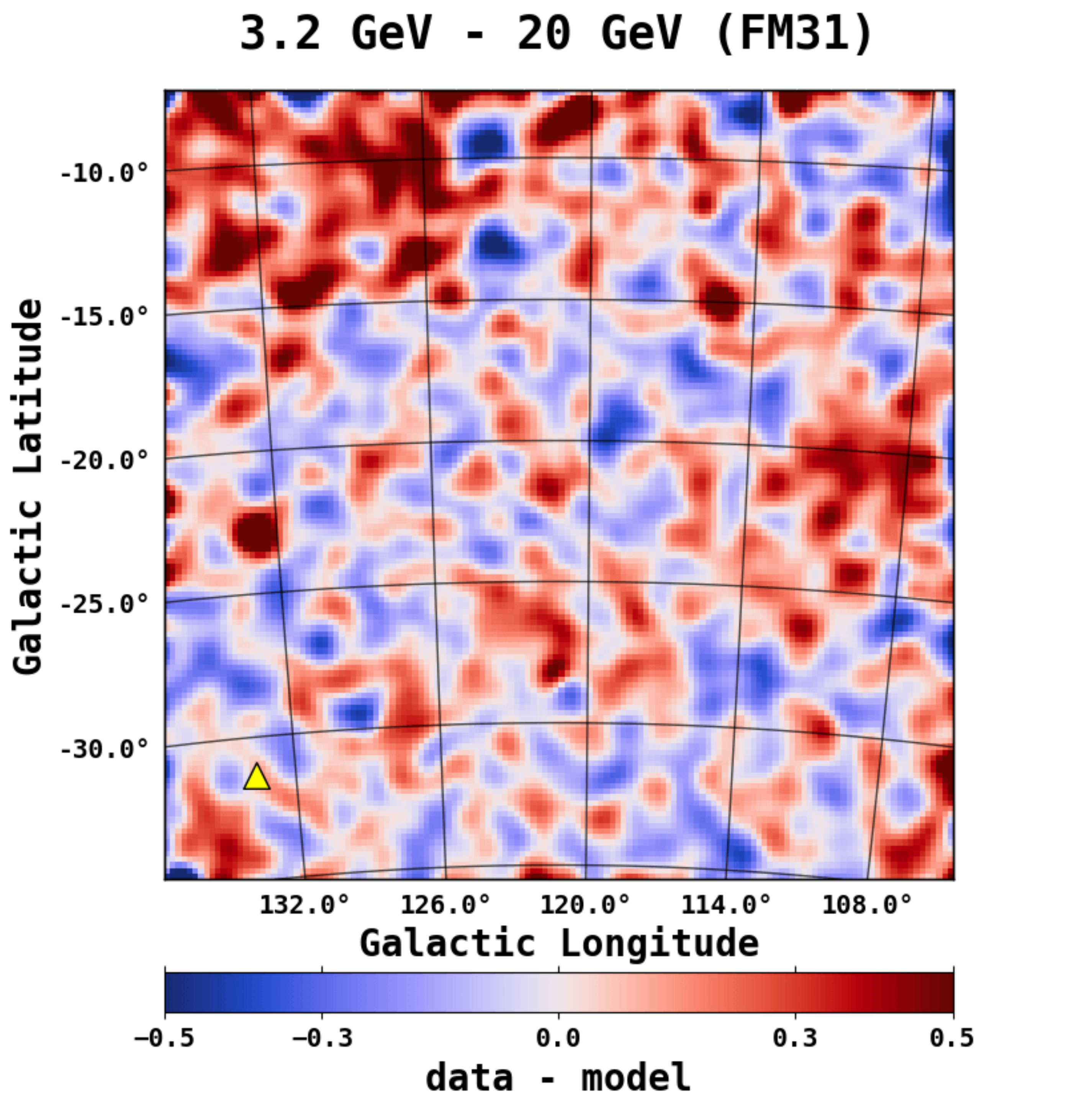}
\includegraphics[width=0.33\textwidth]{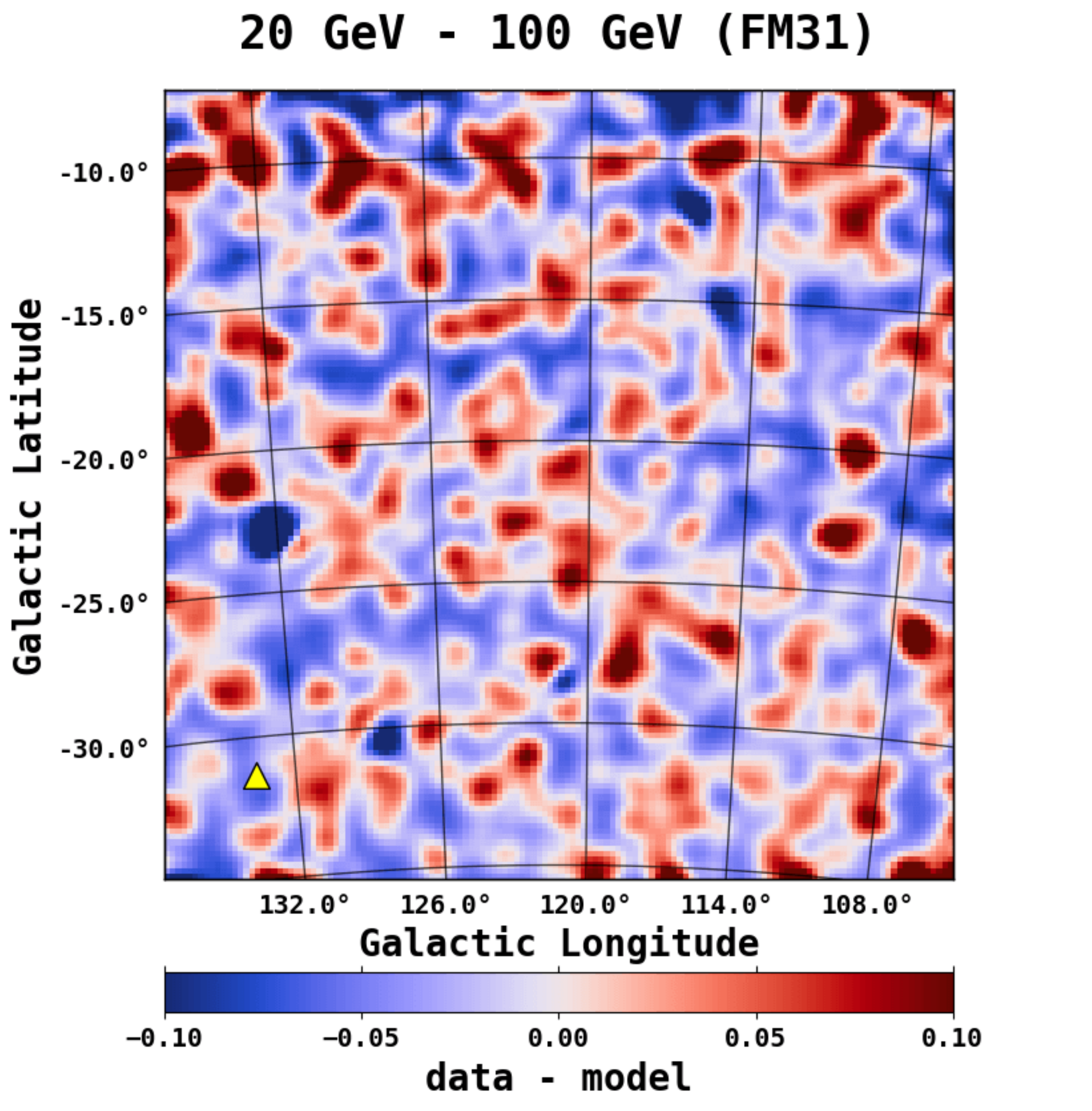}
\caption{Spatial count residuals (data $-$ model) resulting from the fit in FM31 (tuned) for three different energy bands, as indicated above each plot. The energy bins are chosen to coincide with the excess observed in the fractional residuals. The color scale corresponds to counts/pixel, and the pixel size is $0.2^\circ \times 0.2^\circ$. The images are smoothed using a $1^\circ$ Gaussian kernel. This value corresponds to the PSF (68\% containment angle) of \textit{Fermi}-LAT, which at 1 GeV is $\sim$$1^\circ$. For reference, the position of M33, $(l,b) = (133.61^\circ, -31.33^\circ)$, is shown with a yellow triangle.}
\label{fig:spatial_residuals_FM31_tuned}
\end{figure*}

\begin{figure*}[tbh!]
\centering
\includegraphics[width=0.33\textwidth]{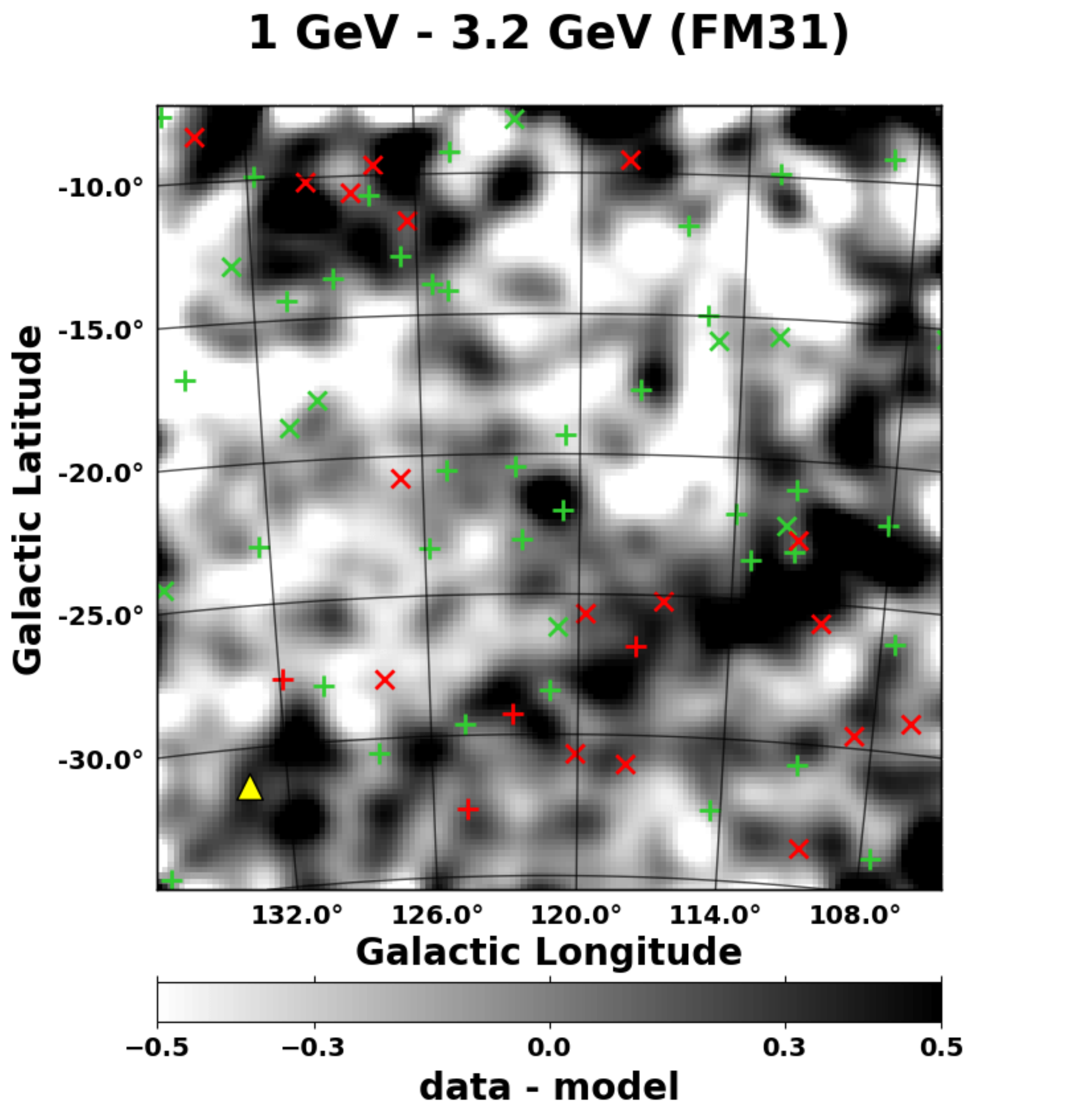}
\includegraphics[width=0.33\textwidth]{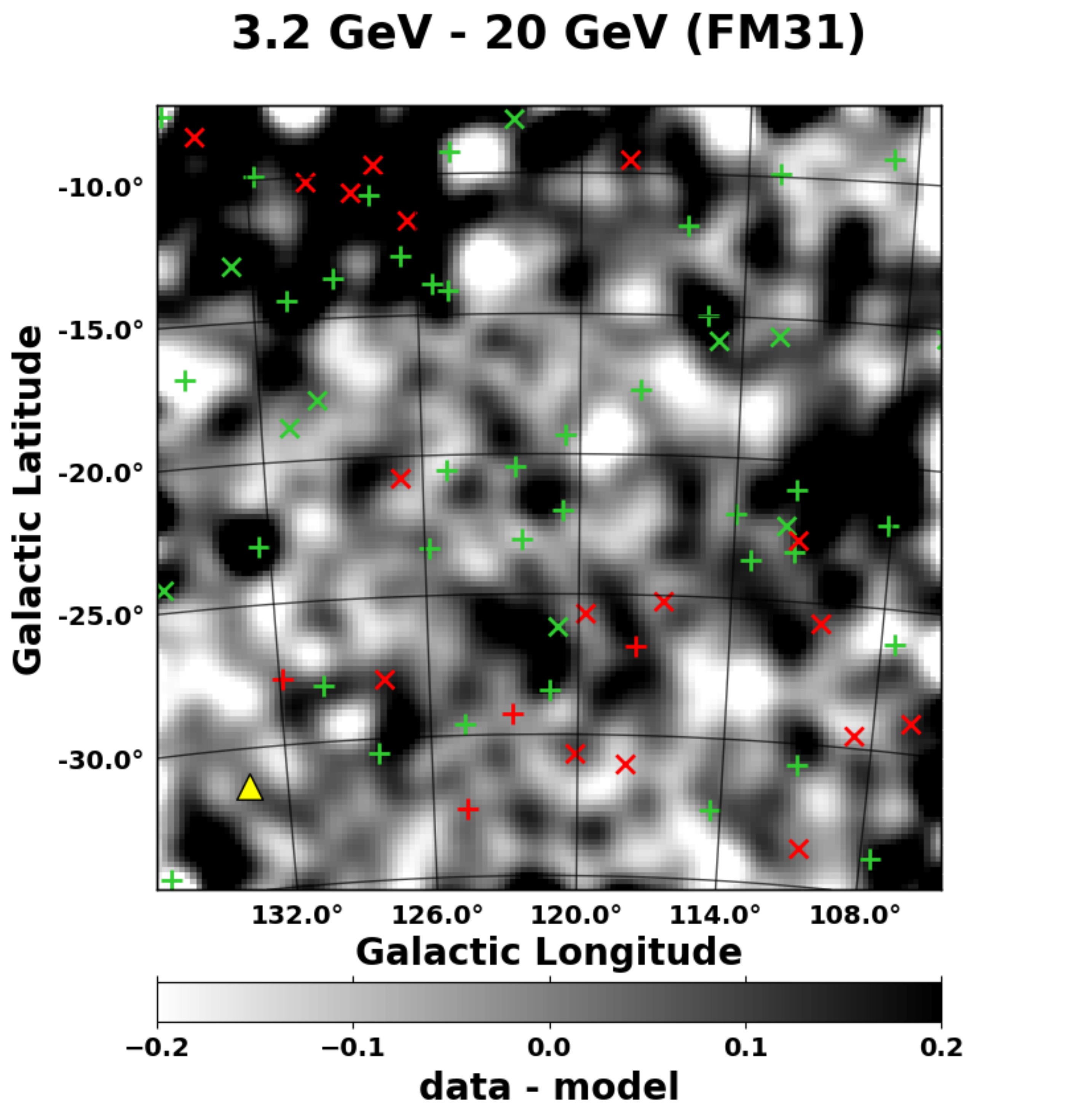}
\includegraphics[width=0.33\textwidth]{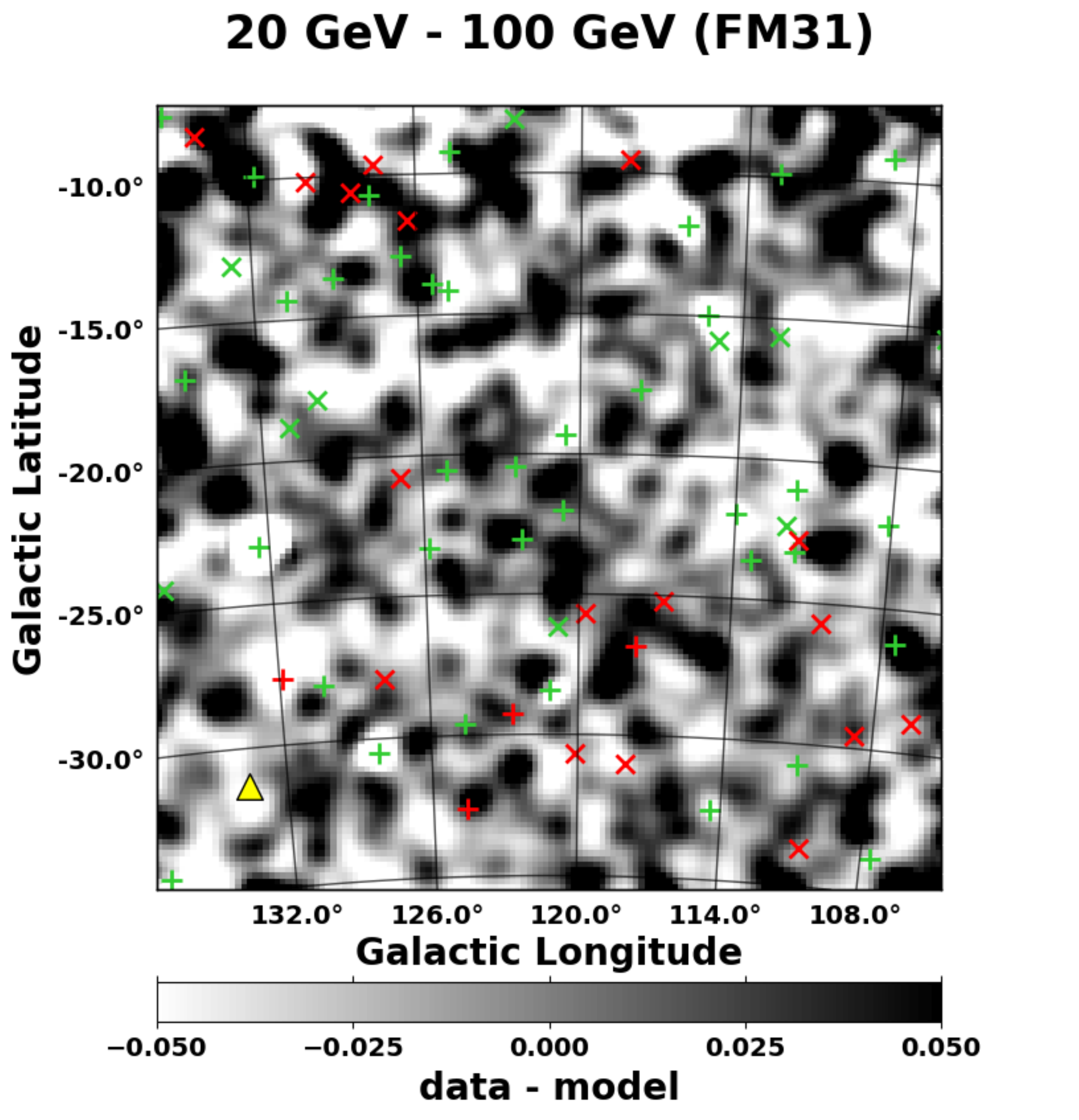}
\caption{Same residual maps as shown in Figure~\ref{fig:spatial_residuals_FM31_tuned}. Here we show the maps in gray scale, and intentionally saturate the images to bring out weaker features. Overlaid are the point sources in the region. Crosses show sources with TS$\geq$25 and slanted crosses show sources with 9$\leq$TS$<$25. Fermi 3FGL sources are shown in green, and new sources found in this analysis are shown in red.}
\label{fig:spatial_residuals_gray_FM31_tuned}
\end{figure*}

The spatial count residuals (data $-$ model) resulting from the baseline fit are shown in Figures~\ref{fig:spatial_residuals_FM31_tuned} and~\ref{fig:spatial_residuals_gray_FM31_tuned}. The residuals are integrated in three different energy bins, as indicated above each plot. The energy bins are chosen to coincide with the positive residual emission observed in the fractional energy residuals. The residuals show structured excesses and deficits. In the first energy bin a large arc structure is observed. The upper-left corner shows bright excess emission, which extends around the field towards the projected position of M33. This structure is similar to what is seen in the TS map (Figure~\ref{fig:TS_map}). Positive residual emission is also observed at the position of the M31 disk. In addition, the first energy bin shows deep over-modeling towards the top of the map and around the M31 disk. The second energy bin shows positive residual emission which is roughly uniform throughout the field, although the arc structure is also visible. In the third energy bin some holes can be seen corresponding to poorly modeled 3FGL sources, but otherwise no obvious structures can be identified. 

Figure~\ref{fig:spatial_residuals_gray_FM31_tuned} shows the same spatial residuals in gray scale, intentionally saturated in order to bring out weaker features. Overlaid are the point sources in the region, both 3FGL (green markers) and additional sources found in this analysis (red markers). Most of the additional sources are correlated with the arc structure. A majority of the 3FGL sources are AGN and are modeled with power-law (PL) spectra. We attempted to optimize the 3FGL spectra by fitting with a LogParabola spectral model, but this did not significantly change the positive residual emission, as discussed further in Appendix~\ref{sec:different_IEMs}.

\begin{deluxetable}{lccc}[tbh!]
\tablecolumns{4}
\tablewidth{0mm}
\tablecaption{Baseline Values for the IEM Components in FM31 (Tuned) \label{tab:norm_FM31_Tuned}}
\tablehead{
\colhead{Component} &
\colhead{Normalization} &
\colhead{Flux  ($\times 10^{-9})$}&
\colhead{Intensity ($\times 10^{-8})$}\\
& 
&
\colhead{(ph cm$^{-2}$ s$^{-1}$)} &
\colhead{(ph cm$^{-2}$ s$^{-1}$ sr$^{-1}$)}
}
\startdata
\hi\ $\pi^0$, A5 &0.82 \p 0.01&149.7 \p 2.5 &63.6 \p 1.1 \\
\hi\ $\pi^0$, A6 &0.1 \p 0.2&1.1 \p 2.4 & 0.5 \p 1.0 \\
\hi\ $\pi^0$, A7 &3.2  \p 0.4&17.1 \p 2.0 & 7.3 \p 0.9 \\
\htwo\ $\pi^0$, A5 &2.9 \p 0.3& 3.9 \p 0.4&1.7 \p 0.2 \\
IC, A8 & 61.3 \p 13.0 &11.3 \p 2.4& 4.8 \p 1.0
\enddata
\tablecomments{The normalizations of the isotropic and IC components (A5 and A6 -- A7) are held fixed to their best-fit values obtained in the TR. The normalizations of the $\pi^0$-related (\hi\ and \htwo) components are fit to the $\gamma$-ray data in FM31. Note that the top of FM31 has contribution from IC A8, and its normalization is also freely scaled. We also fit all 3FGL sources within $20^\circ$ of M31, as well as additional point sources which we find using our point source finding procedure. Intensities are calculated by using the total area of FM31, which is 0.2352 sr. Note that the reported errors are 1$\rm{\sigma}$ statistical only (and likewise for all tables).}
\end{deluxetable}


\begin{figure*}[tbh!]
\centering
\includegraphics[width=0.33\textwidth]{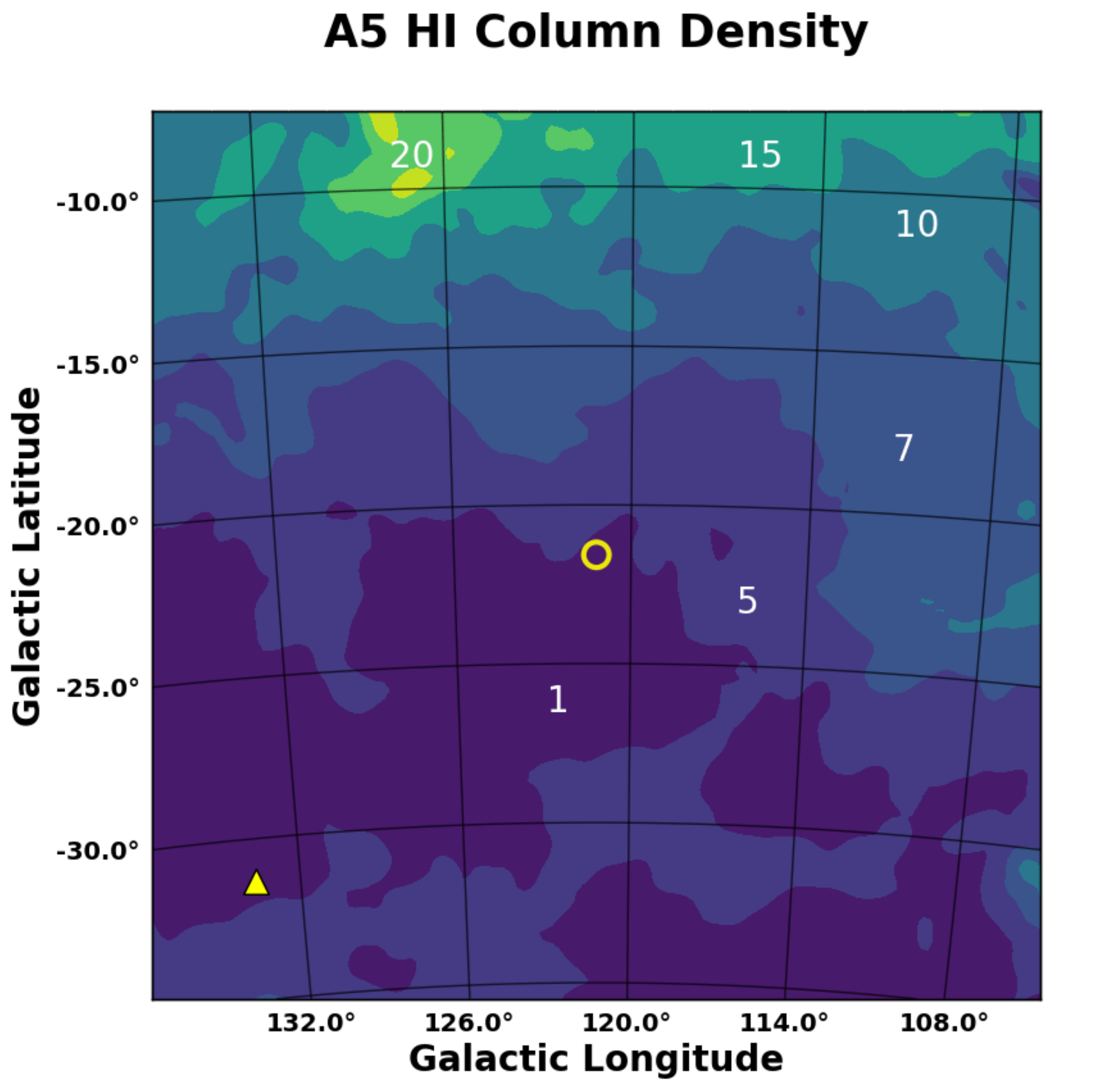}
\includegraphics[width=0.33\textwidth]{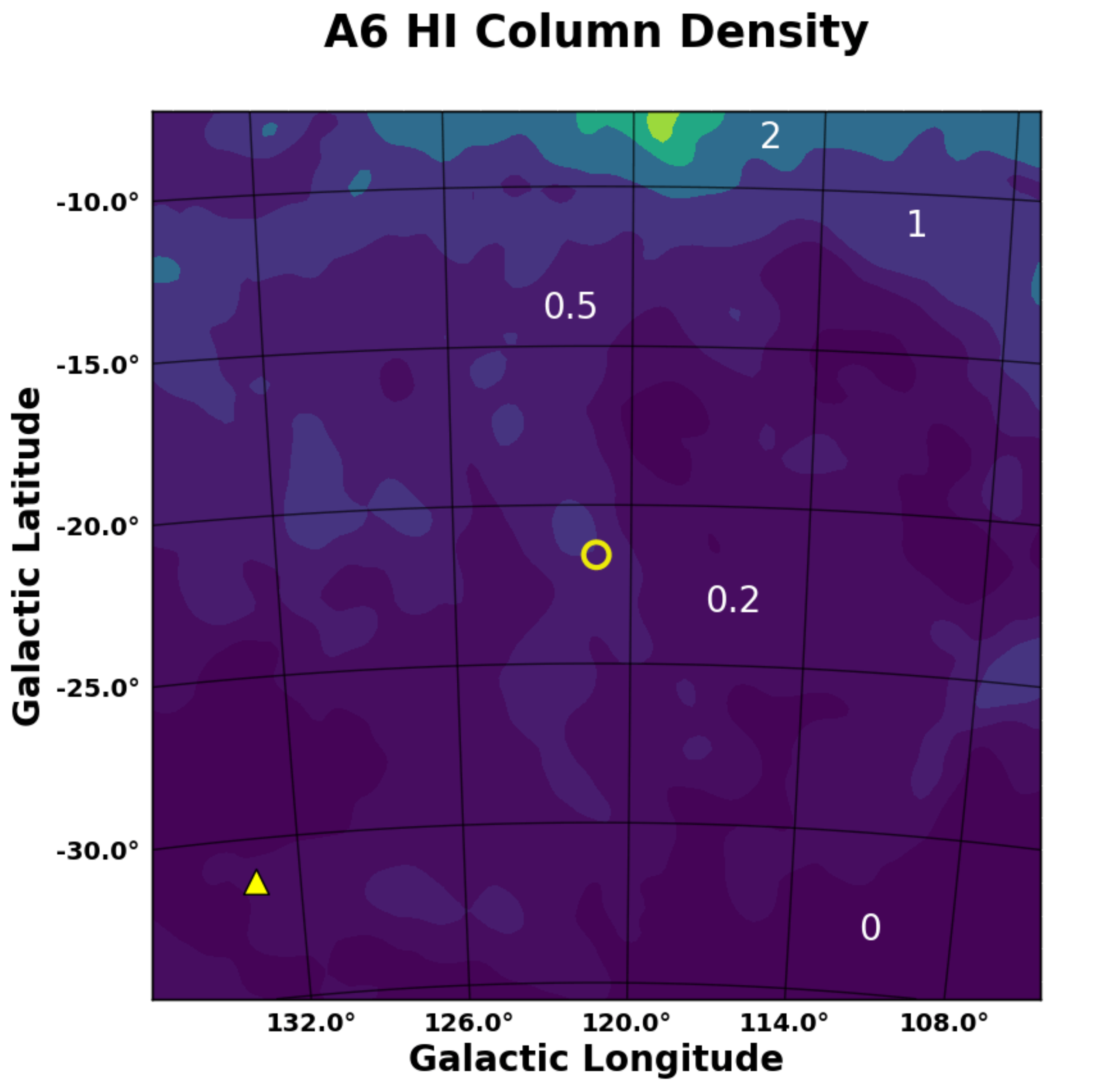}
\includegraphics[width=0.33\textwidth]{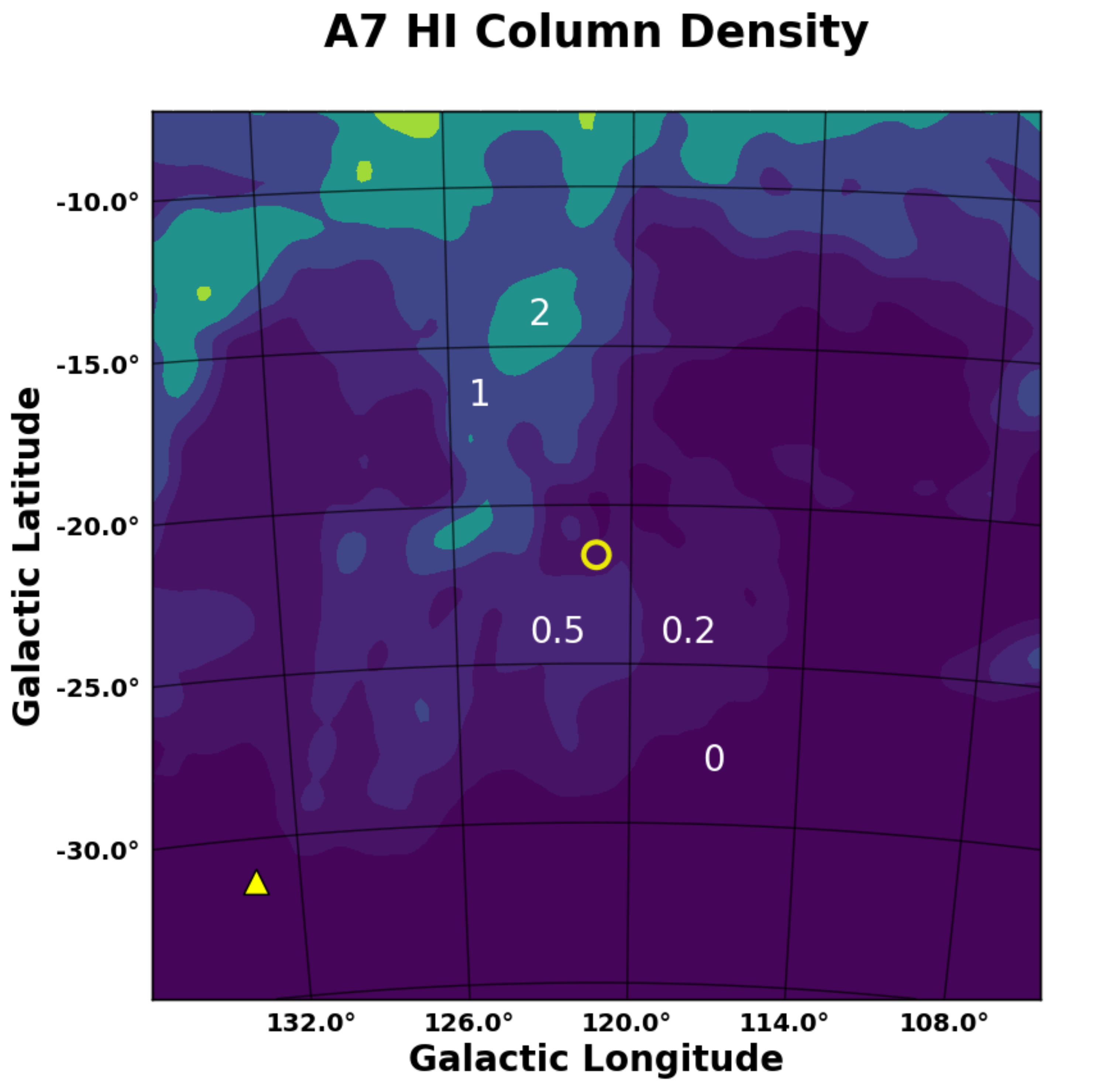}
\includegraphics[width=0.33\textwidth]{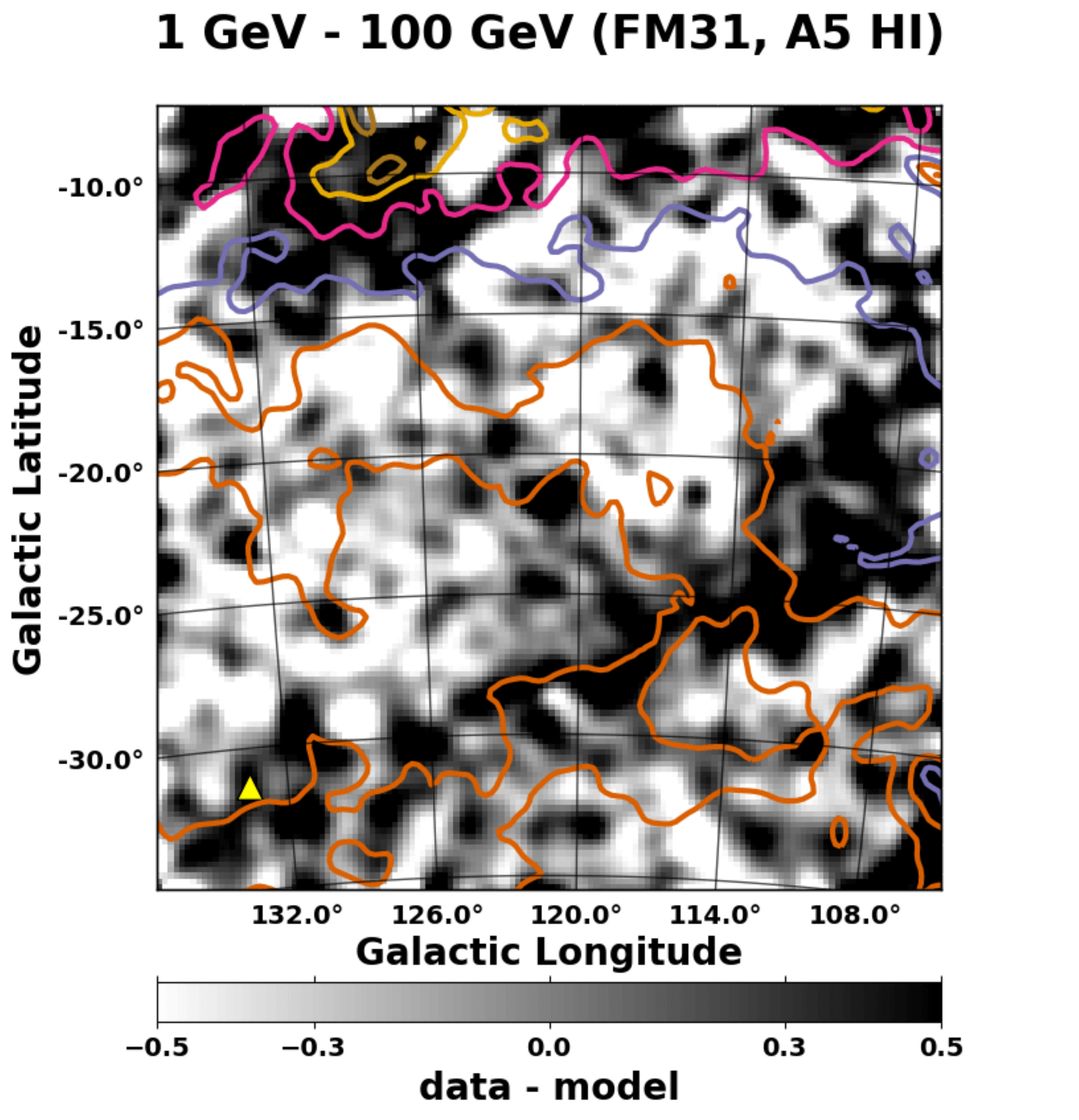}
\includegraphics[width=0.33\textwidth]{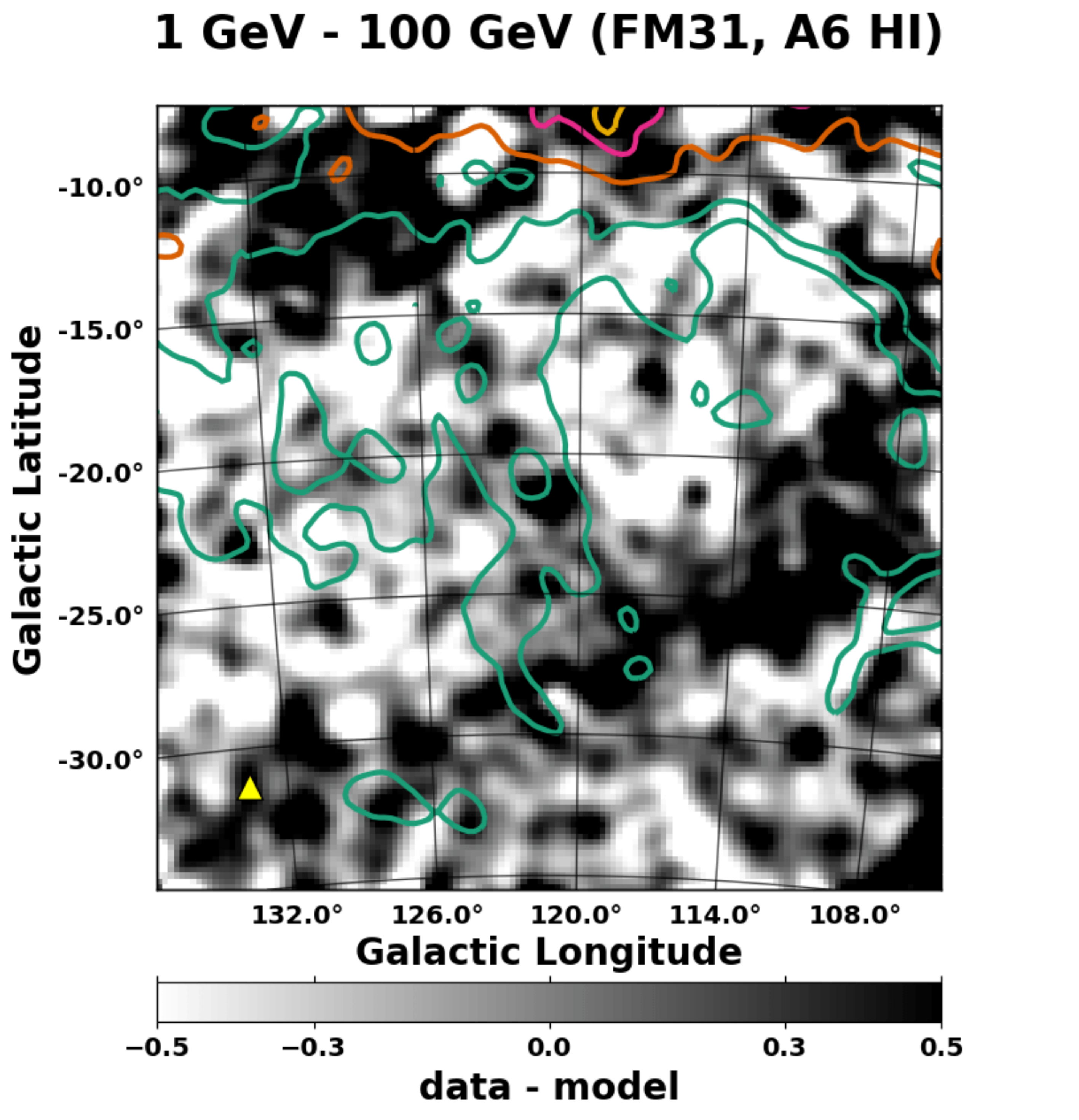}
\includegraphics[width=0.33\textwidth]{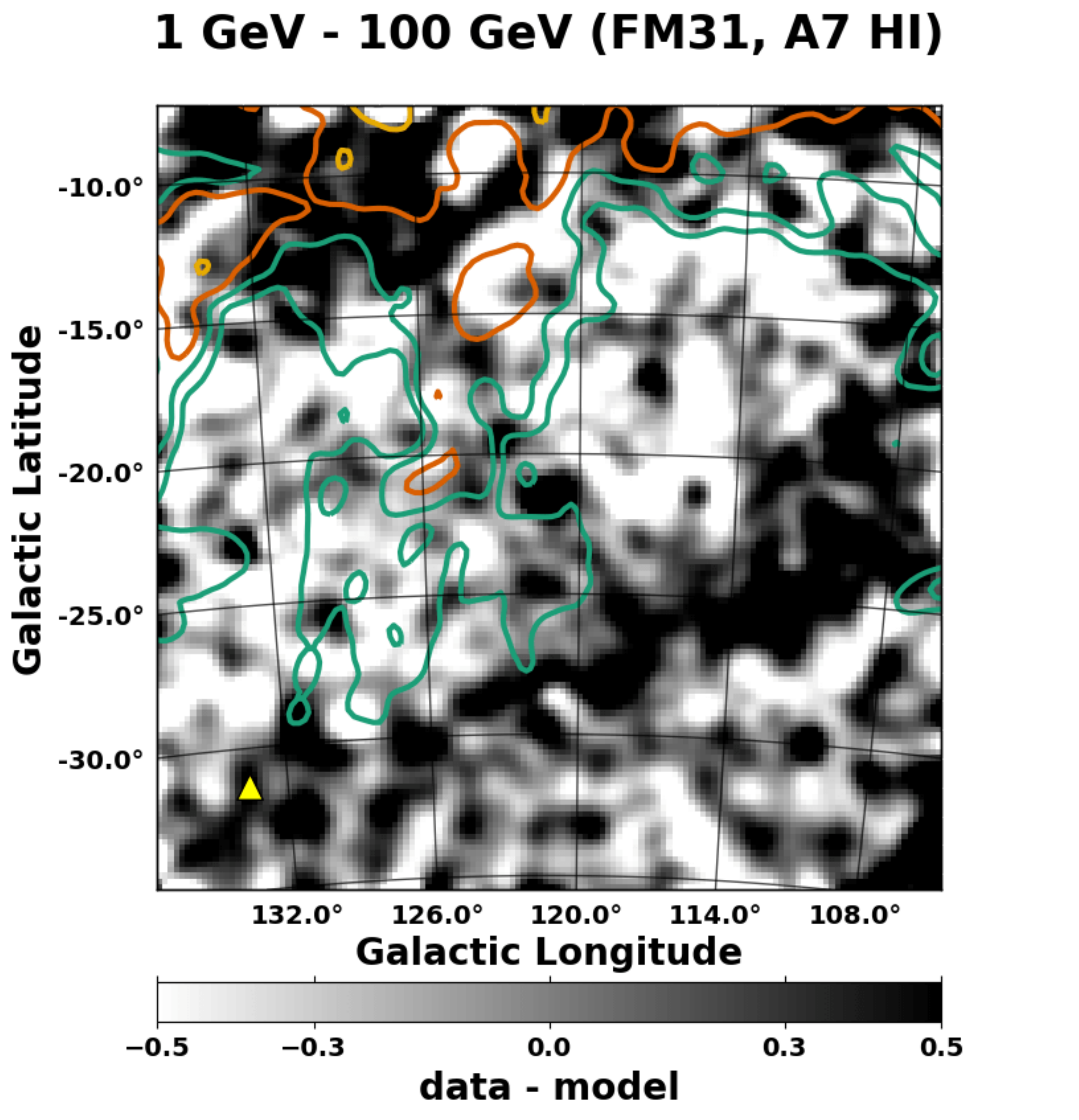}
\includegraphics[width=0.33\textwidth]{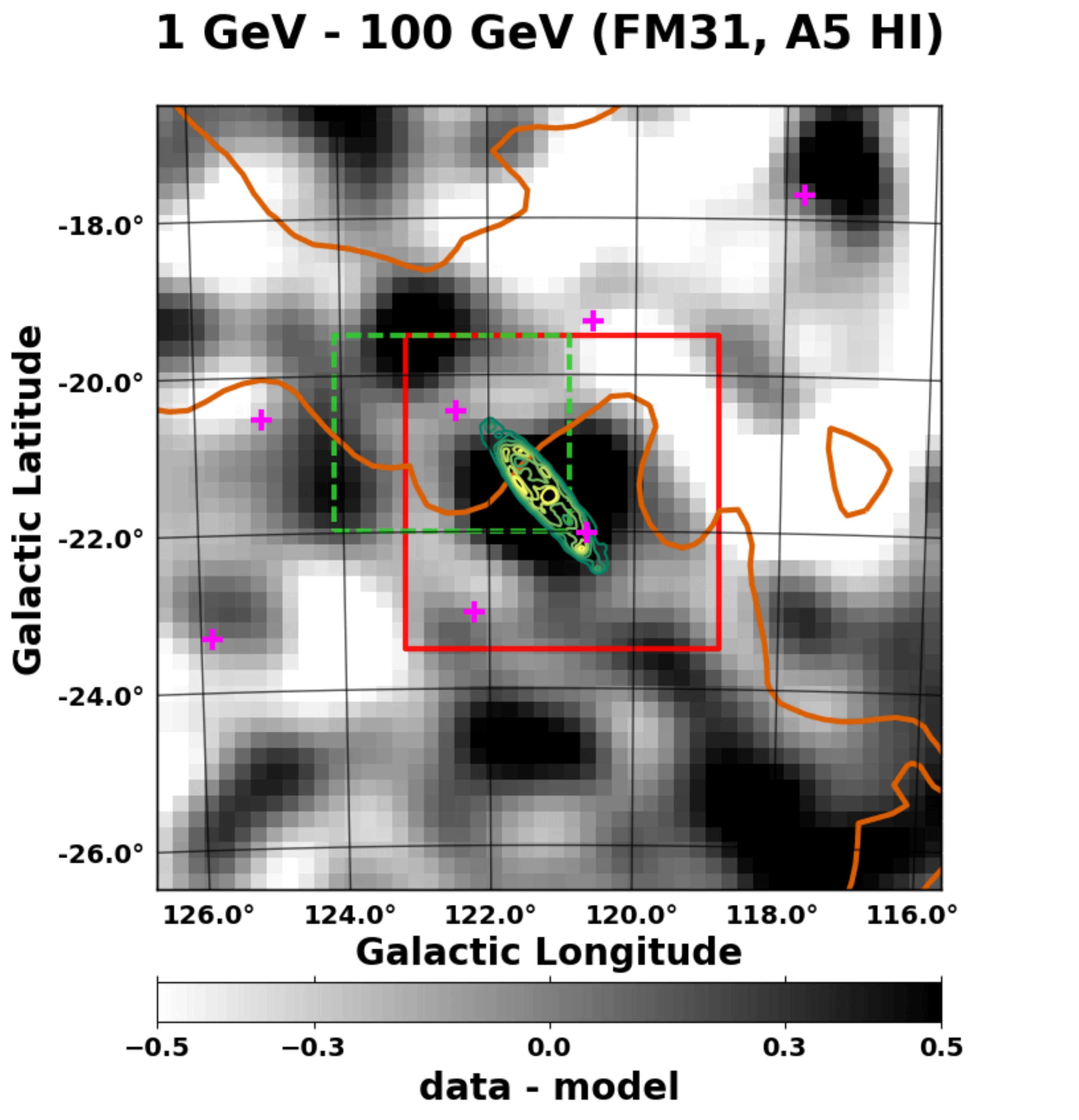}
\includegraphics[width=0.33\textwidth]{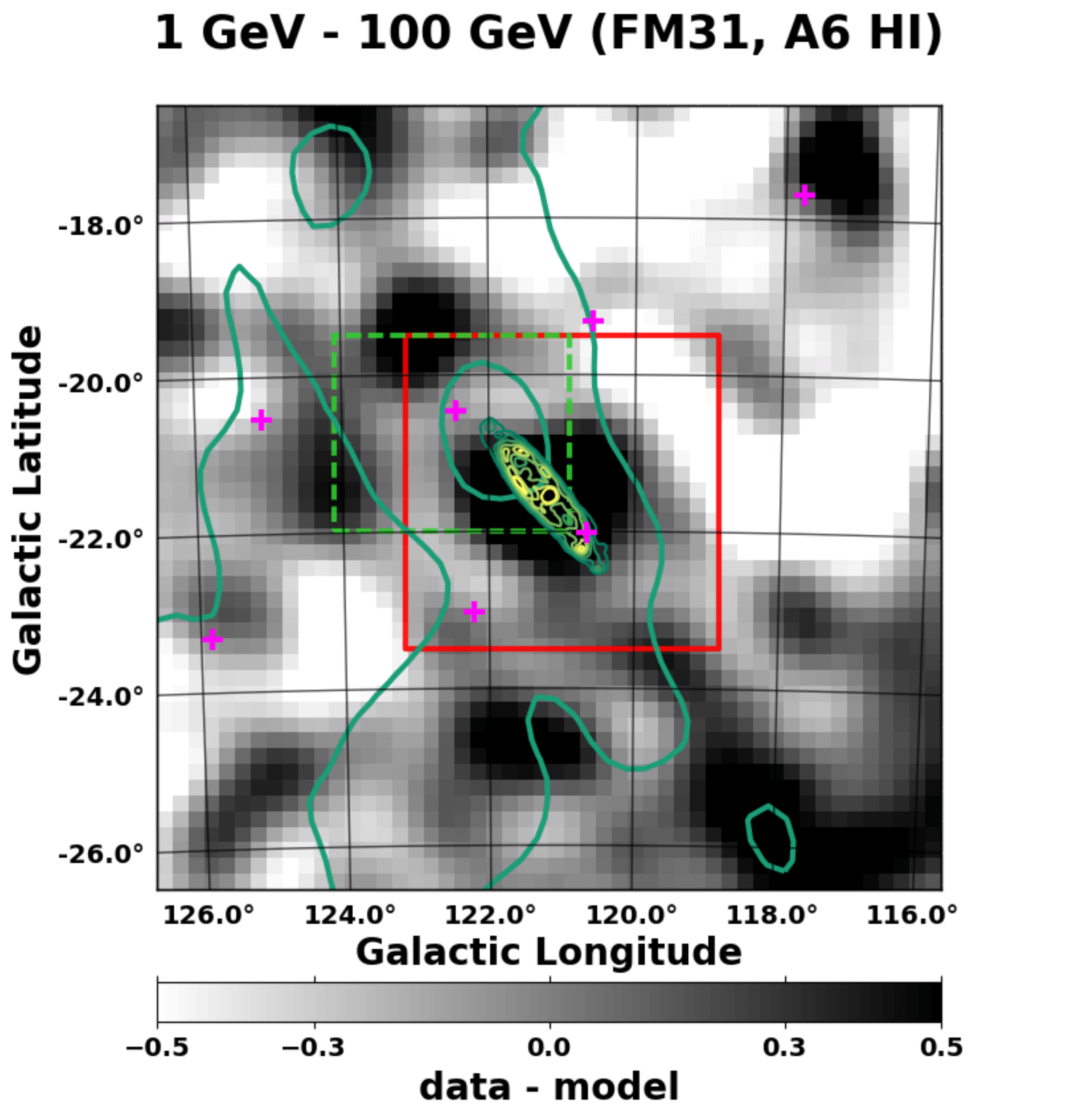}
\includegraphics[width=0.33\textwidth]{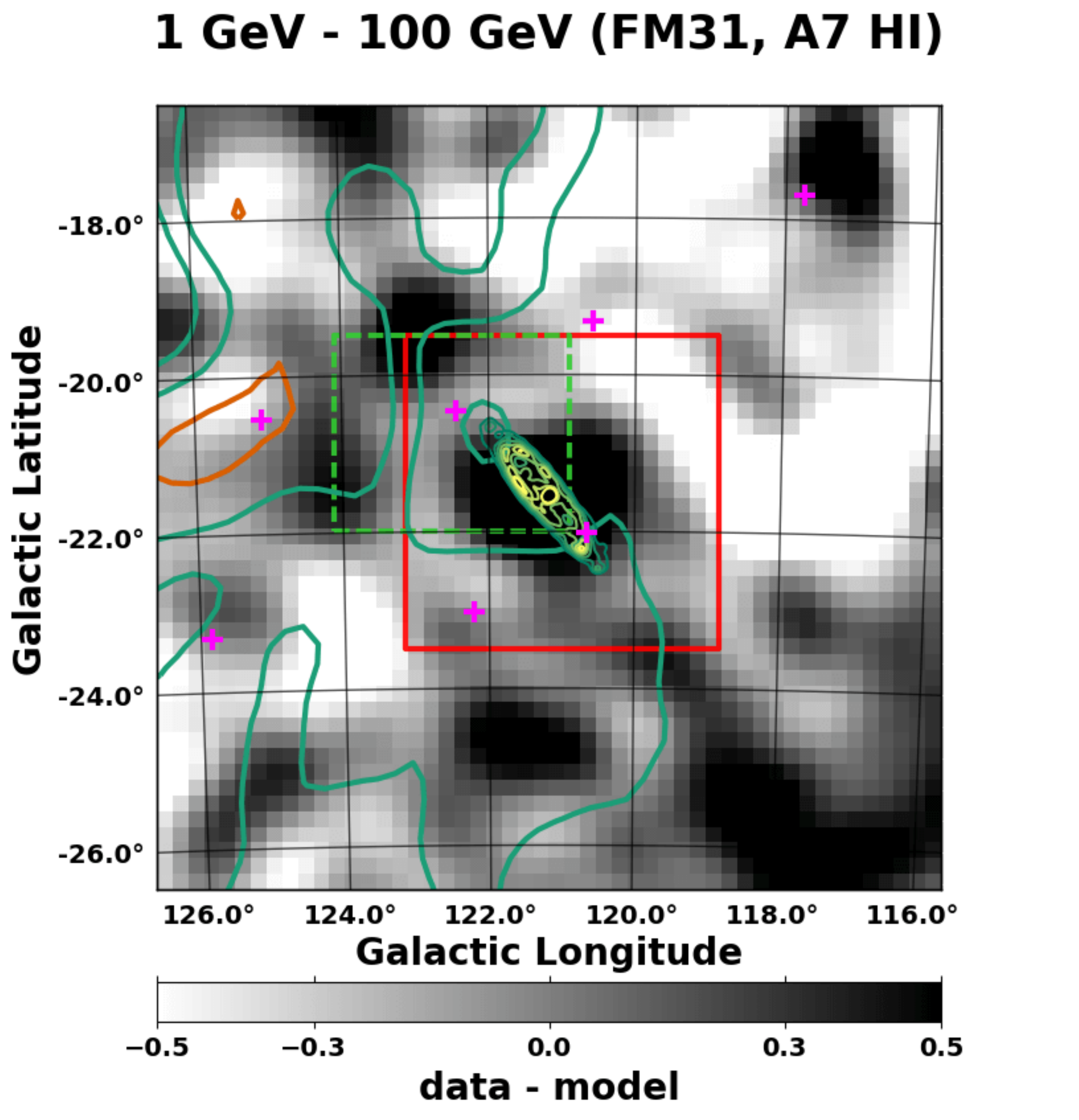}

\caption{\textbf{Top Row:} \hi\ column density contours for A5, A6, and A7, as indicated above each plot. For reference, a yellow circle ($0.4^\circ$) centered at M31 is overlaid, and a yellow triangle is overlaid at the position of M33. The units are $10^{20} \ \mathrm{cm^{-2}}$, and the levels are indicated on the maps. \textbf{Middle Row:} The same \hi\ column density contours are overlaid on the residual maps for FM31. The maps are integrated over the entire energy range 1--100 GeV. The residual emission is observed to be correlated with the column densities. In addition, the column densities of A6 and A7 are observed to be correlated with the major axis of M31 (the position angle of M31 is 38$^\circ$).  \textbf{Bottom Row:} The same maps as for the middle row but for a $5^\circ$ radius centered at M31. Contours for the IRIS 100 $\mu$m map of M31 are overlaid. The levels shown range from 6--22 MJy sr$^{-1}$. Also overlaid are the regions corresponding to the two main cuts (space and velocity) which are made on the underlying gas maps when constructing the MW IEM, as detailed in the text. Lastly, we overlay the 3FGL sources (magenta crosses) in the region with TS$\geq$25. In particular, we consider the two point sources located closest to the M31 disk, since we are interested in the true morphology of the M31 emission. The source located to the right of the disk (3FGL J0040.3+4049) is a blazar candidate and has an association. The source located to the left of the disk (3FGL J0049.0+4224) is unassociated.}
\label{fig:gas_column_densities}
\end{figure*}

\subsection{Analysis of the Galactic \hi-related Emission in FM31} \label{sec:Galactic_gas_analysis}

\begin{figure}[tbh!]
\centering
\includegraphics[width=0.49\textwidth]{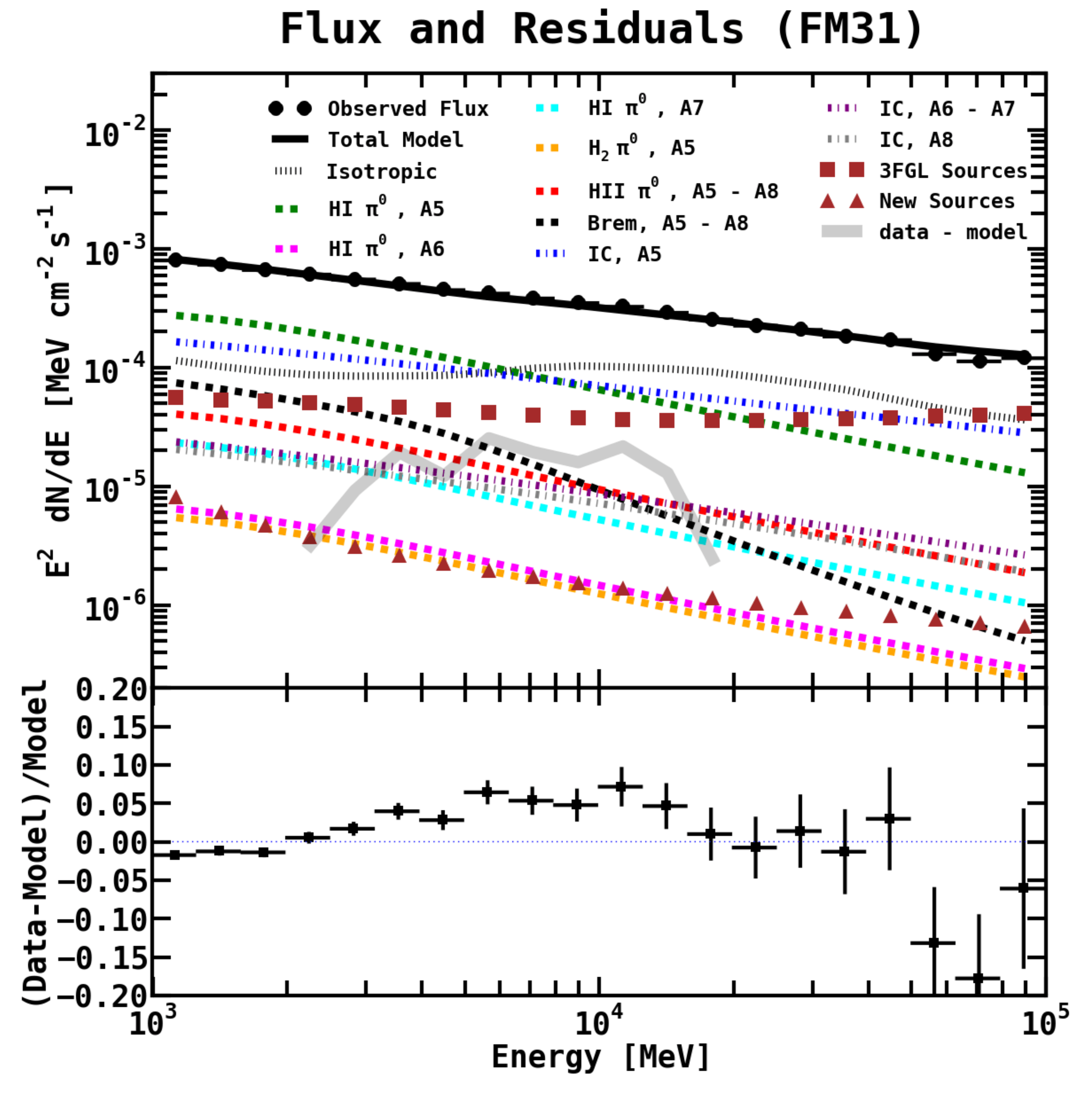}
\caption{Additional freedom is given to the baseline fit. The IC components are fit simultaneously with the other contributing diffuse components and point sources. The isotropic component remains fixed to its value obtained in the TR (1.06).}
\label{fig:flux_and_residuals_true}
\end{figure}

The structured excesses and deficits are an indication that the foreground emission may not be accurately modeled. In particular, the large arc structure observed in the first energy bin points to poorly modeled \hi\ gas in the line of sight. The \hi-related $\gamma$-ray emission depends on the column density of the gas, which in turn depends on the spin temperature. For this analysis the spin temperature is assumed to have a uniform value of 150 K, however, in reality it may vary over the region. 

To further investigate the systematic uncertainty relating to the characterization of \hi\ in the line of sight, we first compare the residual maps to the column densities for A5--A7, as shown in Figure~\ref{fig:gas_column_densities}. For visual clarity, the top row shows the column density filled contour maps. The units are $10^{20} \ \mathrm{cm}^{-2}$, and the levels are indicated on the maps. The second row shows the \hi\ contours overlaid to the residual map integrated between 1--100 GeV. The residual emission is observed to be correlated with the column densities. In addition, the column densities of A6 and A7 are observed to be correlated with the major axis of M31 (the position angle of M31 is 38$^\circ$). 

The last row shows the same maps as the middle row, but for a $5^\circ$ radius centered at M31. The IRIS 100 $\mu$m map of M31 is overlaid. Also overlaid are the regions corresponding to the two main spatial cuts which are made on the underlying \hi\ maps when constructing the MW IEM. The spatial cuts correspond to cuts in velocity space, where the velocity is defined relative to the local standard of rest (LSR). Here we summarize all of the pertinent cuts made to the underlying \hi\ gas maps:
\begin{itemize}
	\item[$\circ$] M31 cut (solid red box in Figure~\ref{fig:gas_column_densities}):\\ 
	$119^\circ \leq l \leq 123^\circ$, $-23.5^\circ \leq b \leq -19.5^\circ$,\\ 
	$V_{\rm LSR} < -120 \ \mathrm{km \ s^{-1}}$;
	
	\item[$\circ$] M31 cut (dashed green box in Figure~\ref{fig:gas_column_densities}):\\ 
	$121^\circ \leq l \leq 123^\circ$, $-22^\circ \leq b \leq -19.5^\circ$,\\
	$-120 \ \mathrm{km \ s^{-1}} < V_{\rm LSR} < -50 \ \mathrm{km \ s^{-1}}$;
	
	\item[$\circ$] M33 cut:\\ $132.5^\circ < l < 134.5^\circ, -33 < b < -30$ \\
	 $-460 \ \mathrm{km \ s^{-1}} \leq V_{\rm LSR} \leq -60 \ \mathrm{km \ s^{-1}}$;

	\item[$\circ$] Anything above a given height $z$ is assumed to be local gas (A5). The height is 1 kpc for $R$$<$8 kpc, but then increases linearly with $R$ with a slope of 0.5 kpc/kpc. The cut is applied after determining the radial distance with the rotation curve and obtaining an estimate of $z$;

	\item[$\circ$] Everything with $|V_{\rm LSR}| > 170 \ \mathrm{km \ s^{-1} \ and} \ |b|>5^\circ$ is considered to be extragalactic;

	\item[$\circ$] Everything with $V_{\rm LSR} < -100 \ \mathrm{km \ s^{-1} \ and} \ |b|>30^\circ$ is considered to be extragalactic.
\end{itemize}

Note that these are the same cuts which are made for the official FSSC IEM. It was pointed out in~\citet{Ackermann:2017nya} that for $-50$ km s$^{-1}$$ < V_{\rm LSR} < -30$ km s$^{-1}$, foreground emission from the MW blends with the remaining signal from M31 at the north-eastern\footnote{For all directions relating to M31, north is up, and east is to the left.} tip of M31, and it is estimated that on some lines of sight in this direction up to $\sim$40\% of the M31 signal might have been incorporated in the MW IEM. Besides, there may be additional \hi\ gas in M31's outer regions which is wrongfully assigned to the MW, as discussed further in Section~\ref{sec:gas_related_emission}. Overall, the cuts (velocity and space) made to the underlying \hi\ maps may be introducing systematics in the morphology of the extended M31 emission. 

Also shown in Figure~\ref{fig:gas_column_densities} are the 3FGL sources in the region with TS$\geq$25. In particular, we consider the two point sources located closest to the M31 disk, since we are ultimately interested in ascertaining the true morphology of the M31 emission. The source located to the right of the disk (3FGL J0040.3+4049) is a blazar candidate and has an association. The source located to the left of the disk (3FGL J0049.0+4224) is unassociated. We identify this source as potentially spurious, in that it may actually be part of a larger diffuse structure.

Because of the poor data--model agreement and the poor description of the \hi-related components, we allowed for additional freedom in the fit by also scaling the IC components (A5 and A6--A7) in FM31. The fit is performed just as for the baseline fit. Figure~\ref{fig:flux_and_residuals_true} shows the resulting flux and residuals, and the corresponding best-fit normalizations are reported in Table~\ref{tab:norm_FM31_True}. Overall, a better fit is obtained. The likelihood value is $-\log L = 143268$, compared to the tuned fit which is $-\log L = 143302$. 

\begin{figure}[tbh!]
\centering
\includegraphics[width=0.45\textwidth]{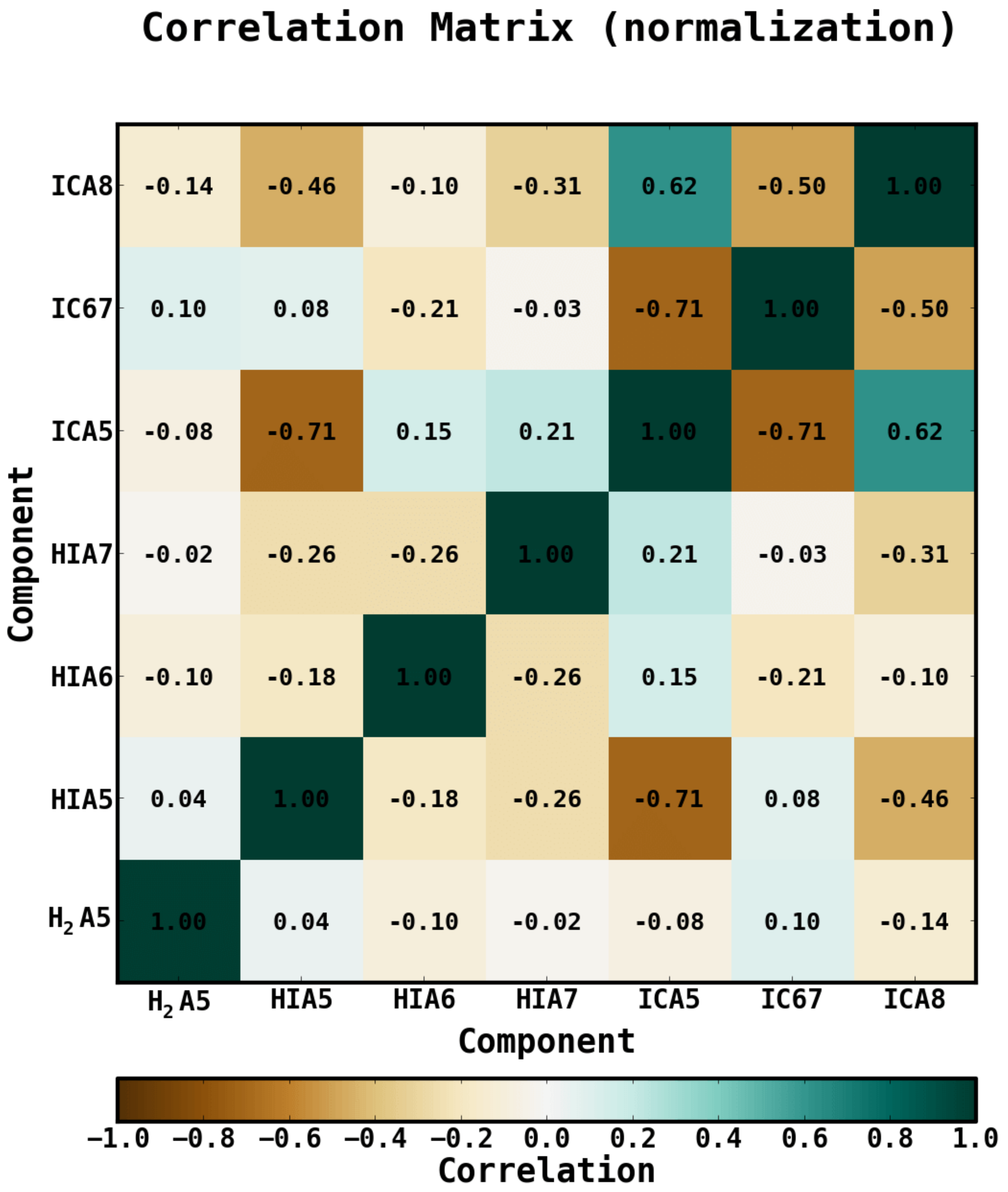}
\caption{Correlation matrix for the FM31 baseline fit with the IC components scaled.}
\label{fig:baseline_true_correlation}
\end{figure}

The \hi\ A5 component obtains a normalization of 1.04, which is comparable to the value obtained in the TR, and close to the GALPROP model prediction. The normalization of \hi\ A6 is still low at $\sim$40\% of the model prediction. We note that the \hi\ A6 flux is less than that of \hi\ A7, which is due to the fact that the radial extension of A6 is 1.5 kpc, compared to A7 which has a radial extension of 5 kpc. The normalization of IC A5 is consistent with the value obtained in the TR. On the other hand, the normalization of IC A6--A7 has a value of 0.9 \p 0.3, compared to the TR value of 3.5 \p 0.4. The normalization of IC A8 is very high, but this is a weak component with contribution only towards the top of the field. Note that because IC A8 only contributes at the very top of the field, it is not well constrained, and this allows its normalization to get high, but its overall effect on the residuals remains subdominant. Despite the additional freedom the model is unable to flatten the positive residual emission between $\sim$3--20 GeV, and it actually becomes slightly more pronounced. The spatial residuals for this fit are qualitatively consistent with the residuals in Figure~\ref{fig:spatial_residuals_FM31_tuned}. The correlation matrix for the fit is given in Figure~\ref{fig:baseline_true_correlation}.

As already discussed, the \hi\ column density depends on the value of the spin temperature, which is used to convert the observed 21-cm brightness temperature to column densities. In general the spin temperature may have some spatial variation. The CR density may also vary over the field, and likewise for the ISRF density.  To account for these possibilities we divide FM31 into three equal subregions: top, middle, and bottom. Each subregion is then further divided equally into right and left. In each subregion we rescale the diffuse components. The point sources remain fixed to the best-fit values obtained in the baseline fit (with IC scaled).  

\begin{figure*}[tbh!]
\centering
\includegraphics[width=0.33\textwidth]{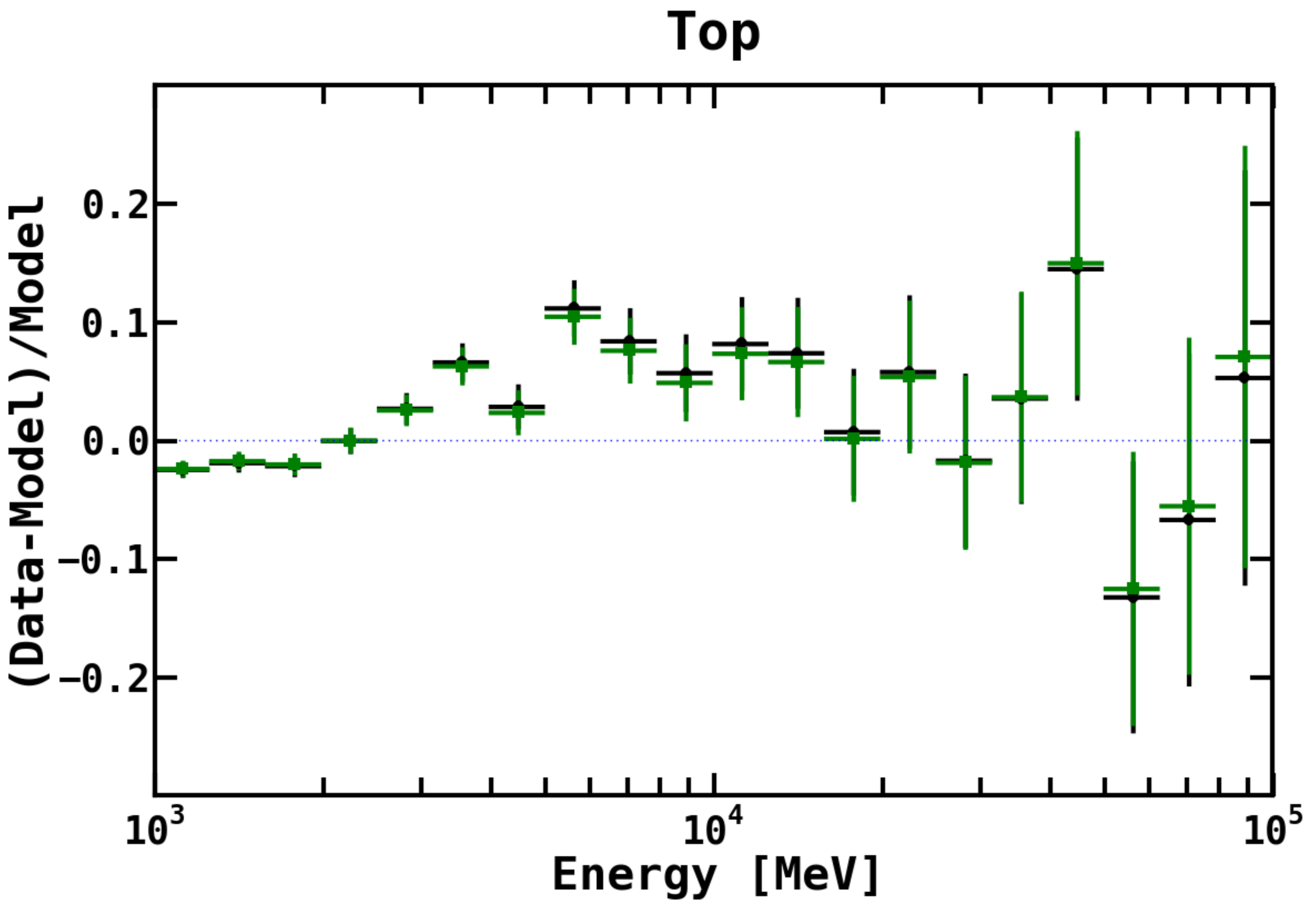}
\includegraphics[width=0.33\textwidth]{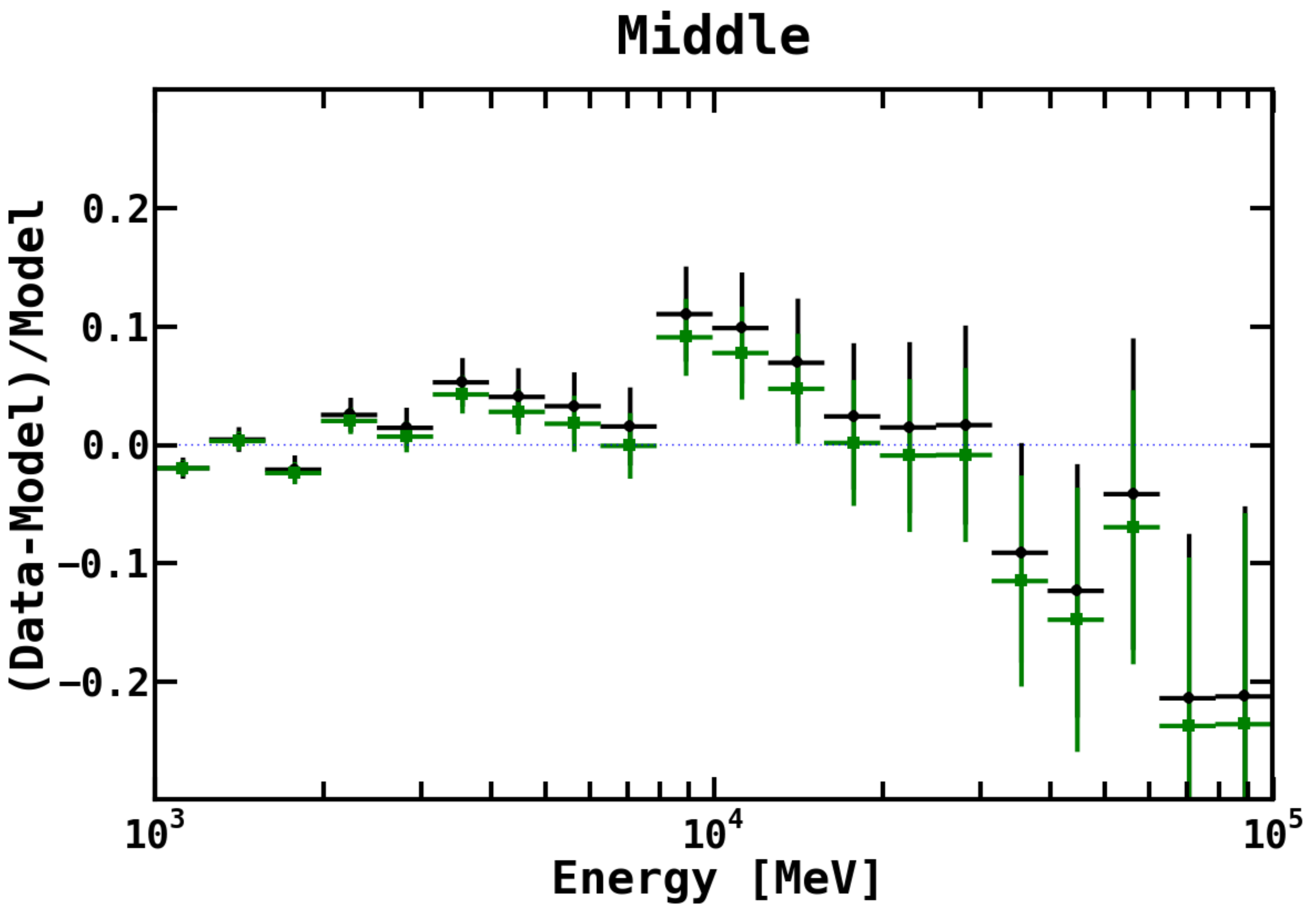}
\includegraphics[width=0.33\textwidth]{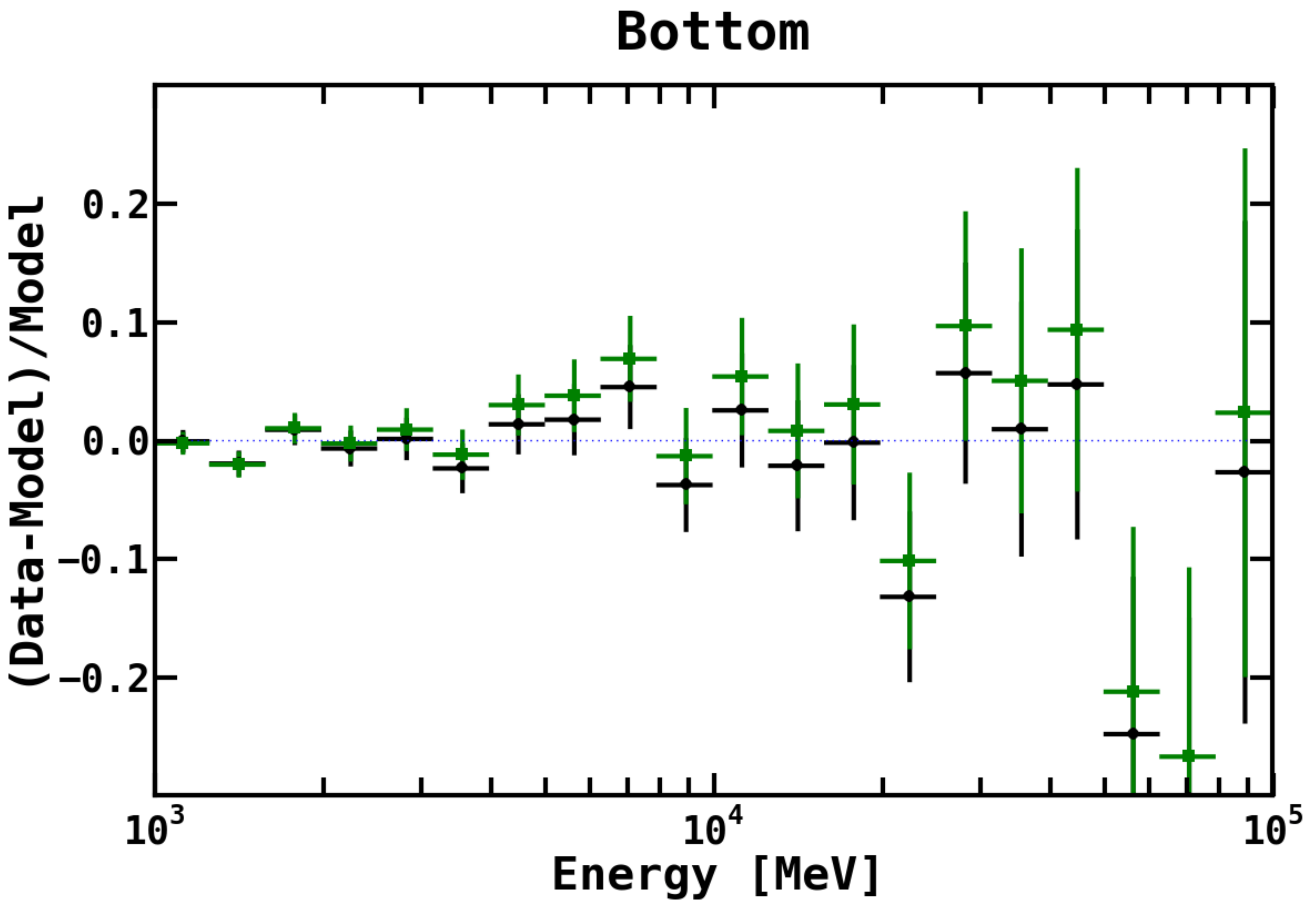}
\includegraphics[width=0.33\textwidth]{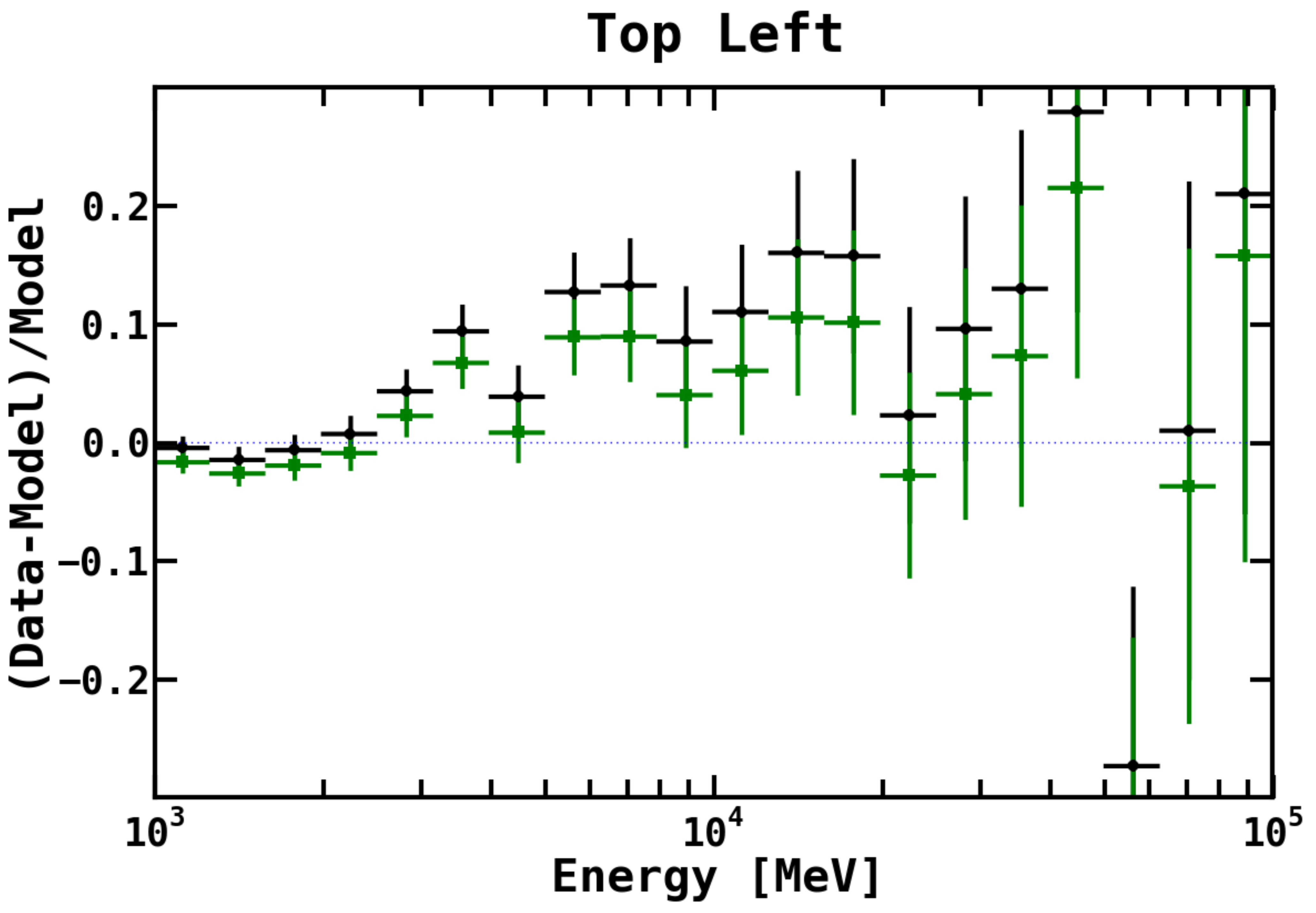}
\includegraphics[width=0.33\textwidth]{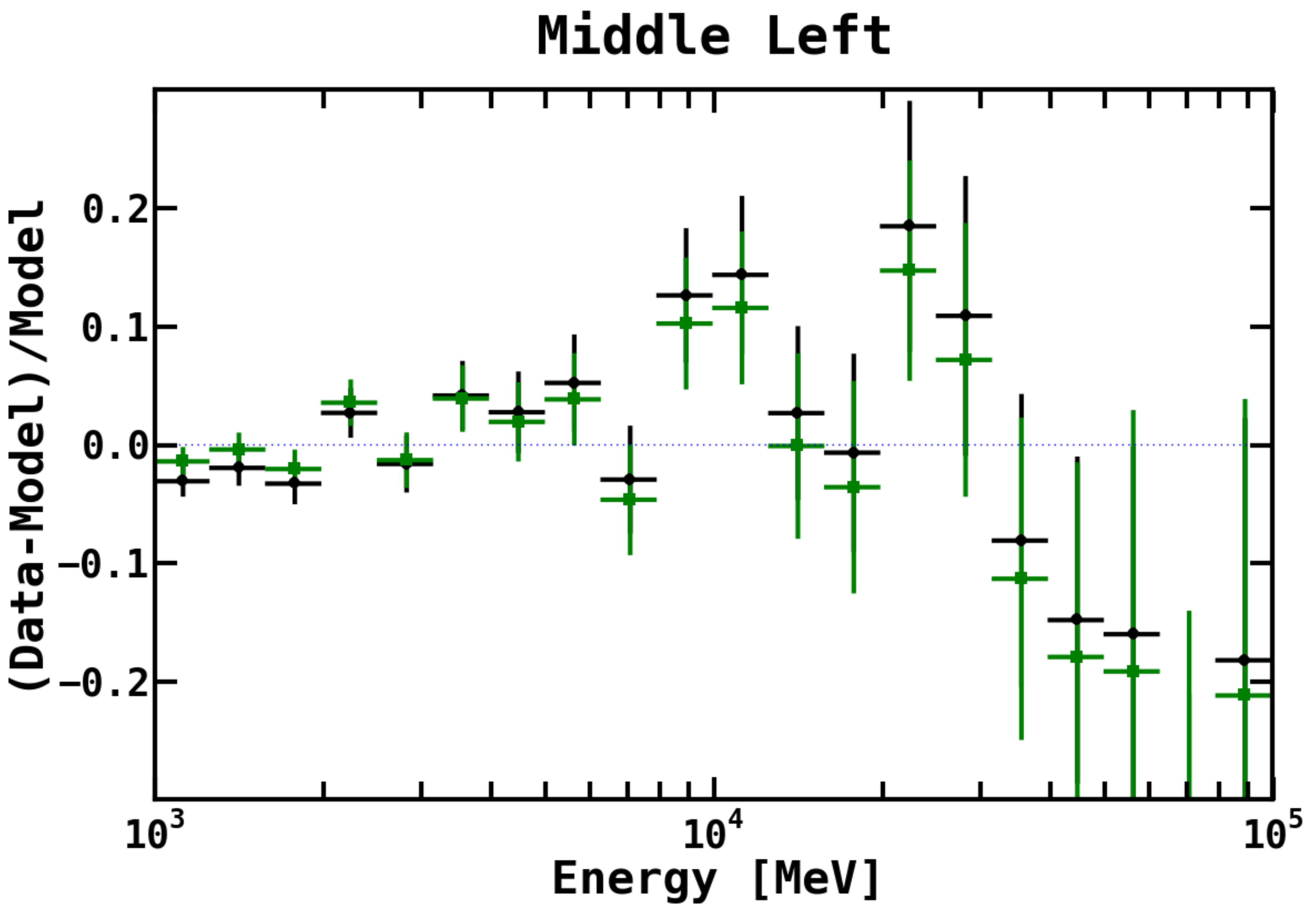}
\includegraphics[width=0.33\textwidth]{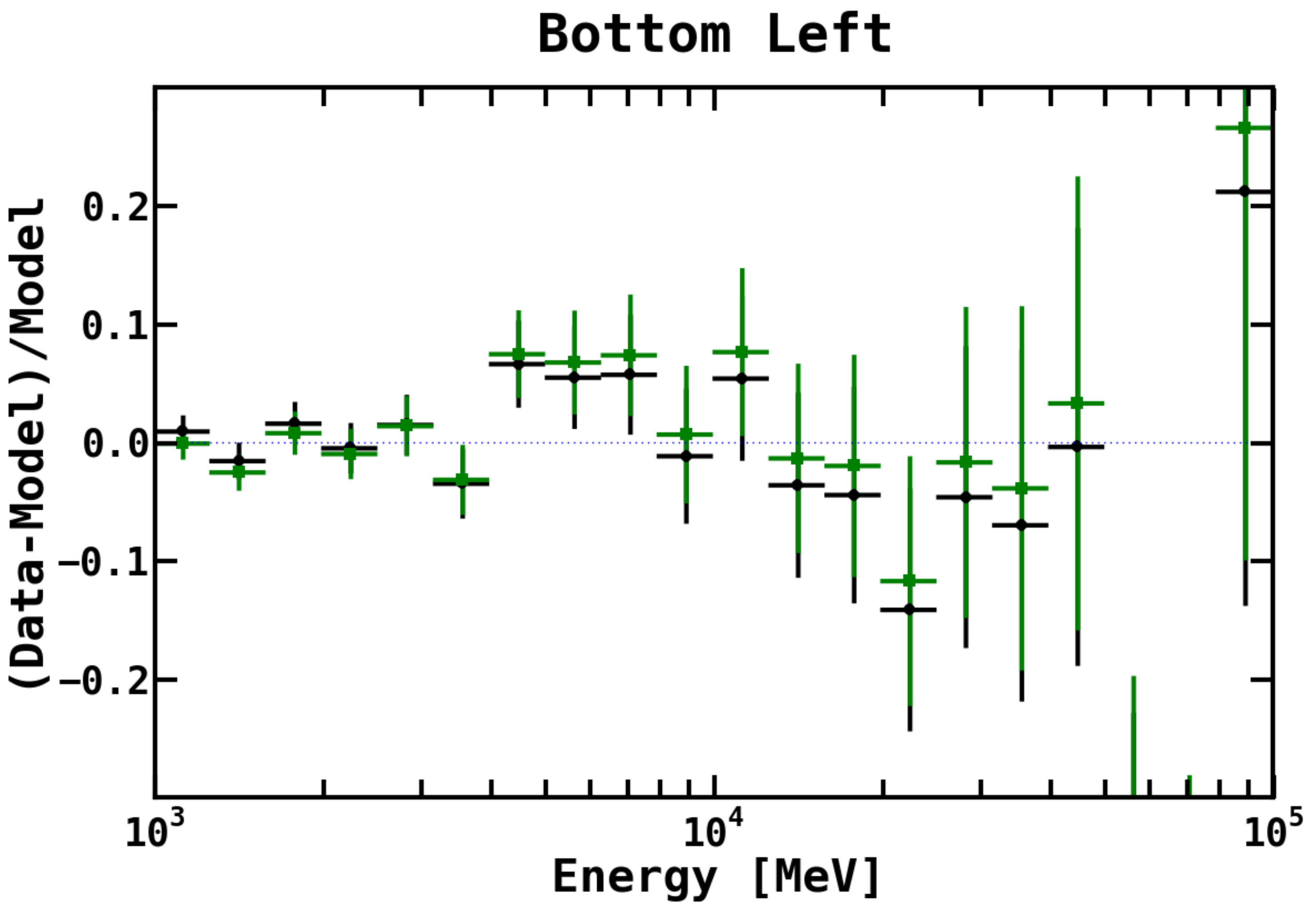}
\includegraphics[width=0.33\textwidth]{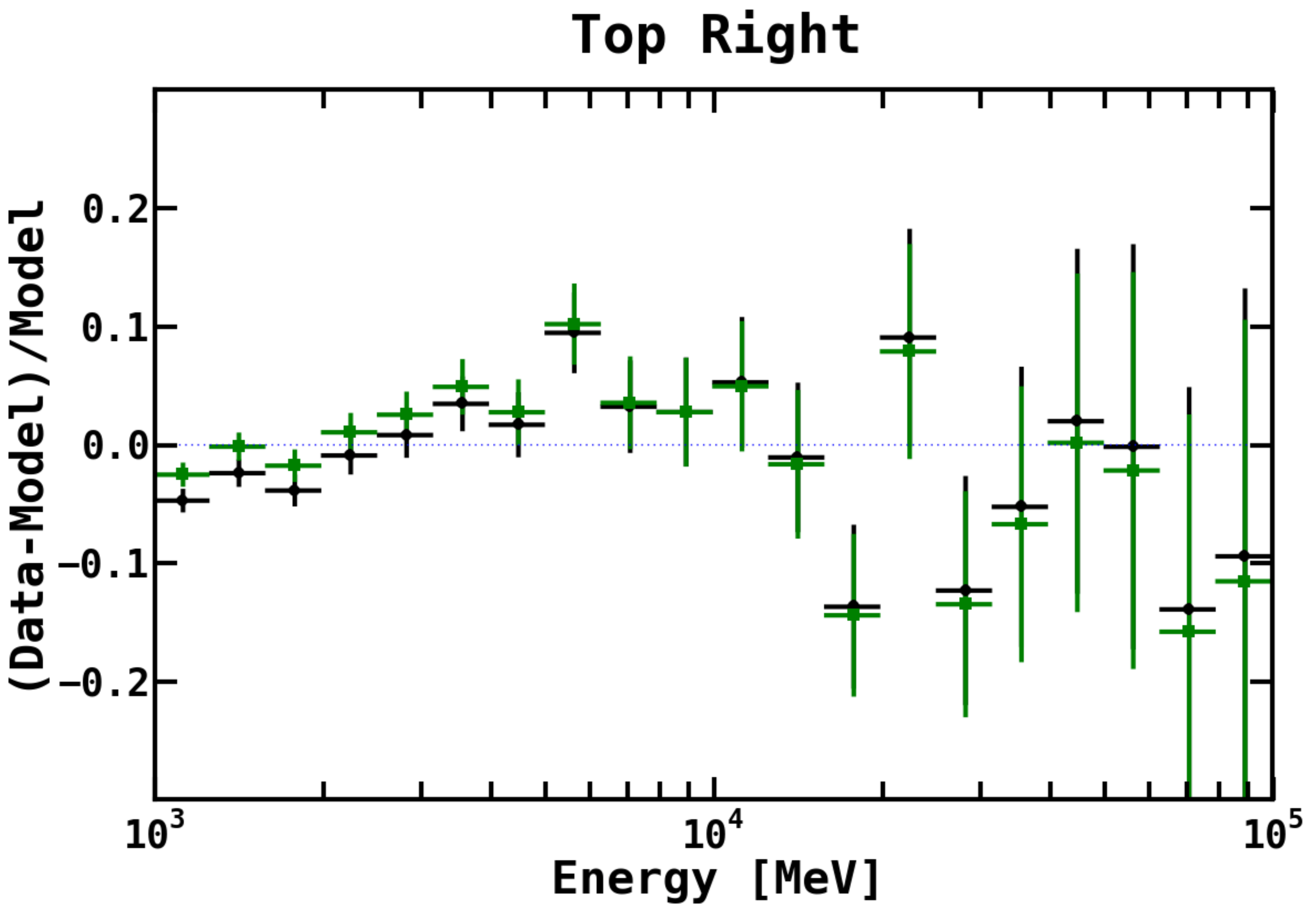}
\includegraphics[width=0.33\textwidth]{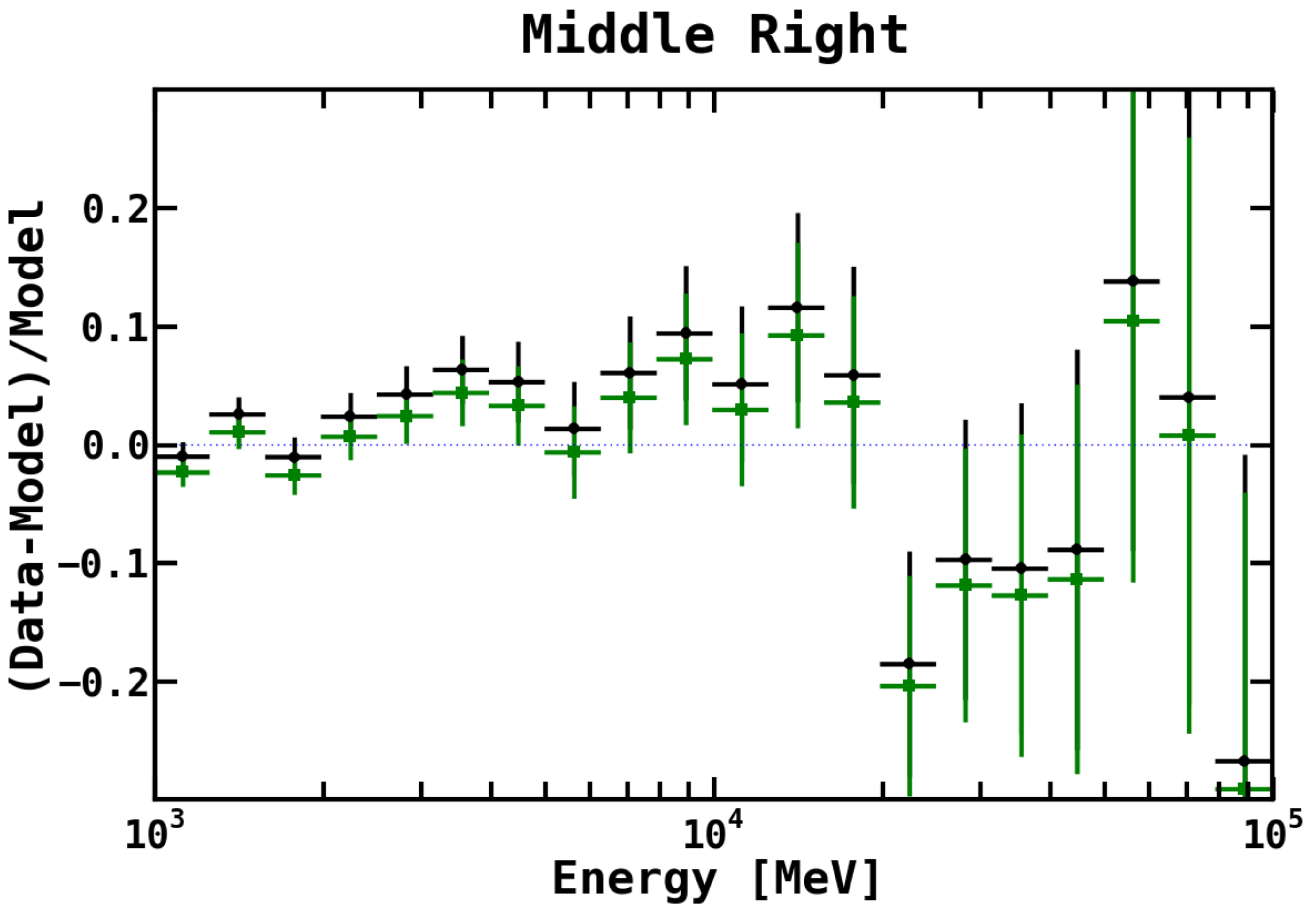}
\includegraphics[width=0.33\textwidth]{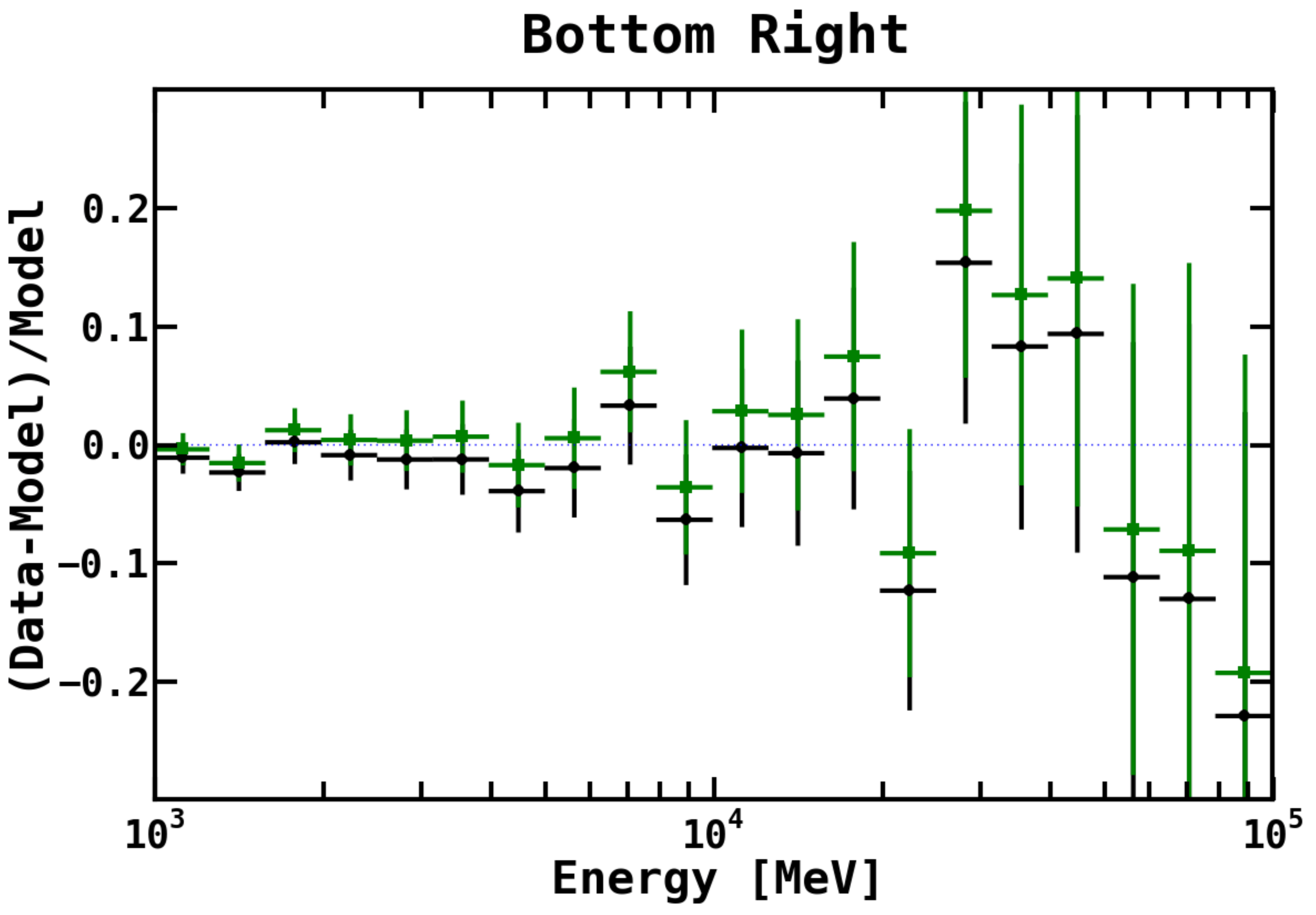}
\caption{Fractional residuals calculated in different spatial regions. The field is evenly divided into top, middle, and bottom. Each slice is then further divided into right and left. The regions are indicated above each plot. Black data points show the residuals resulting from the baseline fit (which is over the entire field, with IC scaled in addition to the other contributing components). We then rescale the diffuse components in the different subregions, masking the rest of the region, and keeping the point sources fixed to their baseline values (green data points). This is done to allow for a spatially varying spin temperature and/or CR and ISRF densities, which would in turn change the normalizations of the $\gamma$-ray components. Even in these smaller regions the diffuse components are unable to flatten the residuals, with the exception of the bottom right, which is pretty flat.}
\label{fig:top_middle_bottom}
\end{figure*}

The fractional energy residuals that result from this rescaling are shown in Figure~\ref{fig:top_middle_bottom}. The black data points show the residuals resulting from the baseline fit (over the entire field) calculated in the given subregion. The top row shows the residuals for the fit performed in the top, middle, and bottom regions, respectively. The second and third rows show the results for rescaling the normalizations in the regions which are further divided into right and left. 

Even with these smaller subregions the model is unable to flatten the positive residual emission between $\sim$3--20 GeV. Note that for many of these subregions the best-fit normalizations of the diffuse components resulting from the rescaling are not very physical, as some of the components go to zero, since they are not very well constrained and the fit simply tries to optimize the likelihood. Nevertheless, the model is still unable to fully flatten the residuals.

\begin{deluxetable}{lcccccc}[tbh!]
\tablecolumns{7}
\tablewidth{0mm}
\tablecaption{Baseline Values for the IEM Components in FM31 (IC scaled) \label{tab:norm_FM31_True}}
\tablehead{
\colhead{Component} &
\colhead{Normalization} &
\colhead{Flux  ($\times 10^{-9})$}&
\colhead{Intensity ($\times 10^{-8})$}\\
& 
&
\colhead{(ph cm$^{-2}$ s$^{-1}$)} &
\colhead{(ph cm$^{-2}$ s$^{-1}$ sr$^{-1}$)}
}
\startdata
\hi\ $\pi^0$, A5 &1.04 \p 0.04&189.3 \p 6.9 &80.5 \p 2.9 \\
\hi\ $\pi^0$, A6 &0.4 \p 0.2  &4.4 \p 2.5 &1.9 \p 1.0 \\
\hi\ $\pi^0$, A7 &2.9 \p 0.4  &15.8 \p 2.1 &6.7 \p 8.8 \\
\htwo\ $\pi^0$, A5 &2.7 \p 0.3&3.7 \p 0.4 &1.6 \p 0.2 \\
IC, A5 &2.4 \p 0.1  &125.0 \p7.0 &53.1 \p 3.0 \\
IC, A6 -- A7&0.9 \p 0.3  &17.3 \p 6.4 &7.3 \p 2.7\\
IC, A8 & 80.5 \p 16.4 &14.8 \p 3.0&6.3 \p 1.3
 \enddata
\tablecomments{The isotropic component is held fixed to the best-fit value obtained in the TR (1.06). All other diffuse sources and point sources are freely scaled in FM31, including the IC components. This is in contrast to the FM31 tuned fit, where the IC components are held fixed to the best-fit values obtained in the TR. Intensities are calculated by using the total area of FM31, which is 0.2352 sr.}
\end{deluxetable}

\begin{figure*}[t]
\centering
\includegraphics[width=0.33\textwidth]{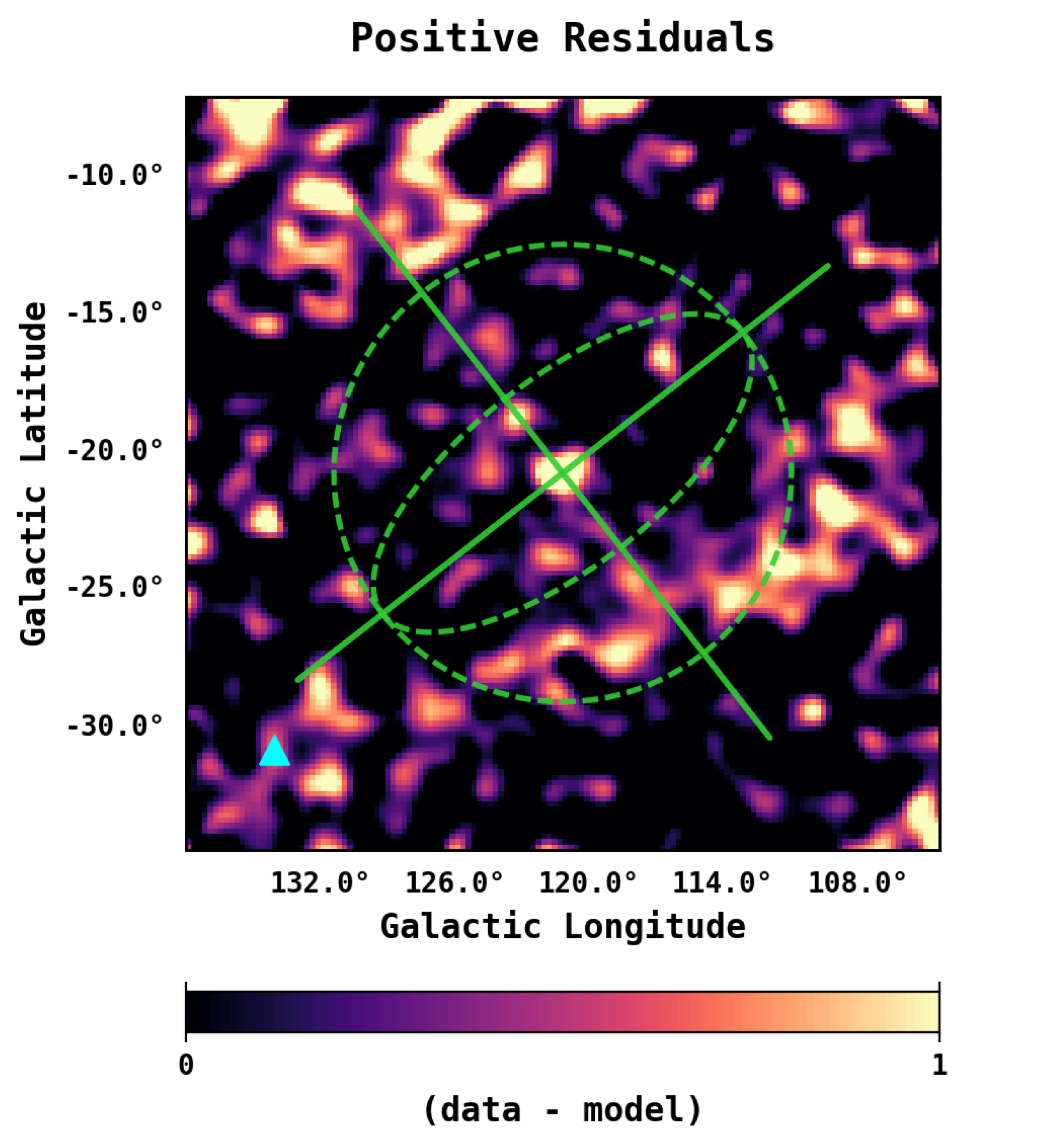}
\includegraphics[width=0.33\textwidth]{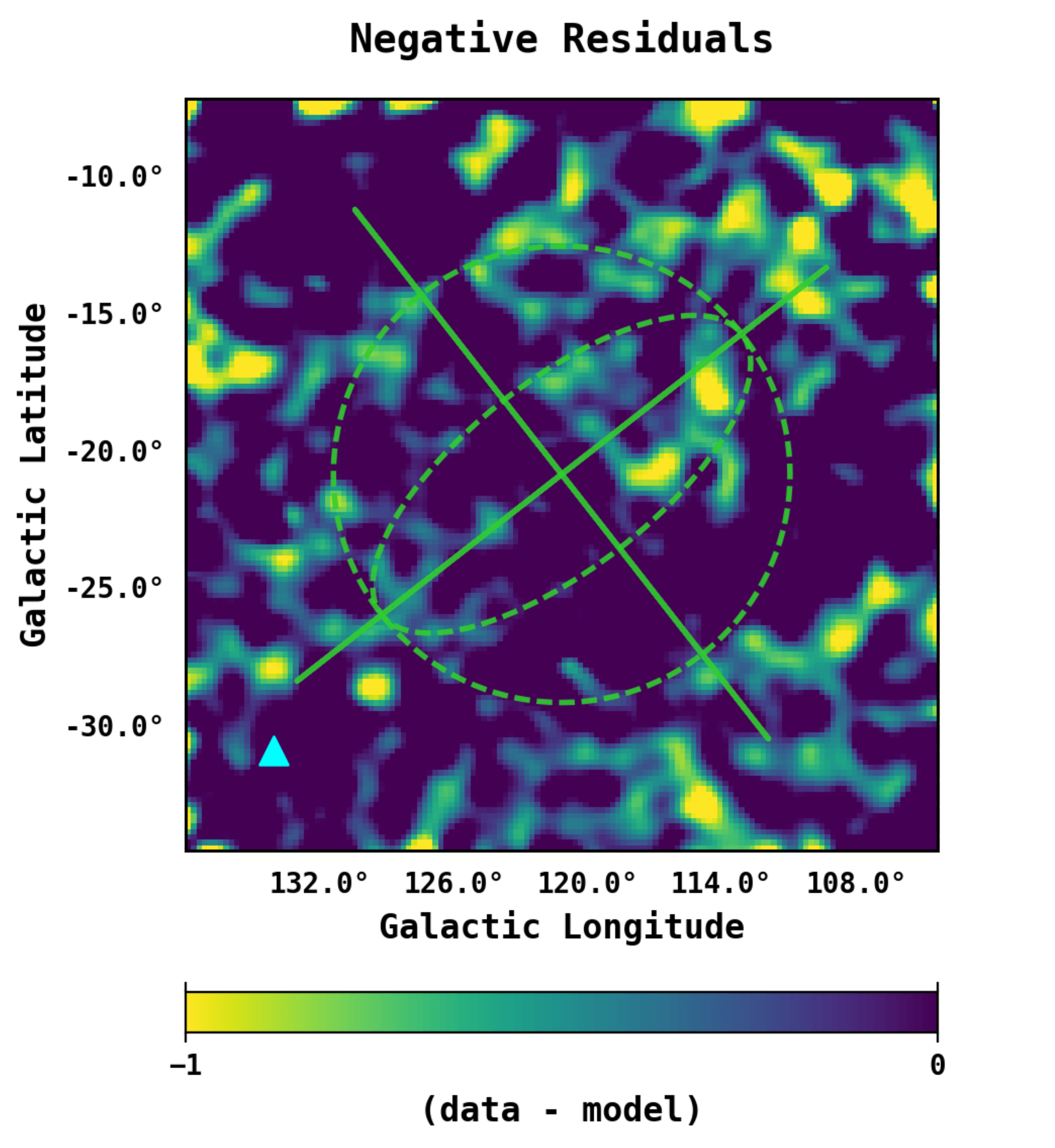}
\includegraphics[width=0.33\textwidth]{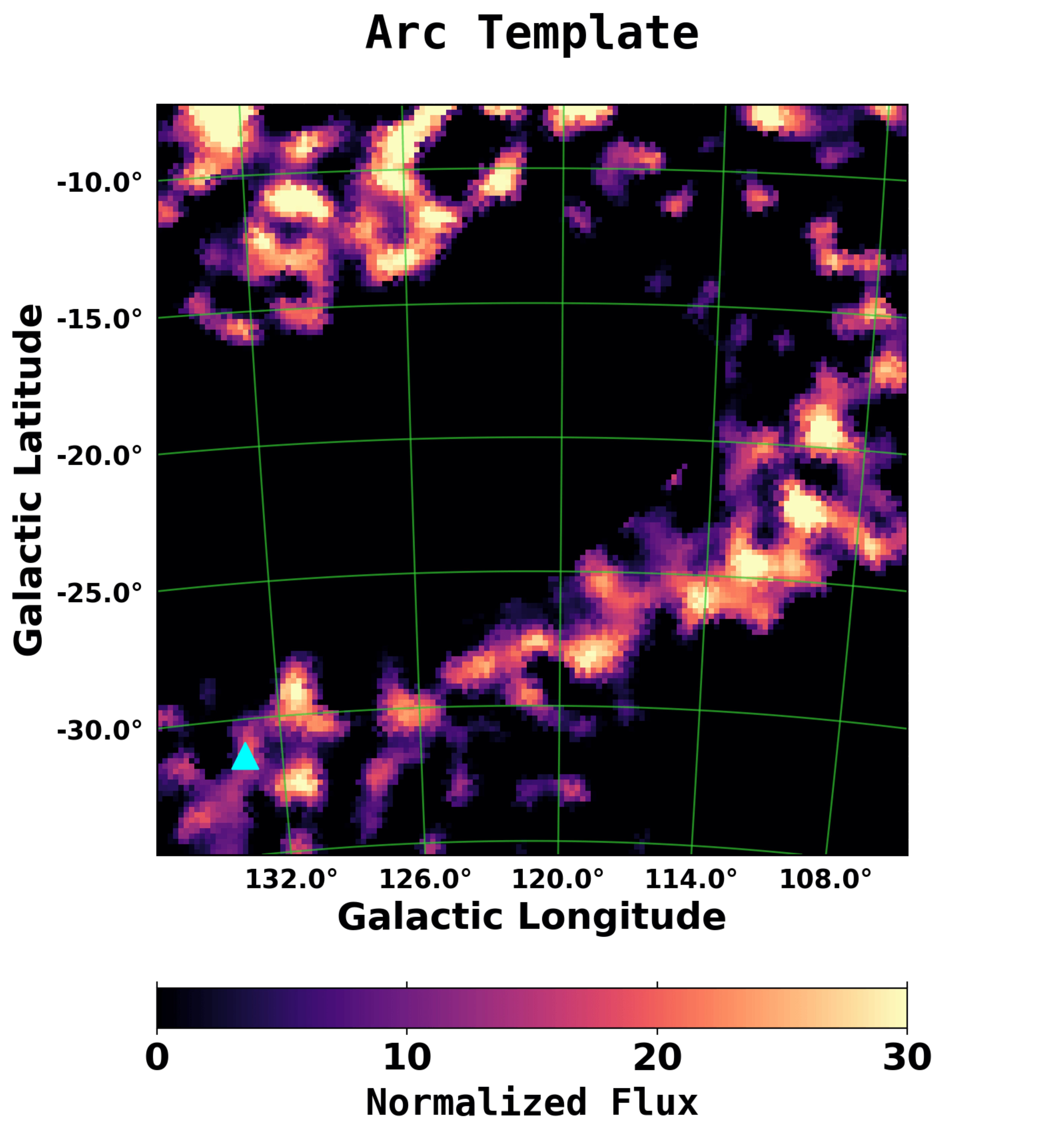}
\caption{The first two panels show the spatial count residuals integrated between 1--100 GeV, resulting from the baseline fit (see Figure~\ref{fig:flux_and_residuals_true}). In order to construct a template for the large arc extending from the top left corner to the projected position of M33 (arc template), we divide the total residual map into positive residuals (left) and negative residuals (middle). The maps show the geometry used to help facilitate the template construction (the green axes, circle, and ellipse), as detailed in the text. The corresponding geometrical parameters are given in Table~\ref{tab:template_parameters}. The resulting arc template is shown in the far right panel. In addition to fitting the full arc template, we also perform a variation of the fit in which the arc template is divided into a north component (arc north: $b > -16.5^\circ$) and a south component (arc south: $b \leq -16.5^\circ$), where the spectral parameters of each component are allowed to vary independently. The cut is made right below the bright emission in the upper-left corner, and it allows the north component to be at a different distance along the line of sight than the south component, as discussed in the text. The cyan triangle shows the projected position of M33.}
\label{fig:template_geometry}
\end{figure*}

Meanwhile, the residuals do start to become a bit more uniformly distributed. For example, when performing the fit over the entire field, the residuals in the top left are much more pronounced than the top right. For the rescaling in the different subregions, the top left residuals are decreased (between $\sim$3--20 GeV), whereas the top right residuals become a bit more pronounced. The same general trend can be seen in most of the subregions. The residuals are fairly flat in the bottom right, however, the bottom left (which contains M33) shows positive residual emission.

\subsection{Arc Template} \label{sec:arc_template}

Thus far the model has been unable to flatten the positive residual emission observed between $\sim$3--20 GeV. Furthermore, the spatial residuals show structured excess and deficits. It may be due to some foreground MW gas that is not well traced by the 21-cm emission. On the other hand, or in addition, the positive residual emission may be related to the M31 system, for which no model components are currently included. We note that the residuals behave qualitatively the same even when masking the inner region of the M31 disk (0.4$^\circ$). 

Our ultimate goal is to test for a $\gamma$-ray signal exhibiting spherical symmetry with respect to the center of M31, since there are numerous physical motivations for such a signal. However, before adding these components to the model, we employ a template approach to account for the arc-like feature observed in the spatial residuals, which may be related to foreground MW emission, and is not obviously related to the M31 system.

The first two panels in Figure~\ref{fig:template_geometry} show the spatial residuals integrated between 1--100 GeV, resulting from the baseline fit (see Figure~\ref{fig:flux_and_residuals_true}). In order to construct a template for the large arc extending from the top left corner to the projected position of M33 (arc template), we divide the total residual map into positive residuals (left) and negative residuals (middle). Overlaid is the geometry used to help facilitate the template construction. All geometry is plotted based on the general equation of an ellipse, which can be written as
\begin{equation}
\begin{split}
&a^{-2}\left\{(x-h)\cos\phi + (y-k)\sin\phi\right\}^2  \\
&+ \ b^{-2}\left\{(x-h)\sin\phi + (y-k)\cos\phi\right\}^2 = 1,
\end{split}
\end{equation}
where the center is given by $(h,k)$, $a$ and $b$ are the major and minor axes, respectively, and $\phi$ is the orientation angle of the ellipse. All geometrical parameters are given in Table~\ref{tab:template_parameters}. Note that the geometry corresponds to the $\gamma$-ray emission as observed in the stereographic projection, with the pole of the projection centered at M31. The plotted coordinate system (solid axes) is centered at M31 and oriented with respect to the position angle of the M31 disk ($38^\circ$). The large dashed green circle has a radius of $8.5^\circ$ ($R_{\rm tan}=117 \ \mathrm{kpc}$). The corresponding border facilitates the cut for the north-east side, and the radius is determined by the bright emission in the upper-left corner. The inner ellipse is used to facilitate the cut on the south-west side. This cut follows the natural curvature of the arc. Any emission not connected to the large arc is removed.

\begin{deluxetable}{lcccc}[tbh!]
\tablecolumns{5}
\tablewidth{0mm}
\tablecaption{Geometrical Parameters for the Arc Template\label{tab:template_parameters}}
\tablehead{
\colhead{Component} &
\colhead{2$a$ [deg]}&
\colhead{2$b$ [deg]}&
\colhead{$\phi$ [deg]}
}
\startdata
M31 position angle axis &25&0 &38\\
M31 perpendicular axis&25&0&128 \\
Dashed circle &17&17 &38\\
Dashed ellipse&17&7&38
\enddata
\tablecomments{M31 geometry is centered at $(h,k) = (121.17^\circ, -21.57^\circ)$. Angles are defined with respect to the positive $x$-axis (Cartesian plane), and they correspond to the major axis of the ellipse. Note that the geometry corresponds to the $\gamma$-ray emission as observed in the stereographic projection, with the pole of the projection centered at M31.}
\end{deluxetable}

\begin{figure*}[tbh!]
\centering
\includegraphics[width=0.4\textwidth]{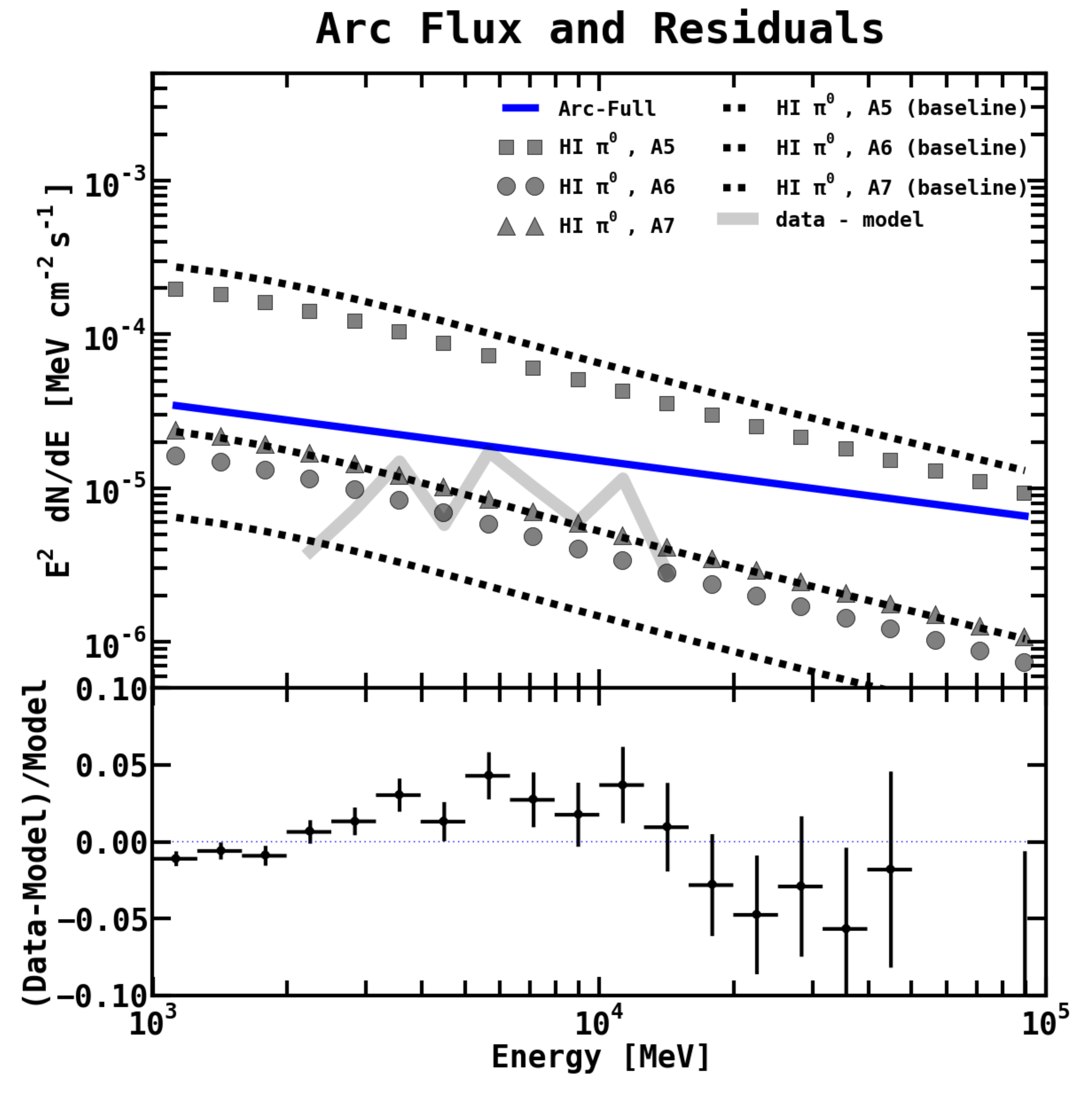}
\includegraphics[width=0.4\textwidth]{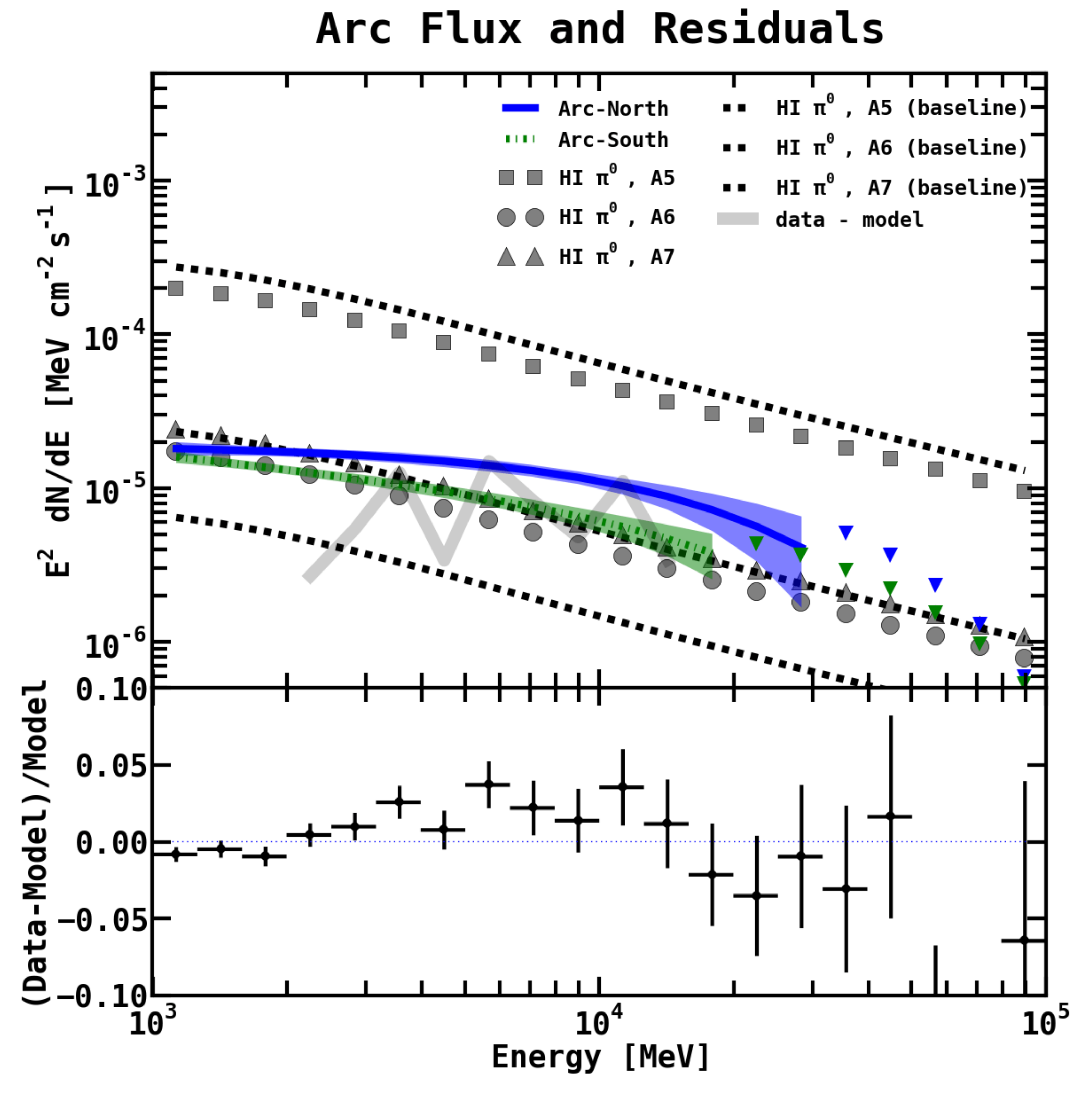}
\caption{Spectra and fractional energy residuals resulting from the arc fit. \textbf{Left:} The full arc component is given a PL spectrum, and the normalization and index are fit simultaneously with the other components in the region, just as for the baseline fit. Black dashed lines show the \hi\ A5 (top),  A6 (bottom), and A7 (middle) components from the baseline fit (not the arc fit). Note that A7 has a greater radial extension than that of A6, and likewise it has a greater overall flux. Correspondingly, the gray markers (squares, circles, and triangles) show the \hi\ A5--A7 spectra resulting from the arc fit. The blue solid line is the best-fit spectrum for the arc template. The bottom panel shows the remaining fractional residuals. For reference, the residuals (data -- model) are also plotted in the upper panel (faint gray band). \textbf{Right:} The arc template is given additional freedom by dividing it into north and south components. The arc components are given PLEXP spectral models, and the spectral parameters (normalization, index, and cutoff) are freely scaled with the other components. Downward pointing blue and green triangles give upper-limits. Bands give the 1$\sigma$ error. The arc template is unable to flatten the excess between $\sim$3--20 GeV.}
\label{fig:M31_Arc_flux_and_Residuals}
\end{figure*}

\begin{deluxetable*}{lcccc}[tbh!]
\tablecolumns{5}
\tablewidth{0mm}
\tablecaption{Normalizations of the Diffuse Components, Integrated Flux, and Likelihoods for the Arc Fits \label{tab:arc_fit_normalizations}}
\tablehead{
\colhead{Component} &
\colhead{Arc Full (PL)} &
\colhead{Arc North and South (PLEXP)}&
\colhead{Flux ($\times 10^{-9})$}&
\colhead{Intensity ($\times 10^{-8})$}\\
&
&
&
\colhead{(ph cm$^{-2}$ s$^{-1}$)} &
\colhead{(ph cm$^{-2}$ s$^{-1}$ sr$^{-1}$)} 
}
\startdata
\hi\ $\pi^0$, A5 &0.74 \p 0.04 &0.75 \p 0.04& 137.3 \p 8.0 & 58.4 \p 3.4\\
\hi\ $\pi^0$, A6 &1.1 \p 0.2 &1.2 \p 0.2 &11.7 \p 2.5&5.0 \p 1.1\\
\hi\ $\pi^0$, A7 &3.0 \p 0.4&3.0 \p 0.4&16.2 \p 2.1 &6.9 \p 0.9\\
\htwo\ $\pi^0$, A5 &2.6 \p 0.3 &2.7 \p 0.3 &3.7 \p 0.4 &1.6 \p 0.2 \\
IC, A5 &2.5 \p 0.1 &2.6 \p 0.1&134.2 \p 7.4&57.1 \p 3.1 \\
IC, A6 -- A7&1.6 \p 0.3 &1.5 \p 0.3&28.5 \p 6.4& 12.1 \p 2.7\\
IC, A8 &92.0 \p 17.0 &62.0 \p 18.2 &11.4 \p 3.3& 4.8 \p 1.4\\
$-\log L$&142972&142954&\nodata&\nodata
 \enddata
\tablecomments{Columns 2--3 give the best fit normalizations for the diffuse components.  The last two columns report the total integrated flux and intensity between 1--100 GeV for the arc north and south fit, which is the fit with the best likelihood. Note that the normalizations for the diffuse components are comparable for both variations of the fit. The bottom row gives the resulting likelihood for each respective fit. Intensities are calculated by using the total area of FM31, which is 0.2352 sr. }
\end{deluxetable*}

%
\begin{deluxetable*}{lccccccc}[tbh!]
\tablecolumns{8}
\tablewidth{0mm}
\tablecaption{Results for the Arc Templates\label{tab:arc_fit_params}}
\tablehead{
\colhead{Template} &
\colhead{area} &
\colhead{TS} &
\colhead{Flux ($\times 10^{-9})$}&
\colhead{Intensity ($\times 10^{-8})$}&
\colhead{Counts}&
\colhead{Index}&
\colhead{Cutoff,  $E_c$}\\
&
\colhead{(sr)}&
&
\colhead{(ph cm$^{-2}$ s$^{-1}$)} &
\colhead{(ph cm$^{-2}$ s$^{-1}$ sr$^{-1}$)} &
&
$\alpha$&
\colhead{(GeV)}
}
\startdata
Arc Full (PL) 			  &0.080232	& 651&26.0 \p 1.4&32.4 \p 1.7  &6872& 2.38 \p 0.05&\nodata	\\

Arc North (PLEXP) 			&0.033864& 457		&15.7 \p 1.4		& 46.4 \p 4.1	&4071& 2.0 \p 0.2&18.3 \p 14.8	\\

Arc South (PLEXP) &0.046368	&416	&12.0 \p 1.0			& 25.9 \p 2.2	&3210& 2.3 \p 0.1& 24.6 \p 19.7		
\enddata
\tablecomments{The TS is defined as $-2\Delta\log L$, and it is the value reported by {\it pylikelihood} (a fitting routine from the \fermilat{} ScienceTools package), without refitting. Fits are made with a power-law spectral model $dN/dE\propto E^{-\alpha}$ and with a model with exponential cut off $dN/dE\propto E^{-\alpha} \exp{(-E/E_c)}.$}
\end{deluxetable*}

The resulting normalized template is shown in the far right panel of Figure~\ref{fig:template_geometry}. By adding the arc template to the model we obtain a cleaner view towards M31's outer halo, and we are able to make inferences regarding the origin of the arc structure. We test two variations of the fit. In one variation we add a single template for the full arc. The arc is given a PL spectral model and the spectral parameters (normalization and index) are fit simultaneously with the other components in the region, just as for the baseline fit. In the second variation of the fit, the arc template is divided into a north component (arc north: $b > -16.5^\circ$) and a south component (arc south: $b \leq -16.5^\circ$). The cut is made right below the bright emission in the upper-left corner. Both components are given PLEXP spectral models (power law function with exponential cutoff), and the spectral parameters (normalization, index, and cutoff) of each component are allowed to vary independently. This allows the north component to be at a different distance along the line of sight than the south component, since different distances may correspond to different spectral parameters. Note that we also tried a number of different variations to the arc fit, and they all gave similar results as the two variations that we show here. 

\begin{figure}[tbh!]
\centering
\includegraphics[width=0.45\textwidth]{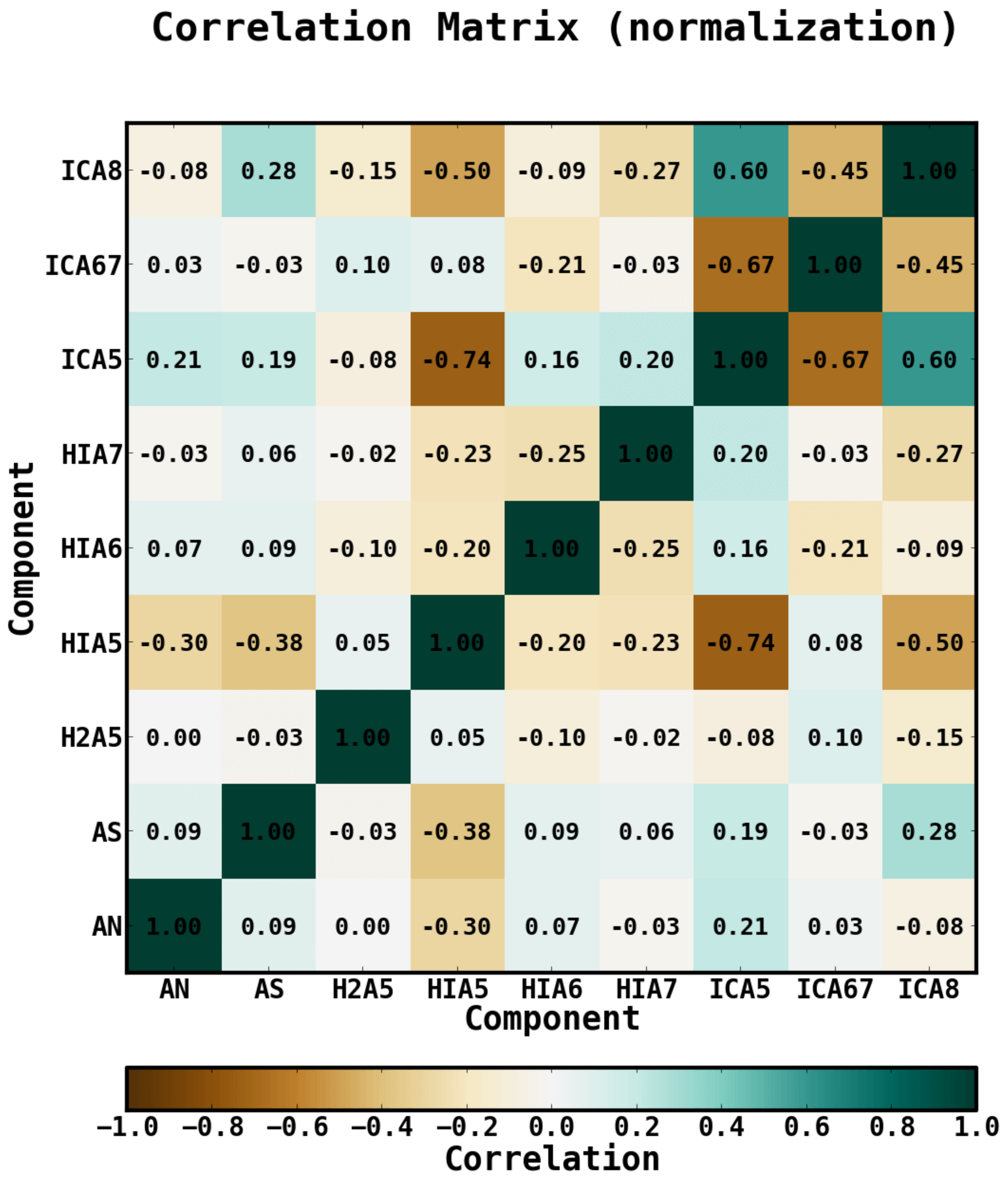}
\caption{The correlation matrix for the arc north (AN) and south (AS) fit.}
\label{fig:Arc_correlation}
\end{figure}

Results for the fits are given in Figure~\ref{fig:M31_Arc_flux_and_Residuals}. The top panels show best-fit spectra, and bottom panels show the remaining fractional residuals. For comparison, black dashed lines show the best-fit \hi\ spectra that result from the baseline fit, as shown in Figure~\ref{fig:flux_and_residuals_true}. For visual clarity, we show just the arc template and gas-related components. Spectra for the other components are qualitatively consistent with the results shown in Figure~\ref{fig:flux_and_residuals_true}. The arc template is unable to flatten the positive residual emission between $\sim$3--20 GeV, but the split arc fit with PLEXP spectral models does provide flatter residuals above $\sim$20 GeV. The correlation matrix for the arc north and south fit is shown in Figure~\ref{fig:Arc_correlation}.

Table~\ref{tab:arc_fit_normalizations} gives the best-fit normalizations for the diffuse components for both fits, as well as the overall likelihoods. Note that the normalizations are comparable for both fit variations. The last two columns report the total integrated flux and intensity for the arc north and south fit, which has the best likelihood. The corresponding best-fit parameters for the arc template components are reported in Table~\ref{tab:arc_fit_params}. For the baseline fit (Figure~\ref{fig:flux_and_residuals_true}) the total integrated flux for \hi\ A5 is (189.3 \p 6.9) $\times 10^{-9}$ ph cm$^{-2}$ s$^{-1}$. For the arc north and south fit the total integrated flux for \hi\ A5 plus the arc flux is (165.0 \p 10.4) $\times 10^{-9}$ ph cm$^{-2}$ s$^{-1}$. Thus with the arc template the total \hi\ A5 flux is decreased by $\sim$13\%. The flux is later increased when adding the M31-related components to the model, in addition to the arc template, as discussed in Section~\ref{sec:M31_components}. With the arc template the \hi\ A6 normalization has a value close to the GALPROP prediction. The normalization for IC A8 remains high, but this is a weak component with contribution only towards the top of the field.

\begin{figure*}[tbh!]
\centering
\includegraphics[width=0.3\textwidth]{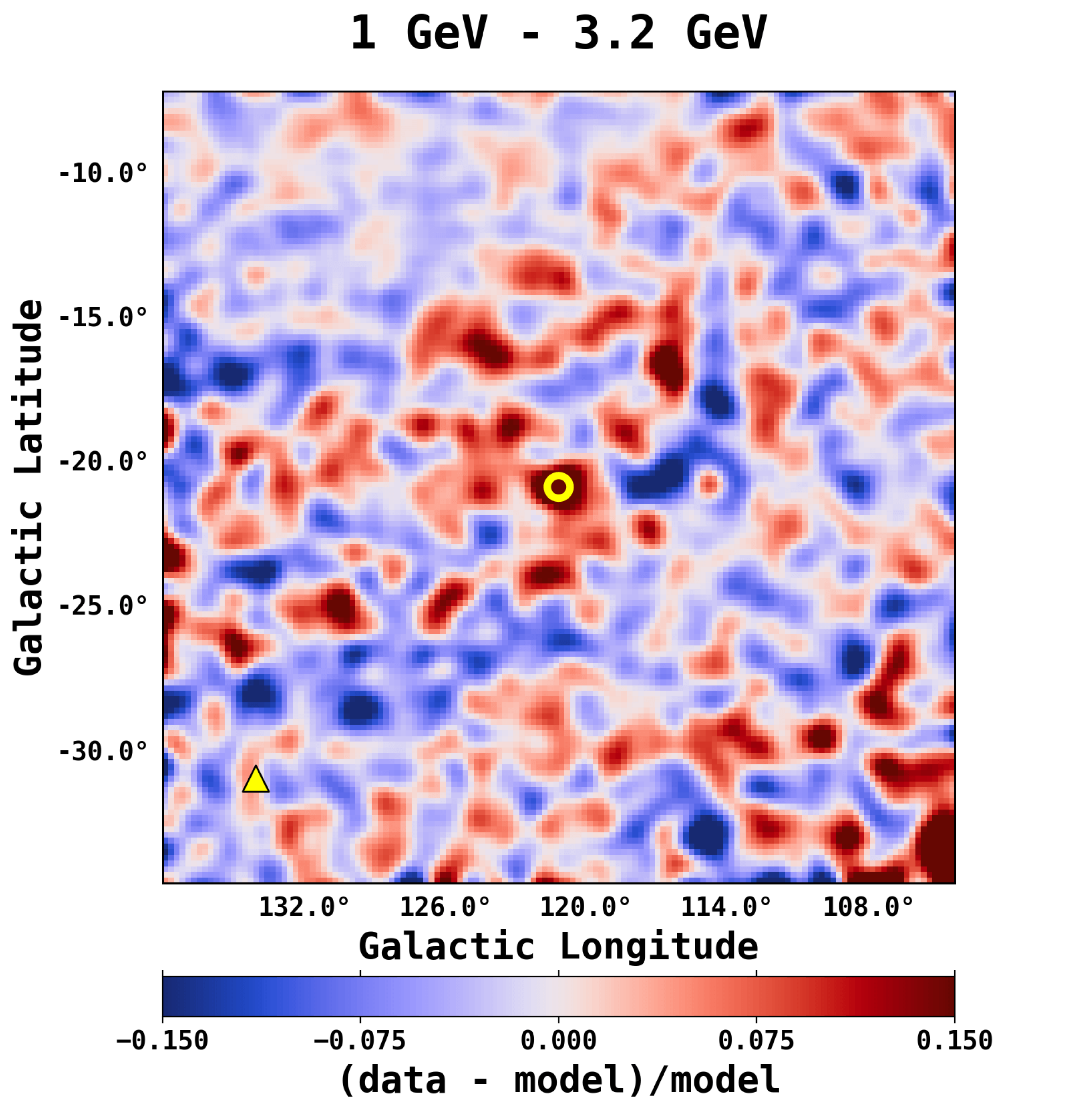}
\includegraphics[width=0.3\textwidth]{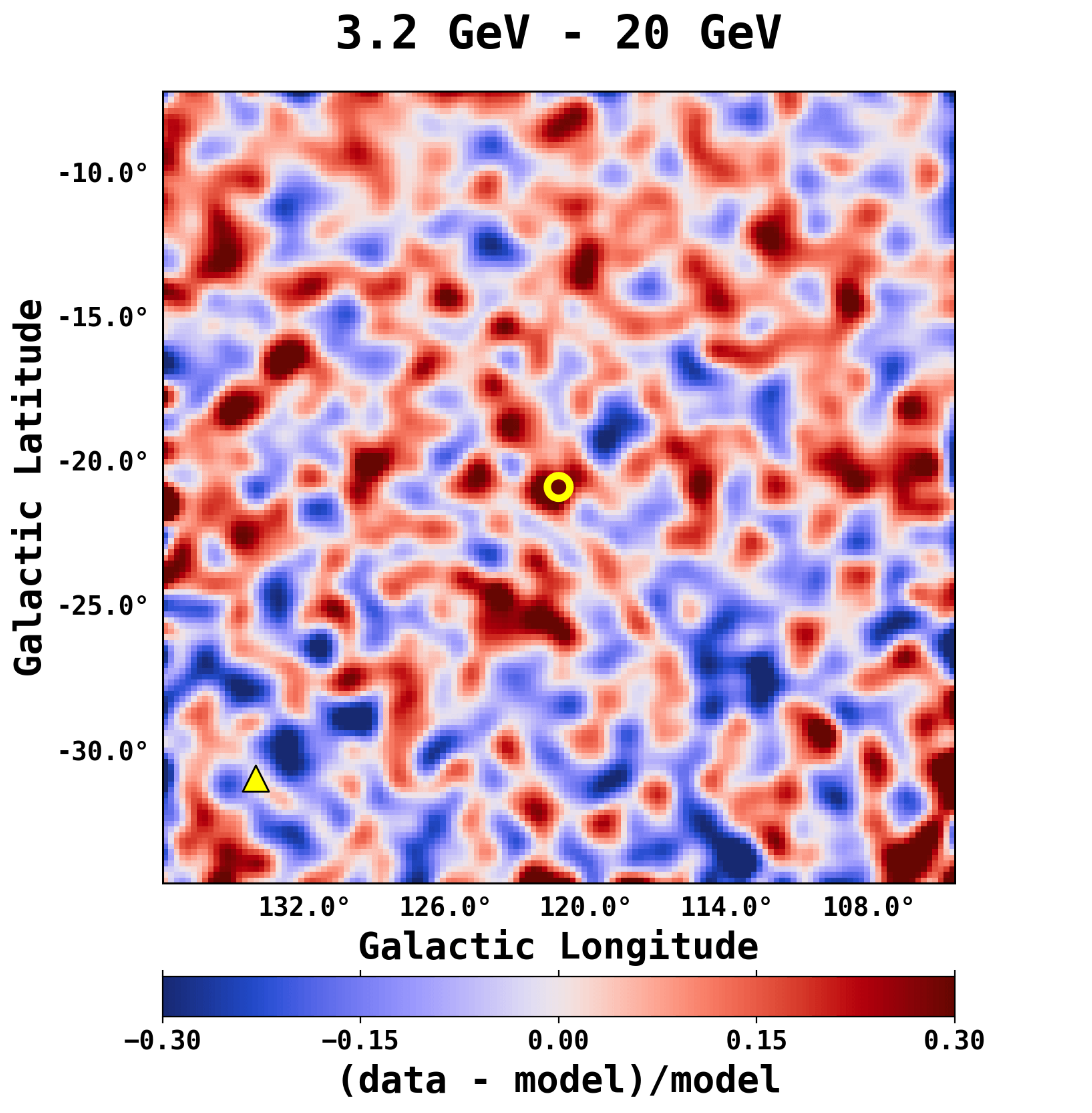}
\includegraphics[width=0.3\textwidth]{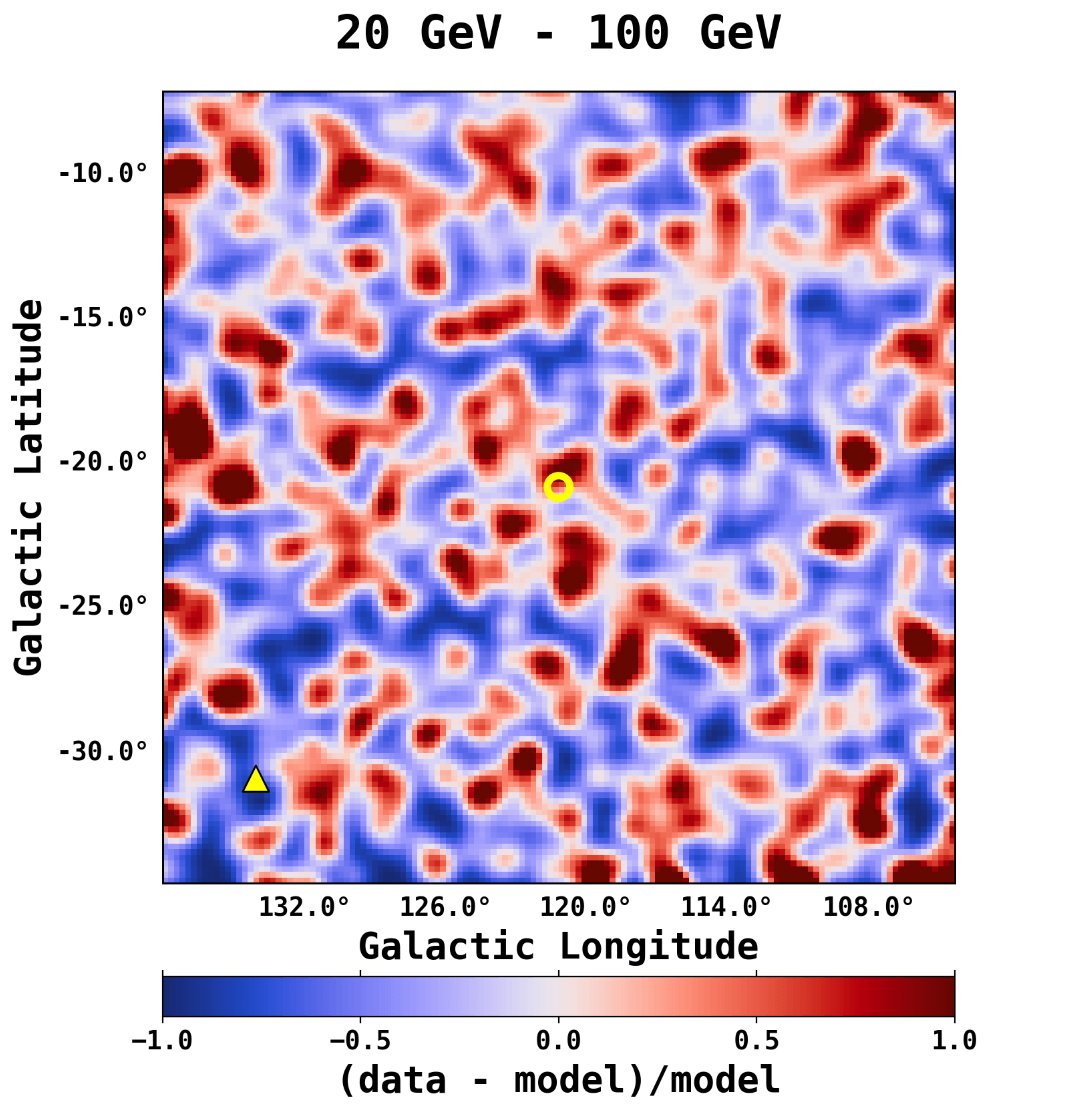}
\caption{Spatial count residuals resulting from the arc fit. To give a sense of the deviations, here we show the fractional residuals, where we divide by the model counts for each pixel. The residuals are integrated in three energy bins, just as for the residuals in Figure~\ref{fig:spatial_residuals_FM31_tuned}. We show residuals from the arc north and south fit, with PLEXP spectral model. Residuals for the full arc fit with PL spectral model are very similar. The arc structure no longer dominates the residuals, as expected. The position of M33 is indicated with a yellow triangle, and the center of M31 is indicated with a $0.4^\circ$ open circle.}
\label{fig:M31_Arc_spatial_residuals}
\end{figure*}

\begin{figure}[tbh!]
\centering
\includegraphics[width=0.45\textwidth]{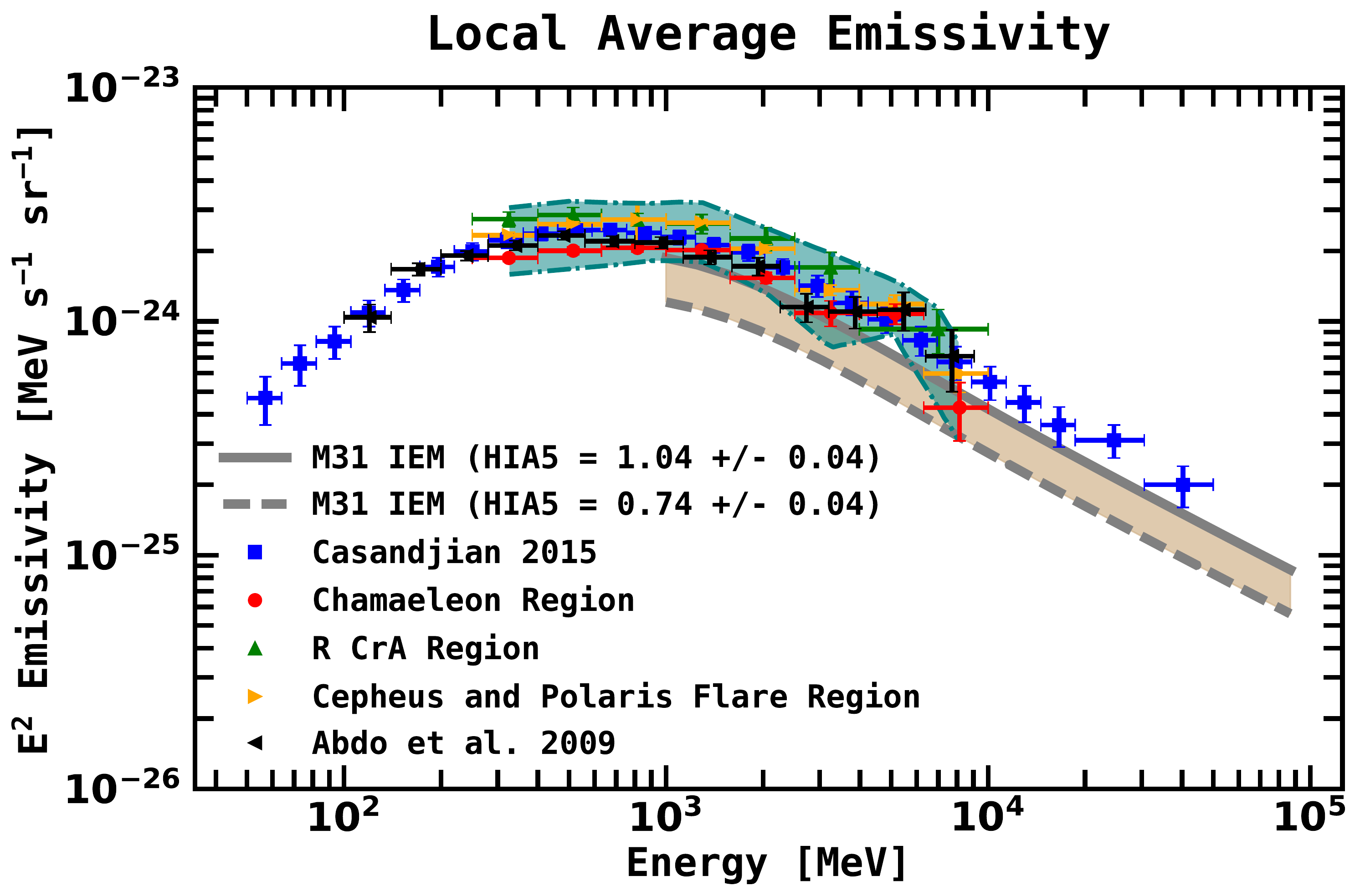}
\caption{The average local (A5) emissivity per H atom. The solid gray curve comes from the baseline fit with IC scaled, and it gives the proper estimate of the emissivity in FM31. The dashed gray curve comes from the arc fit with PL spectral model, and it only includes the contribution from the \hi\ A5 component, but not the emission associated with the arc. The blue data points (squares) are from \citet{Casandjian:2015hja}, and the corresponding error bars are systematic$+$statistical. The fit includes absolute latitudes between $10^\circ$--$70^\circ$.  The data points for the different regions (red circles, green upward-pointing triangles, and yellow rightward-pointing triangles) are from~\citet{ackermann2012fermi}, and the corresponding error bars are statistical only (1$\sigma$). The teal band shows the total uncertainty (statistical$+$systematic) from the same analysis (from the erratum). The different regions are among the nearest molecular cloud complexes, within $\sim$300 pc from the solar system. We also plot the measurements from~\citet{abdo2009fermi} (black leftward-pointing triangles), as determined from a mid-latitude region in the third Galactic quadrant.}
\label{fig:emissivity}
\end{figure}

\begin{figure}[tbh!]
\centering
\includegraphics[width=0.37\textwidth]{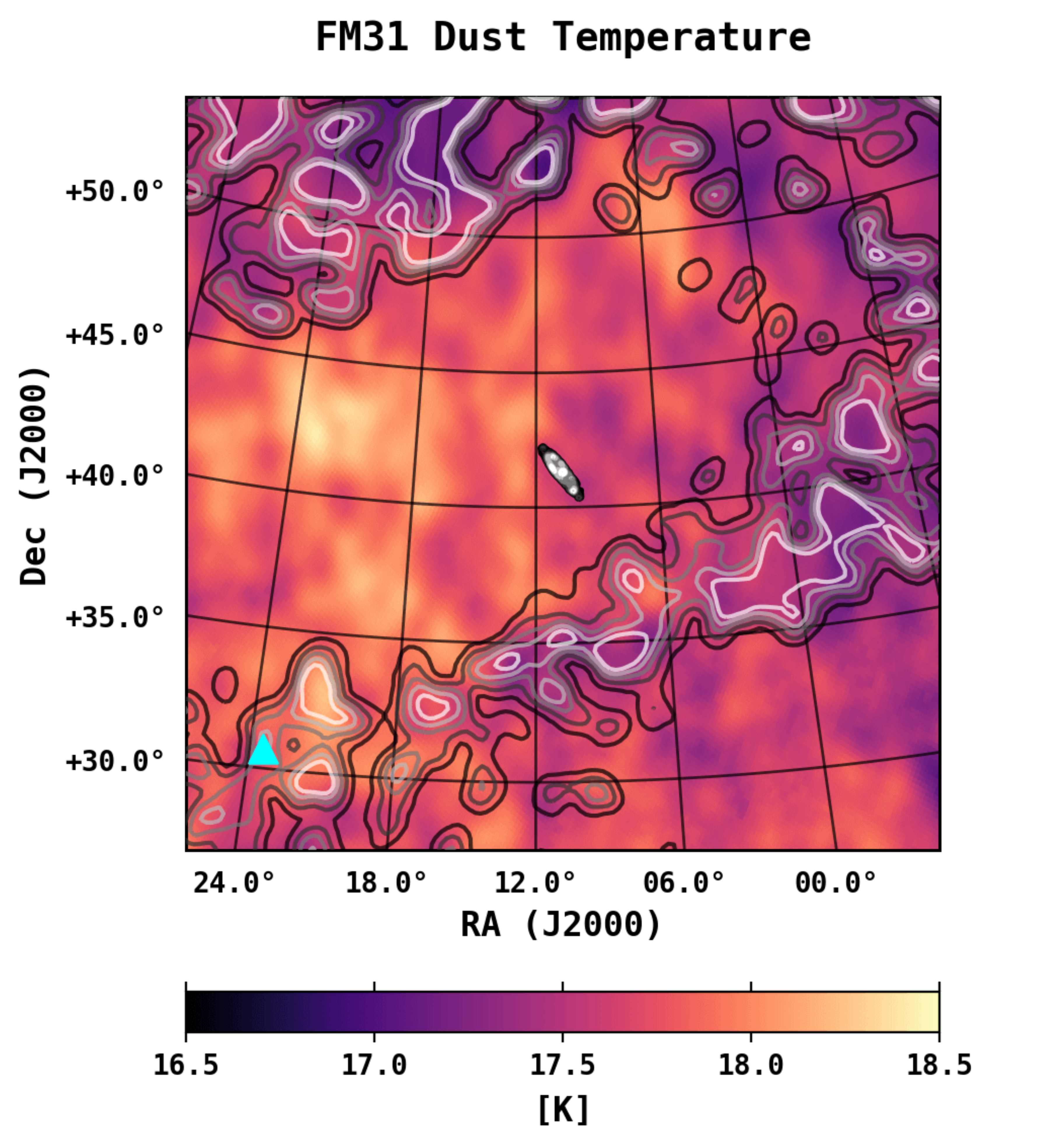}
\includegraphics[width=0.37\textwidth]{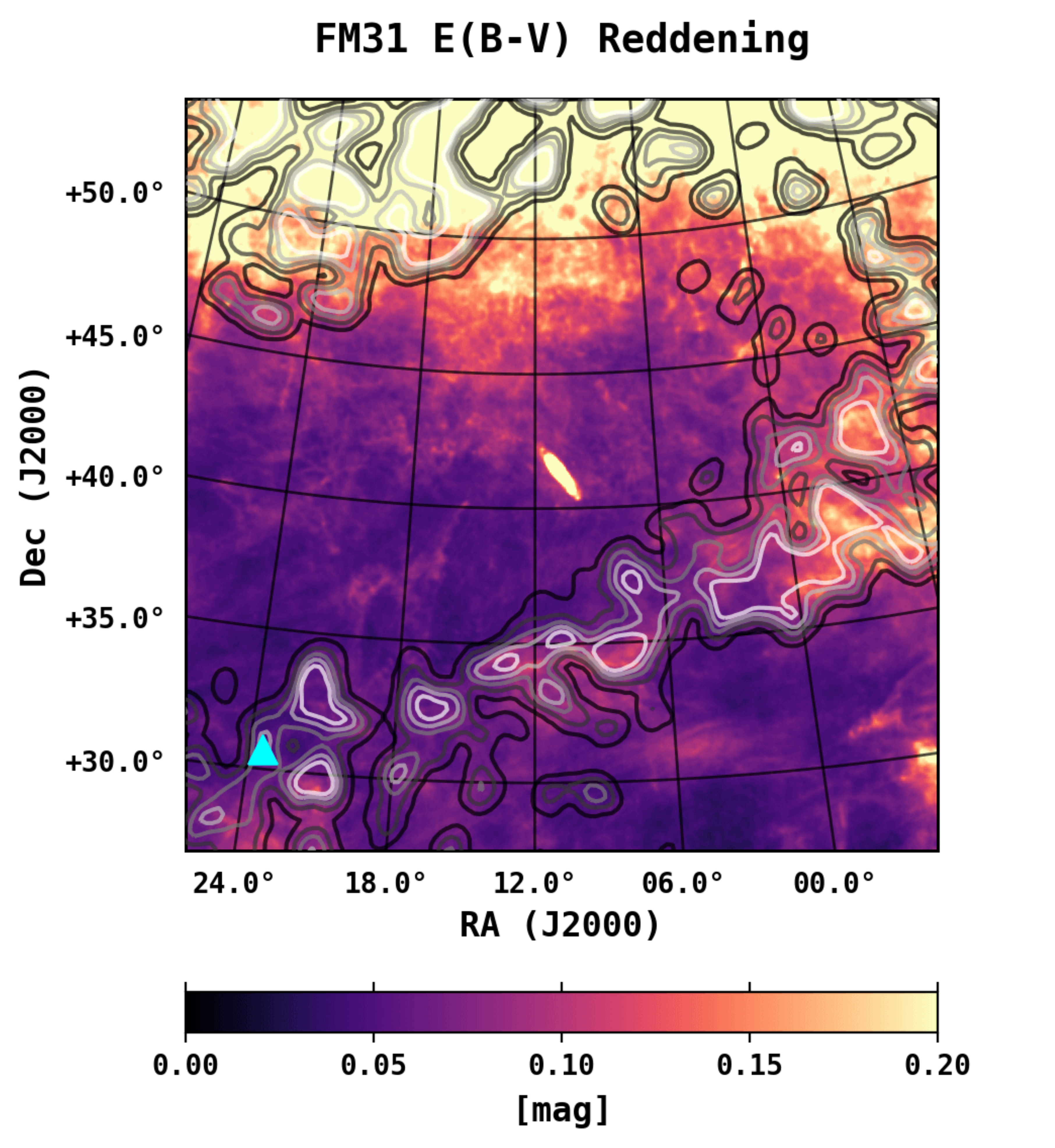}
\caption{Top panel shows the dust temperature map for FM31, and the bottom panel shows the dust reddening map, from~\citet{schlegel1998maps}, as discussed in the text. Overlaid are contours for the arc template. Contours for the IRIS 100 $\mu$m map of M31 are also overlaid in the top panel. The cyan triangle shows the (projected) position of M33.}
\label{fig:FM31_dust}
\end{figure}

Spatial residuals resulting from the arc north and south fit are shown in Figure~\ref{fig:M31_Arc_spatial_residuals}. Results for the full arc fit are very similar. To give a sense of the deviations, we show the fractional residuals, where we divide by the model counts for each pixel. The residuals are divided into three energy bins, just as for the residuals in Figure~\ref{fig:spatial_residuals_FM31_tuned}. The arc structure no longer dominates the residuals, as expected. In the first energy bin bright emission can be seen at the center of the map, corresponding to the inner galaxy of M31. In addition, the residuals in the first bin still show structured excesses and deficits, possibly associated with emission from M31's outer disk and halo. The second energy bin coincides with the positive residual emission observed in the fractional energy residuals. The spatial distribution of the emission is roughly uniform throughout the field, although small-scale structures can be observed. The third energy bin is roughly uniform with no obvious features. The distribution of the residual emission in FM31 is further quantified in Section~\ref{sec:symmetry}, where we consider the symmetry of the excess.

In Figure~\ref{fig:emissivity} we plot the measured local average emissivity per H atom, resulting from all fits in FM31. The solid gray curve comes from the baseline fit with IC scaled, and gives the proper estimate of the emissivity in FM31. The dashed gray curve comes from the arc fit with PL spectral model, and it only includes the contribution from the \hi\ A5 component, but not the emission associated with the arc. The best-fit normalizations are listed in the legend. Also plotted is the corresponding measurement made in~\citet{Casandjian:2015hja}, which is determined from a fit including absolute latitudes between $10^\circ$--$70^\circ$. Additionally, we plot the results from~\citet{ackermann2012fermi}, for which the emissivity is determined from different nearby molecular cloud complexes, within $\sim$300 pc from the solar system. Lastly, we plot the measurements from~\citet{abdo2009fermi}, as determined from a mid-latitude region in the third Galactic quadrant, i.e.\ $200^\circ < l < 260^\circ$ and $22^\circ < |b| < 60^\circ$. The local emissivity as determined from FM31 is slightly lower (referring to the baseline normalization of 1.04), but it is consistent within 1$\sigma$ with these other measurements. This is not surprising since the analysis by \citet{ackermann2012fermi} is based on observations of the well-defined gas clouds residing within $\sim$300 pc from the solar system. Meanwhile, our ``local ring'' is 2 kpc thick (Table~\ref{tab:GALPROP_parameters}), while FM31 is projected toward the outer Galaxy where the CR density is predictably low.

As we see, inclusion of the arc template into the fit improves its quality significantly. Meanwhile, the origin of the arc itself remains unknown. As we show below, the arc is most likely associated with the interstellar gas, its under-predicted column density, and/or with particles whose spectrum is distinctly flatter than the rest of CRs.

\begin{figure*}[tbh!]
\centering
\includegraphics[width=0.4\textwidth]{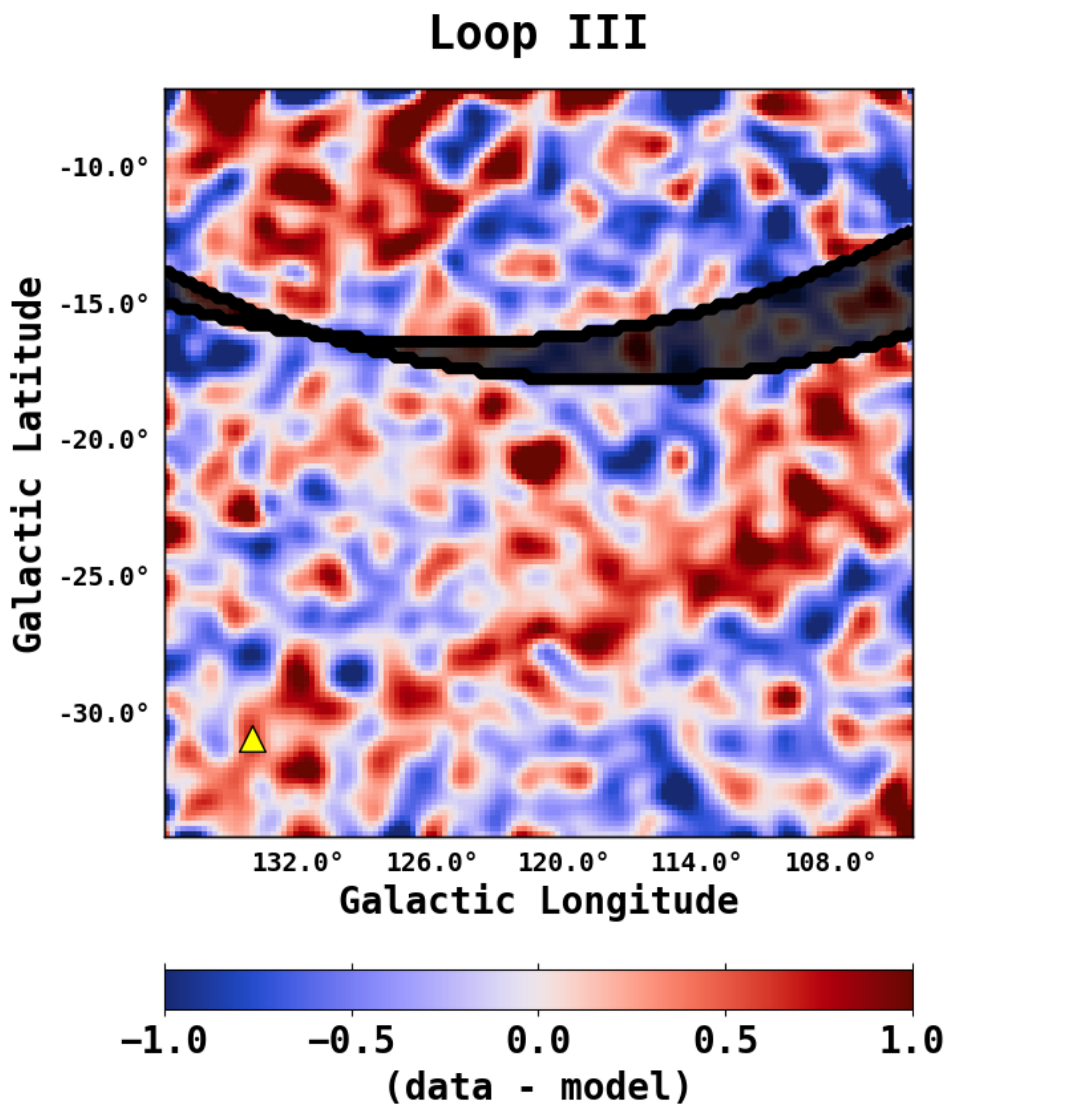}
\includegraphics[width=0.4\textwidth]{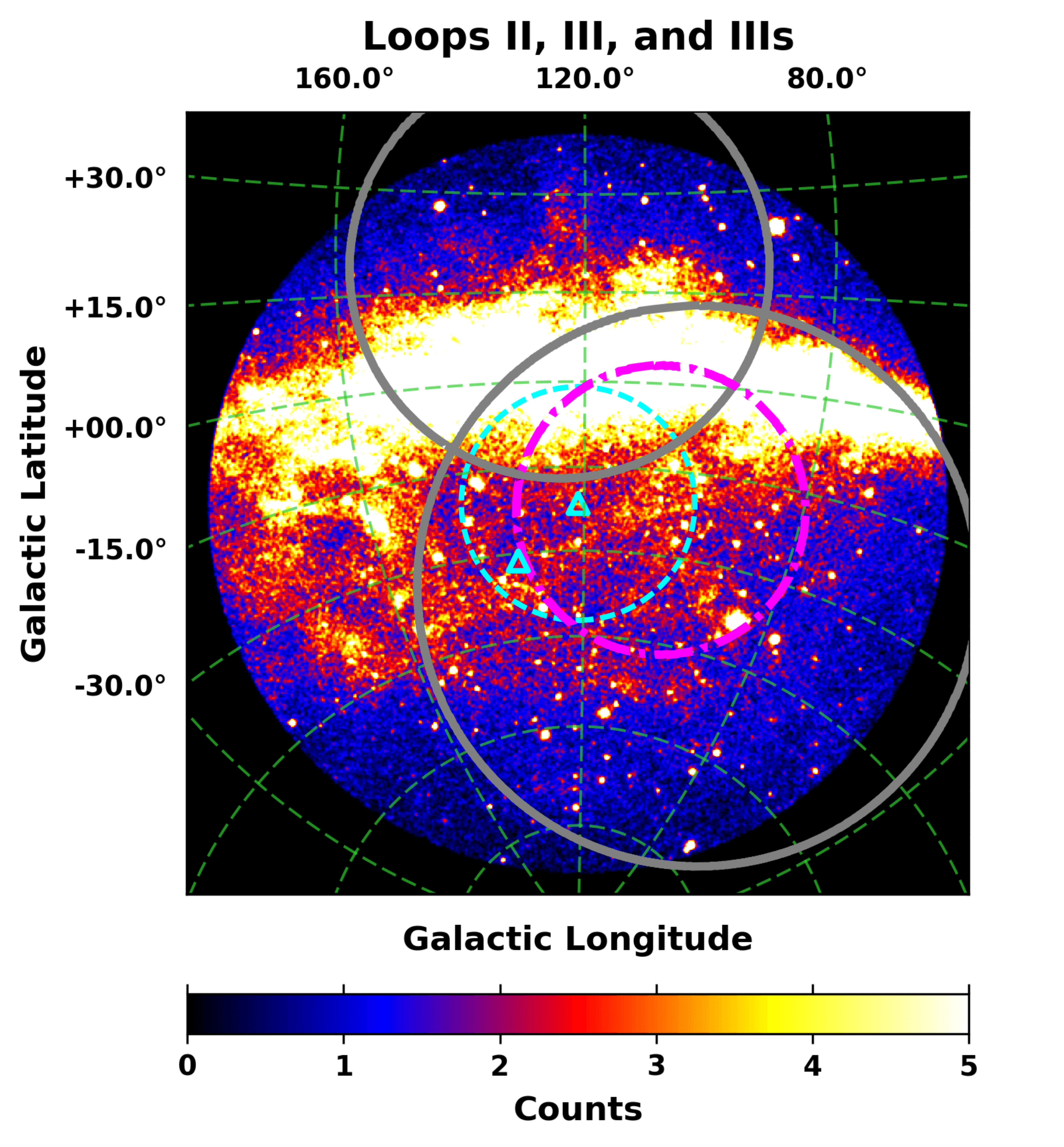}
\caption{{\bf Left:} FM31 residuals from the baseline fit (with IC scaled) with the Loop III shell plotted over it. The two lines correspond to two somewhat different positions and radii obtained from continuum and polarization observations \citep{2015MNRAS.452..656V}. The shell radius is approximate and the shell itself can be several degrees thick. The shaded area gives an idea of the error associated with the parameters of the shell. {\bf Right:} M31's virial radius (300 kpc) is shown with a cyan dashed circle, and cyan triangles show the positions of M31 and M33. The gray circles show Loop III at the top and Loop II at the bottom. Loop IIIs (which is only visible in polarization) is shown with a dash-dot magenta circle.}
\label{fig:LoopIII}
\end{figure*}

In Figure~\ref{fig:FM31_dust} we show the dust temperature map and the $\rm{E(B-V)}$ reddening map for FM31 from~\citet{schlegel1998maps}. Overlaid are contours for the arc template. The levels correspond to the normalized flux, and they range from 1--20 in increments of 5. The dust temperature serves as a possible proxy for the gas temperature. In this analysis we have assumed a uniform spin temperature of 150 K, but as can be seen in the top panel of Figure~\ref{fig:FM31_dust}, much of the arc template correlates with cold regions in the dust, indicating that at least part of the corresponding residuals may be caused by an under prediction of the \hi\ column density.

As can be seen in Figure~\ref{fig:FM31_dust}, much of the arc template closely correlates with the foreground dust, and likewise it correlates with the local \hi\ column density, as seen in Figure~\ref{fig:gas_column_densities}, indicating that the corresponding emission is most likely due to inaccuracies in the foreground model. Although our model already corrects for the DNM, the method is full sky and may use an incorrect gas to dust ratio for this particular region. In addition, the method also assumes a linear conversion between gas and dust, which may not actually be the case. Also, we note that while the spatial correlation between the arc template and properties of the dust is clearly visible towards the Galactic plane and the extended arm at the far right of the map, the region associated closest with M33 (in projection) and its general vicinity is not as obviously correlated. 
 
The analysis described in this section clearly shows that the arc is associated with the gas, but its components have the spectral index of $\sim$2.0--2.4, noticeably flatter than the rest of the \hi\ gas $\sim$2.75 in the ROI (Figure~\ref{fig:M31_Arc_flux_and_Residuals}). This may imply that the spectrum of CR particles interacting with gas in this direction is flatter than the spectrum of the old component of CRs that is altered by the long propagation history. Indeed, radio observations and sometimes X-rays and \gray{s} reveal structures that cover a considerable area of the sky and are often referred to as ``radio loops''. The most well-known is Loop I, which has a prominent part of its shell aligned with the North Polar Spur, but other circular structures and filaments also become visible in polarization skymaps. There are, at least, 17 known structures \citep[for details see][and references therein]{2015MNRAS.452..656V} with the radii of tens of degrees that can be as large as $\sim$$80^\circ$ for the Loop XI. The spectral indices of these structures indicate a non-thermal (synchrotron) origin for the radio emission, but the origin of the loops is not completely clear. One of the major limitations is the lack of precise measurements of their distances. The current explanations include old and nearby SNR, bubbles/shells powered by OB associations, and some others.
  
It turns out that a part of the shell of Loop III seems to be associated with the north part of the arc (Figure~\ref{fig:LoopIII}) and Loops II and IIIs are covering the entire ROI. Presence of accelerated electrons associated with the Loop III shell hints that protons with a flat spectrum can also be present there. This may explain the distinctly different spectral index of the arc template and an exponential cutoff significantly below 50 GeV (Figure~\ref{fig:M31_Arc_flux_and_Residuals} right) that corresponds to the ambient particle energies below $\sim$1 TeV. Here we are not speculating further if the whole arc or only a part of it is associated with the Loop III shell or with other Loops, leaving a detailed analysis for a followup paper.

 \subsection{M31 Components} \label{sec:M31_components}

The baseline model seems unable to account for the total emission in FM31. We now proceeded to add to the model M31-related components, for which we make the simplifying assumption of spherical symmetry with respect to the center of M31. For the inner galaxy we add a uniform disk with a radius of 0.4$^\circ$, consistent with the best-fit morphology in~\citet{Ackermann:2017nya}. We add a second uniform template centered at M31 with a radial extension of $0.4^\circ < r \leq 8.5^\circ$. This is the geometry as determined in Figure~\ref{fig:template_geometry}, which was used to help facilitate the construction of the arc template. We note that although the outer radius was set by the bright residual emission in the upper-left corner, it also happens to encompass a large \hi\ cloud centered in projection on M31, possibly associated with the M31 system (i.e.\ the M31 cloud), as well as a majority of M31's globular cluster population and stellar halo, which will be further discussed in Section~\ref{sec:gas_related_emission}. The radial extension corresponds to a projected radius of 117 kpc. We label this component as FM31 spherical halo. 

Lastly, we add a third uniform template with a radial extension of $r > 8.5^\circ$, covering the remaining extent of the field. This corresponds to M31's far outer halo, and likewise it begins to approach the MW plane towards the top of the field. This is the template that suffers most from Galactic confusion. We label this component as FM31 far outer halo. 

All of the M31-related component are given PLEXP spectral models, and the spectral parameters (normalization, index, cutoff) are fit with the arc template and the other baseline components. We note that the spectra of the M31 components have also been fit with a power law per every other energy band, as well as a standard power law, and the results are consistent with the PLEXP model (see Section~\ref{sec:FSSC_IEM}).
The fit is performed in the standard way just as for the baseline fit. We perform two main variations of the fit, amounting to different variations of the arc template. For one variation we use the full arc template with PL spectral model. For the second variation we use the north and south arc templates with PLEXP spectral models. 

\begin{figure*}[tbh!]
\centering
\includegraphics[width=0.49\textwidth]{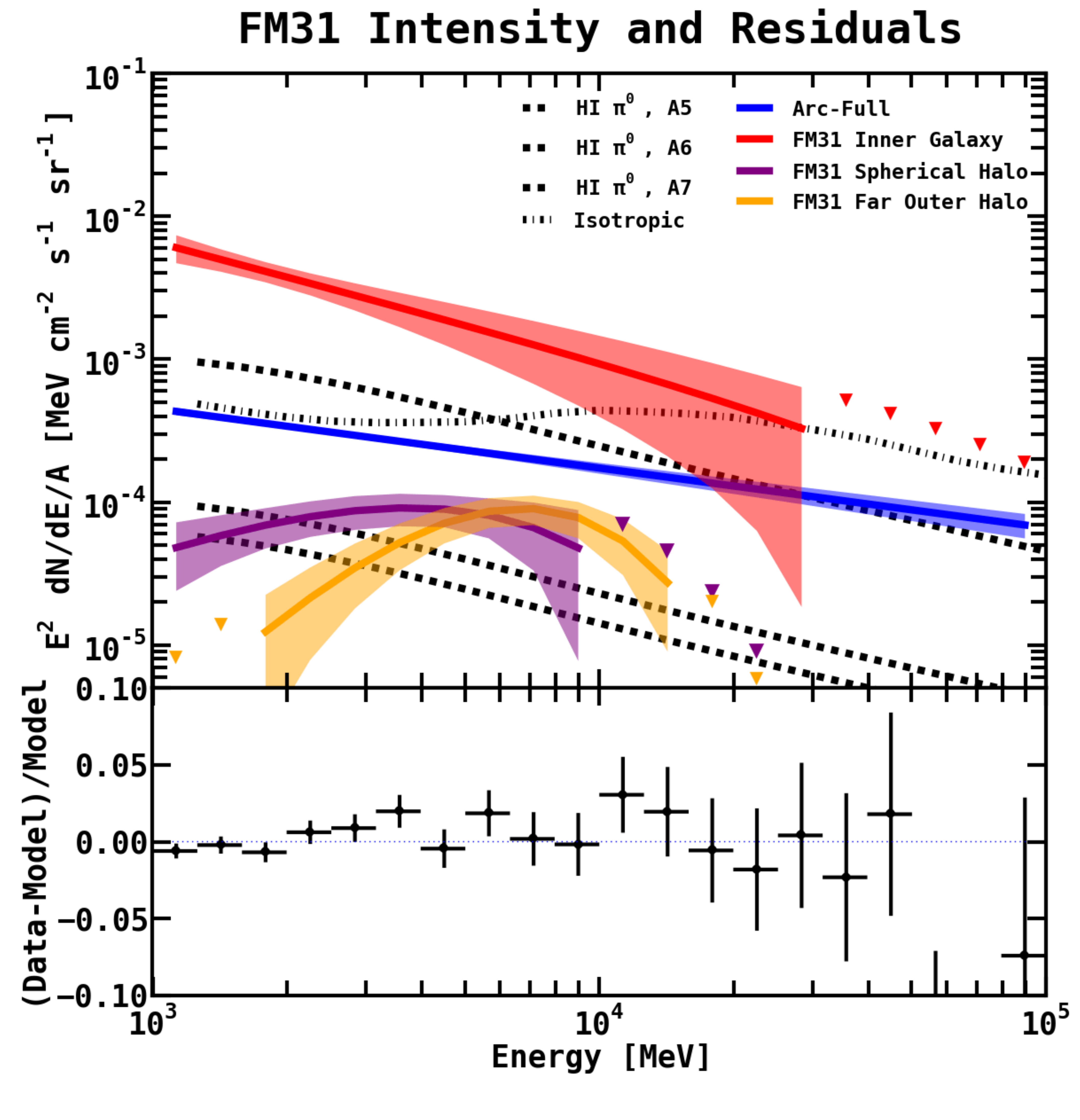}
\includegraphics[width=0.49\textwidth]{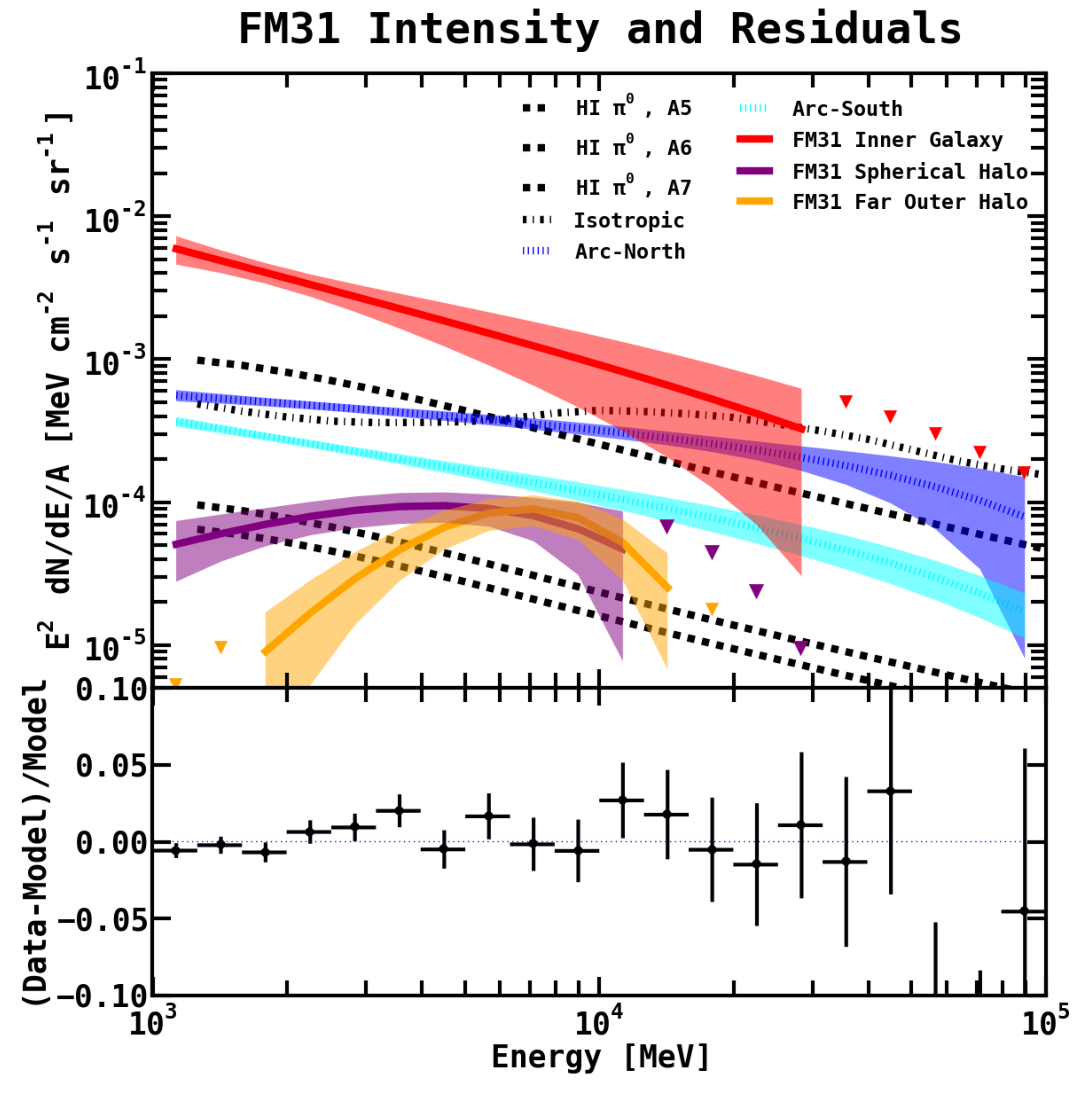}
\caption{M31-related components are added to the model, in addition to the arc template, and standard baseline components. The left panel is for the full arc template with PL spectral model, and the right panel is for the north and south arc templates with PLEXP spectral model, just as in Figure~\ref{fig:M31_Arc_flux_and_Residuals}. Black dashed lines show the best-fit spectra for the \hi\ A5 (top),  A6 (bottom), and A7 (middle) components. The black dashed-dot line shows the isotropic component, which remains fixed to its best-fit value obtained in the tuning region, just as for all other fits. The best-fit spectra of the remaining components are similar to that shown in Figure~\ref{fig:flux_and_residuals_true}, and are left out here for visual clarity. Downward pointing triangles give upper-limits. Bands give the 1$\sigma$ error. The bottom panel shows the remaining fractional residuals, which are fairly flat over the entire energy range, and likewise show a normal distribution with a mean of zero.}
\label{fig:M31_components}
\end{figure*}

\begin{deluxetable*}{lcccc}[tbh!]
\tablecolumns{5}
\tablewidth{0mm}
\tablecaption{Normalizations of the Diffuse Components, Integrated Flux, and Likelihoods for the Arc Fits with M31 Components\label{tab:arc_and_M31_norms}}
\tablehead{
\colhead{Component} &
\colhead{Arc Full (PL)} &
\colhead{Arc North and South}&
\colhead{Flux ($\times 10^{-9})$}&
\colhead{Intensity ($\times 10^{-8})$}\\
&
&
&
\colhead{(ph cm$^{-2}$ s$^{-1}$)} &
\colhead{(ph cm$^{-2}$ s$^{-1}$ sr$^{-1}$)} 
}
\startdata
\hi\ $\pi^0$, A5 &0.85 \p 0.05&0.88 \p 0.05&159.8 \p 9.1 & 67.9 \p 3.9 \\
\hi\ $\pi^0$, A6 &0.9 \p 0.2&1.0 \p 0.2 & 10.3 \p 2.5 & 4.4 \p 1.1 \\
\hi\ $\pi^0$, A7 &2.8 \p 0.4&2.9 \p 0.4& 15.3 \p 2.1 & 6.5 \p 0.9 \\
\htwo\ $\pi^0$, A5 &2.7 \p 0.3&2.7 \p 0.3&3.7 \p 0.4& 1.6 \p 0.2  \\
IC, A5 &2.2 \p 0.2&2.2 \p 0.2& 115.2 \p 8.6 & 49.0 \p 3.7  \\
IC, A6 -- A7&1.2 \p 0.4&1.0 \p 0.4&20.1 \p 7.0 & 8.6 \p 3.0 \\
IC, A8 &88.5 \p 19.0&59.7 \p 20.2&11.0 \p 3.6 & 4.7 \p 1.5 \\
$-\log L$&142933&142919&\nodata&\nodata
 \enddata
\tablecomments{Columns 2 and 3 give the best fit normalizations for the diffuse components.  The last two columns report the total integrated flux and intensity between 1--100 GeV for the arc north and south fit. The bottom row gives the resulting likelihood for each respective fit. Intensities are calculated by using the total area of FM31, which is 0.2352 sr.}
\end{deluxetable*}

The intensities and residuals resulting from the fits with the arc template and M31 components are shown in Figure~\ref{fig:M31_components}. The left panel is for the full arc template with PL spectral model. The right panel is for the north and south arc templates with PLEXP spectral model. Black dashed lines show the best-fit spectra for the \hi\ A5 (top),  A6 (bottom), and A7 (middle) components. The black dashed-dot line shows the isotropic component, which remains fixed to its best-fit value obtained in the tuning region, just as for all other fits. The best-fit spectra of the remaining components are similar to that shown in Figure~\ref{fig:flux_and_residuals_true}, and are left out here for visual clarity. The bottom panel shows the remaining fractional residuals, which are fairly flat over the entire energy range, and likewise show a normal distribution with a mean of zero. The best fit normalizations and flux for the diffuse components are reported in Table~\ref{tab:arc_and_M31_norms}, as well as the fit likelihood. Best-fit parameters for the arc template and M31-related components are reported in Tables~\ref{tab:arc_M31_fit_params_PL} and~\ref{tab:arc_M31_fit_params_North_and_South}.

We note that for the M31-related components the TS is defined as $-2\Delta\log L$, and it is the value reported by {\it pylikelihood} (a fitting routine from the \fermilat{} ScienceTools package), without refitting. In order to obtain a more conservative estimate of the statistical significance of the M31-related components, and in particular, the components corresponding to the outer halo, we make the following calculation. We define the null model as consisting of the standard components (point sources and diffuse), arc template (north and south), and M31 inner galaxy component. Then for the alternative model we also include the spherical halo and far outer halo components. We find that the alternative model is preferred at the confidence level of roughly 8$\sigma$ ($-2\Delta\log L$=63). 

The total integrated flux for the \hi\ A5 component plus the arc north and south components is (185.6 \p 12.9) $\times 10^{-9}$ ph cm$^{-2}$ s$^{-1}$, consistent with that of the baseline fit (with IC scaled). The normalization of  the \hi\ A6 component is consistent with the GALPROP prediction. The normalization of the \hi\ A7 component is still a bit high (2.8 \p 0.4). The normalizations of the IC A5 and A6-A7 components are consistent with the all-sky average obtained in the isotropic calculation (Table~\ref{tab:norm_isotropic}). The intensity of the arc south component at $\sim$10 GeV is at the same level as that of the M31-related components, and its spectrum is softer than the spectrum of the north component.

%
\begin{deluxetable*}{lcccccccc}[tbh!]
\tablecolumns{9}
\tablewidth{0mm}
\tablecaption{Results for the Arc Template (Full, PL) and M31 Components\label{tab:arc_M31_fit_params_PL}}
\tablehead{
\colhead{Template} &
\colhead{area} &
\colhead{TS} &
\colhead{Flux ($\times 10^{-9})$}&
\colhead{Energy Flux ($\times 10^{-12})$}&
\colhead{Intensity ($\times 10^{-8})$}&
\colhead{Counts}&
\colhead{Index}&
\colhead{Cutoff, $E_c$}\\
&
\colhead{(sr)}&
&
\colhead{(ph cm$^{-2}$ s$^{-1}$)} &
\colhead{(erg cm$^{-2}$ s$^{-1}$)} &
\colhead{(ph cm$^{-2}$ s$^{-1}$ sr$^{-1}$)} &
&
$\alpha$&
\colhead{(GeV)}
}
\startdata

Arc Full (PL)  			&0.080232&616&25.5 \p 1.4&118.5 \p 7.0&31.8 \p 1.7 &6739&2.42 \p 0.05&\nodata	\\
	
FM31 Inner Galaxy 	&0.000144&55&0.5 \p 0.1&1.7 \p 0.4&347.2 \p 69.4 &141&2.8 \p 0.3&96.4 \p 151.6\\ 

FM31 Spherical Halo 	&0.0684&34&4.2 \p 1.6&19.4 \p 6.2&6.1 \p 2.3 &1158&0.7 \p 1.1&2.9 \p 2.9 \\

FM31 Far Outer Halo 	&0.166656&32&4.3 \p 1.9 &33.8 \p 9.0 &2.6 \p 1.1&1142&--1.4 \p 1.2&2.0 \p 0.7 
\enddata
\tablecomments{The TS is defined as $-2\Delta\log L$, and it is the value reported by pylikelihood, without refitting. Fits are made with a power-law spectral model $dN/dE\propto E^{-\alpha}$ and with a model with exponential cut off $dN/dE\propto E^{-\alpha} \exp{(-E/E_c)}.$}
\end{deluxetable*}

%
\begin{deluxetable*}{lcccccccc}[tbh!]
\tablecolumns{9}
\tablewidth{0mm}
\tablecaption{Results for the Arc Template (North and South, PLEXP) and M31 Components\label{tab:arc_M31_fit_params_North_and_South}}
\tablehead{
\colhead{Template} &
\colhead{area} &
\colhead{TS} &
\colhead{Flux ($\times 10^{-9})$}&
\colhead{Energy Flux ($\times 10^{-12})$}&
\colhead{Intensity ($\times 10^{-8})$}&
\colhead{Counts}&
\colhead{Index}&
\colhead{Cutoff, $E_c$}\\
&
\colhead{(sr)}&
&
\colhead{(ph cm$^{-2}$ s$^{-1}$)} &
\colhead{(erg cm$^{-2}$ s$^{-1}$)} &
\colhead{(ph cm$^{-2}$ s$^{-1}$ sr$^{-1}$)} &
&
$\alpha$&
\colhead{(GeV)}
}
\startdata

Arc North &0.033864 & 438 &15.5 \p 1.3 &78.9 \p 6.4 &45.8 \p 3.8& 4027 &2.2 \p 0.1 &84.5 \p 100.4	\\

Arc South &0.046368& 395 &11.8 \p 0.7 &47.8 \p 4.1 &25.4 \p 1.5 & 3155& 2.5 \p 0.1 &100.0 \p 6.6 	\\	

FM31 Inner Galaxy 	&0.000144 & 53 &0.5 \p 0.08 &1.7 \p 0.4&347.2 \p 55.6 & 139 & 2.8 \p 0.3&100.0 \p 10.6 \\ 

FM31 Spherical Halo 	&0.0684 & 39  &4.5 \p 1.2 & 22.0 \p 6.4  &6.6 \p 1.8 & 1223 &0.9 \p 0.8 &4.0 \p 3.6  \\

FM31 Far Outer Halo 	&0.166656 & 30 &3.8 \p 1.3 & 31.6 \p 8.7  &2.3 \p 0.8 & 1020 &--1.8 \p 1.3 & 1.8 \p 0.6 
\enddata
\tablecomments{The TS is defined as $-2\Delta\log L$, and it is the value reported by pylikelihood, without refitting. Fits are made with a model with exponential cut off $dN/dE\propto E^{-\alpha} \exp{(-E/E_c)}.$}
\end{deluxetable*}

%
\begin{deluxetable*}{lcccccccc}[tbh!]
\tablecolumns{9}
\tablewidth{0mm}
\tablecaption{Results for the Symmetry Test\label{tab:symmetry_test}}
\tablehead{
\colhead{Template} &
\colhead{area} &
\colhead{TS} &
\colhead{Flux ($\times 10^{-9})$}&
\colhead{Energy Flux ($\times 10^{-12})$}&
\colhead{Intensity ($\times 10^{-8})$}&
\colhead{Counts}&
\colhead{Index}&
\colhead{Cutoff, $E_c$}\\
&
\colhead{(sr)}&
&
\colhead{(ph cm$^{-2}$ s$^{-1}$)} &
\colhead{(erg cm$^{-2}$ s$^{-1}$)} &
\colhead{(ph cm$^{-2}$ s$^{-1}$ sr$^{-1}$)} &
&
$\alpha$&
\colhead{(GeV)}
}
\startdata

Spherical Halo North &0.0342 &89 &5.1 \p 1.3&22.4 \p 5.2&14.9 \p 3.8 &1388&1.2 \p 0.6 & 4.2 \p 3.3	\\

Spherical Halo South &0.0342 &28 &2.7 \p 1.2&11.9 \p 5.1 &7.9 \p 3.5&743&1.9 \p 0.5 &11.6 \p 15.0 	\\	

Far Outer Halo North &0.0833 &89 &6.8 \p 2.1 &47.6 \p 9.6  &8.2 \p 2.5&1805&--0.6 \p 0.8&2.4 \p 0.8  	\\ 

Far Outer Halo South &0.0833 &31 &4.7 \p 2.4&16.9 \p 11.6  &5.6 \p 2.9&1233  &2.7 \p 0.4&97.5 \p 21.9

\enddata
\tablecomments{The TS is defined as $-2\Delta\log L$, and it is the value reported by pylikelihood, without refitting. Fits are made with a model with exponential cut off $dN/dE\propto E^{-\alpha} \exp{(-E/E_c)}.$}
\end{deluxetable*}

In Appendix~\ref{sec:different_IEMs} we perform additional systematic checks. Using the M31 IEM we allow for extra freedom in the fit. We also repeat the analysis with two alternative IEMs, namely, the IG IEM and FSSC IEM. Each alternative IEM has its own self-consistently derived isotropic spectrum and additional point sources. Full details of these tests are given in Appendix~\ref{sec:different_IEMs}. Here we summarize the main findings.

Using the M31 IEM we allow for extra freedom in the fit by varying the index of the IC components with a PL scaling. In this case the IC components show a spectral hardening towards the outer Galaxy, for both the TR and FM31. However, this is unable to flatten the excess in FM31, and the properties of the excess remain qualitatively consistent with the results presented above.

Using the M31 IEM we also vary the index of the \hi-related components using a PL scaling. In the TR the local annulus shows no change in the index. However, in FM31 there is a hardening of the index for the local annulus, with a significantly increasing hardening towards the outer Galaxy. This result is in direct contrast to the gradual softening which has been reported by other studies~\citep{Acero:2016qlg,yang2016radial}. FM31 clearly shows an anomaly with respect to these other measurements, as well as an anomaly with respect to the results in the TR and the GALPROP predictions (see Section \ref{sec:M31_IEM_extra}). The anomaly is most clearly evident for the outer Galaxy rings, A6 and A7, and it is also these rings which are found to be partially correlated with the M31 system, as is clearly seen in Figure~\ref{fig:gas_column_densities}. In particular, the \hi\ A7 component obtains a best-fit index $\Delta\alpha$ of  --0.39 \p 0.11, which corresponds to an effective index of 2.37, compared to its GALPROP prediction of 2.76. This result further supports the conclusion that there is some significant anomaly in FM31. This particular fit is also able to do a better job at flattening the excess in the fractional energy residuals, however, some excess emission still remains. To quantify the remaining excess we fit the M31-related components. In this case the spherical halo is still detected at $\sim$3--4$\sigma$ and the spectral properties are qualitatively consistent with the main results.

For the IG IEM the spectrum of the isotropic component is determined at high latitudes ($|b|>50^\circ$), and the normalization is held fixed to its nominal value (1.0). This is in contrast to the M31 IEM, for which we use the all-sky isotropic spectrum, with the normalization determined in a tuning region directly below FM31. The fit is otherwise performed in the standard way. The residuals are qualitatively consistent with what we find for the M31 IEM.

\begin{figure*}[tbh!]
\centering
\includegraphics[width=0.7\textwidth]{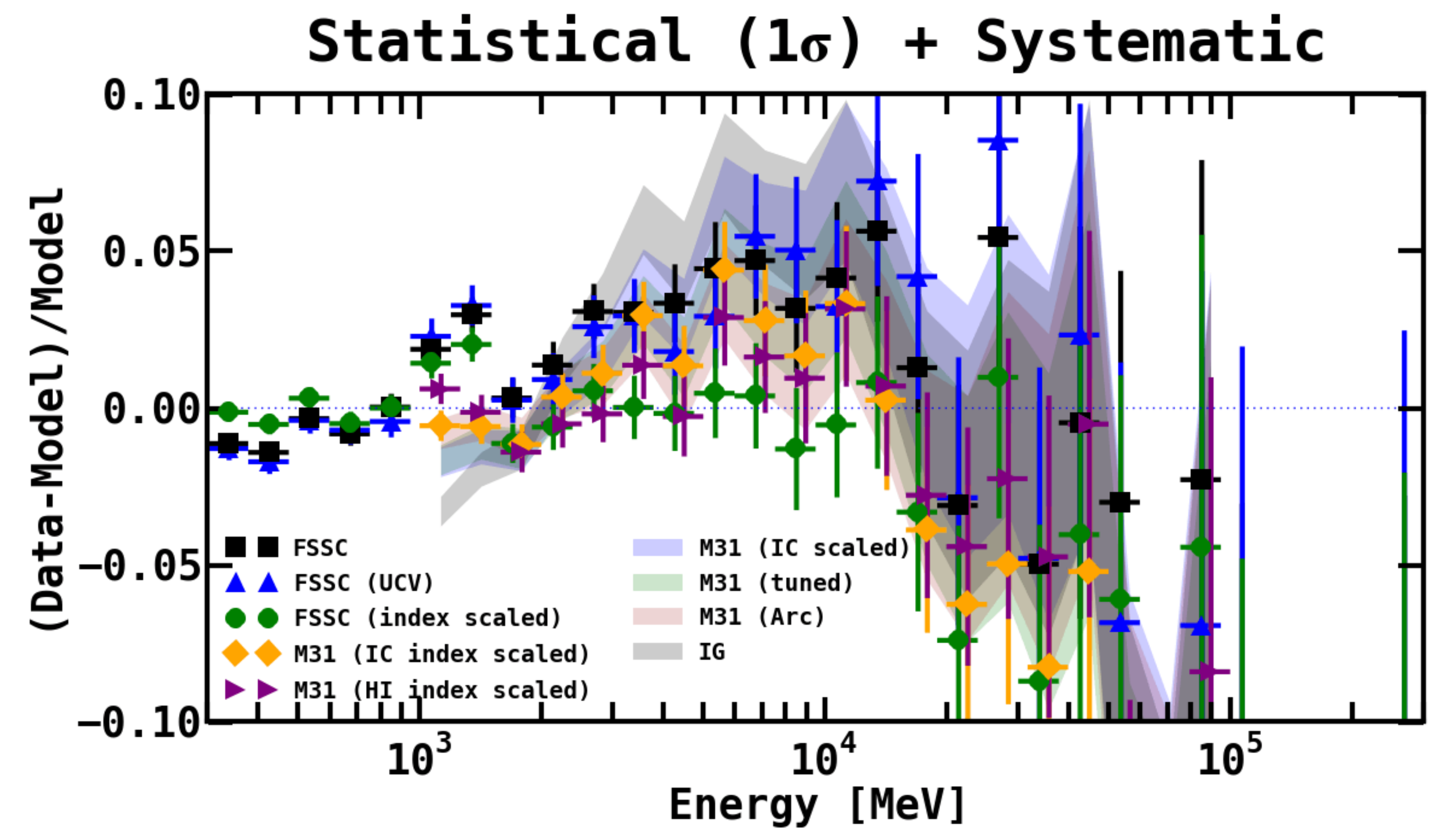}
\caption{A systematic excess can be observed between $\sim$3--20 GeV at the level of $\sim$3--5\%. Systematic over-modeling is also present above and below this range. We note that there is one model for which the signal can be flattened (shown with green circles). This results from using the FSSC IEM (intended for point source analysis) and fitting both the isotropic and Galactic diffuse (including the index) in the signal region. The FSSC IEM is not intended for extended source analysis, and this result illustrates how the application of an improper IEM for analysis of largely extended emission can alter the physical results. The M31 IEM is our benchmark model. The different models are as follows: \textbf{black squares:} FSSC IEM, fitting the isotropic and Galactic diffuse (with index fixed) in the signal region, using Clean data, corresponding to the fit in Figure~\ref{fig:flux_and_residuals_FSSC}; \textbf{blue upward-pointing triangles:} same as for the black squares but using UltraCleanVeto (UCV) data, see Section \ref{sec:FSSC_IEM} for details; \textbf{green circles:} same as for the black squares but also freeing the index of the Galactic diffuse; \textbf{orange diamonds:} M31 IEM baseline fit, varying the index of the IC components A5-A8 using a power law scaling, corresponding to the fit in Figure~\ref{fig:IC_index_scaled}; \textbf{purple rightward-pointing triangles:} M31 IEM baseline fit, varying the index of the \hi-related components A5--A8 using a power law scaling, corresponding to the fit in Figure~\ref{fig:HI_index_scaled}. Note that in this case FM31 shows a significant anomaly in the index of the gas-related emission towards the outer Galaxy, as is clearly shown in Figure~\ref{fig:HI_index_scaled}. \textbf{blue band:} M31 IEM baseline fit, corresponding to the fit in Figure~\ref{fig:flux_and_residuals_true}; \textbf{green band:} M31 IEM tuned fit, corresponding to the fit in Figure~\ref{fig:flux_and_residuals_FM31_Tuned}; \textbf{pink band:} M31 IEM arc fit, corresponding to the fit in Figure~\ref{fig:M31_Arc_flux_and_Residuals} (this is our primary model); \textbf{black band:} inner Galaxy (IG) IEM, corresponding to the fit in Figure~\ref{fig:flux_and_residuals_IG}.}
\label{fig:residuals_all}
\end{figure*}

We also repeat the fit using the FSSC IEM. We fit both the isotropic component and the Galactic diffuse component in the signal region, as well as the point sources. We perform the fit with and without freeing the index of the Galactic diffuse component. In the latter case the excess remains qualitatively consistent with what we find for the M31 IEM (both the fractional count residuals and the spatial residuals). However, in the former case the IEM is able to flatten the excess in the fractional count residuals (the spatial residuals remain qualitatively the same). This illustrates how the application of an improper IEM for analysis of largely extended emission can alter the physical results. 

We note that as a test we have also performed the fit with the M31 IEM by freely scaling the isotropic component in FM31, along with the other diffuse components and point sources. In this case the isotropic component obtains a normalization of 1.46 \p 0.06, and the excess in the fractional count residuals remains qualitatively the same. We do not consider this to be a proper procedure for our analysis, but nevertheless this test shows that even with an increase in the normalization of the isotropic emission upwards of 46\% the residual is still observed.

A summary of the excess in the fractional count residuals for all fit variations tested in this analysis is shown in Figure~\ref{fig:residuals_all}. We conclude that a systematic excess is present between $\sim$3--20 GeV at the level of $\sim$3--5\%. The signal is only flattened with the FSSC IEM (intended for point source analysis), when fitting all components in the signal region (including the index of the Galactic diffuse component), whereas all other fits result in an excess. Our benchmark model is the M31 IEM.

\subsection{Symmetry of the Residual Emission in FM31} \label{sec:symmetry}

In this section we further test the symmetry of the residual emission in FM31. We divide the spherical halo and far outer halo templates into north and south components. The cut is made at the midpoint of FM31 along the horizontal direction (parallel to the Galactic plane), corresponding to a latitude of $\sim$$-21.5^\circ$. This allows for deviation from spherical symmetry, as well as a gradient with respect to the Galactic plane. 

\begin{figure*}
\centering
\includegraphics[width=0.33\textwidth]{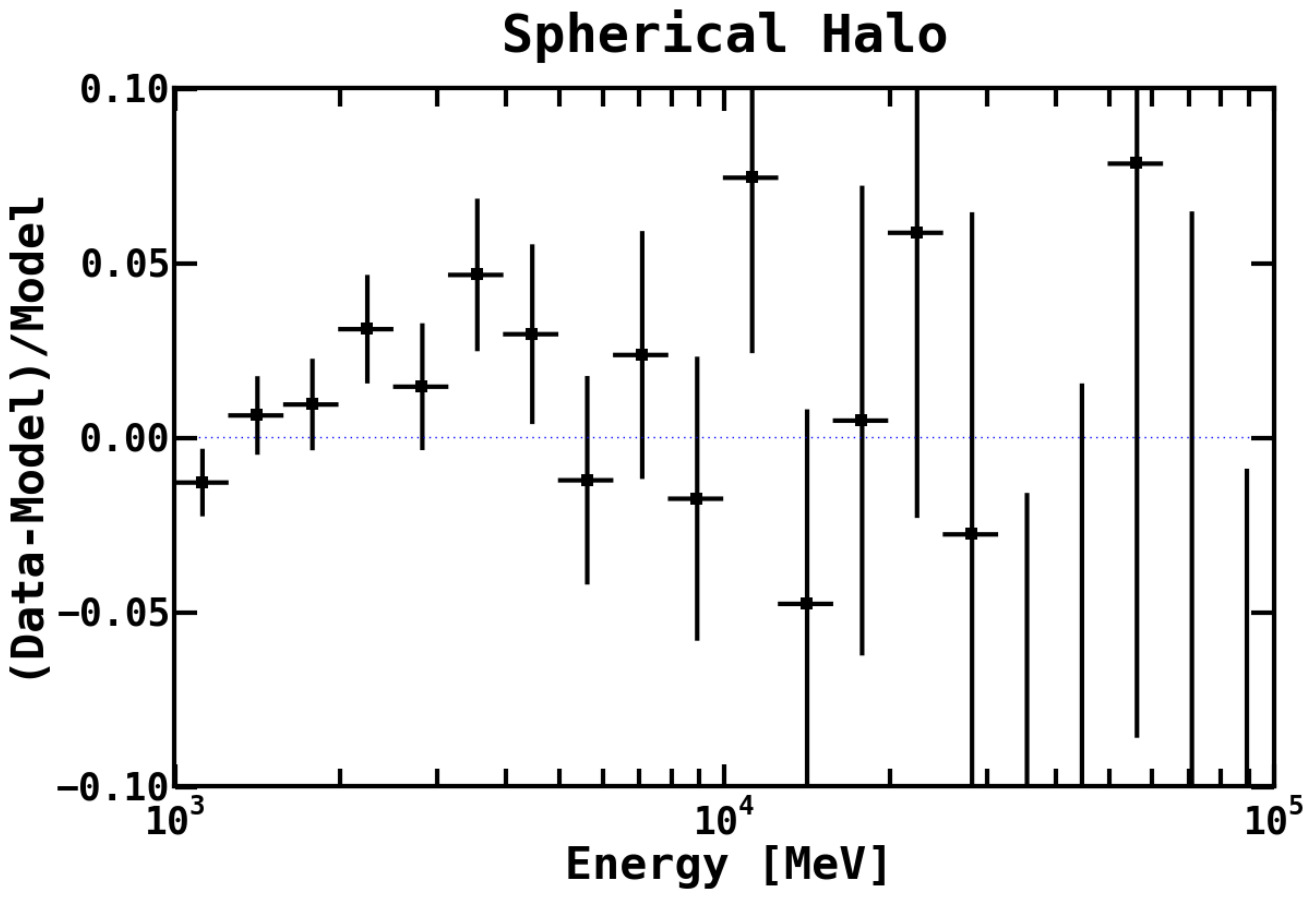}
\includegraphics[width=0.33\textwidth]{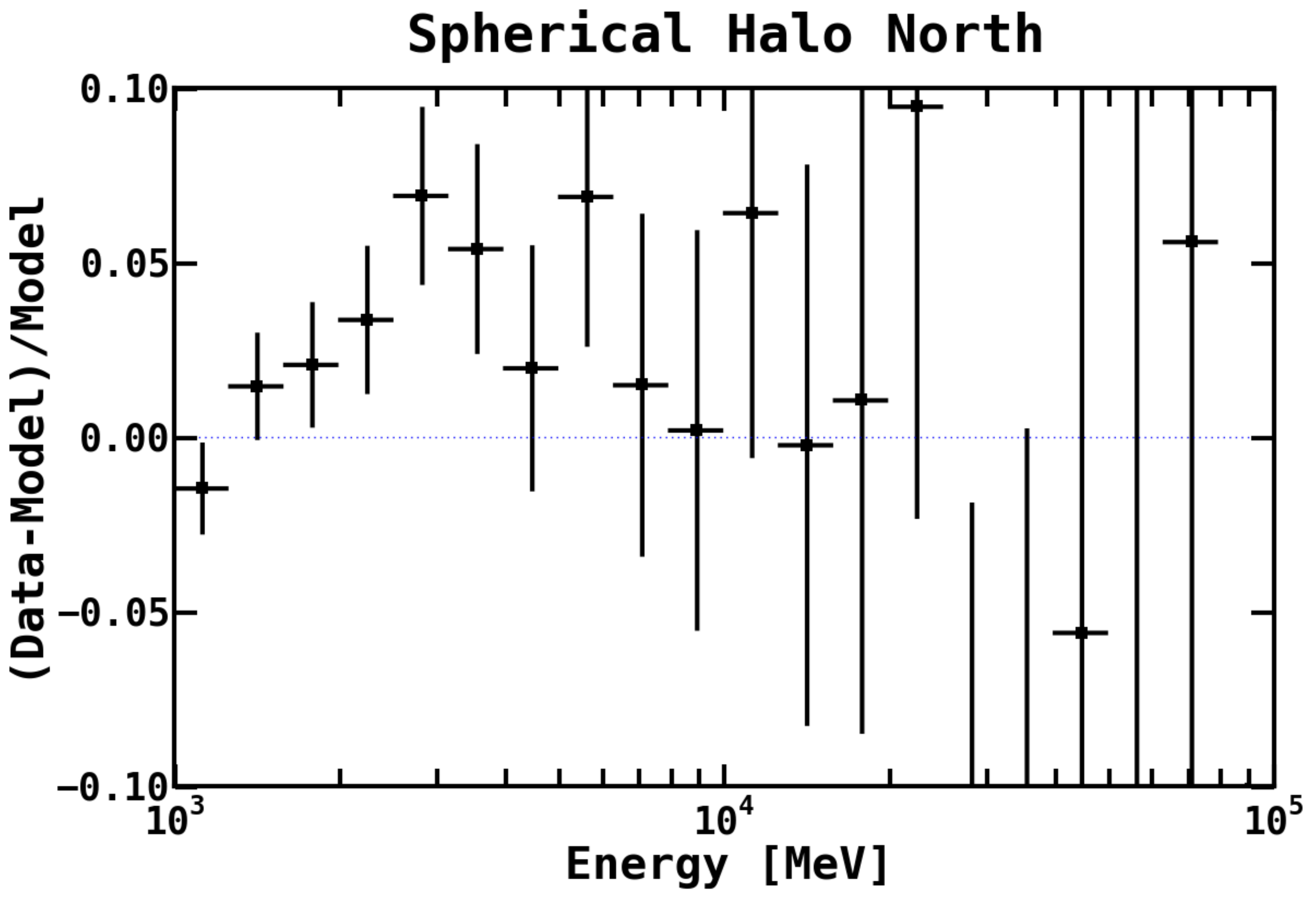}
\includegraphics[width=0.33\textwidth]{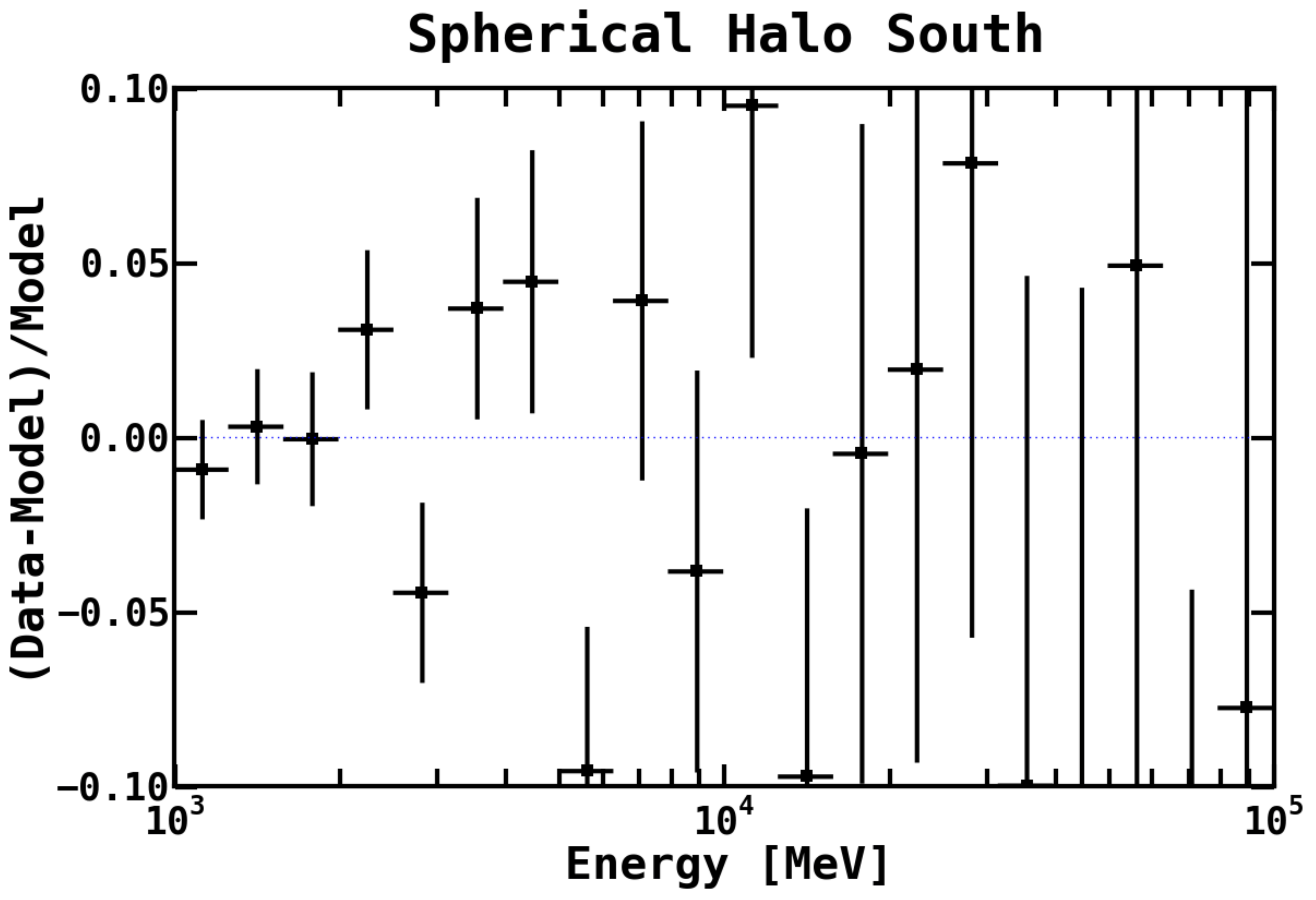}
\includegraphics[width=0.33\textwidth]{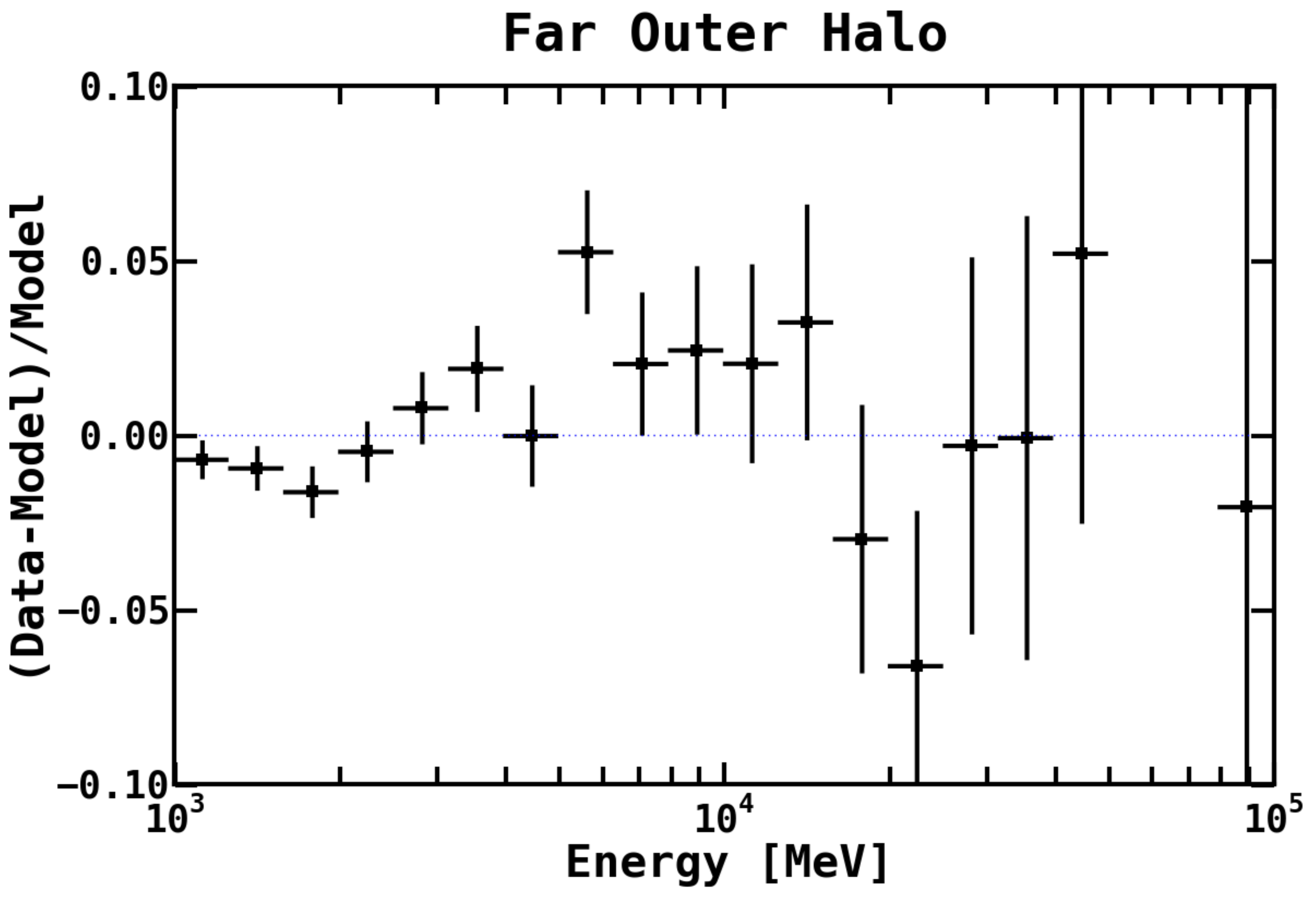}
\includegraphics[width=0.33\textwidth]{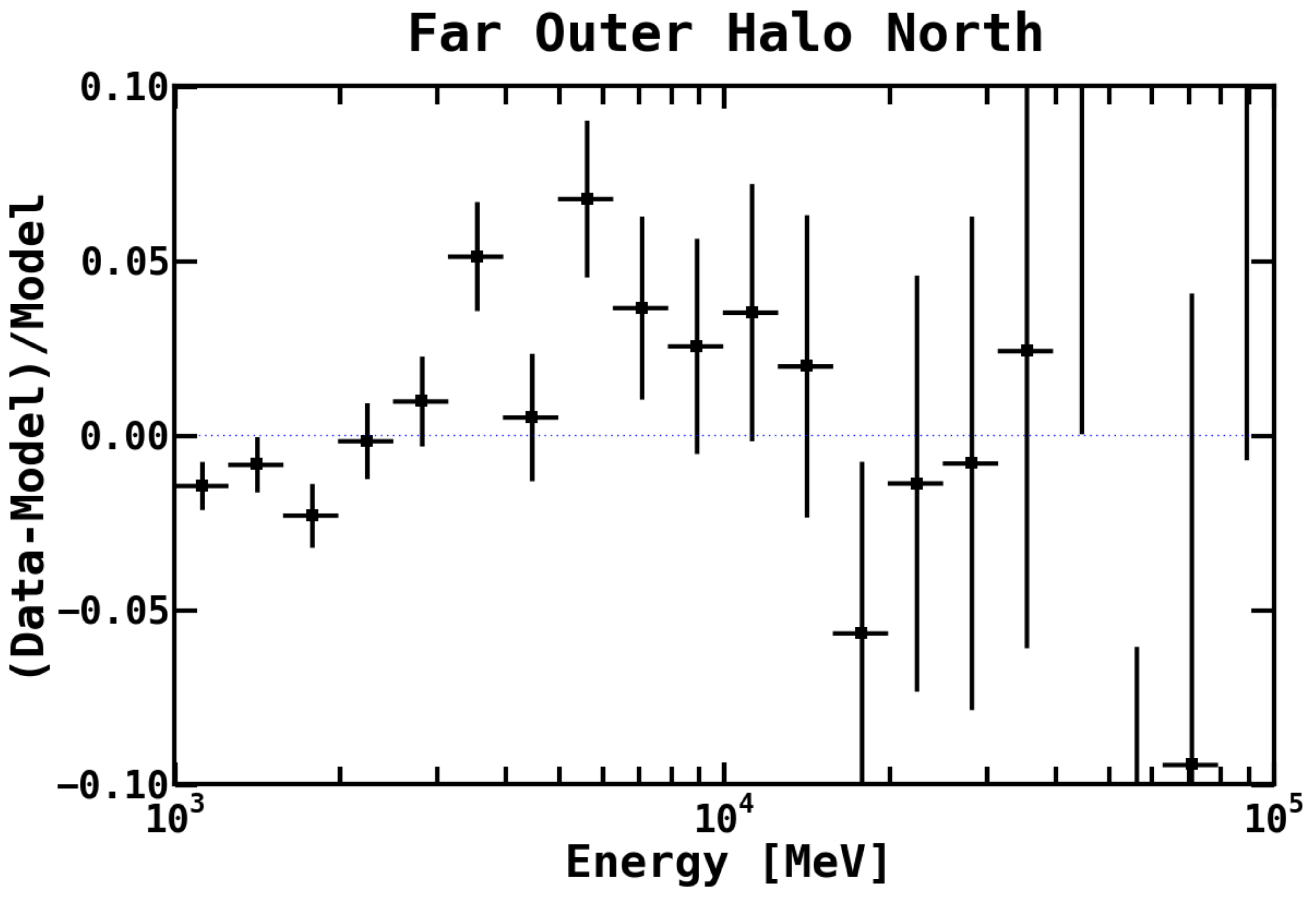}
\includegraphics[width=0.33\textwidth]{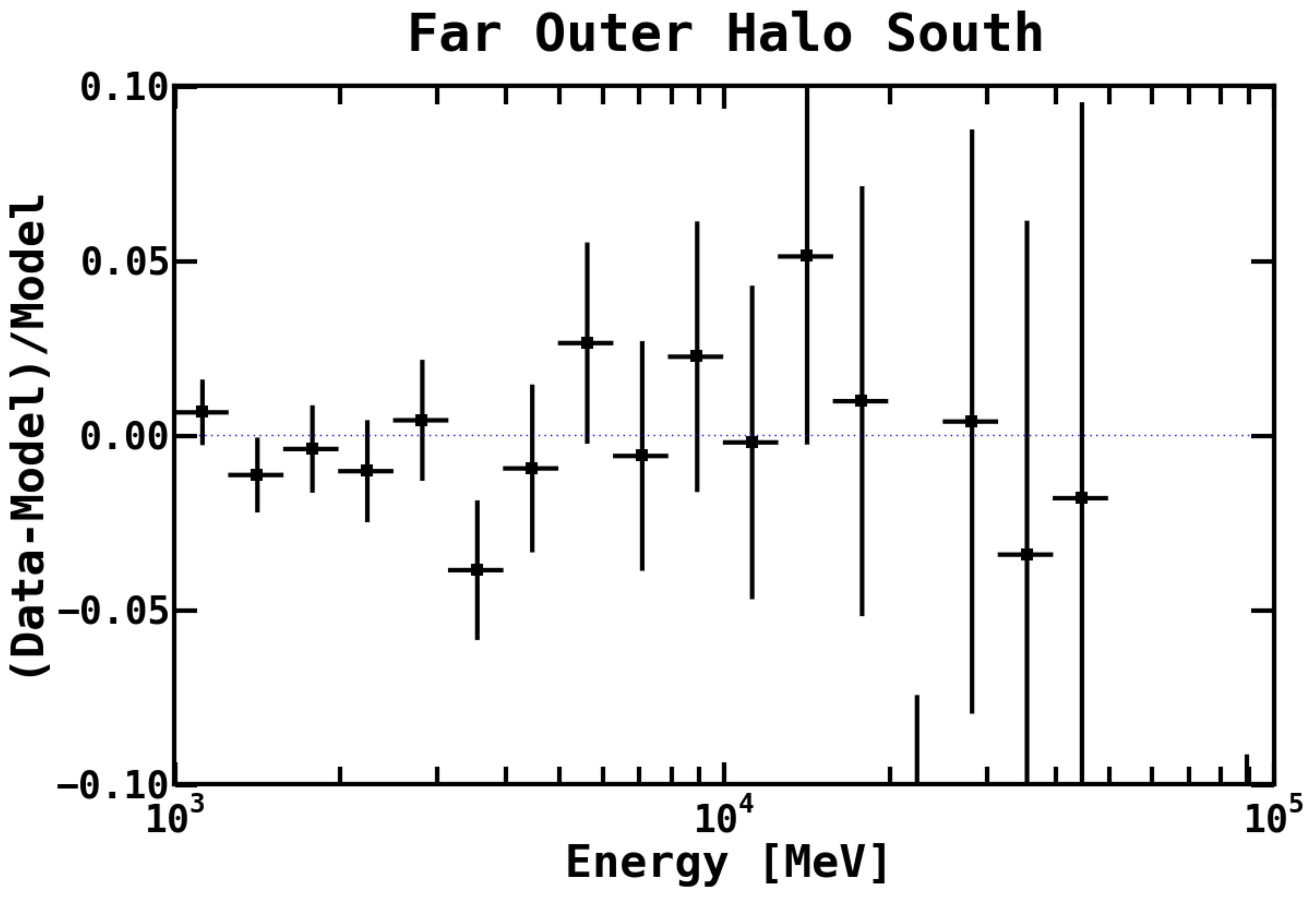}
\caption{The fractional count residuals calculated over the different spatial regions corresponding to the spherical halo and far outer halo components, as indicated above each plot. Note that these are the residuals before adding the M31-related components, and they correspond to the spatial residuals shown in Figure~\ref{fig:M31_Arc_spatial_residuals}, resulting from the baseline fit with the arc north and south templates. The goal here is to further examine the symmetry of the residual emission associated with the M31-related  components. We consider the northern and southern regions of the templates, where the cut is made at the midpoint of FM31 along the horizontal direction (parallel to the Galactic plane), corresponding to a latitude of $-21.5^\circ$. The first column shows the residuals calculated over the entire region, for the spherical halo and far outer halo, respectively. The second column shows the residuals in the north, and the third column shows the residuals in the south.}
\label{fig:M31_fractional_residuals}
\end{figure*}

We first calculate the fractional count residuals in the different regions without fitting any of the M31-related templates. These results are shown in Figure~\ref{fig:M31_fractional_residuals}, and they correspond to the spatial residuals shown in Figure~\ref{fig:M31_Arc_spatial_residuals}, resulting from the baseline fit with the arc north and south templates. The excess can be seen for both the spherical halo and far outer halo regions. For the spherical halo region, the excess appears to be more prominent in the north compared to the south, although it is present in both. For the far outer halo region, the excess is prominent in the north, whereas the residuals in the south are pretty flat. 

We quantify the symmetry of the residual emission by fitting templates for the different regions simultaneously with the other components of the IEM. The M31-related components include the inner galaxy and the northern and southern regions of the spherical halo and far outer halo (5 components in total). Each component is given a PLEXP spectral model, and the spectral parameters are allowed to vary independently (although the components are fit simultaneously). The fit also includes the arc north and south components. Lastly, we scale the diffuse components and point sources in the standard way. 

\begin{figure*}
\centering
\includegraphics[width=0.33\textwidth]{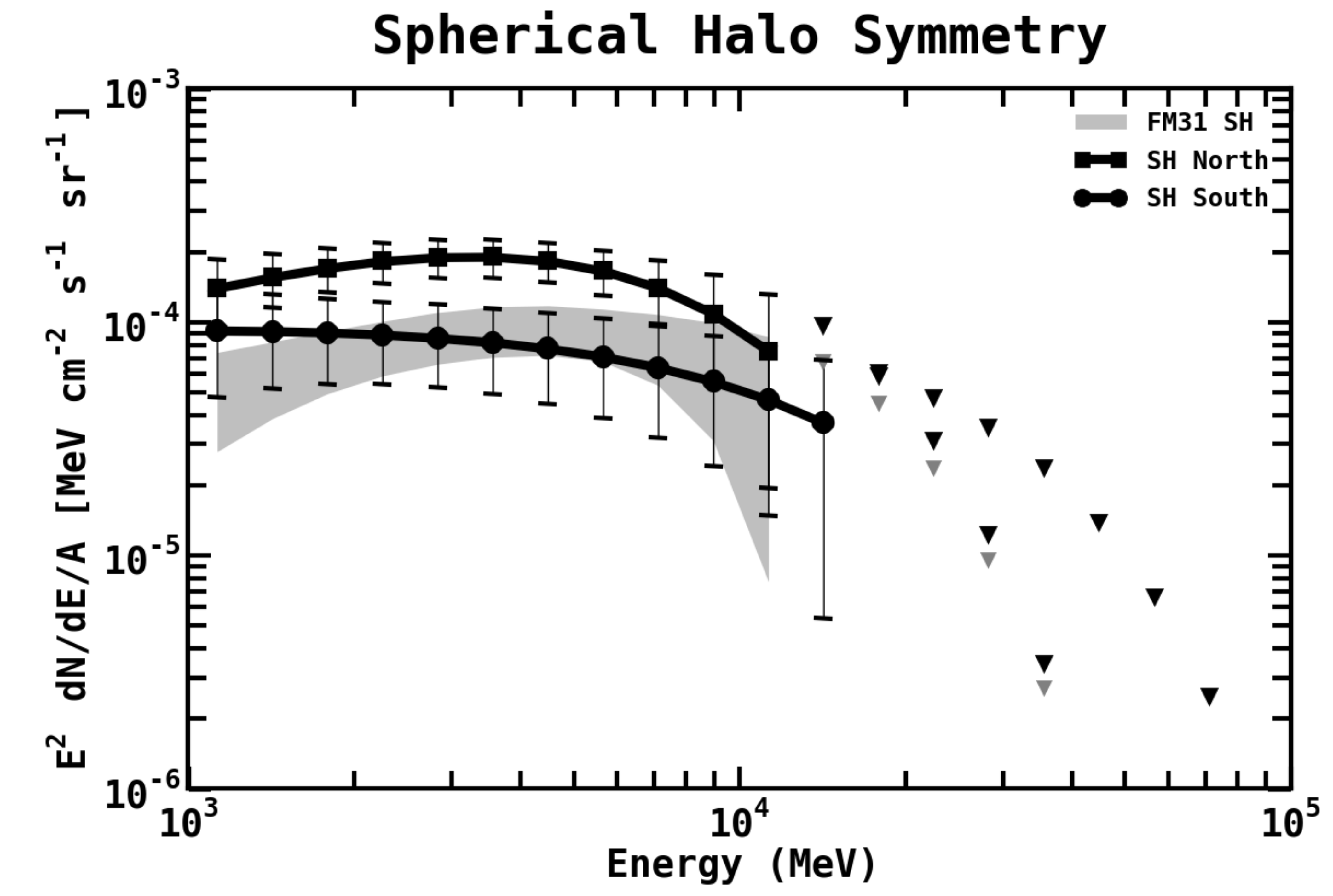}
\includegraphics[width=0.33\textwidth]{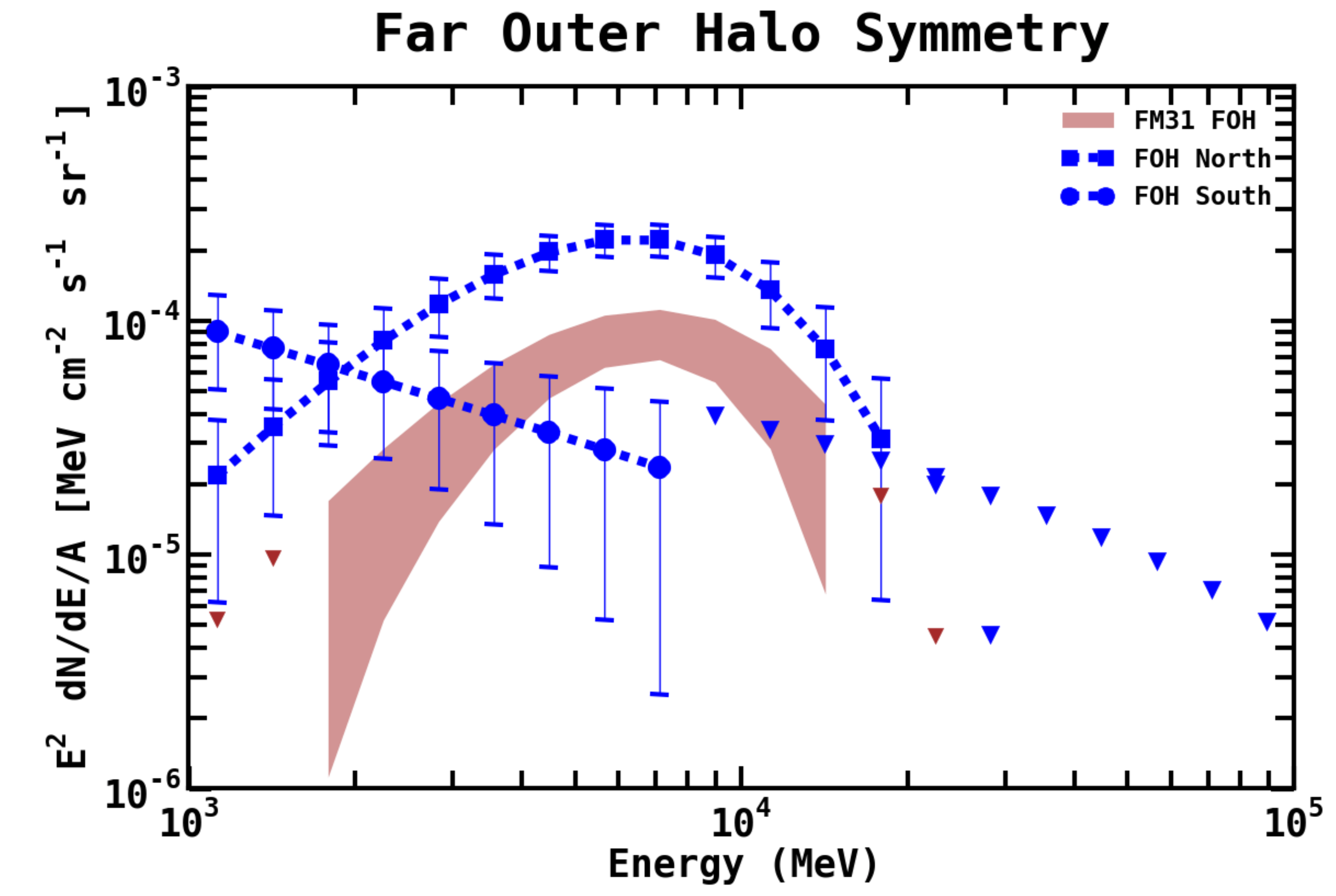}
\includegraphics[width=0.33\textwidth]{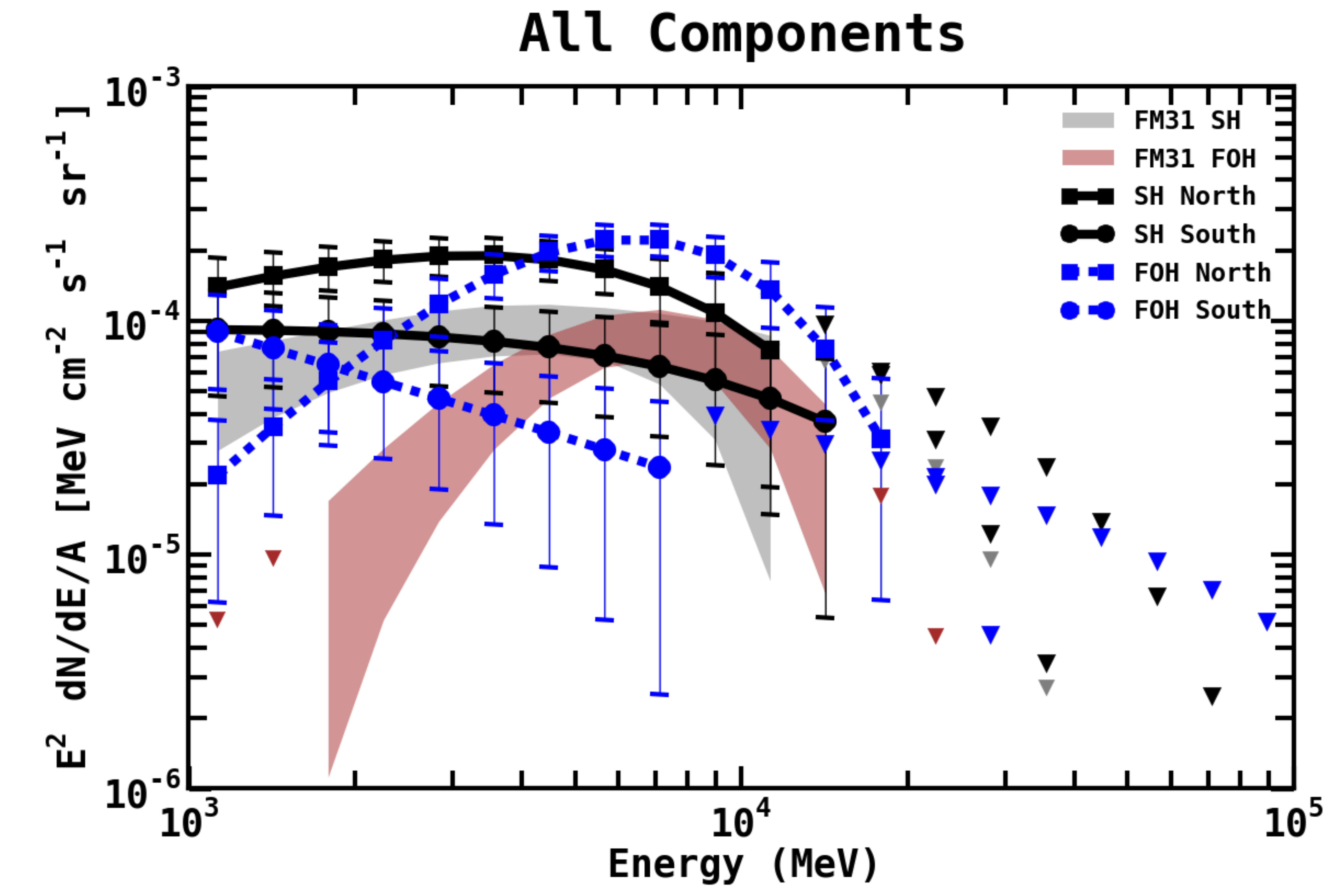}
\caption{The best-fit spectra resulting from the symmetry test fit, where the spherical halo and far outer halo templates are divided into north and south components, and the spectral parameters for each component are allowed to vary independently. The cut is made at the midpoint of FM31 along the horizontal direction (parallel to the Galactic plane), corresponding to a latitude of $-21.5^\circ$. The northern components are shown with square markers, and the southern components are shown with circle markers. Downward pointing triangles give upper limits. Also overlaid are the spectra for the full component fit (with arc north and south), as shown in Figure~\ref{fig:M31_components}.}
\label{fig:M31_symmetry}
\end{figure*}

\begin{figure}
\centering
\includegraphics[width=0.45\textwidth]{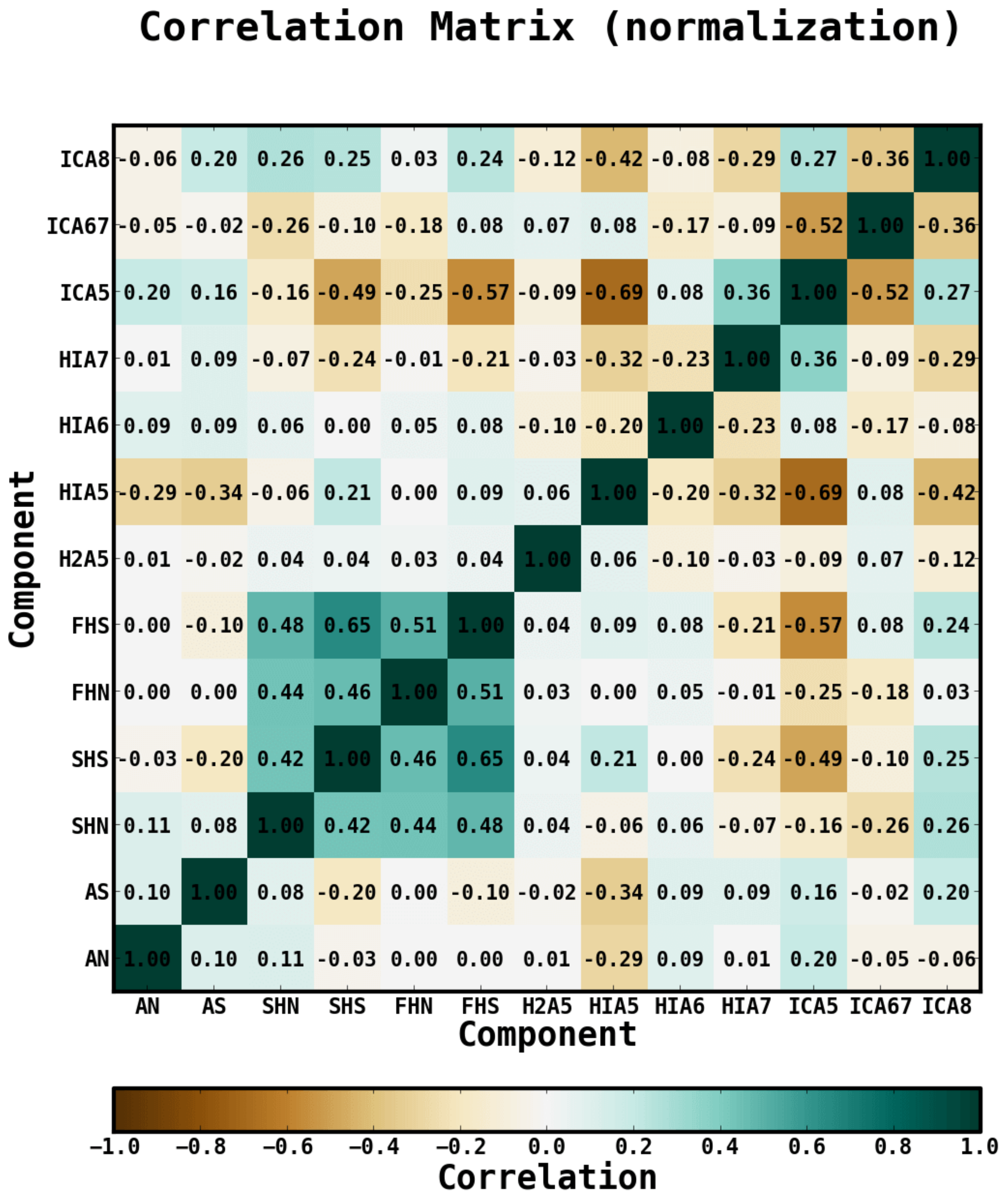}
\caption{Correlation matrix for the symmetry test fit. In addition to the standard components, the fit includes components for the arc north and south (AN and AS), inner galaxy (not shown here), spherical halo north and south (SHN and SHS), and the far outer halo north and south (FHN and FHS).}
\label{fig:symmetry_correlation}
\end{figure}

The resulting spectra for the northern and southern regions of the spherical halo and far outer halo are shown in Figure~\ref{fig:M31_symmetry}. For reference, we also overlay the spectra for the full M31-related components (from Figure~\ref{fig:M31_components}). The spectra for the arc components are very similar to the results shown in Figure~\ref{fig:M31_components}, and therefore we do not show them here. The corresponding best-fit parameters for the halo components are reported in Table~\ref{tab:symmetry_test}. All components are significantly detected (with a significance $>5\sigma$).  

The spherical halo region is slightly brighter in the north than the south. The best-fit spectra for the two components have similar spectral shapes and are qualitatively consistent with that of the full template. We note that we have elected to define north and south with respect to the plane of the MW. However, if the spherical halo component is in fact physically associated with the M31-system, then it may be just as well to cut the two halves with respect to the major axis of M31 ($38^\circ$), which may increase the symmetry between north and south. However, our primary objective here is to simply quantify the gross properties of the residual emission, and a more detailed determination of the morphology is left for a follow-up study. 

The far outer halo region shows a significant spectral variation between the north and south. The northern component has a high spectral curvature, identical to the spectral shape that results when fitting the full template, and is generally brighter than the southern component.

The correlation matrix for the fit is shown in Figure~\ref{fig:symmetry_correlation}. The southern components for both the spherical halo and far outer halo have a stronger anti-correlation with the IEM, compared to the northern components. In particular, the southern components have relatively strong anti-correlations with ICA5 and \hi\ A7. We also note that the southern component of the spherical halo has some anti-correlation with the arc template, whereas the northern component does not. The normalizations of the diffuse components are mostly in agreement with those obtained for the fit with the full M31-related templates. However, the IC A6-A7 component obtains a best-fit normalization of 0.42 \p 0.38, which may not be very physical, and is in contrast to the values obtained for the other fits in FM31. These results highlight one major shortcoming of this test; that is, the northern and southern regions correlate differently with the IEM, and this can potentially lead to inaccuracies regarding the actual symmetry of the tentative signal. This is especially problematic for the excess in FM31, since the corresponding emission lies well below the foreground/background emission.  

The fit with the north and south M31-related templates further shows the importance of the MW modeling and also that the excess is likely to contain a significant MW component. In particular, the excess emission associated with the far outer halo is likely to be related to the MW. Indeed, the Galactic disk region directly above FM31 has many complications, and it is known to contain extended excess $\gamma$-ray emission of unknown origin~\citep{Acero:2016qlg}. In addition, the region (in projection) also contains an extended high-velocity cloud known as Complex H~\citep{hulsbosch1975studies,blitz1999high,Lockman:2003zs,Simon:2005vh}, which has been postulated to be either a dark galaxy of the Local Group or an example of a cold accretion flow onto the MW~\citep{Simon:2005vh}. Here we only point out a couple of these associated difficulties, but our primary goal is to quantify the rough properties of the excess emission. 

A portion of the excess emission is also likely related to the M31 system, and in particular, the emission associated with the spherical halo region. We note that of the four halo components, the overall intensity is highest for the northern spherical halo. Given the significant modeling uncertainties, we make the simplifying conclusion that the excess emission in FM31 is significantly detected and has a total radial extension upwards of $\sim$120--200 kpc from the center of M31. The lower limit corresponds to the boundary of the spherical halo, and the upper limit corresponds to the boundary of the far outer halo. This conclusion encapsulates the possibility that the excess emission may have contributions from both M31 and the MW, and it also refers to the emission associated with the arc template, the nature of which remains unclear. 

\section{The Smooth Component of the Residual Emission in FM31 and Dark Matter} \label{sec:smooth_residual_emission}

The dominant component of the residual emission in FM31 has a total radial extension upwards of $\sim$120--200 kpc from the center of M31, corresponding to the excess between $\sim$3--20 GeV in the fractional count residuals. It is plausible that a portion of the signal may be related to M31's DM halo. In general, the exact properties of M31's DM halo remain highly uncertain, i.e.\ the geometry, extent, and substructure content. Here we make some simplifying assumptions to get a rough sense of the consistency between the observed signal and a possible DM interpretation. In particular, we check for consistency with the DM interpretation of the excess $\gamma$-ray emission observed in the Galactic center~\citep{Goodenough:2009gk, Hooper:2010mq,Hooper:2011ti, Abazajian:2012pn,Hooper:2013rwa,Gordon:2013vta,Huang:2013pda,Abazajian:2014fta,Zhou:2014lva,Calore:2014xka,Abazajian:2014hsa,Calore:2014nla,Huang:2015rlu,TheFermi-LAT:2015kwa,Daylan:2014rsa,Carlson:2016iis,Fermi-LAT:2017yoi,Karwin:2016tsw,TheFermi-LAT:2017vmf,agrawal2017point}. This by no means encompasses all possibilities, and more detailed evaluations are left for future studies. 

In addition to M31's DM halo, we also consider the contribution from the MW's DM halo along the line of sight, since this component has not been explicitly accounted for in our analysis. If such a component actually exists, then it may be at least partially absorbed by the isotropic component, as well as the other components of the IEM, but it will not necessarily be fully absorbed, and a portion of such a signal could be contained in the M31-related components.

\begin{figure*}[tbh!]
\centering
\includegraphics[width=0.49\textwidth]{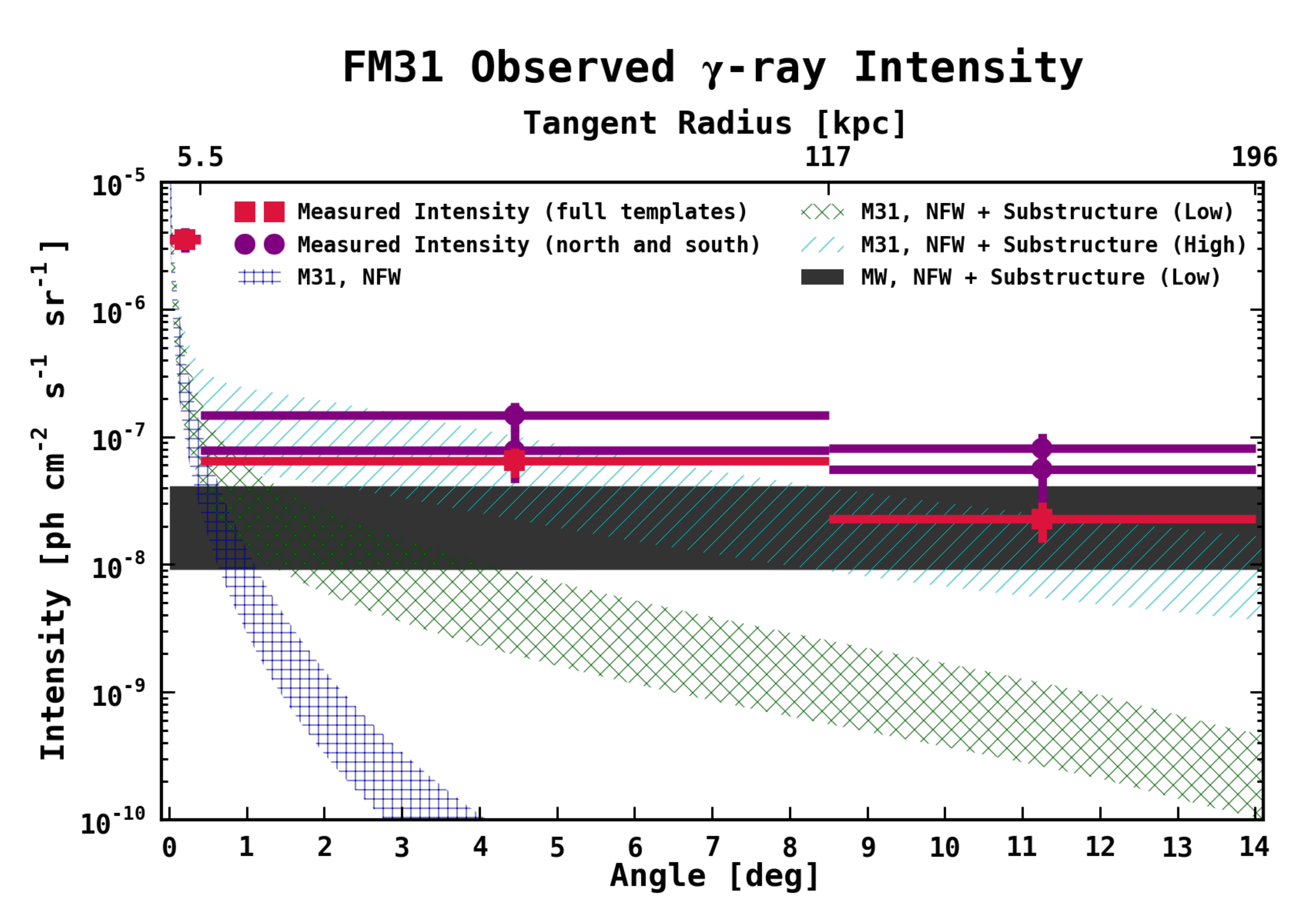}
\includegraphics[width=0.49\textwidth]{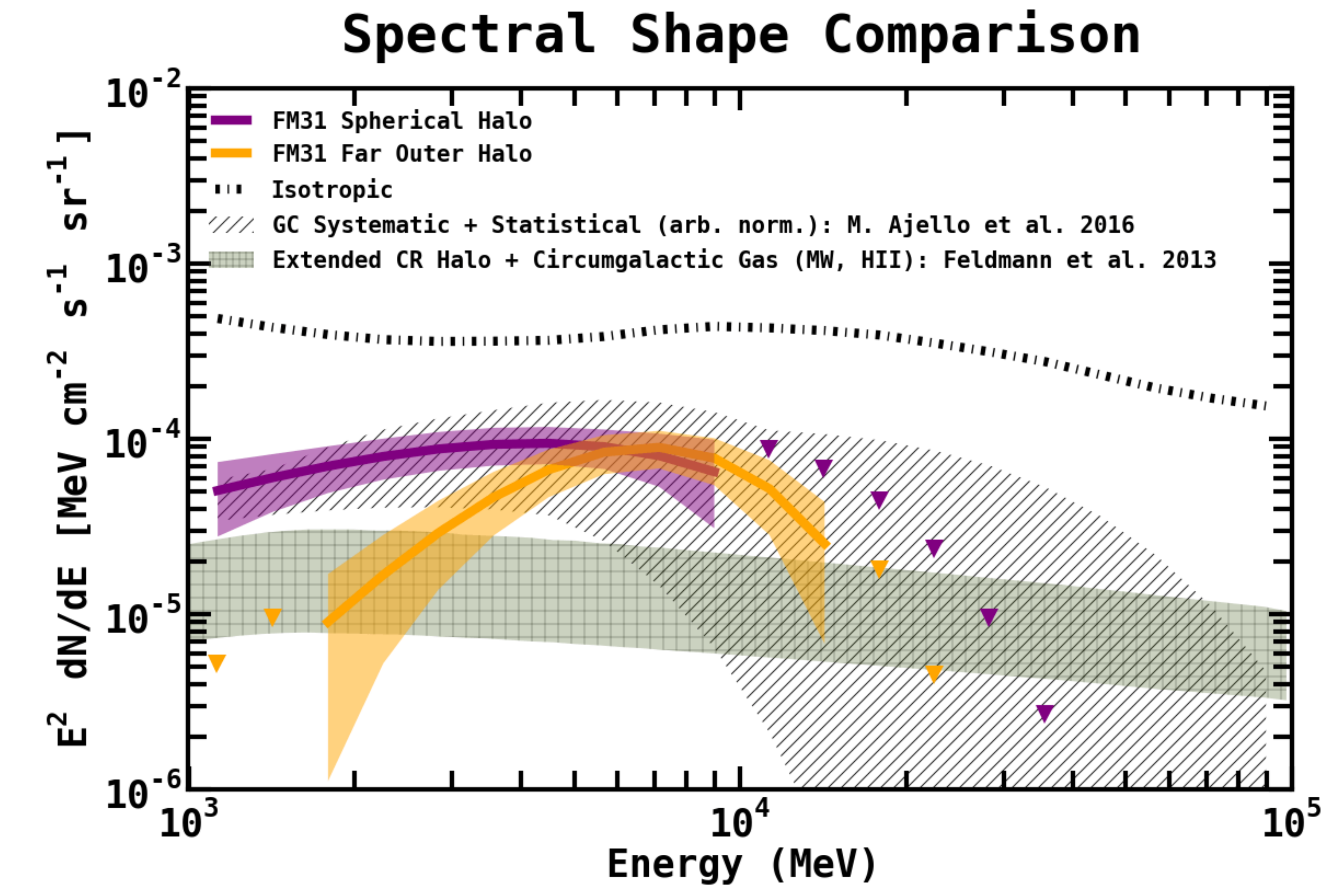}
\caption{\textbf{Left:} Radial intensity profile for the M31-related components. Red square markers show the results from the north and south arc template with PLEXP. The profiles for the PL arc fit are basically the same. Purple circle markers show the results from the fit with the M31-related templates divided into north and south components (from Figure~\ref{fig:M31_symmetry}). For reference, we compare the radial profile to expectations for DM annihilation in the line of sight. Note that this also includes the contribution from the MW's DM halo in the line of sight, which has not been accounted for in our analysis, and may be at least partially embedded in the isotropic component and Galactic diffuse components. Likewise, the M31-related components may contain a significant contribution from the MW's extended halo. Details regarding the DM profiles are given in Appendix~\ref{sec:DM}. \textbf{Right:} Spectral shape comparison to the Galactic center excess (for an arbitrary normalization), as observed in~\citet{TheFermi-LAT:2015kwa}. Also shown is a prediction for CRs interacting with the ionized gas of the circumgalactic medium from~\citet{Feldmann:2012rx}. Note that the prediction is for a MW component, but we are primarily interested in a spectral shape comparison. 
}
\label{fig:M31_radial_profile}
\end{figure*}

The left panel of Figure~\ref{fig:M31_radial_profile} shows the radial profile of the $\gamma$-ray intensity for the M31-related components. Red square markers show the fit with the full M31-related templates, including the arc north and south with PLEXP. Purple circle markers show the fit with the M31-related templates divided into north and south components (from Figure~\ref{fig:M31_symmetry}). The individual intensities of the divided north and south components are a bit higher than the intensity of the combined template because of the different correlation of the tentative signal in these regions with the IEM components (see Figure~\ref{fig:symmetry_correlation} and the corresponding discussion in Section~\ref{sec:symmetry}). The intensity of the M31-related emission is far less in the outer regions than it is towards the inner galaxy. Furthermore, the signal is not detected in the TR. This is consistent with the hypothesis that the emission originates (at least partially) from the M31 system. 

In the figure, we compare the radial dependence of the observed intensity to the predicted intensity for a DM signal. Plots of the corresponding $J$-factors and a description of all parameters for the predicted $\gamma$-ray flux due to DM annihilation are given in Appendix~\ref{sec:DM}. For the \emph{DM attribute quantity}, Eq.~(\ref{eq1}),
we use the best-fit values as determined from the GC excess in~\citet{Karwin:2016tsw}. The uncertainty bands for each of the three intensity profiles come from the uncertainty in the \emph{DM attribute quantity} (as described in Appendix~\ref{sec:DM}). The black band shows the corresponding intensity profile for the MW DM component along the line of sight. Note that in general there is also expected to be an additional contribution from the local DM filament between M31 and the MW.

We find that the radial intensity profile of the positive residual emission in FM31 is roughly consistent with a cold DM scenario that includes a large boost factor due to substructures. Granted, however, the exact partitioning of individual contributions to the signal remains unclear, i.e.\ primary emission from M31's DM halo, secondary emission in M31, emission from the local DM filament between M31 and the MW, and emission from the MW's DM halo along the line of sight. We note that for the radial intensity profile in Figure~\ref{fig:M31_radial_profile} we have not included a MW prediction for the high substructure model. Our main intention here is just to get a rough sense of the consistency with a DM interpretation, but in general the MW substructure high prediction would also be relevant, and it would imply that a significant portion of the MW halo signal would need to be almost fully absorbed by the isotropic component and other components of the IEM.

We again stress that the properties of the excess emission observed towards the outer halo have a strong dependence on the modeling of the IEM. This is partially reflected by the large uncertainty in the radial profile between the two different fit variations, as can be seen in the left panel of Figure~\ref{fig:M31_radial_profile}. We also stress that the excess  in FM31 is likely to contain a significant contribution from the MW. In particular, the emission associated with the far outer halo is more likely to be related to the MW than the M31 system. But still, the nature of this emission remains unclear.  

The right panel in Figure~\ref{fig:M31_radial_profile} shows a spectral shape comparison with the excess emission observed in the Galactic center~\citep{TheFermi-LAT:2015kwa}. The band for the Galactic center excess shows the systematic $+$ statistical uncertainty (although it is dominated by the systematics), and it is shown for an arbitrary normalization. We find that the spectra of the M31-related components are qualitatively consistent with the uncertainty band of the Galactic center excess. We note that the spectrum of the far outer halo component has a higher curvature at low energies. If this is indeed a real feature of the signal (and not just a systematic effect), then it could be related to secondary processes. If the DM produces some fraction of leptons, then the leptons may generate secondary $\gamma$-ray emission from IC and Bremsstrahlung, due to interactions with the interstellar radiation fields and gas~\citep{Cirelli:2013mqa,Lacroix:2014eea,Abazajian:2014hsa}. For M31, the secondary emission may have a dependence on the radial distance from the center of M31, since the stellar halo and gaseous halo also have a radial dependence. However, this possibility would need to be quantified to get a better sense of the effect.

Also plotted in the right panel of Figure~\ref{fig:M31_radial_profile} is the isotropic component. The intensity of the M31-related components is below that of the isotropic component by a factor of $\sim$5. There is a bump in the isotropic spectrum around $\sim$10 GeV (as is more clearly seen in Figure~\ref{fig:Isotropic_Sytematics}), and this energy also somewhat corresponds to the peak emission of the M31-related components. This might suggest that the isotropic emission may include a contribution that originates from similar processes in the extended halo of the MW. As it pertains to DM in particular, this issue is significantly complicated and is beyond the scope of this work, but related discussions can be found in \citet{Cuoco:2010jb}, \citet{Cholis:2013ena}, \citet{Fornasa:2015qua}, \citet{Ajello:2015mfa}, and \citet{Ackermann:2015tah}.

\begin{figure*}[tbh!]
\centering
\includegraphics[width=0.33\textwidth]{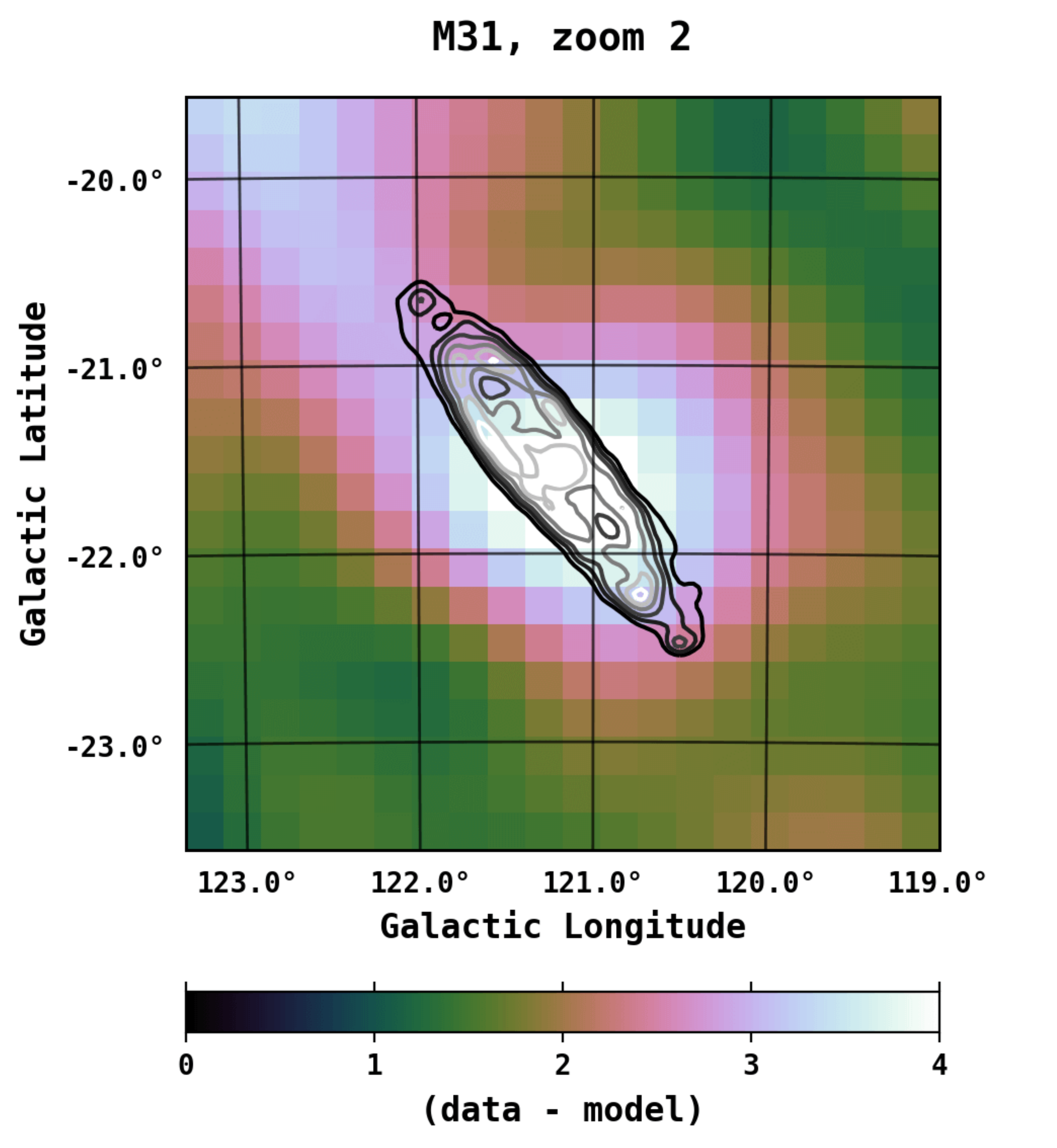}
\includegraphics[width=0.33\textwidth]{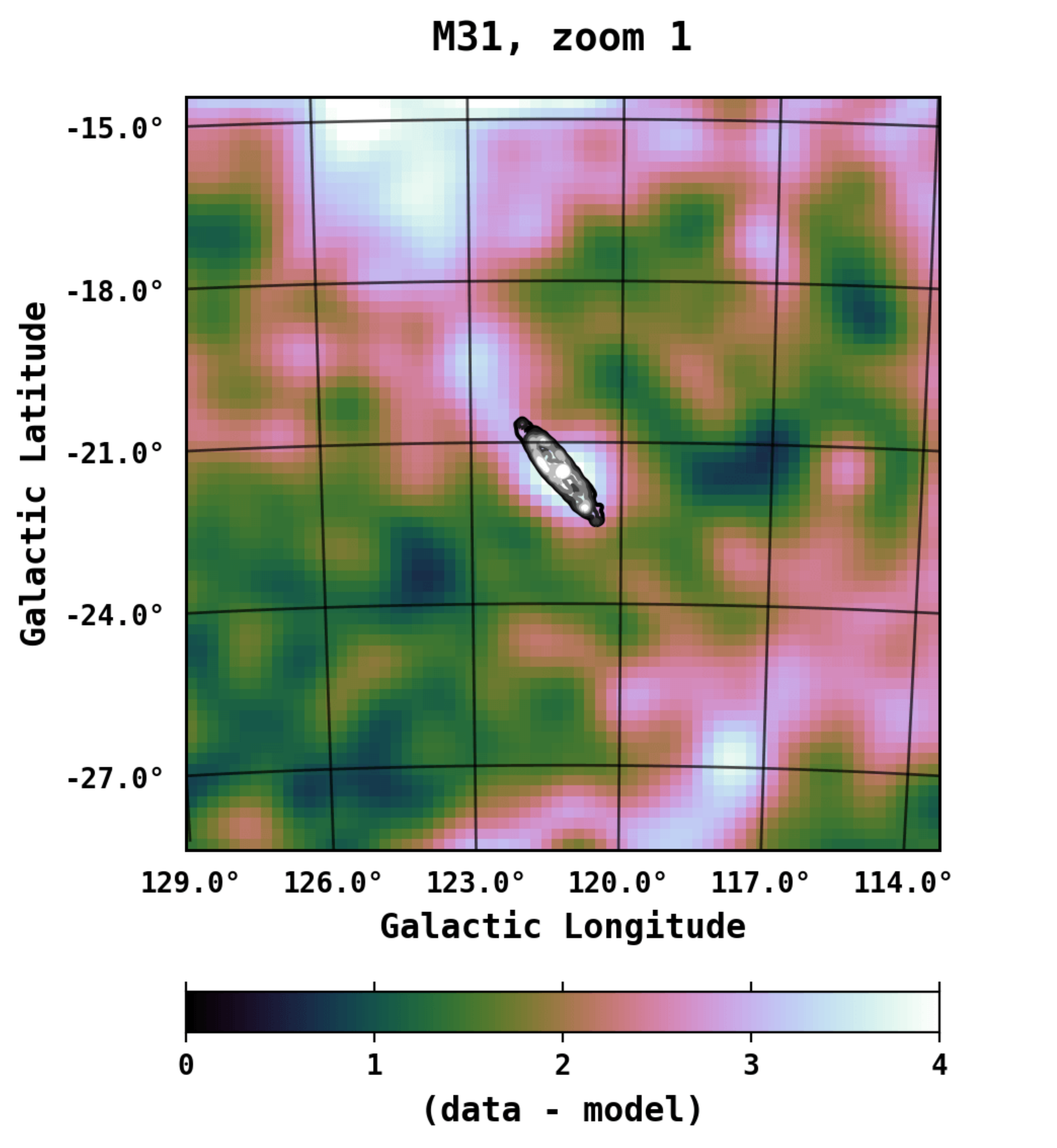}
\includegraphics[width=0.33\textwidth]{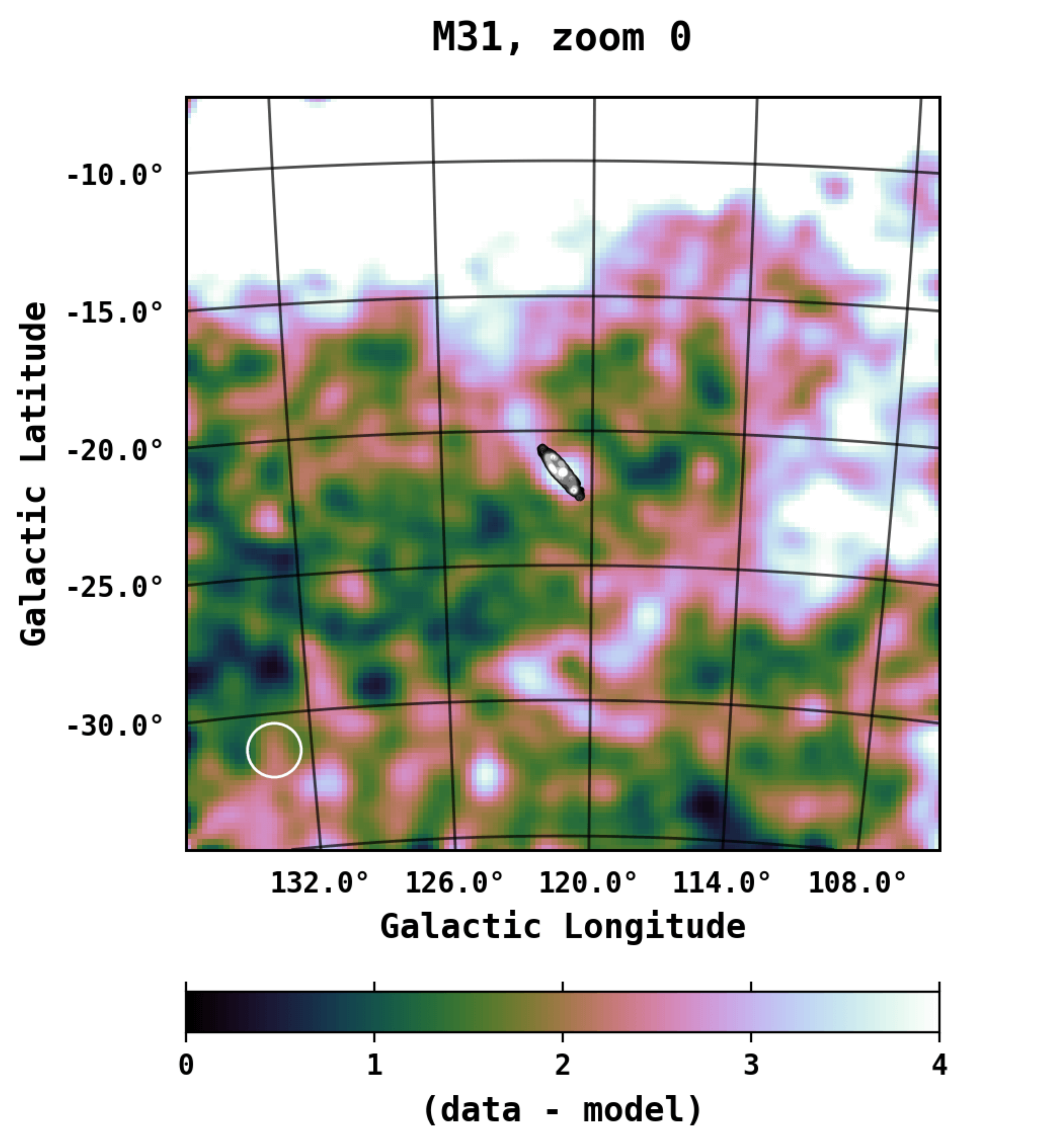}
\includegraphics[width=0.33\textwidth]{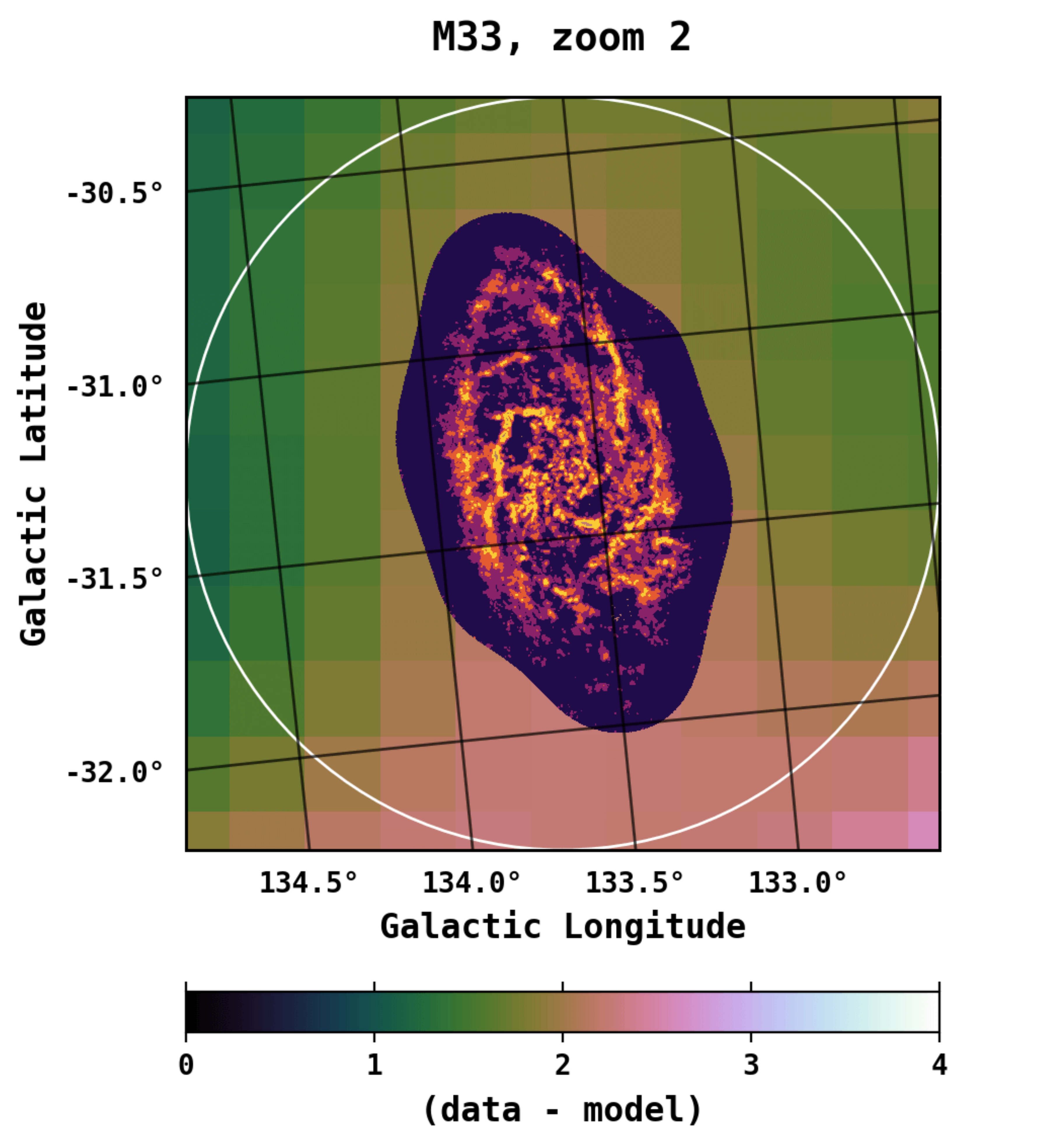}
\includegraphics[width=0.33\textwidth]{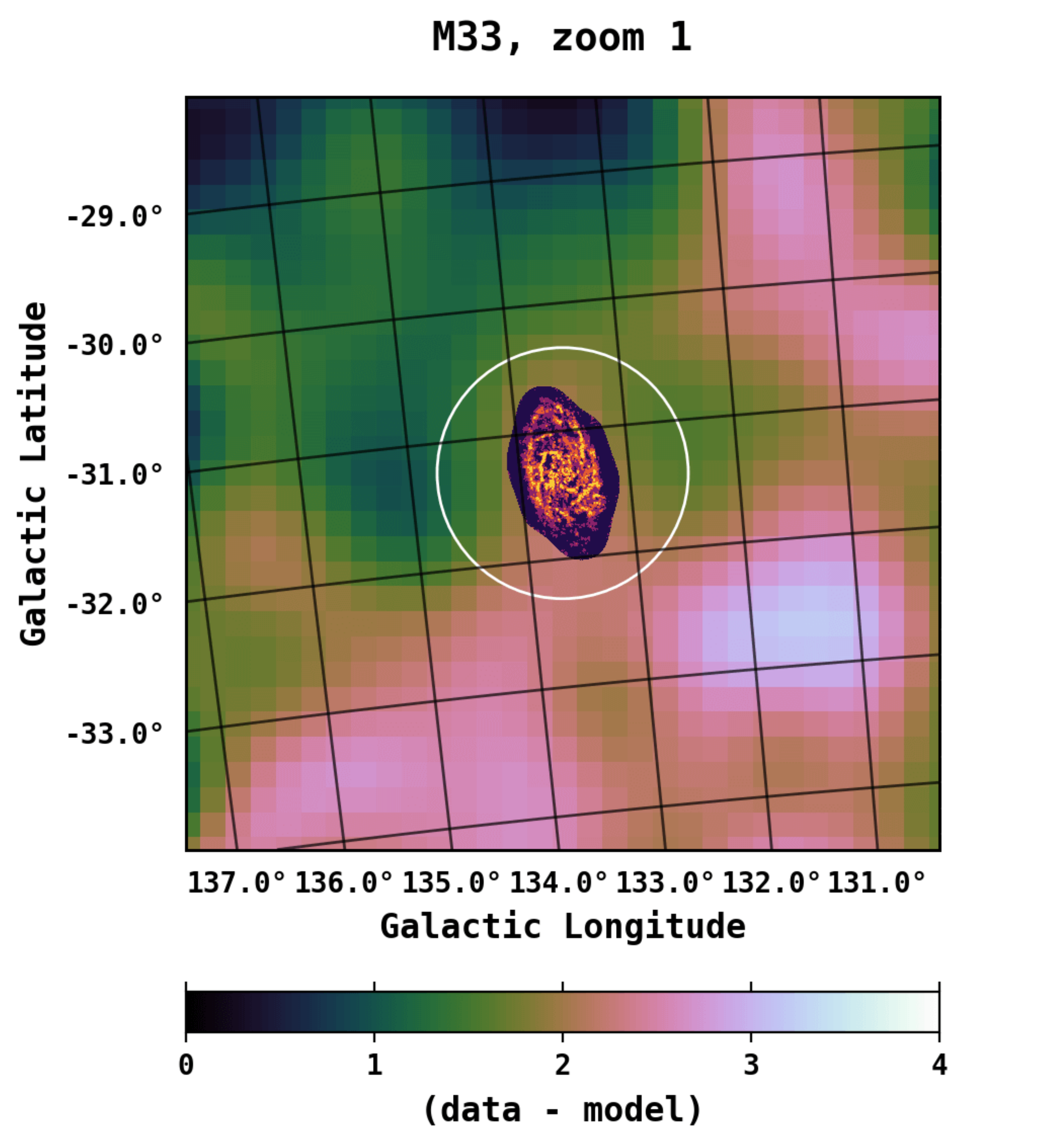}
\includegraphics[width=0.33\textwidth]{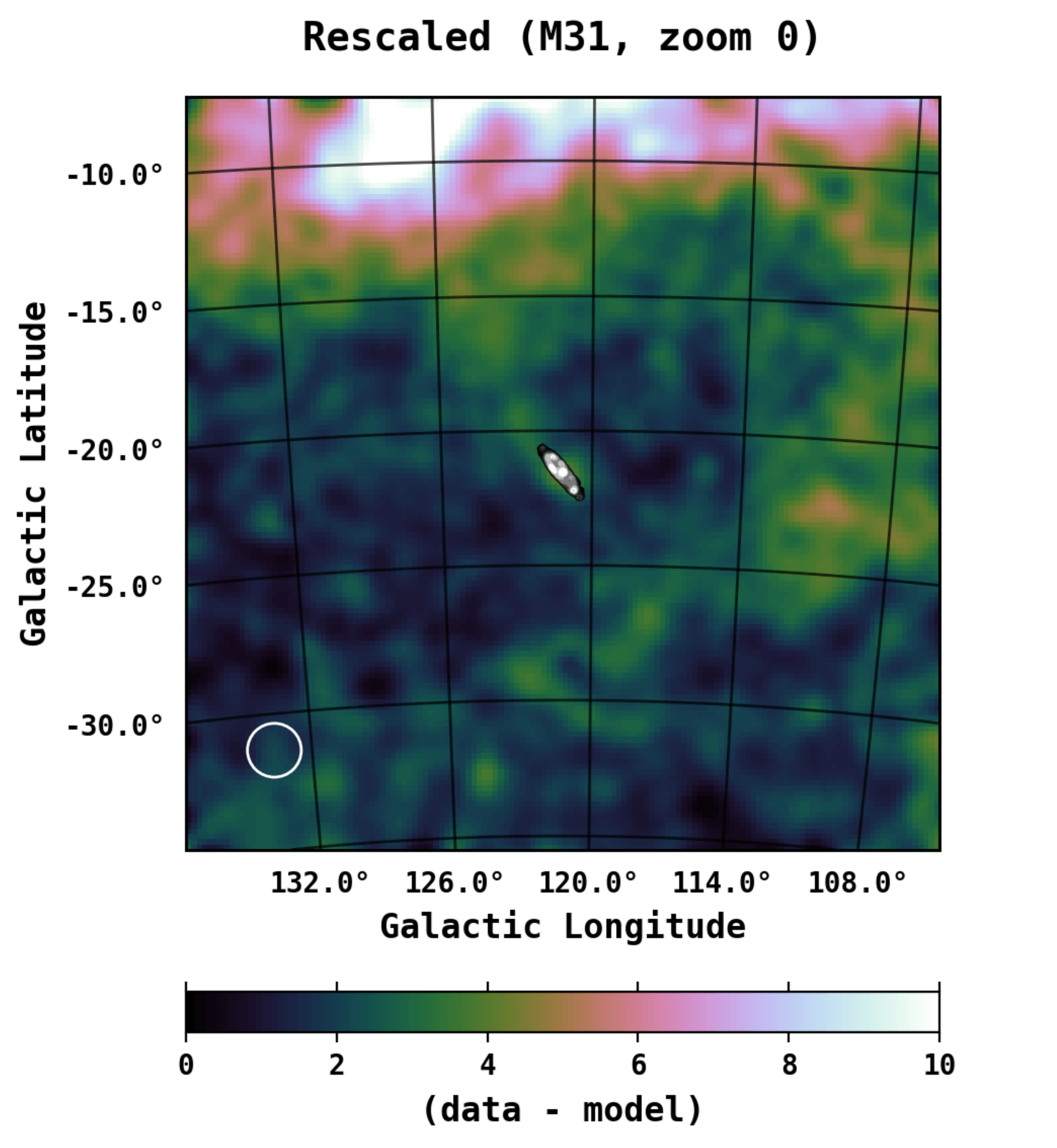}
\caption{Residual maps showing the structured emission integrated in the energy range 1--100 GeV. The color scale corresponds to counts/pixel, and the pixel size is $0.2^\circ \times 0.2^\circ$. The images are smoothed using a $1^\circ$ Gaussian kernel. This value corresponds to the PSF (68\% containment angle) of \textit{Fermi}-LAT, which at 1 GeV is $\sim$$1^\circ$. Maps are shown in the cubehelix color scheme~\citep{green2011colour}. In the top row contours for the IRIS 100 $\mu$m map of M31 are overlaid, and three zoom levels ($2^\circ$, $7^\circ$, full field) centered at M31 are shown. The white circle ($1^\circ$) shows the position of M33. The bottom row shows two zoom levels ($1^\circ$, $3^\circ$) centered at M33, and the \hi\ integrated intensity map (units of K) of M33 is overlaid. In the third panel we show the M31 zoom 0 map rescaled, in order to provide a sense of the relative intensity towards the MW disk. \textbf{We stress that these maps have not subtracted any Galactic \hi-related emission.}}
\label{fig:positive_residuals_full}
\end{figure*}

\begin{figure}[tbh!]
\centering
\includegraphics[width=0.45\textwidth]{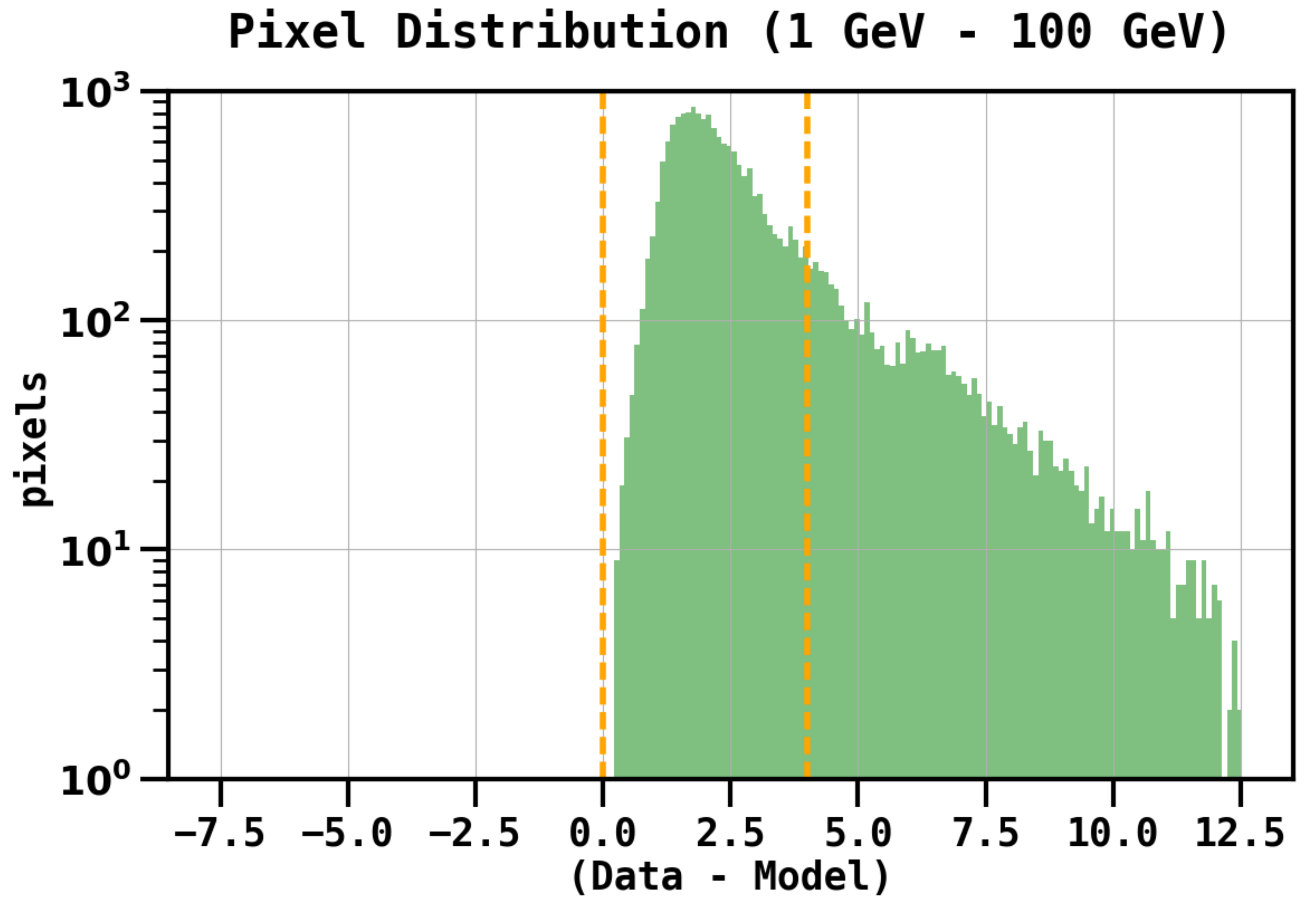}
\caption{Pixel distribution of the smoothed residual map (1 GeV -- 100 GeV) after removing the \hi-related components, as shown in Figure~\ref{fig:positive_residuals_full}. The yellow dashed lines are at 0 and 4 counts.}
\label{fig:map_detatils}
\end{figure}

\begin{figure}[tb!]
\centering
\includegraphics[width=0.49\textwidth]{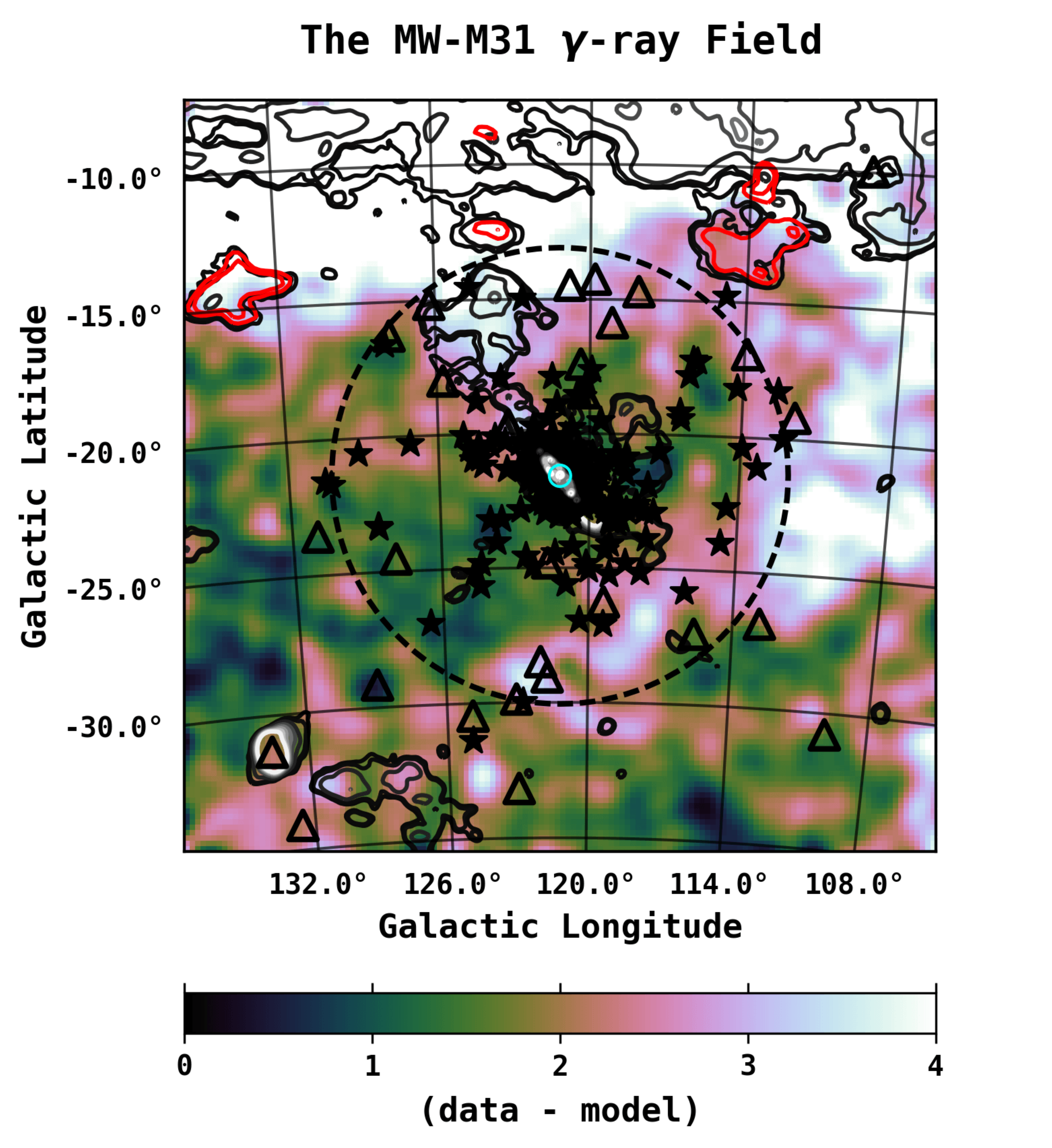}
\caption{To the structured $\gamma$-ray emission in FM31 we overlay some M31-related observations from other wavelengths. We stress that this is only done as a qualitative gauge of M31's outer halo. In the figure we have not subtracted any Galactic \hi-related emission, and we do not expect the M31-related observations to outshine the MW emission, as discussed in the text. Contours for the IRIS 100 $\mu$m map of M31 are overlaid. The solid cyan circle ($0.4^\circ$) shows the boundary of the FM31 inner galaxy component, and the black dashed circle ($8.5^\circ$) shows the outer boundary of the FM31 spherical halo component, as detailed in Section~\ref{sec:M31_components}. \hi\ emission contours from the HI4PI all-sky survey (based on EBHIS and GASS)~\citep{bekhti2016hi4pi}, integrated over the velocity range $-600\ \mathrm{km \ s^{-1}} \leq V_{\rm LSR} \leq -95\ \mathrm{km \ s^{-1}}$ are overlaid. M31's confirmed globular clusters are shown with black stars. M31's population of dwarf galaxies is shown with open black triangles. The M31 cloud can be seen (although obscured by globular clusters). We note the serendipitous enclosure by the spherical halo of the M31 cloud, as well as a majority of M31's globular cluster population and dwarf galaxies. \hi\ contours corresponding to M33 can be seen in the lower-left corner. The hook-shaped gas cloud to the right of M33 is Wright's cloud. The red gas contours towards the top of the map are clouds of Complex H. The black \hi\ contours towards the top of the field correspond to the plane of the MW, and likewise for the bright (white) $\gamma$-ray emission. To the far right of the field a bright arm of emission extends to higher latitudes. Although not considered when making the overlay, the M31-related observations can be seen to trace the left boundary of the arm. This may be an observational bias, due to foreground gas and dust. 
  \textbf{We stress that these maps have not subtracted any Galactic \hi-related emission.}}
\label{fig:FM31_rich}
\end{figure}

We note that the DM could be decaying (see~\citet{Blanco:2018esa} and references therein). In this case the $\gamma$-ray signal would be morphologically more consistent with the excess observed in FM31 without requiring a large boost from substructures, since it scales as the DM density, as opposed to the square of the density for annihilation. Here we restrict the interpretation to annihilating DM, also in the context of the GC excess. We leave a more complete DM study, including decaying DM, to a followup work.

We also note that aside from DM, another possible interpretation of the signal, if it truly originates from the M31 system, would be that it arises from CR interactions with the ionized gas of M31's circumgalactic medium. We do not rule out this possibility; however, if the emission is dominated by CR interactions with the ionized gas, then this would imply that the CR spectrum and distribution are significantly different in M31's outer galaxy than that measured locally in the MW. 

Additionally, the observed intensity of the M31-related components would imply a relatively high emissivity in M31's outer regions compared to the local MW measurements. However, from a study of the $\gamma$-ray emission from a sample of high velocity clouds and intermediate velocity clouds in the halo of the MW, \citet{Tibaldo:2015ooa} concluded that the $\gamma$-ray emissivity per H atom of the clouds decreases as a function of distance from the disk, with indications of a $\sim$50\%-80\% decline of the CR density within a few kpc. 

Likewise, from an analytical study of the MW, \citet{Feldmann:2012rx} estimate that the CR density in the outer halo may be up to 10\% of that found in the disk. Their predicted \gray\ spectrum is shown in Figure~\ref{fig:M31_radial_profile}, right panel, with a green forward-hatch band. Note that the predicted intensity level in their model is based on the prediction for a MW signal, but we are mostly interested in a spectral shape comparison. The study in~\citet{Feldmann:2012rx} uses a distribution of \hii\ gas derived using a high resolution hydrodynamical simulation, along with reasonable estimates for the distribution of CRs in the outer halo of the MW. The spatial extent of the CR halo is the greatest modeling uncertainty. The two CR distributions used in their calculation fall to half of their density (not including the density within the disk itself) by 60 kpc and 360 kpc, respectively. These distributions define their uncertainty band in the figure. 

Considering the radial extent, spectral shape, and intensity of the M31-related components, it is seemingly unlikely that the corresponding emission is dominated by CR interactions with the ionized gas of M31's circumgalactic medium.

\section{The Structured $\gamma$-ray Emission in FM31 and Complementary M31-related Observations} \label{sec:gas_related_emission}

Although the M31-related components are detected with high statistical significance and for multiple IEMs (Appendix~\ref{sec:different_IEMs}), the corresponding intensity lies below that of the isotropic emission, and therefore the signal has a strong dependence on the systematic uncertainties of the isotropic component. In addition, our analysis has demonstrated that the characterization of \hi\ along the line of sight is a significant systematic uncertainty for analysis of the M31 field, including the contribution from the DNM. Overall, $\gamma$-ray observations of M31's outer halo are significantly complicated by confusion with the Galactic and isotropic emission, due to the halo's large extension on the sky. 

To gauge the full extent of the uncertainty pertaining to the \hi-related components, and to help mitigate the uncertainty pertaining to the isotropic component, in this section we supplement our analysis by observing the structured $\gamma$-ray emission in FM31 in a (semi) model-independent way. As a qualitative gauge, we also compare this emission to some of the main tracers of M31's outer disk and halo. 

We observe the $\gamma$-ray emission in a (semi) model-independent way by removing the \hi-related A5--A8 components from the model (including the Bremsstrahlung component). In addition, we remove the two point sources closest to the M31 disk (3FGL J0040.3+4049 and 3FGL J0049.0+4224), and we remove the new point sources that we find with our point source finding procedure, since most of these sources are found to correlate with the diffuse structures in the residuals (see Figure~\ref{fig:TS_map}). All other sources are held fixed to their best-fit values obtained in the baseline fit (with IC scaled). This effectively amounts to removing only the known smooth diffuse sources and point sources from the data, or equivalently, observing only the structured emission.

The resulting count residuals (data $-$ model) integrated between 1--100 GeV are shown in Figure \ref{fig:positive_residuals_full}. The color scale corresponds to counts/pixel, and the pixel size is $0.2^\circ \times 0.2^\circ$. The images are smoothed using a $1^\circ$ Gaussian kernel. This value roughly corresponds to the PSF (68\% containment angle) of \textit{Fermi}-LAT, which at 1 GeV is $\sim$$1^\circ$. The corresponding pixel distribution is shown in Figure~\ref{fig:map_detatils}. All of the pixels have positive counts, which is why we set the lower limit of the plot range to zero. Maps are shown in the cubehelix color scheme~\citep{green2011colour}. Contours for the disk regions of M31 and M33~\citep{gratier2010molecular} are overlaid. Bright emission corresponding to M31's inner galaxy can be observed. The emission can be seen to extend continuously along M31's major axis in the north-east\footnote{For M31-related directions, north points up, and east points to the left.} direction, which then continues to extend upward until blending with the bright emission of the MW plane. This feature is lopsided, as the south-west side shows a more distinct cutoff away from the inner galaxy. The large arc feature observed in the residuals is also clearly visible in the emission. 

We have found that the M31-related components are roughly consistent with arising from DM annihilation. Since there is still a high level of uncertainty regarding the actual nature of DM, especially on galactic scales, we cannot rule out the possibility that the smooth residual emission may in fact have a DM origin. The same also applies for some of the structured emission in FM31. We, therefore, consider the main tracers of M31's outer disk and halo, since these are some of the few observational handles available when searching for a DM signal from the outer regions of the M31 system.

In Figure~\ref{fig:FM31_rich} we overlay the boundaries for the M31 inner galaxy (solid cyan circle) and spherical halo (dashed black circle) components. We also overlay the M31 disk, the M31 cloud \citep{blitz1999high,kerp2016survey}, M33, Wright's cloud \citep{wright1979tail}, M31's population of globular clusters \citep{galleti20042mass,huxor2008globular,peacock2010m31,Mackey:2010ix,veljanoski2014outer,huxor2014outer}, M31's population of satellite galaxies~\citep{McConnachie:2012vd,martin2013pandas,collins2013kinematic,Ibata:2013rh,pawlowski2013dwarf}, and clouds of Complex H \citep{hulsbosch1975studies,blitz1999high,Lockman:2003zs,Simon:2005vh}. The spherical halo component is found to enclose 61\% (22/36) of M31's dwarf galaxy population, which increases to 72\% (26/36) if including the dwarfs which are within $\sim$$1^\circ$ of the spherical halo boundary. We stress that this is only done as a qualitative gauge of M31's outer halo. We do not expect these systems to outshine the local MW emission. In particular, we do not expect to detect the individual M31 dwarfs, since they are mostly undetected in the MW. We also do not expect to detect the individual globular clusters. We do note, however, that we find features in the data that are positionally coincident with some of these tracers, and most prominently with the M31 cloud. Further investigation is left for a follow-up study.

\section{Summary, Discussion, and Conclusion} \label{sec:fianl}

The goal of this work is to search for extended $\gamma$-ray emission originating beyond the galactic disk of M31, and to examine the implications for CRs and DM. There are two primary motivations for this search. First, CR interactions with M31's circumgalactic medium and/or stellar halo could generate a detectable signal in $\gamma$-rays. Secondly, M31's DM halo has a large extension on the sky and could produce a detectable signal within currently allowed DM scenarios, which would be complementary to other targets, and specifically, the Galactic center. Our primary field of interest (FM31) is a $28^\circ \times 28^\circ$ square region, which amounts to a projected radius of $\sim$200 kpc from the center of M31. Our study complements previously published results on M31~\citep{Fermi-LAT:2010kib,ogelman2010discovery,Pshirkov:2015hda,Pshirkov:2016qhu,Ackermann:2017nya} and is the first to explore the farthest reaches of the M31 system in $\gamma$-rays.  

Because of the extended nature of the signal we are investigating, modeling the bright foreground of the MW is the biggest challenge in performing this analysis. The IEM provided by the FSSC cannot be used as a primary foreground model for this study, as it \emph{is not} intended for the analysis of extended sources\textsuperscript{\ref{caveats}} 
\citep{Acero:2016qlg}. We construct specialized interstellar emission models for the analysis of FM31 by employing the CR propagation code GALPROP, including a self-consistent determination of the isotropic component. Additionally, we use a template approach to account for inaccuracies in the foreground model relating to the neutral gas along the line of sight. 

The parameters of the GALPROP model are tuned to the measured local interstellar spectra of CRs, including the latest AMS-02 measurements. We have adopted the best-fit parameters from the tuning procedure performed in~\citet{Boschini:2017fxq,Boschini:2018zdv}, where GALPROP and HelMod are implemented in an iterative manner, thereby accounting for solar modulation in a physically motivated way when fitting to the local CR measurements. 

The total interstellar emission model consists of individual components for $\pi^0$-decay, IC, and Bremsstrahlung, and the components are defined in Galactocentric annuli. In total there are 8 annuli, but for FM31 only annulus 5 (the local annulus) and beyond contribute to the foreground emission. FM31 has a significant emission associated with \hi\ gas, but there is very little emission from \htwo{} gas. A uniform spin temperature of 150 K is assumed for the baseline IEM. The foreground emission from \hii\ and Bremsstrahlung are subdominant. Our model also accounts for the DNM. The anisotropic formalism is employed for the calculation of the IC component. To model the point sources in the region, we employ the 3FGL as a starting point, and because of the larger statistics of our data set, we account for additional point sources self-consistently with the M31 IEM by implementing a point source finding procedure, which is based on a wavelet transform algorithm.

We calculate the isotropic component self-consistently with the M31 IEM. The main calculation is performed over the full sky in the following region: $|b| \geq 30^\circ, \ 45^\circ \leq  l \leq 315^\circ$. To better determine the normalization of the isotropic component we use a tuning region (TR) directly below FM31, outside of the virial radius. The best-fit normalization is found to be 1.06 \p 0.04, and this remains fixed for all other fits with the M31 IEM. The isotropic component anti-correlates with the IC components, and we also use the TR to initially constrain the normalizations of the IC components (A5 and A6-A7) for the fit in FM31. The fit in the TR yields a model that describes the data well across the entire region and at all energies. The best-fit normalizations of the IEM components in the TR are all in reasonable agreement with the GALPROP predictions.

For the initial baseline fit in FM31 we freely scale the normalizations of the \hi\ and \htwo{} $\pi^0$-related components concurrently with the point sources. The normalizations of the isotropic and IC components (A5 and A6-A7) remain fixed to their best-fit values obtained in the TR. The top of FM31 has a minor contribution from IC A8, and it is also freely scaled in the fit. Lastly, the \hii\ and Bremsstrahlung components remain fixed to their GALPROP predictions. Note that the Bremsstrahlung component possesses a normalization of 1.0 \p 0.6 in the TR, consistent with the GALPROP prediction. 

The baseline fit in FM31 results in positive residual emission in the fractional count residuals between $\sim$3--20 GeV. The residual emission in this corresponding energy range is fairly smooth and extends over the entire field. The spatial residuals also show structured excesses and deficits, primarily at lower energies ($\sim$1--3 GeV). Because of this poor data-model agreement, additional freedom is given to the fit, including freely scaling the IC components in FM31 and rescaling the diffuse components in smaller subregions. The latter fit is performed in order to allow for any un-modeled spatial variation in the CR density, ISRF density, and/or spin temperature. We find that the general features of the residual emission persist even with these variations. 

A significant fraction of the structured excess emission in FM31 is found to be spatially correlated with the \hi\ column density and the foreground dust, including regions where the dust is relatively cold. This may be indicative of a spatially varying spin temperature, which is not properly accounted for by the rescaling in the smaller subregions. Correspondingly, the structured residual emission may be related to inaccuracies in the modeling of the DNM, which in general is determined as part of an all-sky procedure. A part of the shell of Loop III is also present in FM31, while Loops II and IIIs cover it completely. This may imply that some of the gas-related emission in the region is produced by a population of particles with the spectrum that is harder than that of the old CR population. Note that the \hi\ $\pi^0$-related $\gamma$-ray component is dominant in FM31 for energies below $\sim$5 GeV. 

We, therefore, refine the baseline IEM by constructing a template to account for potential mis-modeling of these components. The template is obtained by selecting the excess emission in FM31 that correlates with \hi\ tracers. We refer to this as the arc template. This procedure accounts for any un-modeled \hi\ (or other Galactic gas), as well as any mis-modeling in its line of sight distance, spin temperature, and spectral index variations. 

We find that the specialized IEMs for the analysis of FM31, both the baseline model and the baseline model with the arc template, yield an extended excess at the level of $\sim$3--5\% in the $\sim$3--20 GeV energy range. We have also tested a number of additional systematic variations to the fit. With the M31 IEM we allowed for additional freedom by varying the index of the IC components and the \hi-related components using a PL scaling. The fit was also performed with two alternative IEMs, namely, the IG and FSSC IEMs. Each alternative IEM has its own self-consistently derived isotropic component and additional point sources. In addition, we tested systematic variations to the spectra of 3FGL sources (although the point sources are not a major uncertainty for this analysis). In total we perform 9 main variations of the fit (see Figure~\ref{fig:residuals_all}), using 3 different IEMs (although all IEMs share similar underlying \hi\ maps). The excess is observed for all of the physically motivated IEMs intended for extended source analysis.

Using our benchmark model (the M31 IEM) we have demonstrated that the excess is robust against the systematic studies of the MW foreground emission that we have considered, and that it significantly decreases outside of FM31 (as evidenced by the lack of a similar excess in the TR). This indicates that the excess originates at least partially from outside of the MW and it is significant towards M31. However, we do not rule out the possibility that the signal may also include a MW component, as discussed below. 

We note that apart from the structured residual emission correlated with the foreground gas and dust, which is accounted for with the arc template, other structured excesses and deficits in FM31 are found to be correlated with the major axis of the M31 disk. Likewise, a portion of the \hi\ column densities in the outer Galaxy (A6 and A7) are found to be correlated with M31's major axis as well. This is an indication that some of the gas which is currently assigned to the MW may actually reside in the M31 system, as was also pointed out in~\citet{Ackermann:2017nya}. This will be fully addressed in a forthcoming work. 

A component of the residual emission in FM31 is observed to be positionally coincident with the projected position of M33, and a portion of this emission may have an actual physical association; however, further investigation has been left for future studies. Aside from the structured excesses and deficits, which are observed primarily in the lower energy range ($\sim$1--3 GeV), the majority of the excess emission is roughly uniformly distributed across FM31, corresponding to the positive residual emission observed in the fractional count residuals between $\sim$3--20 GeV.

To determine whether the excess presents a spherically symmetric gradient about the center of M31, which would lend support to the hypothesis that it originates from there, we perform a further fit in FM31 by including three symmetric uniform templates centered at M31. This also allows us to quantify the spectrum and gradient of the positive residual emission. The templates are fit concurrently with the other components of the baseline IEM, including the arc template. 

The inner disk (inner galaxy) has a radial extension of 0.4$^\circ$ (5.5 kpc projected radius). This is the best-fit morphology as determined in~\citet{Ackermann:2017nya}, and it corresponds to the bright $\gamma$-ray emission towards M31's inner galaxy. The intermediate ring (spherical halo) has a radial extension from $0.4^\circ < r \leq 8.5^\circ$ (117 kpc projected radius). This extension excludes most of the residual emission associated with the arc template, while also enclosing a majority of M31's globular cluster population and stellar halo, as well as the M31 cloud. The outer ring (far outer halo) covers the remaining extent of FM31, corresponding to a total projected radius of $\sim$200 kpc, and likewise it begins to approach the MW plane towards the top of the field.  We find that all templates are significantly detected (with a significance of $\geq 5 \sigma$). Furthermore, the M31-related components are able to flatten the positive residual emission in the fractional count residuals. 

For the fit with the arc template and M31-related components, the best-fit normalizations of the IEM components are overall in good agreement with the GALPROP predictions, and they also agree with the best-fit normalizations obtained for the all-sky fit in the determination of the isotropic component. The total integrated flux for the \hi\ A5 component plus the arc north and south components is 185.6 \p 12.9 ph cm$^{-2}$ s$^{-1}$, consistent with that of the baseline fit (with IC scaled). In turn, the corresponding local average emissivity is consistent with the measurements made in \citet{abdo2009fermi}, \citet{ackermann2012fermi}, and \citet{Casandjian:2015hja}. 

The normalization of  the \hi\ A6 component is consistent with the GALPROP prediction. The normalization of the \hi\ A7 component is a bit high at 2.8 \p 0.4 (as for all fits in FM31), but this component may contain a fraction of gas that actually resides in the M31 system, as was already discussed, and will be further discussed below. The normalizations of the IC A5 and A6-A7 components are consistent with the all-sky average obtained in the isotropic calculation (Table~\ref{tab:norm_isotropic}). The normalization of the IC A8 component is high, which is true for all fits in FM31, but this component is subdominant and only contributes along the top of the field, corresponding to the Galactic plane.

The spectrum and intensity for the inner galaxy are consistent with previously published results. We note however that the spectrum derived between 1--100 GeV is softer than that derived between 300 MeV -- 300 GeV (although consistent within errors). This is due to the energy range used for the calculation. The spherical halo and far outer halo have intensities that are much dimmer than the inner galaxy, and present a mild intensity gradient, tapering off with distance from the center of M31. Their spectra are significantly different from all the other extended components in FM31. They peak between $\sim$5--10 GeV, and drop off below and above these energies more steeply than all other contributions. We find it difficult to reconcile these spectra with the possibility that the excess emission originates solely within the MW, further setting it apart from known Galactic sources. Beyond these general features, the spectra for the two outer annuli differ from each other with the far outer halo presenting a harder spectrum at low energies.  

To further test the symmetry of the residual emission in FM31, we also perform a fit in which we divide the spherical halo and far outer halo templates into north and south components, allowing the spectral parameters of each component to vary independently (although all components are fit simultaneously). The cut is made at the midpoint of FM31 along the horizontal direction (parallel to the Galactic plane), corresponding to a latitude of $-21.5^\circ$. The fit is otherwise performed just as for the fit with the full M31-related templates (including the arc north and south). We find that all components are significantly detected (with a significance $>5\sigma$). The results for this test further demonstrate the importance of the MW modeling and that the excess is likely to have a significant MW component. In particular, the emission associated with the far outer halo is more likely to be related to the MW than the M31 system. But even still, the nature of this emission remains unclear. 
 
Given the approximately uniform spatial distribution of the excess emission (as most clearly indicated by the fit with the full M31-related templates), understanding its interplay with the isotropic component is crucial. We have investigated this issue and concluded that the excess emission is robust within the systematic uncertainties in the isotropic component we have considered. Our treatment of the isotropic component can primarily be found in Section~\ref{sec:tuning}, Figure~\ref{fig:Isotropic_Sytematics}, Appendix~\ref{sec:IG_IEMs}, and Appendix~\ref{sec:FSSC_IEM}. We note, however, that the isotropic emission has a bump-like feature in the energy range that somewhat overlaps with the peak in the spectrum of the M31-related components (as is most clearly seen in Figure~\ref{fig:Isotropic_Sytematics}). This might suggest that the isotropic emission may include a component that originates from similar processes in the extended halo of the MW.

These results show that if the excess emission originates from the M31 system (at least partially), its extension reaches a distance upwards of $\sim$120--200 kpc from the center of M31. This is consistent with the expectation for a DM signal, as the virial radius for the DM halo extends at least this far. To test this interpretation, we compare these results with the predictions for a DM signal that originates from the M31 halo, with a spectrum and annihilation cross-section consistent with a DM interpretation of the GC excess. We also consider the contribution from the MW's DM halo along the line of sight, since this component has not been explicitly accounted for in this analysis. If such a component actually exists, then it may be at least partially embedded in the isotropic component, as well as the other components of the IEM, but it will not necessarily be fully absorbed. Note that in general there is also expected to be some contribution from the local DM filament between M31 and the MW. 

We consider different assumptions for the amount of DM substructure in M31 (and the MW), and we find that if a cold DM scenario is assumed that includes a large boost factor due to substructures, the observed excess emission is consistent with this interpretation. Granted, however, the exact partitioning of individual contributions to the signal remains unclear, i.e.\ primary emission from M31's DM halo, secondary emission in M31, emission from the local DM filament between M31 and the MW, and emission from the MW's DM halo along the line of sight.

This is an intriguing finding, however, its implications are far reaching, and better understanding the MW foreground is crucial before drawing any stronger conclusions. Another crucial aspect is complementarity with other DM targets. Although these results are consistent with other observations in $\gamma$-rays, namely the GC excess and the constraints from dwarf spheroidal galaxies, they imply that a large boost factor from substructures would contribute to a DM signal from the MW halo. As already stated, this contribution has not been accounted for in this analysis and might be at least partially embedded in the isotropic component as well as other components of the MW foreground. Likewise, the M31-related components might contain some contribution from the MW DM halo along the line of sight, as well as some contribution from the local DM filament between M31 and the MW. From our substructure calculations we estimate that the intensity of a MW DM contribution in FM31 may be on the order of $\sim$1--10\% of the isotropic intensity. Investigating this possibility in more detail requires a dedicated analysis which is beyond the scope of this work. 

The CR halo of M31 might extend tens to hundreds of  kpc from the center of M31. It is possible that some of the emission in FM31 results from CR interactions with the ionized gas of M31's circumgalactic medium and/or stellar halo, which also extend well beyond the galactic disk. However, based on the radial extent, spectral shape, and intensity of the M31-related components, it is seemingly unlikely that the corresponding emission is dominated by these types of CR interactions. 

We have also investigated the structured residual emission in FM31, as well as the emission correlated with the \hi\ $\gamma$-ray maps, and compared them to different tracers of M31's outer disk and halo. These tracers include the M31 cloud, and M31's populations of globular clusters and satellite galaxies. We find features in the data that are positionally coincident with some of these tracers, and most prominently with the M31 cloud. This is a further indication that some of the structured emission observed in FM31 originates from M31 rather than the MW. This in turn implies that the total $\gamma$-ray emission from the M31 system extends well beyond the inner regions of the galactic disk. The M31 system is very rich, and further analysis of these findings is beyond the scope of this paper. Our primary focus in this analysis is the more significant smoother component of the signal.

In summary, we present the first search for extended emission from M31 in $\gamma$-rays out to a distance of $\sim$200 kpc from its center. We find evidence for an extended excess that appears to be distinct from the conventional MW foreground, having a total radial extension upwards of 120--200 kpc from the center of M31. We discuss plausible interpretations for the excess emission but emphasize that uncertainties in the MW foreground, and in particular modeling of the \hi-related components, have not been fully explored and may impact the results. The results also have a close link with the isotropic component (and likewise the IC components), which may be inevitable considering the nature of the signal under investigation. We find that a DM interpretation provides a good description of the observed emission and is consistent with the GC excess DM interpretation. However, better understanding of the systematics, and complementarity with other DM searches, as discussed in the paper, is critical to settle the issue.

\section*{Acknowledgements}
The authors thank Tsunefumi Mizuno, Gulli J\'ohannesson, Alex Drlica-Wagner, and Troy Porter for many useful comments made at the preparation stage of the manuscript. The authors are also pleased to acknowledge conversations with Ketron Mitchell-Wynne, Sean Fillingham, Tim Tait, Philip Tanedo, Mike Cooper, James Bullock, Manoj Kaplinghat, Kevork N. Abazajian, Sebastian Trojanowski, Ferdinand Badescu, Volodymyr Takhistov, Deano Farinella, and Dan Hooper. A majority of the data analysis has been performed on UCI's HPC, and CK thanks Harry Mangalam for his assistance on numerous occasions. CK also thanks James Chiang for his assistance with the Fermi Science Tools. The work of CK and SM is supported in part by Department of Energy grant DESC0014431. SSC is supported by National Science Foundation Grant PHY-1620638 and a McCue Fellowship. IM acknowledges partial support from NASA grant NNX17AB48G.

The \textit{Fermi}-LAT Collaboration acknowledges generous ongoing support from a number of agencies and institutes that have supported both the development and the operation of the LAT as well as scientific data analysis. These include the National Aeronautics and Space Administration and the Department of Energy in the United States; the Commissariat \`a l'Energie Atomique and the Centre National de la Recherche Scientifique/Institut National de Physique Nucl\'eaire et de Physique des Particules in France; the Agenzia Spaziale Italiana and the Istituto Nazionale di Fisica Nucleare in Italy; the Ministry of Education, Culture, Sports, Science, and Technology (MEXT); the High Energy Accelerator Research Organization (KEK) and the Japan Aerospace Exploration Agency (JAXA) in Japan; and the K.~A.~Wallenberg Foundation, the Swedish Research Council, and the Swedish National Space Board in Sweden. 

Additional support for science analysis during the operations phase is gratefully acknowledged from the Istituto Nazionale di Astrofisica in Italy and the Centre National d'\'Etudes Spatiales in France. This work performed in part under DOE 
Contract DE-AC02-76SF00515.

\bibliography{citations}

\begin{thebibliography}{}
\expandafter\ifx\csname natexlab\endcsname\relax\def\natexlab#1{#1}\fi
\providecommand{\url}[1]{\href{#1}{#1}}
\providecommand{\dodoi}[1]{doi:~\href{http://doi.org/#1}{\nolinkurl{#1}}}
\providecommand{\doeprint}[1]{\href{http://ascl.net/#1}{\nolinkurl{http://ascl.net/#1}}}
\providecommand{\doarXiv}[1]{\href{https://arxiv.org/abs/#1}{\nolinkurl{https://arxiv.org/abs/#1}}}

\bibitem[{Abazajian {et~al.}(2014)Abazajian, Canac, Horiuchi, \&
  Kaplinghat}]{Abazajian:2014fta}
Abazajian, K.~N., Canac, N., Horiuchi, S., \& Kaplinghat, M. 2014, PhRvD, 90,
  023526, \dodoi{10.1103/PhysRevD.90.023526}

\bibitem[{Abazajian {et~al.}(2015)Abazajian, Canac, Horiuchi, Kaplinghat, \&
  Kwa}]{Abazajian:2014hsa}
Abazajian, K.~N., Canac, N., Horiuchi, S., Kaplinghat, M., \& Kwa, A. 2015,
  JCAP, 1507, 013, \dodoi{10.1088/1475-7516/2015/07/013}

\bibitem[{Abazajian \& Kaplinghat(2012)}]{Abazajian:2012pn}
Abazajian, K.~N., \& Kaplinghat, M. 2012, PhRvD, 86, 083511,
  \dodoi{10.1103/PhysRevD.86.083511, 10.1103/PhysRevD.87.129902}

\bibitem[{Abdo {et~al.}(2009)Abdo, Ackermann, Ajello, Atwood, Axelsson,
  Baldini, Ballet, Barbiellini, Bastieri, Baughman, {et~al.}}]{abdo2009fermi}
Abdo, A., Ackermann, M., Ajello, M., {et~al.} 2009, ApJ, 703, 1249

\bibitem[{{Abdo} {et~al.}(2009){Abdo}, {Ackermann}, {Ajello}, {Ampe},
  {Anderson}, {Atwood}, {Axelsson}, {Bagagli}, {Baldini}, {Ballet}, \&
  et~al.}]{Abdo:2009gy}
{Abdo}, A.~A., {Ackermann}, M., {Ajello}, M., {et~al.} 2009, Astroparticle
  Physics, 32, 193, \dodoi{10.1016/j.astropartphys.2009.08.002}

\bibitem[{{Abdo} {et~al.}(2010){Abdo}, {Ackermann}, {Ajello}, {Allafort},
  {Atwood}, {Baldini}, {Ballet}, {Barbiellini}, {Bastieri}, {Bechtol},
  {Bellazzini}, {Berenji}, {Blandford}, {Bloom}, {Bonamente}, {Borgland},
  {Bouvier}, {Brandt}, {Bregeon}, {Brigida}, {Bruel}, {Buehler}, {Burnett},
  {Buson}, {Caliandro}, {Cameron}, {Cannon}, {Caraveo}, {Casandjian}, {Cecchi},
  {{\c C}elik}, {Charles}, {Chekhtman}, {Chiang}, {Ciprini}, {Claus},
  {Cohen-Tanugi}, {Conrad}, {Dermer}, {de Angelis}, {de Palma}, {Digel},
  {Silva}, {Drell}, {Drlica-Wagner}, {Dubois}, {Favuzzi}, {Fegan}, {Fortin},
  {Frailis}, {Fukazawa}, {Funk}, {Fusco}, {Gargano}, {Germani}, {Giglietto},
  {Giordano}, {Giroletti}, {Glanzman}, {Godfrey}, {Grenier}, {Grondin},
  {Guiriec}, {Gustafsson}, {Hadasch}, {Harding}, {Hayashi}, {Hayashida},
  {Hays}, {Healey}, {Jean}, {J{\'o}hannesson}, {Johnson}, {Johnson}, {Johnson},
  {Kamae}, {Katagiri}, {Kataoka}, {Kerr}, {Kn{\"o}dlseder}, {Kuss}, {Lande},
  {Latronico}, {Lee}, {Lemoine-Goumard}, {Longo}, {Loparco}, {Lott},
  {Lovellette}, {Lubrano}, {Madejski}, {Makeev}, {Martin}, {Mazziotta},
  {Mehault}, {Michelson}, {Mitthumsiri}, {Mizuno}, {Moiseev}, {Monte},
  {Monzani}, {Morselli}, {Moskalenko}, {Murgia}, {Naumann-Godo}, {Nolan},
  {Norris}, {Nuss}, {Ohsugi}, {Okumura}, {Omodei}, {Orlando}, {Ormes}, {Ozaki},
  {Paneque}, {Panetta}, {Parent}, {Pepe}, {Persic}, {Pesce-Rollins}, {Piron},
  {Porter}, {Rain{\`o}}, {Rando}, {Razzano}, {Reimer}, {Reimer}, {Ritz},
  {Romani}, {Sadrozinski}, {Saz Parkinson}, {Sgr{\`o}}, {Siskind}, {Smith},
  {Smith}, {Spandre}, {Spinelli}, {Strickman}, {Strigari}, {Strong}, {Suson},
  {Takahashi}, {Takahashi}, {Tanaka}, {Thayer}, {Thompson}, {Tibaldo},
  {Torres}, {Tosti}, {Tramacere}, {Uchiyama}, {Usher}, {Vandenbroucke},
  {Vianello}, {Vilchez}, {Vitale}, {Waite}, {Wang}, {Winer}, {Wood}, {Yang}, \&
  {Ziegler}}]{Fermi-LAT:2010kib}
---. 2010, \aap, 523, L2, \dodoi{10.1051/0004-6361/201015759}

\bibitem[{{Abeysekara} {et~al.}(2014){Abeysekara}, {Alfaro}, {Alvarez},
  {{\'A}lvarez}, {Arceo}, {Arteaga-Vel{\'a}zquez}, {Ayala Solares}, {Barber},
  {Baughman}, {Bautista-Elivar}, {Becerra Gonzalez}, {Belmont}, {BenZvi},
  {Berley}, {Bonilla Rosales}, {Braun}, {Caballero-Lopez}, {Caballero-Mora},
  {Carrami{\~n}ana}, {Castillo}, {Cotti}, {Cotzomi}, {de la Fuente}, {De
  Le{\'o}n}, {DeYoung}, {Diaz Hernandez}, {Diaz-Cruz}, {D{\'{\i}}az-V{\'e}lez},
  {Dingus}, {DuVernois}, {Ellsworth}, {Fiorino}, {Fraija}, {Galindo},
  {Garfias}, {Gonz{\'a}lez}, {Goodman}, {Grabski}, {Gussert}, {Hampel-Arias},
  {Harding}, {Hui}, {H{\"u}ntemeyer}, {Imran}, {Iriarte}, {Karn}, {Kieda},
  {Kunde}, {Lara}, {Lauer}, {Lee}, {Lennarz}, {Le{\'o}n Vargas}, {Linares},
  {Linnemann}, {Longo}, {Luna-Garcia}, {Marinelli}, {Martinez}, {Martinez},
  {Mart{\'{\i}}nez-Castro}, {Matthews}, {McEnery}, {Mendoza Torres},
  {Miranda-Romagnoli}, {Moreno}, {Mostaf{\'a}}, {Nellen}, {Newbold},
  {Noriega-Papaqui}, {Oceguera-Becerra}, {Patricelli}, {Pelayo},
  {P{\'e}rez-P{\'e}rez}, {Pretz}, {Rivi{\`e}re}, {Rosa-Gonz{\'a}lez}, {Ryan},
  {Salazar}, {Salesa}, {Sanchez}, {Sandoval}, {Schneider}, {Silich}, {Sinnis},
  {Smith}, {Sparks Woodle}, {Springer}, {Taboada}, {Toale}, {Tollefson},
  {Torres}, {Ukwatta}, {Villase{\~n}or}, {Weisgarber}, {Westerhoff}, {Wisher},
  {Wood}, {Yodh}, {Younk}, {Zaborov}, {Zepeda}, {Zhou}, {Abazajian}, \&
  {Milagro Collaboration}}]{Abeysekara:2014ffg}
{Abeysekara}, A.~U., {Alfaro}, R., {Alvarez}, C., {et~al.} 2014, \prd, 90,
  122002, \dodoi{10.1103/PhysRevD.90.122002}

\bibitem[{{Acero} {et~al.}(2016){Acero}, {Ackermann}, {Ajello}, {Albert},
  {Baldini}, {Ballet}, {Barbiellini}, {Bastieri}, {Bellazzini}, {Bissaldi},
  {Bloom}, {Bonino}, {Bottacini}, {Brandt}, {Bregeon}, {Bruel}, {Buehler},
  {Buson}, {Caliandro}, {Cameron}, {Caragiulo}, {Caraveo}, {Casandjian},
  {Cavazzuti}, {Cecchi}, {Charles}, {Chekhtman}, {Chiang}, {Chiaro}, {Ciprini},
  {Claus}, {Cohen-Tanugi}, {Conrad}, {Cuoco}, {Cutini}, {D'Ammando}, {de
  Angelis}, {de Palma}, {Desiante}, {Digel}, {Di Venere}, {Drell}, {Favuzzi},
  {Fegan}, {Ferrara}, {Focke}, {Franckowiak}, {Funk}, {Fusco}, {Gargano},
  {Gasparrini}, {Giglietto}, {Giordano}, {Giroletti}, {Glanzman}, {Godfrey},
  {Grenier}, {Guiriec}, {Hadasch}, {Harding}, {Hayashi}, {Hays}, {Hewitt},
  {Hill}, {Horan}, {Hou}, {Jogler}, {J{\'o}hannesson}, {Kamae}, {Kuss},
  {Landriu}, {Larsson}, {Latronico}, {Li}, {Li}, {Longo}, {Loparco},
  {Lovellette}, {Lubrano}, {Maldera}, {Malyshev}, {Manfreda}, {Martin},
  {Mayer}, {Mazziotta}, {McEnery}, {Michelson}, {Mirabal}, {Mizuno}, {Monzani},
  {Morselli}, {Nuss}, {Ohsugi}, {Omodei}, {Orienti}, {Orlando}, {Ormes},
  {Paneque}, {Pesce-Rollins}, {Piron}, {Pivato}, {Rain{\`o}}, {Rando},
  {Razzano}, {Razzaque}, {Reimer}, {Reimer}, {Remy}, {Renault},
  {S{\'a}nchez-Conde}, {Schaal}, {Schulz}, {Sgr{\`o}}, {Siskind}, {Spada},
  {Spandre}, {Spinelli}, {Strong}, {Suson}, {Tajima}, {Takahashi}, {Thayer},
  {Thompson}, {Tibaldo}, {Tinivella}, {Torres}, {Tosti}, {Troja}, {Vianello},
  {Werner}, {Wood}, {Wood}, {Zaharijas}, \& {Zimmer}}]{Acero:2016qlg}
{Acero}, F., {Ackermann}, M., {Ajello}, M., {et~al.} 2016, \apjs, 223, 26,
  \dodoi{10.3847/0067-0049/223/2/26}

\bibitem[{{Ackermann} {et~al.}(2015{\natexlab{a}}){Ackermann}, {Ajello},
  {Albert}, {Baldini}, {Barbiellini}, {Bastieri}, {et~al.}}]{Ackermann:2015tah}
{Ackermann}, M., {Ajello}, M., {Albert}, A., {et~al.} 2015{\natexlab{a}}, JCAP,
  1509, 008, \dodoi{10.1088/1475-7516/2015/09/008}

\bibitem[{{Ackermann} {et~al.}(2012{\natexlab{a}}){Ackermann}, {Ajello},
  {Albert}, {Allafort}, {Atwood}, {Axelsson}, {Baldini}, {Ballet},
  {Barbiellini}, {Bastieri}, {Bechtol}, {Bellazzini}, {Bissaldi}, {Blandford},
  {Bloom}, {Bogart}, {Bonamente}, {Borgland}, {Bottacini}, {Bouvier}, {Brandt},
  {Bregeon}, {Brigida}, {Bruel}, {Buehler}, {Burnett}, {Buson}, {Caliandro},
  {Cameron}, {Caraveo}, {Casandjian}, {Cavazzuti}, {Cecchi}, {{\c C}elik},
  {Charles}, {Chaves}, {Chekhtman}, {Cheung}, {Chiang}, {Ciprini}, {Claus},
  {Cohen-Tanugi}, {Conrad}, {Corbet}, {Cutini}, {D'Ammando}, {Davis}, {de
  Angelis}, {DeKlotz}, {de Palma}, {Dermer}, {Digel}, {Silva}, {Drell},
  {Drlica-Wagner}, {Dubois}, {Favuzzi}, {Fegan}, {Ferrara}, {Focke}, {Fortin},
  {Fukazawa}, {Funk}, {Fusco}, {Gargano}, {Gasparrini}, {Gehrels}, {Giebels},
  {Giglietto}, {Giordano}, {Giroletti}, {Glanzman}, {Godfrey}, {Grenier},
  {Grove}, {Guiriec}, {Hadasch}, {Hayashida}, {Hays}, {Horan}, {Hou}, {Hughes},
  {Jackson}, {Jogler}, {J{\'o}hannesson}, {Johnson}, {Johnson}, {Johnson},
  {Kamae}, {Katagiri}, {Kataoka}, {Kerr}, {Kn{\"o}dlseder}, {Kuss}, {Lande},
  {Larsson}, {Latronico}, {Lavalley}, {Lemoine-Goumard}, {Longo}, {Loparco},
  {Lott}, {Lovellette}, {Lubrano}, {Mazziotta}, {McConville}, {McEnery},
  {Mehault}, {Michelson}, {Mitthumsiri}, {Mizuno}, {Moiseev}, {Monte},
  {Monzani}, {Morselli}, {Moskalenko}, {Murgia}, {Naumann-Godo}, {Nemmen},
  {Nishino}, {Norris}, {Nuss}, {Ohno}, {Ohsugi}, {Okumura}, {Omodei},
  {Orienti}, {Orlando}, {Ormes}, {Paneque}, {Panetta}, {Perkins},
  {Pesce-Rollins}, {Pierbattista}, {Piron}, {Pivato}, {Porter}, {Racusin},
  {Rain{\`o}}, {Rando}, {Razzano}, {Razzaque}, {Reimer}, {Reimer}, {Reposeur},
  {Reyes}, {Ritz}, {Rochester}, {Romoli}, {Roth}, {Sadrozinski}, {Sanchez},
  {Saz Parkinson}, {Sbarra}, {Scargle}, {Sgr{\`o}}, {Siegal-Gaskins},
  {Siskind}, {Spandre}, {Spinelli}, {Stephens}, {Suson}, {Tajima}, {Takahashi},
  {Tanaka}, {Thayer}, {Thayer}, {Thompson}, {Tibaldo}, {Tinivella}, {Tosti},
  {Troja}, {Usher}, {Vandenbroucke}, {Van Klaveren}, {Vasileiou}, {Vianello},
  {Vitale}, {Waite}, {Wallace}, {Winer}, {Wood}, {Wood}, {Wood}, {Yang}, \&
  {Zimmer}}]{Ackermann:2012kna}
---. 2012{\natexlab{a}}, \apjs, 203, 4, \dodoi{10.1088/0067-0049/203/1/4}

\bibitem[{{Ackermann} {et~al.}(2012{\natexlab{b}}){Ackermann}, {Ajello},
  {Atwood}, {Baldini}, {Ballet}, {Barbiellini}, {Bastieri}, {Bechtol},
  {Bellazzini}, {Berenji}, {Blandford}, {Bloom}, {Bonamente}, {Borgland},
  {Brandt}, {Bregeon}, {Brigida}, {Bruel}, {Buehler}, {Buson}, {Caliandro},
  {Cameron}, {Caraveo}, {Cavazzuti}, {Cecchi}, {Charles}, {Chekhtman},
  {Chiang}, {Ciprini}, {Claus}, {Cohen-Tanugi}, {Conrad}, {Cutini}, {de
  Angelis}, {de Palma}, {Dermer}, {Digel}, {Silva}, {Drell}, {Drlica-Wagner},
  {Falletti}, {Favuzzi}, {Fegan}, {Ferrara}, {Focke}, {Fortin}, {Fukazawa},
  {Funk}, {Fusco}, {Gaggero}, {Gargano}, {Germani}, {Giglietto}, {Giordano},
  {Giroletti}, {Glanzman}, {Godfrey}, {Grove}, {Guiriec}, {Gustafsson},
  {Hadasch}, {Hanabata}, {Harding}, {Hayashida}, {Hays}, {Horan}, {Hou},
  {Hughes}, {J{\'o}hannesson}, {Johnson}, {Johnson}, {Kamae}, {Katagiri},
  {Kataoka}, {Kn{\"o}dlseder}, {Kuss}, {Lande}, {Latronico}, {Lee},
  {Lemoine-Goumard}, {Longo}, {Loparco}, {Lott}, {Lovellette}, {Lubrano},
  {Mazziotta}, {McEnery}, {Michelson}, {Mitthumsiri}, {Mizuno}, {Monte},
  {Monzani}, {Morselli}, {Moskalenko}, {Murgia}, {Naumann-Godo}, {Norris},
  {Nuss}, {Ohsugi}, {Okumura}, {Omodei}, {Orlando}, {Ormes}, {Paneque},
  {Panetta}, {Parent}, {Pesce-Rollins}, {Pierbattista}, {Piron}, {Pivato},
  {Porter}, {Rain{\`o}}, {Rando}, {Razzano}, {Razzaque}, {Reimer}, {Reimer},
  {Sadrozinski}, {Sgr{\`o}}, {Siskind}, {Spandre}, {Spinelli}, {Strong},
  {Suson}, {Takahashi}, {Tanaka}, {Thayer}, {Thayer}, {Thompson}, {Tibaldo},
  {Tinivella}, {Torres}, {Tosti}, {Troja}, {Usher}, {Vandenbroucke},
  {Vasileiou}, {Vianello}, {Vitale}, {Waite}, {Wang}, {Winer}, {Wood}, {Wood},
  {Yang}, {Ziegler}, \& {Zimmer}}]{Ackermann:2012pya}
{Ackermann}, M., {Ajello}, M., {Atwood}, W.~B., {et~al.} 2012{\natexlab{b}},
  \apj, 750, 3, \dodoi{10.1088/0004-637X/750/1/3}

\bibitem[{{Ackermann} {et~al.}(2012{\natexlab{c}}){Ackermann}, {Ajello},
  {Allafort}, {Baldini}, {Ballet}, {Barbiellini}, {Bastieri}, {Bechtol},
  {Bellazzini}, {Berenji}, {Blandford}, {Bloom}, {Bonamente}, {Borgland},
  {Bottacini}, {Brandt}, {Bregeon}, {Brigida}, {Bruel}, {Buehler}, {Busetto},
  {Buson}, {Caliandro}, {Cameron}, {Caraveo}, {Casandjian}, {Cecchi},
  {Charles}, {Chekhtman}, {Chiang}, {Ciprini}, {Claus}, {Cohen-Tanugi},
  {Conrad}, {D'Ammando}, {de Angelis}, {de Palma}, {Dermer}, {Digel}, {Silva},
  {Drell}, {Drlica-Wagner}, {Falletti}, {Favuzzi}, {Fegan}, {Ferrara}, {Focke},
  {Fukazawa}, {Fukui}, {Funk}, {Fusco}, {Gargano}, {Gasparrini}, {Germani},
  {Giglietto}, {Giordano}, {Giroletti}, {Glanzman}, {Godfrey}, {Grenier},
  {Grondin}, {Grove}, {Guiriec}, {Hadasch}, {Hanabata}, {Harding}, {Hayashi},
  {Horan}, {Hou}, {Hughes}, {Itoh}, {Jackson}, {J{\'o}hannesson}, {Johnson},
  {Kamae}, {Katagiri}, {Kataoka}, {Kn{\"o}dlseder}, {Kuss}, {Lande}, {Larsson},
  {Lee}, {Lemoine-Goumard}, {Longo}, {Loparco}, {Lovellette}, {Lubrano},
  {Martin}, {Mazziotta}, {McEnery}, {Mehault}, {Michelson}, {Mitthumsiri},
  {Mizuno}, {Moiseev}, {Monte}, {Monzani}, {Morselli}, {Moskalenko}, {Murgia},
  {Naumann-Godo}, {Nemmen}, {Nishino}, {Norris}, {Nuss}, {Ohno}, {Ohsugi},
  {Okumura}, {Omodei}, {Orlando}, {Ormes}, {Ozaki}, {Paneque}, {Panetta},
  {Parent}, {Pesce-Rollins}, {Pierbattista}, {Piron}, {Pivato}, {Porter},
  {Rain{\`o}}, {Rando}, {Razzano}, {Reimer}, {Reimer}, {Romoli}, {Roth},
  {Sada}, {Sadrozinski}, {Sanchez}, {Sbarra}, {Sgr{\`o}}, {Siskind}, {Spandre},
  {Spinelli}, {Strong}, {Suson}, {Takahashi}, {Takahashi}, {Tanaka}, {Thayer},
  {Thayer}, {Thompson}, {Tibaldo}, {Tibolla}, {Tinivella}, {Torres}, {Tosti},
  {Tramacere}, {Troja}, {Uchiyama}, {Uehara}, {Usher}, {Vandenbroucke},
  {Vasileiou}, {Vianello}, {Vitale}, {Waite}, {Wang}, {Winer}, {Wood},
  {Yamamoto}, {Yang}, \& {Zimmer}}]{ackermann2012fermi}
{Ackermann}, M., {Ajello}, M., {Allafort}, A., {et~al.} 2012{\natexlab{c}},
  \apj, 755, 22, \dodoi{10.1088/0004-637X/755/1/22}

\bibitem[{{Ackermann} {et~al.}(2015{\natexlab{b}}){Ackermann}, {Albert},
  {Anderson}, {Atwood}, {Baldini}, {Barbiellini}, {Bastieri}, {Bechtol},
  {Bellazzini}, {Bissaldi}, {Blandford}, {Bloom}, {Bonino}, {Bottacini},
  {Brandt}, {Bregeon}, {Bruel}, {Buehler}, {Caliandro}, {Cameron}, {Caputo},
  {Caragiulo}, {Caraveo}, {Cecchi}, {Charles}, {Chekhtman}, {Chiang}, {Chiaro},
  {Ciprini}, {Claus}, {Cohen-Tanugi}, {Conrad}, {Cuoco}, {Cutini}, {D'Ammando},
  {de Angelis}, {de Palma}, {Desiante}, {Digel}, {Di Venere}, {Drell},
  {Drlica-Wagner}, {Essig}, {Favuzzi}, {Fegan}, {Ferrara}, {Focke},
  {Franckowiak}, {Fukazawa}, {Funk}, {Fusco}, {Gargano}, {Gasparrini},
  {Giglietto}, {Giordano}, {Giroletti}, {Glanzman}, {Godfrey}, {Gomez-Vargas},
  {Grenier}, {Guiriec}, {Gustafsson}, {Hays}, {Hewitt}, {Horan}, {Jogler},
  {J{\'o}hannesson}, {Kuss}, {Larsson}, {Latronico}, {Li}, {Li}, {Llena Garde},
  {Longo}, {Loparco}, {Lubrano}, {Malyshev}, {Mayer}, {Mazziotta}, {McEnery},
  {Meyer}, {Michelson}, {Mizuno}, {Moiseev}, {Monzani}, {Morselli}, {Murgia},
  {Nuss}, {Ohsugi}, {Orienti}, {Orlando}, {Ormes}, {Paneque}, {Perkins},
  {Pesce-Rollins}, {Piron}, {Pivato}, {Porter}, {Rain{\`o}}, {Rando},
  {Razzano}, {Reimer}, {Reimer}, {Ritz}, {S{\'a}nchez-Conde}, {Schulz},
  {Sehgal}, {Sgr{\`o}}, {Siskind}, {Spada}, {Spandre}, {Spinelli}, {Strigari},
  {Tajima}, {Takahashi}, {Thayer}, {Tibaldo}, {Torres}, {Troja}, {Vianello},
  {Werner}, {Winer}, {Wood}, {Wood}, {Zaharijas}, {Zimmer}, \& {Fermi-LAT
  Collaboration}}]{Ackermann:2015zua}
{Ackermann}, M., {Albert}, A., {Anderson}, B., {et~al.} 2015{\natexlab{b}},
  Physical Review Letters, 115, 231301, \dodoi{10.1103/PhysRevLett.115.231301}

\bibitem[{{Ackermann} {et~al.}(2017{\natexlab{a}}){Ackermann}, {Ajello},
  {Albert}, {Baldini}, {Ballet}, {Barbiellini}, {Bastieri}, {Bellazzini},
  {Bissaldi}, {Bloom}, {Bonino}, {Bottacini}, {Brandt}, {Bregeon}, {Bruel},
  {Buehler}, {Cameron}, {Caputo}, {Caragiulo}, {Caraveo}, {Cavazzuti},
  {Cecchi}, {Charles}, {Chekhtman}, {Chiaro}, {Ciprini}, {Costanza}, {Cutini},
  {D'Ammando}, {de Palma}, {Desiante}, {Digel}, {Di Lalla}, {Di Mauro}, {Di
  Venere}, {Favuzzi}, {Funk}, {Fusco}, {Gargano}, {Giglietto}, {Giordano},
  {Giroletti}, {Glanzman}, {Green}, {Grenier}, {Guillemot}, {Guiriec},
  {Hayashi}, {Hou}, {J{\'o}hannesson}, {Kamae}, {Kn{\"o}dlseder}, {Kong},
  {Kuss}, {La Mura}, {Larsson}, {Latronico}, {Li}, {Longo}, {Loparco},
  {Lubrano}, {Maldera}, {Malyshev}, {Manfreda}, {Martin}, {Mazziotta},
  {Michelson}, {Mirabal}, {Mitthumsiri}, {Mizuno}, {Monzani}, {Morselli},
  {Moskalenko}, {Negro}, {Nuss}, {Ohsugi}, {Omodei}, {Orlando}, {Ormes},
  {Paneque}, {Persic}, {Pesce-Rollins}, {Piron}, {Porter}, {Principe},
  {Rain{\`o}}, {Rando}, {Razzano}, {Reimer}, {S{\'a}nchez-Conde}, {Sgr{\`o}},
  {Simone}, {Siskind}, {Spada}, {Spandre}, {Spinelli}, {Tanaka}, {Tibaldo},
  {Torres}, {Troja}, {Uchiyama}, {Wang}, {Wood}, {Wood}, {Zaharijas}, \&
  {Zhou}}]{Ackermann:2017nya}
{Ackermann}, M., {Ajello}, M., {Albert}, A., {et~al.} 2017{\natexlab{a}}, \apj,
  836, 208, \dodoi{10.3847/1538-4357/aa5c3d}

\bibitem[{{Ackermann} {et~al.}(2017{\natexlab{b}}){Ackermann}, {Ajello},
  {Albert}, {Atwood}, {Baldini}, {Ballet}, {Barbiellini}, {Bastieri},
  {Bellazzini}, {Bissaldi}, {Blandford}, {Bloom}, {Bonino}, {Bottacini},
  {Brandt}, {Bregeon}, {Bruel}, {Buehler}, {Burnett}, {Cameron}, {Caputo},
  {Caragiulo}, {Caraveo}, {Cavazzuti}, {Cecchi}, {Charles}, {Chekhtman},
  {Chiang}, {Chiappo}, {Chiaro}, {Ciprini}, {Conrad}, {Costanza}, {Cuoco},
  {Cutini}, {D'Ammando}, {de Palma}, {Desiante}, {Digel}, {Di Lalla}, {Di
  Mauro}, {Di Venere}, {Drell}, {Favuzzi}, {Fegan}, {Ferrara}, {Focke},
  {Franckowiak}, {Fukazawa}, {Funk}, {Fusco}, {Gargano}, {Gasparrini},
  {Giglietto}, {Giordano}, {Giroletti}, {Glanzman}, {Gomez-Vargas}, {Green},
  {Grenier}, {Grove}, {Guillemot}, {Guiriec}, {Gustafsson}, {Harding}, {Hays},
  {Hewitt}, {Horan}, {Jogler}, {Johnson}, {Kamae}, {Kocevski}, {Kuss}, {La
  Mura}, {Larsson}, {Latronico}, {Li}, {Longo}, {Loparco}, {Lovellette},
  {Lubrano}, {Magill}, {Maldera}, {Malyshev}, {Manfreda}, {Martin},
  {Mazziotta}, {Michelson}, {Mirabal}, {Mitthumsiri}, {Mizuno}, {Moiseev},
  {Monzani}, {Morselli}, {Negro}, {Nuss}, {Ohsugi}, {Orienti}, {Orlando},
  {Ormes}, {Paneque}, {Perkins}, {Persic}, {Pesce-Rollins}, {Piron},
  {Principe}, {Rain{\`o}}, {Rando}, {Razzano}, {Razzaque}, {Reimer}, {Reimer},
  {S{\'a}nchez-Conde}, {Sgr{\`o}}, {Simone}, {Siskind}, {Spada}, {Spandre},
  {Spinelli}, {Suson}, {Tajima}, {Tanaka}, {Thayer}, {Tibaldo}, {Torres},
  {Troja}, {Uchiyama}, {Vianello}, {Wood}, {Wood}, {Zaharijas}, {Zimmer}, \&
  {Fermi LAT Collaboration}}]{TheFermi-LAT:2017vmf}
---. 2017{\natexlab{b}}, \apj, 840, 43, \dodoi{10.3847/1538-4357/aa6cab}

\bibitem[{{Ade} {et~al.}(2015){Ade}, {Aghanim}, {Arnaud}, Ashdown, {Aumont},
  {et~al.}}]{Ade:2014bjw}
{Ade}, P. A.~R., {Aghanim}, N., {Arnaud}, M., {et~al.} 2015, A\&A, 582, A28,
  \dodoi{10.1051/0004-6361/201424643}

\bibitem[{Agrawal \& Randall(2017)}]{agrawal2017point}
Agrawal, P., \& Randall, L. 2017, JCAP, 2017, 019

\bibitem[{Aguilar {et~al.}(2014)Aguilar, Aisa, Alpat, Alvino, Ambrosi, Andeen,
  Arruda, Attig, Azzarello, Bachlechner, Barao, Barrau, Barrin, Bartoloni,
  Basara, Battarbee, Battiston, Bazo, Becker, Behlmann, Beischer, Berdugo,
  Bertucci, Bigongiari, Bindi, Bizzaglia, Bizzarri, Boella, de~Boer, Bollweg,
  Bonnivard, Borgia, Borsini, Boschini, Bourquin, Burger, Cadoux, Cai, Capell,
  Caroff, Casaus, Cascioli, Castellini, Cernuda, Cervelli, Chae, Chang, Chen,
  Chen, Cheng, Chen, Cheng, Chikanian, Chou, Choumilov, Choutko, Chung, Clark,
  Clavero, Coignet, Consolandi, Contin, Corti, Coste, Crispoltoni, Cui, Dai,
  Delgado, Della~Torre, Demirk\"oz, Derome, Di~Falco, Di~Masso, Dimiccoli,
  D\'{\i}az, von Doetinchem, Donnini, Du, Duranti, D'Urso, Eline, Eppling,
  Eronen, Fan, Farnesini, Feng, Fiandrini, Fiasson, Finch, Fisher, Galaktionov,
  Gallucci, Garc\'{\i}a, Garc\'{\i}a-L\'opez, Gargiulo, Gast, Gebauer, Gervasi,
  Ghelfi, Gillard, Giovacchini, Goglov, Gong, Goy, Grabski, Grandi, Graziani,
  Guandalini, Guerri, Guo, Habiby, Haino, Han, He, Heil, Hoffman, Hsieh, Huang,
  Huh, Incagli, Ionica, Jang, Jinchi, Kanishev, Kim, Kim, Kirn, Kossakowski,
  Kounina, Kounine, Koutsenko, Krafczyk, Kunz, La~Vacca, Laudi, Laurenti,
  Lazzizzera, Lebedev, Lee, Lee, Leluc, Li, Li, Li, Li, Li, Li, Li, Li, Li,
  Lim, Lin, Lipari, Lippert, Liu, Liu, Lomtadze, Lu, Lu, Luebelsmeyer, Luo,
  Luo, Lv, Majka, Malinin, Ma\~n\'a, Mar\'{\i}n, Martin, Mart\'{\i}nez, Masi,
  Maurin, Menchaca-Rocha, Meng, Mo, Morescalchi, Mott, M\"uller, Ni, Nikonov,
  Nozzoli, Nunes, Obermeier, Oliva, Orcinha, Palmonari, Palomares, Paniccia,
  Papi, Pauluzzi, Pedreschi, Pensotti, Pereira, Pilo, Piluso, Pizzolotto,
  Plyaskin, Pohl, Poireau, Postaci, Putze, Quadrani, Qi, R\"aih\"a, Rancoita,
  Rapin, Ricol, Rodr\'{\i}guez, Rosier-Lees, Rozhkov, Rozza, Sagdeev,
  Sandweiss, Saouter, Sbarra, Schael, Schmidt, Schuckardt, Schulz~von Dratzig,
  Schwering, Scolieri, Seo, Shan, Shan, Shi, Shi, Shi, Siedenburg, Son, Spada,
  Spinella, Sun, Sun, Tacconi, Tang, Tang, Tang, Tao, Tescaro, Ting, Ting,
  Tomassetti, Torsti, T\"urko\ifmmode~\breve{g}\else \u{g}\fi{}lu, Urban,
  Vagelli, Valente, Vannini, Valtonen, Vaurynovich, Vecchi, Velasco, Vialle,
  Wang, Wang, Wang, Wang, Wang, Weng, Whitman, Wienkenh\"over, Wu, Xia, Xie,
  Xie, Xiong, Xin, Xu, Xu, Yan, Yang, Yang, Ye, Yi, Yu, Yu, Zeissler, Zhang,
  Zhang, Zhang, Zhang, Zheng, Zhuang, Zhukov, Zichichi, Zimmermann, Zuccon, \&
  Zurbach}]{PhysRevLett.113.221102}
Aguilar, M., Aisa, D., Alpat, B., {et~al.} 2014, PhRvL, 113, 221102,
  \dodoi{10.1103/PhysRevLett.113.221102}

\bibitem[{Aguilar {et~al.}(2015{\natexlab{a}})Aguilar, Aisa, Alpat, Alvino,
  Ambrosi, Andeen, Arruda, Attig, Azzarello, Bachlechner, Barao, Barrau,
  Barrin, Bartoloni, Basara, Battarbee, Battiston, Bazo, Becker, Behlmann,
  Beischer, Berdugo, Bertucci, Bigongiari, Bindi, Bizzaglia, Bizzarri, Boella,
  de~Boer, Bollweg, Bonnivard, Borgia, Borsini, Boschini, Bourquin, Burger,
  Cadoux, Cai, Capell, Caroff, Casaus, Cascioli, Castellini, Cernuda, Cerreta,
  Cervelli, Chae, Chang, Chen, Chen, Cheng, Chen, Cheng, Chou, Choumilov,
  Choutko, Chung, Clark, Clavero, Coignet, Consolandi, Contin, Corti, Gil,
  Coste, Creus, Crispoltoni, Cui, Dai, Delgado, Della~Torre, Demirk\"oz,
  Derome, Di~Falco, Di~Masso, Dimiccoli, D\'{\i}az, von Doetinchem, Donnini,
  Du, Duranti, D'Urso, Eline, Eppling, Eronen, Fan, Farnesini, Feng, Fiandrini,
  Fiasson, Finch, Fisher, Galaktionov, Gallucci, Garc\'{\i}a,
  Garc\'{\i}a-L\'opez, Gargiulo, Gast, Gebauer, Gervasi, Ghelfi, Gillard,
  Giovacchini, Goglov, Gong, Goy, Grabski, Grandi, Graziani, Guandalini,
  Guerri, Guo, Haas, Habiby, Haino, Han, He, Heil, Hoffman, Hsieh, Huang, Huh,
  Incagli, Ionica, Jang, Jinchi, Kanishev, Kim, Kim, Kirn, Kossakowski,
  Kounina, Kounine, Koutsenko, Krafczyk, La~Vacca, Laudi, Laurenti, Lazzizzera,
  Lebedev, Lee, Lee, Leluc, Levi, Li, Li, Li, Li, Li, Li, Li, Li, Li, Lim, Lin,
  Lipari, Lippert, Liu, Liu, Lolli, Lomtadze, Lu, Lu, Lu, Luebelsmeyer, Luo,
  Lv, Majka, Ma\~n\'a, Mar\'{\i}n, Martin, Mart\'{\i}nez, Masi, Maurin,
  Menchaca-Rocha, Meng, Mo, Morescalchi, Mott, M\"uller, Ni, Nikonov, Nozzoli,
  Nunes, Obermeier, Oliva, Orcinha, Palmonari, Palomares, Paniccia, Papi,
  Pauluzzi, Pedreschi, Pensotti, Pereira, Picot-Clemente, Pilo, Piluso,
  Pizzolotto, Plyaskin, Pohl, Poireau, Postaci, Putze, Quadrani, Qi, Qin, Qu,
  R\"aih\"a, Rancoita, Rapin, Ricol, Rodr\'{\i}guez, Rosier-Lees, Rozhkov,
  Rozza, Sagdeev, Sandweiss, Saouter, Sbarra, Schael, Schmidt, von Dratzig,
  Schwering, Scolieri, Seo, Shan, Shan, Shi, Shi, Shi, Siedenburg, Son, Spada,
  Spinella, Sun, Sun, Tacconi, Tang, Tang, Tang, Tao, Tescaro, Ting, Ting,
  Tomassetti, Torsti, T\"urko\ifmmode~\breve{g}\else \u{g}\fi{}lu, Urban,
  Vagelli, Valente, Vannini, Valtonen, Vaurynovich, Vecchi, Velasco, Vialle,
  Vitale, Vitillo, Wang, Wang, Wang, Wang, Wang, Wang, Weng, Whitman,
  Wienkenh\"over, Wu, Wu, Xia, Xie, Xie, Xiong, Xin, Xu, Xu, Yan, Yang, Yang,
  Ye, Yi, Yu, Yu, Zeissler, Zhang, Zhang, Zhang, Zhang, Zheng, Zhuang, Zhukov,
  Zichichi, Zimmermann, Zuccon, \& Zurbach}]{PhysRevLett.114.171103}
---. 2015{\natexlab{a}}, PhRvL, 114, 171103,
  \dodoi{10.1103/PhysRevLett.114.171103}

\bibitem[{Aguilar {et~al.}(2015{\natexlab{b}})Aguilar, Aisa, Alpat, Alvino,
  Ambrosi, Andeen, Arruda, Attig, Azzarello, Bachlechner, Barao, Barrau,
  Barrin, Bartoloni, Basara, Battarbee, Battiston, Bazo, Becker, Behlmann,
  Beischer, Berdugo, Bertucci, Bindi, Bizzaglia, Bizzarri, Boella, de~Boer,
  Bollweg, Bonnivard, Borgia, Borsini, Boschini, Bourquin, Burger, Cadoux, Cai,
  Capell, Caroff, Casaus, Castellini, Cernuda, Cerreta, Cervelli, Chae, Chang,
  Chen, Chen, Chen, Chen, Cheng, Chou, Choumilov, Choutko, Chung, Clark,
  Clavero, Coignet, Consolandi, Contin, Corti, Gil, Coste, Creus, Crispoltoni,
  Cui, Dai, Delgado, Della~Torre, Demirk\"oz, Derome, Di~Falco, Di~Masso,
  Dimiccoli, D\'{\i}az, von Doetinchem, Donnini, Duranti, D'Urso, Egorov,
  Eline, Eppling, Eronen, Fan, Farnesini, Feng, Fiandrini, Fiasson, Finch,
  Fisher, Formato, Galaktionov, Gallucci, Garc\'{\i}a, Garc\'{\i}a-L\'opez,
  Gargiulo, Gast, Gebauer, Gervasi, Ghelfi, Giovacchini, Goglov, Gong, Goy,
  Grabski, Grandi, Graziani, Guandalini, Guerri, Guo, Haas, Habiby, Haino, Han,
  He, Heil, Hoffman, Hsieh, Huang, Huh, Incagli, Ionica, Jang, Jinchi,
  Kanishev, Kim, Kim, Kirn, Korkmaz, Kossakowski, Kounina, Kounine, Koutsenko,
  Krafczyk, La~Vacca, Laudi, Laurenti, Lazzizzera, Lebedev, Lee, Lee, Leluc,
  Li, Li, Li, Li, Li, Li, Li, Li, Li, Li, Lim, Lin, Lipari, Lippert, Liu, Liu,
  Liu, Lolli, Lomtadze, Lu, Lu, Lu, Luebelsmeyer, Luo, Luo, Lv, Majka,
  Ma\~n\'a, Mar\'{\i}n, Martin, Mart\'{\i}nez, Masi, Maurin, Menchaca-Rocha,
  Meng, Mo, Morescalchi, Mott, M\"uller, Nelson, Ni, Nikonov, Nozzoli, Nunes,
  Obermeier, Oliva, Orcinha, Palmonari, Palomares, Paniccia, Papi, Pauluzzi,
  Pedreschi, Pensotti, Pereira, Picot-Clemente, Pilo, Piluso, Pizzolotto,
  Plyaskin, Pohl, Poireau, Putze, Quadrani, Qi, Qin, Qu, R\"aih\"a, Rancoita,
  Rapin, Ricol, Rodr\'{\i}guez, Rosier-Lees, Rozhkov, Rozza, Sagdeev,
  Sandweiss, Saouter, Schael, Schmidt, von Dratzig, Schwering, Scolieri, Seo,
  Shan, Shan, Shi, Shi, Shi, Siedenburg, Son, Song, Spada, Spinella, Sun, Sun,
  Tacconi, Tang, Tang, Tang, Tao, Tescaro, Ting, Ting, Tomassetti, Torsti,
  T\"urko\ifmmode~\breve{g}\else \u{g}\fi{}lu, Urban, Vagelli, Valente,
  Vannini, Valtonen, Vaurynovich, Vecchi, Velasco, Vialle, Vitale, Vitillo,
  Wang, Wang, Wang, Wang, Wang, Wang, Weng, Whitman, Wienkenh\"over,
  Willenbrock, Wu, Wu, Xia, Xie, Xie, Xiong, Xu, Xu, Yan, Yang, Yang, Yang, Ye,
  Yi, Yu, Yu, Zeissler, Zhang, Zhang, Zhang, Zhang, Zhang, Zhang, Zhang, Zheng,
  Zhuang, Zhukov, Zichichi, Zimmermann, \& Zuccon}]{PhysRevLett.115.211101}
---. 2015{\natexlab{b}}, PhRvL, 115, 211101,
  \dodoi{10.1103/PhysRevLett.115.211101}

\bibitem[{{Ajello} {et~al.}(2017){Ajello}, {Baldini}, {Ballet}, {Barbiellini},
  {et~al.}}]{Fermi-LAT:2017yoi}
{Ajello}, M., {Baldini}, L., {Ballet}, J., {Barbiellini}, G., {et~al.} 2017,
  Submitted to ApJ.
\newblock \doarXiv{1705.00009}

\bibitem[{{Ajello} {et~al.}(2015){Ajello}, {Gasparrini}, {S{\'a}nchez-Conde},
  {Zaharijas}, {Gustafsson}, {Cohen-Tanugi}, {Dermer}, {Inoue}, {Hartmann},
  {Ackermann}, {Bechtol}, {Franckowiak}, {Reimer}, {Romani}, \&
  {Strong}}]{Ajello:2015mfa}
{Ajello}, M., {Gasparrini}, D., {S{\'a}nchez-Conde}, M., {et~al.} 2015, \apjl,
  800, L27, \dodoi{10.1088/2041-8205/800/2/L27}

\bibitem[{{Ajello} {et~al.}(2016){Ajello}, {Albert}, {Atwood}, {Barbiellini},
  {Bastieri}, {Bechtol}, {Bellazzini}, {Bissaldi}, {Blandford}, {Bloom},
  {Bonino}, {Bottacini}, {Brandt}, {Bregeon}, {Bruel}, {Buehler}, {Buson},
  {Caliandro}, {Cameron}, {Caputo}, {Caragiulo}, {Caraveo}, {Cecchi},
  {Chekhtman}, {Chiang}, {Chiaro}, {Ciprini}, {Cohen-Tanugi}, {Cominsky},
  {Conrad}, {Cutini}, {D'Ammando}, {de Angelis}, {de Palma}, {Desiante}, {Di
  Venere}, {Drell}, {Favuzzi}, {Ferrara}, {Fusco}, {Gargano}, {Gasparrini},
  {Giglietto}, {Giommi}, {Giordano}, {Giroletti}, {Glanzman}, {Godfrey},
  {Gomez-Vargas}, {Grenier}, {Guiriec}, {Gustafsson}, {Harding}, {Hewitt},
  {Hill}, {Horan}, {Jogler}, {J{\'o}hannesson}, {Johnson}, {Kamae}, {Karwin},
  {Kn{\"o}dlseder}, {Kuss}, {Larsson}, {Latronico}, {Li}, {Li}, {Longo},
  {Loparco}, {Lovellette}, {Lubrano}, {Magill}, {Maldera}, {Malyshev},
  {Manfreda}, {Mayer}, {Mazziotta}, {Michelson}, {Mitthumsiri}, {Mizuno},
  {Moiseev}, {Monzani}, {Morselli}, {Moskalenko}, {Murgia}, {Nuss}, {Ohno},
  {Ohsugi}, {Omodei}, {Orlando}, {Ormes}, {Paneque}, {Pesce-Rollins}, {Piron},
  {Pivato}, {Porter}, {Rain{\`o}}, {Rando}, {Razzano}, {Reimer}, {Reimer},
  {Ritz}, {S{\'a}nchez-Conde}, {Saz Parkinson}, {Sgr{\`o}}, {Siskind}, {Smith},
  {Spada}, {Spandre}, {Spinelli}, {Suson}, {Tajima}, {Takahashi}, {Thayer},
  {Torres}, {Tosti}, {Troja}, {Uchiyama}, {Vianello}, {Winer}, {Wood},
  {Zaharijas}, \& {Zimmer}}]{TheFermi-LAT:2015kwa}
{Ajello}, M., {Albert}, A., {Atwood}, W.~B., {et~al.} 2016, \apj, 819, 44,
  \dodoi{10.3847/0004-637X/819/1/44}

\bibitem[{Allgood {et~al.}(2006)Allgood, Flores, Primack, Kravtsov, Wechsler,
  Faltenbacher, \& Bullock}]{Allgood:2005eu}
Allgood, B., Flores, R.~A., Primack, J.~R., {et~al.} 2006, MNRAS, 367, 1781,
  \dodoi{10.1111/j.1365-2966.2006.10094.x}

\bibitem[{Arp(1964)}]{arp1964spiral}
Arp, H. 1964, ApJ, 139, 1045

\bibitem[{{Atwood} {et~al.}(2009){Atwood}, {Abdo}, {Ackermann}, {Althouse},
  {Anderson}, {Axelsson}, {Baldini}, {Ballet}, {Band}, {Barbiellini}, \&
  et~al.}]{Atwood:2009ez}
{Atwood}, W.~B., {Abdo}, A.~A., {Ackermann}, M., {et~al.} 2009, \apj, 697,
  1071, \dodoi{10.1088/0004-637X/697/2/1071}

\bibitem[{Babcock(1939)}]{babcock1939rotation}
Babcock, H.~W. 1939, Lick Observatory Bulletin, 19, 41

\bibitem[{Bailin \& Steinmetz(2005)}]{Bailin:2004wu}
Bailin, J., \& Steinmetz, M. 2005, ApJ, 627, 647, \dodoi{10.1086/430397}

\bibitem[{Banerjee \& Jog(2008)}]{Banerjee:2008kt}
Banerjee, A., \& Jog, C.~J. 2008, ApJ, 685, 254, \dodoi{10.1086/591223}

\bibitem[{Banerjee \& Jog(2011)}]{Banerjee:2011rr}
---. 2011, ApJ, 732, L8, \dodoi{10.1088/2041-8205/732/1/L8}

\bibitem[{Barmby {et~al.}(2006)Barmby, Ashby, Bianchi, Engelbracht, Gehrz,
  Gordon, Hinz, Huchra, Humphreys, Pahre, {et~al.}}]{barmby2006dusty}
Barmby, P., Ashby, M., Bianchi, L., {et~al.} 2006, ApJ Letters, 650, L45

\bibitem[{{Bate} {et~al.}(2014){Bate}, {Conn}, {McMonigal}, {Lewis}, {Martin},
  {McConnachie}, {Veljanoski}, {Mackey}, {Ferguson}, {Ibata}, {Irwin},
  {Fardal}, {Huxor}, \& {Babul}}]{Bate:2013jha}
{Bate}, N.~F., {Conn}, A.~R., {McMonigal}, B., {et~al.} 2014, \mnras, 437,
  3362, \dodoi{10.1093/mnras/stt2139}

\bibitem[{Beck \& Gr{\"a}ve(1982)}]{beck1982distribution}
Beck, R., \& Gr{\"a}ve, R. 1982, A\&A, 105, 192

\bibitem[{Bekhti {et~al.}(2016)Bekhti, Fl{\"o}er, Keller, Kerp, Lenz, Winkel,
  Bailin, Calabretta, Dedes, Ford, {et~al.}}]{bekhti2016hi4pi}
Bekhti, N.~B., Fl{\"o}er, L., Keller, R., {et~al.} 2016, A\&A, 594, A116

\bibitem[{Bernal {et~al.}(2016)Bernal, Necib, \& Slatyer}]{Bernal:2016guq}
Bernal, N., Necib, L., \& Slatyer, T.~R. 2016, JCAP, 1612, 030,
  \dodoi{10.1088/1475-7516/2016/12/030}

\bibitem[{Bernard {et~al.}(2015)Bernard, Ferguson, Richardson, Irwin, Barker,
  Hidalgo, Aparicio, Chapman, Ibata, Lewis, {et~al.}}]{bernard2015nature}
Bernard, E.~J., Ferguson, A.~M., Richardson, J.~C., {et~al.} 2015, MNRAS, 446,
  2789

\bibitem[{Bett {et~al.}(2007)Bett, Eke, Frenk, Jenkins, Helly, \&
  Navarro}]{Bett:2006zy}
Bett, P., Eke, V., Frenk, C.~S., {et~al.} 2007, MNRAS, 376, 215,
  \dodoi{10.1111/j.1365-2966.2007.11432.x}

\bibitem[{Binder {et~al.}(2016)Binder, Covi, Kamada, Murayama, Takahashi, \&
  Yoshida}]{Binder:2016pnr}
Binder, T., Covi, L., Kamada, A., {et~al.} 2016, JCAP, 1611, 043,
  \dodoi{10.1088/1475-7516/2016/11/043}

\bibitem[{Bird(2016)}]{Bird:2015npa}
Bird, R. 2016, PoS, ICRC2015, 851.
\newblock \doarXiv{1508.07195}

\bibitem[{Blanco \& Hooper(2019)}]{Blanco:2018esa}
Blanco, C., \& Hooper, D. 2019, JCAP, 1903, 019,
  \dodoi{10.1088/1475-7516/2019/03/019}

\bibitem[{Blitz {et~al.}(1999)Blitz, Spergel, Teuben, Hartmann, \&
  Burton}]{blitz1999high}
Blitz, L., Spergel, D.~N., Teuben, P.~J., Hartmann, D., \& Burton, W.~B. 1999,
  ApJ, 514, 818

\bibitem[{Boehm {et~al.}(2001)Boehm, Fayet, \&
  Schaeffer}]{boehm2001constraining}
Boehm, C., Fayet, P., \& Schaeffer, R. 2001, PhLB, 518, 8

\bibitem[{Boehm \& Schaeffer(2005)}]{boehm2005constraints}
Boehm, C., \& Schaeffer, R. 2005, A \& A, 438, 419

\bibitem[{{Bolatto} {et~al.}(2013){Bolatto}, {Wolfire}, \&
  {Leroy}}]{2013ARA&A..51..207B}
{Bolatto}, A.~D., {Wolfire}, M., \& {Leroy}, A.~K. 2013, \araa, 51, 207,
  \dodoi{10.1146/annurev-astro-082812-140944}

\bibitem[{{Boschini} {et~al.}(2017){Boschini}, {Della Torre}, {Gervasi},
  {Grandi}, {J{\'o}hannesson}, {Kachelriess}, {La Vacca}, {Masi}, {Moskalenko},
  {Orlando}, {Ostapchenko}, {Pensotti}, {Porter}, {Quadrani}, {Rancoita},
  {Rozza}, \& {Tacconi}}]{Boschini:2017fxq}
{Boschini}, M.~J., {Della Torre}, S., {Gervasi}, M., {et~al.} 2017, \apj, 840,
  115, \dodoi{10.3847/1538-4357/aa6e4f}

\bibitem[{{Boschini} {et~al.}(2018{\natexlab{a}}){Boschini}, {Della Torre},
  {Gervasi}, {Grandi}, {J{\'o}hannesson}, {La Vacca}, {Masi}, {Moskalenko},
  {Pensotti}, {Porter}, {Quadrani}, {Rancoita}, {Rozza}, \&
  {Tacconi}}]{Boschini:2018zdv}
---. 2018{\natexlab{a}}, \apj, 854, 94, \dodoi{10.3847/1538-4357/aaa75e}

\bibitem[{{Boschini} {et~al.}(2018{\natexlab{b}}){Boschini}, {Della Torre},
  {Gervasi}, {Grandi}, {J{\'o}hannesson}, {La Vacca}, {Masi}, {Moskalenko},
  {Pensotti}, {Porter}, {Quadrani}, {Rancoita}, {Rozza}, \&
  {Tacconi}}]{2018ApJ...858...61B}
---. 2018{\natexlab{b}}, \apj, 858, 61, \dodoi{10.3847/1538-4357/aabc54}

\bibitem[{Bose {et~al.}(2016)Bose, Hellwing, Frenk, Jenkins, Lovell, Helly, Li,
  Gonzalez-Perez, \& Gao}]{bose2016substructure}
Bose, S., Hellwing, W.~A., Frenk, C.~S., {et~al.} 2016, MNRAS, stw2686

\bibitem[{Bose {et~al.}(2017)Bose, Hellwing, Frenk, Jenkins, Lovell, Helly, Li,
  Gonzalez-Perez, \& Gao}]{Bose2016irl}
---. 2017, MNRAS, 464, 4520, \dodoi{10.1093/mnras/stw2686}

\bibitem[{Braun \& Burton(1999)}]{Braun:1998ik}
Braun, R., \& Burton, W.~B. 1999, A\&A, 341, 437.
\newblock \doarXiv{astro-ph/9810433}

\bibitem[{Braun \& Thilker(2004)}]{Braun:2003ey}
Braun, R., \& Thilker, D. 2004, A\&A, 417, 421,
  \dodoi{10.1051/0004-6361:20034423}

\bibitem[{Braun {et~al.}(2009)Braun, Thilker, Walterbos, \&
  Corbelli}]{braun2009wide}
Braun, R., Thilker, D., Walterbos, R., \& Corbelli, E. 2009, ApJ, 695, 937

\bibitem[{Bringmann {et~al.}(2016)Bringmann, Ihle, Kersten, \&
  Walia}]{bringmann2016suppressing}
Bringmann, T., Ihle, H.~T., Kersten, J., \& Walia, P. 2016, PhRvD, 94, 103529

\bibitem[{Brinks \& Burton(1984)}]{brinks1984high}
Brinks, E., \& Burton, W. 1984, A\&A, 141, 195

\bibitem[{Buckley {et~al.}(2014)Buckley, Zavala, Cyr-Racine, Sigurdson, \&
  Vogelsberger}]{buckley2014scattering}
Buckley, M.~R., Zavala, J., Cyr-Racine, F.-Y., Sigurdson, K., \& Vogelsberger,
  M. 2014, PhRvD, 90, 043524

\bibitem[{Bullock {et~al.}(2001)Bullock, Kolatt, Sigad, Somerville, Kravtsov,
  Klypin, Primack, \& Dekel}]{Bullock:1999he}
Bullock, J.~S., Kolatt, T.~S., Sigad, Y., {et~al.} 2001, MNRAS, 321, 559,
  \dodoi{10.1046/j.1365-8711.2001.04068.x}

\bibitem[{Calore {et~al.}(2015{\natexlab{a}})Calore, Cholis, McCabe, \&
  Weniger}]{Calore:2014nla}
Calore, F., Cholis, I., McCabe, C., \& Weniger, C. 2015{\natexlab{a}}, PhRvD,
  91, 063003, \dodoi{10.1103/PhysRevD.91.063003}

\bibitem[{Calore {et~al.}(2015{\natexlab{b}})Calore, Cholis, \&
  Weniger}]{Calore:2014xka}
Calore, F., Cholis, I., \& Weniger, C. 2015{\natexlab{b}}, JCAP, 1503, 038,
  \dodoi{10.1088/1475-7516/2015/03/038}

\bibitem[{Campbell {et~al.}(2010)Campbell, Dutta, \& Komatsu}]{Campbell:2010xc}
Campbell, S., Dutta, B., \& Komatsu, E. 2010, PhRvD, 82, 095007,
  \dodoi{10.1103/PhysRevD.82.095007}

\bibitem[{Carignan {et~al.}(2006)Carignan, Chemin, Huchtmeier, \&
  Lockman}]{carignan2006extended}
Carignan, C., Chemin, L., Huchtmeier, W.~K., \& Lockman, F.~J. 2006, ApJ
  Letters, 641, L109

\bibitem[{{Carlesi} {et~al.}(2016){Carlesi}, {Sorce}, {Hoffman},
  {Gottl{\"o}ber}, {Yepes}, {Libeskind}, {Pilipenko}, {Knebe}, {Courtois},
  {Tully}, \& {Steinmetz}}]{Carlesi:2016qqp}
{Carlesi}, E., {Sorce}, J.~G., {Hoffman}, Y., {et~al.} 2016, \mnras, 458, 900,
  \dodoi{10.1093/mnras/stw357}

\bibitem[{Carlson {et~al.}(2016)Carlson, Linden, \& Profumo}]{Carlson:2016iis}
Carlson, E., Linden, T., \& Profumo, S. 2016, PhRvD, 94, 063504,
  \dodoi{10.1103/PhysRevD.94.063504}

\bibitem[{Casandjian(2015)}]{Casandjian:2015hja}
Casandjian, J.-M. 2015, ApJ, 806, 240, \dodoi{10.1088/0004-637X/806/2/240}

\bibitem[{{Cherry} {et~al.}(2014){Cherry}, {Friedland}, \&
  {Shoemaker}}]{Cherry:2014xra}
{Cherry}, J.~F., {Friedland}, A., \& {Shoemaker}, I.~M. 2014, arXiv: 1411.1071.
\newblock \doarXiv{1411.1071}

\bibitem[{Cholis {et~al.}(2014)Cholis, Hooper, \& McDermott}]{Cholis:2013ena}
Cholis, I., Hooper, D., \& McDermott, S.~D. 2014, JCAP, 1402, 014,
  \dodoi{10.1088/1475-7516/2014/02/014}

\bibitem[{Cholis {et~al.}(2019)Cholis, Linden, \& Hooper}]{Cholis:2019ejx}
Cholis, I., Linden, T., \& Hooper, D. 2019.
\newblock \doarXiv{1903.02549}

\bibitem[{Cirelli {et~al.}(2013)Cirelli, Serpico, \&
  Zaharijas}]{Cirelli:2013mqa}
Cirelli, M., Serpico, P.~D., \& Zaharijas, G. 2013, JCAP, 1311, 035,
  \dodoi{10.1088/1475-7516/2013/11/035}

\bibitem[{Collins {et~al.}(2013)Collins, Chapman, Rich, Ibata, Martin, Irwin,
  Bate, Lewis, Pe{\~n}arrubia, Arimoto, {et~al.}}]{collins2013kinematic}
Collins, M.~L., Chapman, S.~C., Rich, R.~M., {et~al.} 2013, ApJ, 768, 172

\bibitem[{Colombi {et~al.}(1996)Colombi, Dodelson, \& Widrow}]{Colombi:1995ze}
Colombi, S., Dodelson, S., \& Widrow, L.~M. 1996, ApJ, 458, 1,
  \dodoi{10.1086/176788}

\bibitem[{Conn {et~al.}(2012)Conn, Ibata, Lewis, Parker, Zucker, Martin,
  McConnachie, Irwin, Tanvir, Fardal, {et~al.}}]{conn2012bayesian}
Conn, A.~R., Ibata, R.~A., Lewis, G.~F., {et~al.} 2012, ApJ, 758, 11

\bibitem[{Conn {et~al.}(2013)}]{Conn:2013iu}
Conn, A.~R., {et~al.} 2013, ApJ, 766, 120, \dodoi{10.1088/0004-637X/766/2/120}

\bibitem[{Conn {et~al.}(2016)}]{Conn:2016nnu}
---. 2016, MNRAS, 458, 3282, \dodoi{10.1093/mnras/stw513}

\bibitem[{Conrad {et~al.}(2015)Conrad, Cohen-Tanugi, \&
  Strigari}]{Conrad:2015bsa}
Conrad, J., Cohen-Tanugi, J., \& Strigari, L.~E. 2015, J. Exp. Theor. Phys.,
  121, 1104, \dodoi{10.1134/S1063776115130099}

\bibitem[{{Corbelli} {et~al.}(2010){Corbelli}, {Lorenzoni}, {Walterbos},
  {Braun}, \& {Thilker}}]{2010A&A...511A..89C}
{Corbelli}, E., {Lorenzoni}, S., {Walterbos}, R., {Braun}, R., \& {Thilker}, D.
  2010, A\&A, 511, A89, \dodoi{10.1051/0004-6361/200913297}

\bibitem[{{Cummings} {et~al.}(2016){Cummings}, {Stone}, {Heikkila}, {Lal},
  {Webber}, {J{\'o}hannesson}, {Moskalenko}, {Orlando}, \&
  {Porter}}]{2016ApJ...831...18C}
{Cummings}, A.~C., {Stone}, E.~C., {Heikkila}, B.~C., {et~al.} 2016, \apj, 831,
  18, \dodoi{10.3847/0004-637X/831/1/18}

\bibitem[{Cuoco {et~al.}(2017)Cuoco, Kramer, \& Korsmeier}]{Cuoco:2016eej}
Cuoco, A., Kramer, M., \& Korsmeier, M. 2017, Phys. Rev. Lett., 118, 191102,
  \dodoi{10.1103/PhysRevLett.118.191102}

\bibitem[{Cuoco {et~al.}(2011)Cuoco, Sellerholm, Conrad, \&
  Hannestad}]{Cuoco:2010jb}
Cuoco, A., Sellerholm, A., Conrad, J., \& Hannestad, S. 2011, MNRAS, 414, 2040,
  \dodoi{10.1111/j.1365-2966.2011.18525.x}

\bibitem[{Cyr-Racine {et~al.}(2014)Cyr-Racine, de~Putter, Raccanelli, \&
  Sigurdson}]{cyr2014constraints}
Cyr-Racine, F.-Y., de~Putter, R., Raccanelli, A., \& Sigurdson, K. 2014, PhRvD,
  89, 063517

\bibitem[{Cyr-Racine \& Sigurdson(2013)}]{cyr2013cosmology}
Cyr-Racine, F.-Y., \& Sigurdson, K. 2013, PhRvD, 87, 103515

\bibitem[{Cyr-Racine {et~al.}(2016)Cyr-Racine, Sigurdson, Zavala, Bringmann,
  Vogelsberger, \& Pfrommer}]{cyr2016ethos}
Cyr-Racine, F.-Y., Sigurdson, K., Zavala, J., {et~al.} 2016, PhRvD, 93, 123527

\bibitem[{{Dame} {et~al.}(2001){Dame}, {Hartmann}, \&
  {Thaddeus}}]{2001ApJ...547..792D}
{Dame}, T.~M., {Hartmann}, D., \& {Thaddeus}, P. 2001, \apj, 547, 792,
  \dodoi{10.1086/318388}

\bibitem[{{Damiani} {et~al.}(1997){Damiani}, {Maggio}, {Micela}, \&
  {Sciortino}}]{1997ApJ...483..350D}
{Damiani}, F., {Maggio}, A., {Micela}, G., \& {Sciortino}, S. 1997, ApJ, 483,
  350, \dodoi{10.1086/304217}

\bibitem[{Daylan {et~al.}(2016)Daylan, Finkbeiner, Hooper, Linden, Portillo,
  Rodd, \& Slatyer}]{Daylan:2014rsa}
Daylan, T., Finkbeiner, D.~P., Hooper, D., {et~al.} 2016, PDU, 12, 1,
  \dodoi{10.1016/j.dark.2015.12.005}

\bibitem[{de~Heij {et~al.}(2002)de~Heij, Braun, \& Burton}]{deHeij:2002ne}
de~Heij, V., Braun, R., \& Burton, W.~B. 2002, A\&A, 392, 417,
  \dodoi{10.1051/0004-6361:20020908}

\bibitem[{De~Looze {et~al.}(2012)De~Looze, Baes, Parkin, Wilson, Bendo,
  Boquien, Boselli, Cooray, Cormier, Fritz, {et~al.}}]{de2012herschel}
De~Looze, I., Baes, M., Parkin, T., {et~al.} 2012, MNRAS, 423, 2359

\bibitem[{Demers {et~al.}(2003)Demers, Battinelli, \&
  Letarte}]{demers2003carbon}
Demers, S., Battinelli, P., \& Letarte, B. 2003, AJ, 125, 3037

\bibitem[{Diemand {et~al.}(2007)Diemand, Kuhlen, \& Madau}]{Diemand:2006ik}
Diemand, J., Kuhlen, M., \& Madau, P. 2007, ApJ, 657, 262,
  \dodoi{10.1086/510736}

\bibitem[{Dugger {et~al.}(2010)Dugger, Jeltema, \& Profumo}]{Dugger:2010ys}
Dugger, L., Jeltema, T.~E., \& Profumo, S. 2010, JCAP, 1012, 015,
  \dodoi{10.1088/1475-7516/2010/12/015}

\bibitem[{Faber \& Gallagher(1979)}]{faber1979masses}
Faber, S.~M., \& Gallagher, J. 1979, ARA\&A, 17, 135

\bibitem[{{Falvard} {et~al.}(2004){Falvard}, {Giraud}, {Jacholkowska},
  {Lavalle}, {Nuss}, {Piron}, {Sapinski}, {Salati}, {Taillet}, {Jedamzik}, \&
  {Moultaka}}]{Falvard:2002ny}
{Falvard}, A., {Giraud}, E., {Jacholkowska}, A., {et~al.} 2004, Astroparticle
  Physics, 20, 467, \dodoi{10.1016/j.astropartphys.2003.07.001}

\bibitem[{{Fardal} {et~al.}(2013){Fardal}, {Weinberg}, {Babul}, {Irwin},
  {Guhathakurta}, {Gilbert}, {Ferguson}, {Ibata}, {Lewis}, {Tanvir}, \&
  {Huxor}}]{Fardal:2013asa}
{Fardal}, M.~A., {Weinberg}, M.~D., {Babul}, A., {et~al.} 2013, \mnras, 434,
  2779, \dodoi{10.1093/mnras/stt1121}

\bibitem[{Faria {et~al.}(2007)Faria, Johnson, Ferguson, Irwin, Ibata, Johnston,
  Lewis, \& Tanvir}]{faria2007probing}
Faria, D., Johnson, R.~A., Ferguson, A.~M., {et~al.} 2007, AJ, 133, 1275

\bibitem[{Feldmann {et~al.}(2013)Feldmann, Hooper, \& Gnedin}]{Feldmann:2012rx}
Feldmann, R., Hooper, D., \& Gnedin, N.~Y. 2013, ApJ, 763, 21,
  \dodoi{10.1088/0004-637X/763/1/21}

\bibitem[{Feng {et~al.}(2009)Feng, Kaplinghat, Tu, \& Yu}]{feng2009hidden}
Feng, J.~L., Kaplinghat, M., Tu, H., \& Yu, H.-B. 2009, JCAP, 2009, 004

\bibitem[{Ferguson {et~al.}(2002)Ferguson, Irwin, Ibata, Lewis, \&
  Tanvir}]{Ferguson:2002yi}
Ferguson, A., Irwin, M., Ibata, R., Lewis, G., \& Tanvir, N. 2002, AJ, 124,
  1452, \dodoi{10.1086/342019}

\bibitem[{{Fichtel} {et~al.}(1975){Fichtel}, {Hartman}, {Kniffen}, {Thompson},
  {Ogelman}, {Ozel}, {Tumer}, \& {Bignami}}]{fichtel1974high}
{Fichtel}, C.~E., {Hartman}, R.~C., {Kniffen}, D.~A., {et~al.} 1975, ApJ, 198,
  163

\bibitem[{Fornasa \& S{\'a}nchez-Conde(2015)}]{Fornasa:2015qua}
Fornasa, M., \& S{\'a}nchez-Conde, M.~A. 2015, PhR, 598, 1,
  \dodoi{10.1016/j.physrep.2015.09.002}

\bibitem[{Fornengo {et~al.}(2004)Fornengo, Pieri, \& Scopel}]{Fornengo:2004kj}
Fornengo, N., Pieri, L., \& Scopel, S. 2004, PhRvD, 70, 103529,
  \dodoi{10.1103/PhysRevD.70.103529}

\bibitem[{Funk(2015)}]{Funk:2015ena}
Funk, S. 2015, ARNPS, 65, 245, \dodoi{10.1146/annurev-nucl-102014-022036}

\bibitem[{Gaensler {et~al.}(2008)Gaensler, Madsen, Chatterjee, \&
  Mao}]{gaensler2008vertical}
Gaensler, B., Madsen, G., Chatterjee, S., \& Mao, S. 2008, PASA, 25, 184

\bibitem[{Galleti {et~al.}(2004)Galleti, Federici, Bellazzini, Pecci, \&
  Macrina}]{galleti20042mass}
Galleti, S., Federici, L., Bellazzini, M., Pecci, F.~F., \& Macrina, S. 2004,
  A\&A, 416, 917

\bibitem[{Gao {et~al.}(2012)Gao, Frenk, Jenkins, Springel, \&
  White}]{Gao:2011rf}
Gao, L., Frenk, C.~S., Jenkins, A., Springel, V., \& White, S. D.~M. 2012,
  MNRAS, 419, 1721, \dodoi{10.1111/j.1365-2966.2011.19836.x}

\bibitem[{Garcia {et~al.}(2010)Garcia, Hextall, Baganoff, Galache, Melia,
  Murray, Primini, Sjouwerman, \& Williams}]{Garcia:2009hu}
Garcia, M.~R., Hextall, R., Baganoff, F.~K., {et~al.} 2010, ApJ, 710, 755,
  \dodoi{10.1088/0004-637X/710/1/755}

\bibitem[{Garrison-Kimmel {et~al.}(2017)Garrison-Kimmel, Wetzel, Bullock,
  Hopkins, Boylan-Kolchin, Faucher-Gigu{\`e}re, Kere{\v{s}}, Quataert,
  Sanderson, Graus, {et~al.}}]{garrison2017not}
Garrison-Kimmel, S., Wetzel, A., Bullock, J.~S., {et~al.} 2017, MNRAS, 471,
  1709

\bibitem[{Gaskins(2016)}]{Gaskins:2016cha}
Gaskins, J.~M. 2016, ConPh, 57, 496, \dodoi{10.1080/00107514.2016.1175160}

\bibitem[{G\'enolini {et~al.}(2018)G\'enolini, Maurin, Moskalenko, \&
  Unger}]{PhysRevC.98.034611}
G\'enolini, Y., Maurin, D., Moskalenko, I.~V., \& Unger, M. 2018, PhRvC, 98,
  034611, \dodoi{10.1103/PhysRevC.98.034611}

\bibitem[{{Gil de Paz} {et~al.}(2007){Gil de Paz}, {Boissier}, {Madore},
  {Seibert}, {Joe}, {Boselli}, {Wyder}, {Thilker}, {Bianchi}, {Rey}, {Rich},
  {Barlow}, {Conrow}, {Forster}, {Friedman}, {Martin}, {Morrissey}, {Neff},
  {Schiminovich}, {Small}, {Donas}, {Heckman}, {Lee}, {Milliard}, {Szalay}, \&
  {Yi}}]{GildePaz:2006bw}
{Gil de Paz}, A., {Boissier}, S., {Madore}, B.~F., {et~al.} 2007, \apjs, 173,
  185, \dodoi{10.1086/516636}

\bibitem[{Goodenough \& Hooper(2009)}]{Goodenough:2009gk}
Goodenough, L., \& Hooper, D. 2009, arXiv: 0910.2998.
\newblock \doarXiv{0910.2998}

\bibitem[{Gordon \& Macias(2013)}]{Gordon:2013vta}
Gordon, C., \& Macias, O. 2013, PhRvD, 88, 083521,
  \dodoi{10.1103/PhysRevD.88.083521, 10.1103/PhysRevD.89.049901}

\bibitem[{Gratier {et~al.}(2010)Gratier, Braine, Rodriguez-Fernandez, Schuster,
  Kramer, Xilouris, Tabatabaei, Henkel, Corbelli, Israel,
  {et~al.}}]{gratier2010molecular}
Gratier, P., Braine, J., Rodriguez-Fernandez, N., {et~al.} 2010, A\&A, 522, A3

\bibitem[{Green(2011)}]{green2011colour}
Green, D.~A. 2011, arXiv: 1108.5083

\bibitem[{Grenier {et~al.}(2005)Grenier, Casandjian, \&
  Terrier}]{grenier2005unveiling}
Grenier, I.~A., Casandjian, J.-M., \& Terrier, R. 2005, Sci, 307, 1292

\bibitem[{Gupta {et~al.}(2012)Gupta, Mathur, Krongold, Nicastro, \&
  Galeazzi}]{Gupta:2012rh}
Gupta, A., Mathur, S., Krongold, Y., Nicastro, F., \& Galeazzi, M. 2012, ApJ,
  756, L8, \dodoi{10.1088/2041-8205/756/1/L8}

\bibitem[{Hammer {et~al.}(2010)Hammer, Yang, Wang, Puech, Flores, \&
  Fouquet}]{Hammer:2010ug}
Hammer, F., Yang, Y.~B., Wang, J.~L., {et~al.} 2010, ApJ, 725, 542,
  \dodoi{10.1088/0004-637X/725/1/542}

\bibitem[{Han {et~al.}(2016)Han, Cole, Frenk, \& Jing}]{han2016unified}
Han, J., Cole, S., Frenk, C.~S., \& Jing, Y. 2016, MNRAS, 457, 1208

\bibitem[{{Hartman} {et~al.}(1999){Hartman}, {Bertsch}, {Bloom}, {Chen},
  {Deines-Jones}, {Esposito}, {Fichtel}, {Friedlander}, {Hunter}, {McDonald},
  {Sreekumar}, {Thompson}, {Jones}, {Lin}, {Michelson}, {Nolan}, {Tompkins},
  {Kanbach}, {Mayer-Hasselwander}, {M{\"u}cke}, {Pohl}, {Reimer}, {Kniffen},
  {Schneid}, {von Montigny}, {Mukherjee}, \& {Dingus}}]{Hartman:1999fc}
{Hartman}, R.~C., {Bertsch}, D.~L., {Bloom}, S.~D., {et~al.} 1999, \apjs, 123,
  79, \dodoi{10.1086/313231}

\bibitem[{Hayashi {et~al.}(2007)Hayashi, Navarro, \& Springel}]{Hayashi:2006es}
Hayashi, E., Navarro, J.~F., \& Springel, V. 2007, MNRAS, 377, 50,
  \dodoi{10.1111/j.1365-2966.2007.11599.x}

\bibitem[{Helmi(2004)}]{Helmi:2003pp}
Helmi, A. 2004, MNRAS, 351, 643, \dodoi{10.1111/j.1365-2966.2004.07812.x}

\bibitem[{Henderson(1979)}]{henderson1979model}
Henderson, A. 1979, A\&A, 75, 311

\bibitem[{Hooper {et~al.}(2007)Hooper, Finkbeiner, \&
  Dobler}]{hooper2007possible}
Hooper, D., Finkbeiner, D.~P., \& Dobler, G. 2007, PhRvD, 76, 083012

\bibitem[{Hooper \& Goodenough(2011)}]{Hooper:2010mq}
Hooper, D., \& Goodenough, L. 2011, PhL, B697, 412,
  \dodoi{10.1016/j.physletb.2011.02.029}

\bibitem[{Hooper \& Linden(2011)}]{Hooper:2011ti}
Hooper, D., \& Linden, T. 2011, PhRvD, 84, 123005,
  \dodoi{10.1103/PhysRevD.84.123005}

\bibitem[{Hooper \& Slatyer(2013)}]{Hooper:2013rwa}
Hooper, D., \& Slatyer, T.~R. 2013, PDU, 2, 118,
  \dodoi{10.1016/j.dark.2013.06.003}

\bibitem[{Howk {et~al.}(2017)Howk, Wotta, Berg, Lehner, Lockman, Hafen, Pisano,
  Faucher-Gigu{\`e}re, Wakker, Prochaska, {et~al.}}]{howk2017project}
Howk, J.~C., Wotta, C.~B., Berg, M.~A., {et~al.} 2017, ApJ, 846, 141

\bibitem[{Howley {et~al.}(2008)Howley, Geha, Guhathakurta, Montgomery,
  Laughlin, \& Johnston}]{howley2008darwin}
Howley, K., Geha, M., Guhathakurta, P., {et~al.} 2008, ApJ, 683, 722

\bibitem[{Huang {et~al.}(2013)Huang, Urbano, \& Xue}]{Huang:2013pda}
Huang, W.-C., Urbano, A., \& Xue, W. 2013, arXiv: 1307.6862.
\newblock \doarXiv{1307.6862}

\bibitem[{Huang {et~al.}(2016)Huang, En{\ss}lin, \& Selig}]{Huang:2015rlu}
Huang, X., En{\ss}lin, T., \& Selig, M. 2016, JCAP, 1604, 030,
  \dodoi{10.1088/1475-7516/2016/04/030}

\bibitem[{Hubble(1929)}]{hubble1929spiral}
Hubble, E.~P. 1929, ApJ, 69

\bibitem[{Hulsbosch(1975)}]{hulsbosch1975studies}
Hulsbosch, A.~N. 1975, A\&A, 40, 1

\bibitem[{Huo {et~al.}(2018)Huo, Kaplinghat, Pan, \& Yu}]{huo2018signatures}
Huo, R., Kaplinghat, M., Pan, Z., \& Yu, H.-B. 2018, PhLB

\bibitem[{Huxor {et~al.}(2008)Huxor, Tanvir, Ferguson, Irwin, Ibata, Bridges,
  \& Lewis}]{huxor2008globular}
Huxor, A., Tanvir, N., Ferguson, A., {et~al.} 2008, MNRAS, 385, 1989

\bibitem[{Huxor {et~al.}(2014)Huxor, Mackey, Ferguson, Irwin, Martin, Tanvir,
  Veljanoski, McConnachie, Fishlock, Ibata, {et~al.}}]{huxor2014outer}
Huxor, A., Mackey, A., Ferguson, A., {et~al.} 2014, MNRAS, 442, 2165

\bibitem[{Ibata {et~al.}(2005)Ibata, Chapman, Ferguson, Lewis, Irwin, \&
  Tanvir}]{ibata2005accretion}
Ibata, R., Chapman, S., Ferguson, A., {et~al.} 2005, ApJ, 634, 287

\bibitem[{Ibata {et~al.}(2001)Ibata, Irwin, Lewis, Ferguson, \&
  Tanvir}]{ibata2001giant}
Ibata, R., Irwin, M., Lewis, G., Ferguson, A.~M., \& Tanvir, N. 2001, Natur,
  412, 49

\bibitem[{Ibata {et~al.}(2007)Ibata, Martin, Irwin, Chapman, Ferguson, Lewis,
  \& McConnachie}]{Ibata:2007xz}
Ibata, R., Martin, N.~F., Irwin, M., {et~al.} 2007, ApJ, 671, 1591,
  \dodoi{10.1086/522574}

\bibitem[{Ibata {et~al.}(2013)Ibata, Lewis, Conn, Irwin, McConnachie, Chapman,
  Collins, Fardal, Ferguson, Ibata, {et~al.}}]{Ibata:2013rh}
Ibata, R.~A., Lewis, G.~F., Conn, A.~R., {et~al.} 2013, Nature, 493, 62

\bibitem[{J\'{o}hannesson {et~al.}(2018)J\'{o}hannesson, Porter, \&
  Moskalenko}]{Johannesson:2018bit}
J\'{o}hannesson, G., Porter, T.~A., \& Moskalenko, I.~V. 2018, ApJ, 856, 45,
  \dodoi{10.3847/1538-4357/aab26e}

\bibitem[{{J{\'o}hannesson} {et~al.}(2016){J{\'o}hannesson}, {Ruiz de Austri},
  {Vincent}, {Moskalenko}, {Orlando}, {Porter}, {Strong}, {Trotta}, {Feroz},
  {Graff}, \& {Hobson}}]{Johannesson:2016rlh}
{J{\'o}hannesson}, G., {Ruiz de Austri}, R., {Vincent}, A.~C., {et~al.} 2016,
  \apj, 824, 16, \dodoi{10.3847/0004-637X/824/1/16}

\bibitem[{{Kachelriess} {et~al.}(2014){Kachelriess}, {Moskalenko}, \&
  {Ostapchenko}}]{2014ApJ...789..136K}
{Kachelriess}, M., {Moskalenko}, I.~V., \& {Ostapchenko}, S.~S. 2014, \apj,
  789, 136, \dodoi{10.1088/0004-637X/789/2/136}

\bibitem[{{Kachelrie{\ss}} \& {Ostapchenko}(2012)}]{2012PhRvD..86d3004K}
{Kachelrie{\ss}}, M., \& {Ostapchenko}, S. 2012, PhRvD, 86, 043004,
  \dodoi{10.1103/PhysRevD.86.043004}

\bibitem[{Kalberla {et~al.}(2005)Kalberla, Burton, Hartmann, Arnal, Bajaja,
  Morras, \& P{\"o}ppel}]{kalberla2005leiden}
Kalberla, P.~M., Burton, W., Hartmann, D., {et~al.} 2005, A\&A, 440, 775

\bibitem[{{Kalberla} \& {Kerp}(2009)}]{2009ARA&A..47...27K}
{Kalberla}, P.~M.~W., \& {Kerp}, J. 2009, \araa, 47, 27,
  \dodoi{10.1146/annurev-astro-082708-101823}

\bibitem[{{Kamae} {et~al.}(2006){Kamae}, {Karlsson}, {Mizuno}, {Abe}, \&
  {Koi}}]{2006ApJ...647..692K}
{Kamae}, T., {Karlsson}, N., {Mizuno}, T., {Abe}, T., \& {Koi}, T. 2006, \apj,
  647, 692, \dodoi{10.1086/505189}

\bibitem[{Kamionkowski \& Kinkhabwala(1998)}]{Kamionkowski:1997xg}
Kamionkowski, M., \& Kinkhabwala, A. 1998, PhRvD, 57, 3256,
  \dodoi{10.1103/PhysRevD.57.3256}

\bibitem[{Karwin {et~al.}(2017)Karwin, Murgia, Tait, Porter, \&
  Tanedo}]{Karwin:2016tsw}
Karwin, C., Murgia, S., Tait, T. M.~P., Porter, T.~A., \& Tanedo, P. 2017,
  PhRvD, 95, 103005, \dodoi{10.1103/PhysRevD.95.103005}

\bibitem[{Kerp {et~al.}(2016)Kerp, Kalberla, Bekhti, Fl{\"o}er, Lenz, \&
  Winkel}]{kerp2016survey}
Kerp, J., Kalberla, P., Bekhti, N.~B., {et~al.} 2016, A\& A, 589, A120

\bibitem[{Klypin {et~al.}(2002)Klypin, Zhao, \& Somerville}]{Klypin:2001xu}
Klypin, A., Zhao, H., \& Somerville, R.~S. 2002, ApJ, 573, 597,
  \dodoi{10.1086/340656}

\bibitem[{{Kolpak} {et~al.}(2002){Kolpak}, {Jackson}, {Bania}, \&
  {Dickey}}]{2002ApJ...578..868K}
{Kolpak}, M.~A., {Jackson}, J.~M., {Bania}, T.~M., \& {Dickey}, J.~M. 2002,
  \apj, 578, 868, \dodoi{10.1086/342659}

\bibitem[{Kovesi(2015)}]{kovesi2015good}
Kovesi, P. 2015, arXiv: 1509.03700

\bibitem[{Kraushaar \& Clark(1962)}]{kraushaar1962search}
Kraushaar, W., \& Clark, G. 1962, PhRvL, 8, 106

\bibitem[{Kraushaar {et~al.}(1972)Kraushaar, Clark, Garmire, Borken, Higbie,
  Leong, \& Thorsos}]{kraushaar1972high}
Kraushaar, W., Clark, G., Garmire, G., {et~al.} 1972, ApJ, 177, 341

\bibitem[{Kuhlen {et~al.}(2007)Kuhlen, Diemand, \& Madau}]{Kuhlen:2007ku}
Kuhlen, M., Diemand, J., \& Madau, P. 2007, ApJ, 671, 1135,
  \dodoi{10.1086/522878}

\bibitem[{Lacroix {et~al.}(2014)Lacroix, Boehm, \& Silk}]{Lacroix:2014eea}
Lacroix, T., Boehm, C., \& Silk, J. 2014, Phys. Rev., D90, 043508,
  \dodoi{10.1103/PhysRevD.90.043508}

\bibitem[{{Lauer} {et~al.}(2012){Lauer}, {Bender}, {Kormendy}, {Rosenfield}, \&
  {Green}}]{2012ApJ...745..121L}
{Lauer}, T.~R., {Bender}, R., {Kormendy}, J., {Rosenfield}, P., \& {Green},
  R.~F. 2012, ApJ, 745, 121, \dodoi{10.1088/0004-637X/745/2/121}

\bibitem[{Law {et~al.}(2009)Law, Majewski, \& Johnston}]{Law:2009yq}
Law, D.~R., Majewski, S.~R., \& Johnston, K.~V. 2009, ApJ, 703, L67,
  \dodoi{10.1088/0004-637X/703/1/L67}

\bibitem[{Lehner {et~al.}(2015)Lehner, Howk, \& Wakker}]{lehner2015evidence}
Lehner, N., Howk, J.~C., \& Wakker, B.~P. 2015, ApJ, 804, 79

\bibitem[{{Lewis} {et~al.}(2013){Lewis}, {Braun}, {McConnachie}, {Irwin},
  {Ibata}, {Chapman}, {Ferguson}, {Martin}, {Fardal}, {Dubinski}, {Widrow},
  {Mackey}, {Babul}, {Tanvir}, \& {Rich}}]{Lewis:2012dj}
{Lewis}, G.~F., {Braun}, R., {McConnachie}, A.~W., {et~al.} 2013, \apj, 763, 4,
  \dodoi{10.1088/0004-637X/763/1/4}

\bibitem[{Li {et~al.}(2011)Li, Garcia, Forman, Jones, Kraft, Lal, Murray, \&
  Wang}]{Li:2010kf}
Li, Z., Garcia, M.~R., Forman, W.~R., {et~al.} 2011, ApJ, L10,
  \dodoi{10.1088/2041-8205/728/1/L10}

\bibitem[{Li {et~al.}(2016)Li, Huang, Yuan, \& Xu}]{Li:2013qya}
Li, Z., Huang, X., Yuan, Q., \& Xu, Y. 2016, JCAP, 1612, 028,
  \dodoi{10.1088/1475-7516/2016/12/028}

\bibitem[{Li \& Wang(2007)}]{Li:2007ud}
Li, Z., \& Wang, D. 2007, ApJ Letters, \dodoi{10.1086/522674}

\bibitem[{Lisanti {et~al.}(2018{\natexlab{a}})Lisanti, Mishra-Sharma, Rodd, \&
  Safdi}]{lisanti2018search}
Lisanti, M., Mishra-Sharma, S., Rodd, N.~L., \& Safdi, B.~R.
  2018{\natexlab{a}}, PhRvL, 120, 101101

\bibitem[{Lisanti {et~al.}(2018{\natexlab{b}})Lisanti, Mishra-Sharma, Rodd,
  Safdi, \& Wechsler}]{lisanti2018mapping}
Lisanti, M., Mishra-Sharma, S., Rodd, N.~L., Safdi, B.~R., \& Wechsler, R.~H.
  2018{\natexlab{b}}, PhRvD, 97, 063005

\bibitem[{Lockman(2003)}]{Lockman:2003zs}
Lockman, F.~J. 2003, ApJ Letters, 591, L33, \dodoi{10.1086/376961}

\bibitem[{Lovell {et~al.}(2014)Lovell, Frenk, Eke, Jenkins, Gao, \&
  Theuns}]{lovell2014properties}
Lovell, M.~R., Frenk, C.~S., Eke, V.~R., {et~al.} 2014, MNRAS, 439, 300

\bibitem[{Lucero \& Young(2007)}]{lucero2007radio}
Lucero, D.~M., \& Young, L.~M. 2007, AJ, 134, 2148

\bibitem[{Ludlow {et~al.}(2016)Ludlow, Bose, Angulo, Wang, Hellwing, Navarro,
  Cole, \& Frenk}]{ludlow2016mass}
Ludlow, A.~D., Bose, S., Angulo, R.~E., {et~al.} 2016, MNRAS, 460, 1214

\bibitem[{Macci{\`o} {et~al.}(2012)Macci{\`o}, Paduroiu, Anderhalden,
  Schneider, \& Moore}]{maccio2012cores}
Macci{\`o}, A.~V., Paduroiu, S., Anderhalden, D., Schneider, A., \& Moore, B.
  2012, MNRAS, 424, 1105

\bibitem[{Mack {et~al.}(2008)Mack, Jacques, Beacom, Bell, \&
  Yuksel}]{Mack:2008wu}
Mack, G.~D., Jacques, T.~D., Beacom, J.~F., Bell, N.~F., \& Yuksel, H. 2008,
  PhRvD, 78, 063542, \dodoi{10.1103/PhysRevD.78.063542}

\bibitem[{Mackey {et~al.}(2010)Mackey, Huxor, Ferguson, Irwin, Tanvir,
  McConnachie, Ibata, Chapman, \& Lewis}]{Mackey:2010ix}
Mackey, D., Huxor, A., Ferguson, A., {et~al.} 2010, ApJ, 717, L11,
  \dodoi{10.1088/2041-8205/717/1/L11}

\bibitem[{Marleau {et~al.}(2006)Marleau, Noriega-Crespo, Misselt, Gordon,
  Engelbracht, Rieke, Barmby, Willner, Mould, Gehrz,
  {et~al.}}]{marleau2006mapping}
Marleau, F., Noriega-Crespo, A., Misselt, K., {et~al.} 2006, ApJ, 646, 929

\bibitem[{Martin {et~al.}(2013)Martin, Ibata, McConnachie, Mackey, Ferguson,
  Irwin, Lewis, \& Fardal}]{martin2013pandas}
Martin, N.~F., Ibata, R.~A., McConnachie, A.~W., {et~al.} 2013, ApJ, 776, 80

\bibitem[{{Mattox} {et~al.}(1996){Mattox}, {Bertsch}, {Chiang}, {Dingus},
  {Digel}, {Esposito}, {Fierro}, {Hartman}, {Hunter}, {Kanbach}, {Kniffen},
  {Lin}, {Macomb}, {Mayer-Hasselwander}, {Michelson}, {von Montigny},
  {Mukherjee}, {Nolan}, {Ramanamurthy}, {Schneid}, {Sreekumar}, {Thompson}, \&
  {Willis}}]{1996ApJ...461..396M}
{Mattox}, J.~R., {Bertsch}, D.~L., {Chiang}, J., {et~al.} 1996, \apj, 461, 396,
  \dodoi{10.1086/177068}

\bibitem[{Mayall(1951)}]{mayall1951comparison}
Mayall, N.~U. 1951, Publications of Michigan Observatory, 10

\bibitem[{McConnachie(2012)}]{McConnachie:2012vd}
McConnachie, A.~W. 2012, AJ, 144, 4, \dodoi{10.1088/0004-6256/144/1/4}

\bibitem[{McConnachie {et~al.}(2005)McConnachie, Irwin, Ferguson, Ibata, Lewis,
  \& Tanvir}]{McConnachie:2004dv}
McConnachie, A.~W., Irwin, M.~J., Ferguson, A. M.~N., {et~al.} 2005, MNRAS,
  356, 979, \dodoi{10.1111/j.1365-2966.2004.08514.x}

\bibitem[{{McConnachie} {et~al.}(2009){McConnachie}, {Irwin}, {Ibata},
  {Dubinski}, {Widrow}, {Martin}, {C{\^o}t{\'e}}, {Dotter}, {Navarro},
  {Ferguson}, {Puzia}, {Lewis}, {Babul}, {Barmby}, {Bienaym{\'e}}, {Chapman},
  {Cockcroft}, {Collins}, {Fardal}, {Harris}, {Huxor}, {Mackey},
  {Pe{\~n}arrubia}, {Rich}, {Richer}, {Siebert}, {Tanvir}, {Valls-Gabaud}, \&
  {Venn}}]{McConnachie:2009up}
{McConnachie}, A.~W., {Irwin}, M.~J., {Ibata}, R.~A., {et~al.} 2009, \nat, 461,
  66, \dodoi{10.1038/nature08327}

\bibitem[{McMonigal {et~al.}(2015)McMonigal, Bate, Conn, Mackey, Lewis, Irwin,
  Martin, McConnachie, Ferguson, Ibata, {et~al.}}]{mcmonigal2015major}
McMonigal, B., Bate, N., Conn, A., {et~al.} 2015, MNRAS, 456, 405

\bibitem[{Miville-Deschenes \& Lagache(2005)}]{MivilleDeschenes:2004ci}
Miville-Deschenes, M.-A., \& Lagache, G. 2005, ApJS, 157, 302,
  \dodoi{10.1086/427938}

\bibitem[{Molin{\'e} {et~al.}(2017)Molin{\'e}, S{\'a}nchez-Conde,
  Palomares-Ruiz, \& Prada}]{Moline:2016pbm}
Molin{\'e}, {\'A}., S{\'a}nchez-Conde, M.~A., Palomares-Ruiz, S., \& Prada, F.
  2017, MNRAS, 466, 4974

\bibitem[{Monaco {et~al.}(2009)Monaco, Saviane, Perina, Bellazzini, Buzzoni,
  Federici, Pecci, \& Galleti}]{monaco2009young}
Monaco, L., Saviane, I., Perina, S., {et~al.} 2009, A\&A, 502, L9

\bibitem[{{Moskalenko} {et~al.}(2017){Moskalenko}, {J\'ohannesson}, {Orlando},
  {Porter}, \& {Strong}}]{PoS(ICRC2017)279}
{Moskalenko}, I.~V., {J\'ohannesson}, G., {Orlando}, E., {Porter}, T.~A., \&
  {Strong}, A.~W. 2017, PoS, ICRC2017, 279

\bibitem[{Moskalenko \& Strong(1998)}]{Moskalenko:1997gh}
Moskalenko, I.~V., \& Strong, A.~W. 1998, ApJ, 493, 694, \dodoi{10.1086/305152}

\bibitem[{Moskalenko \& Strong(2000)}]{Moskalenko:1998gw}
---. 2000, ApJ, 528, 357, \dodoi{10.1086/308138}

\bibitem[{Navarro {et~al.}(1997)Navarro, Frenk, \& White}]{Navarro:1996gj}
Navarro, J.~F., Frenk, C.~S., \& White, S. D.~M. 1997, ApJ, 490, 493,
  \dodoi{10.1086/304888}

\bibitem[{Ng {et~al.}(2014)Ng, Laha, Campbell, Horiuchi, Dasgupta, Murase, \&
  Beacom}]{Ng:2013xha}
Ng, K. C.~Y., Laha, R., Campbell, S., {et~al.} 2014, PhRvD, 89, 083001,
  \dodoi{10.1103/PhysRevD.89.083001}

\bibitem[{{\"O}gelman {et~al.}(2011){\"O}gelman, Aksaker, An{\i}lan, Dereli,
  Emraho{\u{g}}lu, \& Yegingil}]{ogelman2010discovery}
{\"O}gelman, H., Aksaker, N., An{\i}lan, S., {et~al.} 2011, AIP Conference
  Proceedings, 1379, 82

\bibitem[{Olive \& Turner(1982)}]{olive1982cosmological}
Olive, K.~A., \& Turner, M.~S. 1982, PhRvD, 25, 213

\bibitem[{Pagels \& Primack(1982)}]{pagels1982supersymmetry}
Pagels, H., \& Primack, J.~R. 1982, PhRvL, 48, 223

\bibitem[{Pawlowski {et~al.}(2017)Pawlowski, Ibata, \&
  Bullock}]{pawlowski2017lopsidedness}
Pawlowski, M.~S., Ibata, R.~A., \& Bullock, J.~S. 2017, ApJ, 850, 132

\bibitem[{Pawlowski {et~al.}(2013)Pawlowski, Kroupa, \&
  Jerjen}]{pawlowski2013dwarf}
Pawlowski, M.~S., Kroupa, P., \& Jerjen, H. 2013, MNRAS, 435, 1928

\bibitem[{Peacock {et~al.}(2010)Peacock, Maccarone, Knigge, Kundu, Waters,
  Zepf, \& Zurek}]{peacock2010m31}
Peacock, M.~B., Maccarone, T.~J., Knigge, C., {et~al.} 2010, MNRAS, 402, 803

\bibitem[{Pease(1918)}]{pease1918rotation}
Pease, F. 1918, Proceedings of the National Academy of Sciences of the United
  States of America, 4, 21

\bibitem[{Peebles(1982)}]{peebles1982large}
Peebles, P. 1982, ApJ, 263, L1

\bibitem[{Pollock {et~al.}(1981)Pollock, Masnou, Bignami, Hermsen, Swanenburg,
  Kanbach, Lichti, \& Wills}]{pollock1981search}
Pollock, A., Masnou, J., Bignami, G., {et~al.} 1981, A\&A, 94, 116

\bibitem[{{Porter} {et~al.}(2017){Porter}, {J{\'o}hannesson}, \&
  {Moskalenko}}]{porter2017high}
{Porter}, T.~A., {J{\'o}hannesson}, G., \& {Moskalenko}, I.~V. 2017, \apj, 846,
  67, \dodoi{10.3847/1538-4357/aa844d}

\bibitem[{{Porter} {et~al.}(2008){Porter}, {Moskalenko}, {Strong}, {Orlando},
  \& {Bouchet}}]{2008ApJ...682..400P}
{Porter}, T.~A., {Moskalenko}, I.~V., {Strong}, A.~W., {Orlando}, E., \&
  {Bouchet}, L. 2008, \apj, 682, 400, \dodoi{10.1086/589615}

\bibitem[{{Porter} \& {Strong}(2005)}]{2005ICRC....4...77P}
{Porter}, T.~A., \& {Strong}, A.~W. 2005, International Cosmic Ray Conference,
  4, 77

\bibitem[{Pshirkov {et~al.}(2016{\natexlab{a}})Pshirkov, Postnov, \&
  Vasiliev}]{Pshirkov:2015hda}
Pshirkov, M., Postnov, K., \& Vasiliev, V. 2016{\natexlab{a}}, PoS, ICRC2015,
  867

\bibitem[{Pshirkov {et~al.}(2016{\natexlab{b}})Pshirkov, Vasiliev, \&
  Postnov}]{Pshirkov:2016qhu}
Pshirkov, M.~S., Vasiliev, V.~V., \& Postnov, K.~A. 2016{\natexlab{b}}, MNRAS,
  459, L76, \dodoi{10.1093/mnrasl/slw045}

\bibitem[{{Ptuskin} {et~al.}(2006){Ptuskin}, {Moskalenko}, {Jones}, {Strong},
  \& {Zirakashvili}}]{2006ApJ...642..902P}
{Ptuskin}, V.~S., {Moskalenko}, I.~V., {Jones}, F.~C., {Strong}, A.~W., \&
  {Zirakashvili}, V.~N. 2006, \apj, 642, 902, \dodoi{10.1086/501117}

\bibitem[{Recchia {et~al.}(2016)Recchia, Blasi, \& Morlino}]{recchia2016radial}
Recchia, S., Blasi, P., \& Morlino, G. 2016, MNRAS: Letters, 462, L88

\bibitem[{Reddy \& Yun(2004)}]{Reddy:2003xn}
Reddy, N.~A., \& Yun, M.~S. 2004, ApJ, 600, 695, \dodoi{10.1086/379871}

\bibitem[{Reinert \& Winkler(2018)}]{Reinert:2017aga}
Reinert, A., \& Winkler, M.~W. 2018, JCAP, 1801, 055,
  \dodoi{10.1088/1475-7516/2018/01/055}

\bibitem[{Richardson {et~al.}(2008)Richardson, Ferguson, Johnson, Irwin,
  Tanvir, Faria, Ibata, Johnston, Lewis, McConnachie,
  {et~al.}}]{richardson2008nature}
Richardson, J., Ferguson, A., Johnson, R., {et~al.} 2008, AJ, 135, 1998

\bibitem[{Roberts(1893)}]{roberts1893selection}
Roberts, I. 1893, A Selection of Photographs of Stars, Star-Clusters and
  Nebulae, together with Information concerning the Instruments and the Methods
  employed in the pursuit of Celestial Photography

\bibitem[{Roberts \& Whitehurst(1975)}]{roberts1975rotation}
Roberts, M.~S., \& Whitehurst, R.~N. 1975, ApJ, 201, 327

\bibitem[{Rubin \& Ford(1970)}]{Rubin:1970zza}
Rubin, V.~C., \& Ford, Jr., W.~K. 1970, ApJ, 159, 379, \dodoi{10.1086/150317}

\bibitem[{Saglia {et~al.}(2010)Saglia, Fabricius, Bender, Montalto, Lee,
  Riffeser, Seitz, Morganti, Gerhard, \& Hopp}]{Saglia:2009tp}
Saglia, R.~P., Fabricius, M., Bender, R., {et~al.} 2010, A\&A, 509, A61,
  \dodoi{10.1051/0004-6361/200912805}

\bibitem[{Saha {et~al.}(2009)Saha, Levine, Jog, \& Blitz}]{Saha:2009dt}
Saha, K., Levine, E.~S., Jog, C.~J., \& Blitz, L. 2009, ApJ, 697, 2015,
  \dodoi{10.1088/0004-637X/697/2/2015}

\bibitem[{S{\'a}nchez-Conde \& Prada(2014)}]{sanchez2014flattening}
S{\'a}nchez-Conde, M.~A., \& Prada, F. 2014, MNRAS, 442, 2271

\bibitem[{Schlegel {et~al.}(1998)Schlegel, Finkbeiner, \&
  Davis}]{schlegel1998maps}
Schlegel, D.~J., Finkbeiner, D.~P., \& Davis, M. 1998, ApJ, 500, 525

\bibitem[{Seigar {et~al.}(2008)Seigar, Barth, \& Bullock}]{seigar2008revised}
Seigar, M.~S., Barth, A.~J., \& Bullock, J.~S. 2008, MNRAS, 389, 1911

\bibitem[{Shull(2014)}]{Shull:2014uia}
Shull, J.~M. 2014, ApJ, 784, 142, \dodoi{10.1088/0004-637X/784/2/142}

\bibitem[{Simon {et~al.}(2006)Simon, Blitz, Cole, Weinberg, \&
  Cohen}]{Simon:2005vh}
Simon, J.~D., Blitz, L., Cole, A.~A., Weinberg, M.~D., \& Cohen, M. 2006, ApJ,
  640, 270, \dodoi{10.1086/499914}

\bibitem[{Slipher(1913)}]{slipher1913radial}
Slipher, V.~M. 1913, Lowell Observatory Bulletin, 2, 56

\bibitem[{Springel {et~al.}(2008)Springel, Wang, Vogelsberger, Ludlow, Jenkins,
  Helmi, Navarro, Frenk, \& White}]{Springel:2008cc}
Springel, V., Wang, J., Vogelsberger, M., {et~al.} 2008, MNRAS, 391, 1685,
  \dodoi{10.1111/j.1365-2966.2008.14066.x}

\bibitem[{Sreekumar {et~al.}(1994)Sreekumar, Bertsch, Dingus, Esposito,
  Fichtel, Hartman, Hunter, Kanbach, Kniffen, Lin,
  {et~al.}}]{sreekumar1994study}
Sreekumar, P., Bertsch, D., Dingus, B., {et~al.} 1994, ApJ, 426, 105

\bibitem[{Stanek \& Garnavich(1998)}]{Stanek:1998cu}
Stanek, K.~Z., \& Garnavich, P.~M. 1998, ApJ, 503, L131, \dodoi{10.1086/311539}

\bibitem[{Strong \& Mattox(1996)}]{strong1996gradient}
Strong, A., \& Mattox, J. 1996, A\&A, 308, L21

\bibitem[{Strong {et~al.}(2004)Strong, Moskalenko, Reimer, Digel, \&
  Diehl}]{strong2004distribution}
Strong, A.~W., Moskalenko, I., Reimer, O., Digel, S., \& Diehl, R. 2004, A\&A,
  422, L47

\bibitem[{Strong \& Moskalenko(1998)}]{Strong:1998pw}
Strong, A.~W., \& Moskalenko, I.~V. 1998, ApJ, 509, 212, \dodoi{10.1086/306470}

\bibitem[{Strong {et~al.}(2007)Strong, Moskalenko, \& Ptuskin}]{Strong:2007nh}
Strong, A.~W., Moskalenko, I.~V., \& Ptuskin, V.~S. 2007, ARNPS, 57, 285,
  \dodoi{10.1146/annurev.nucl.57.090506.123011}

\bibitem[{Strong {et~al.}(2000)Strong, Moskalenko, \& Reimer}]{Strong:1998fr}
Strong, A.~W., Moskalenko, I.~V., \& Reimer, O. 2000, ApJ, 537, 763,
  \dodoi{10.1086/309038}

\bibitem[{{Strong} {et~al.}(2004{\natexlab{a}}){Strong}, {Moskalenko}, \&
  {Reimer}}]{2004ApJ...613..962S}
{Strong}, A.~W., {Moskalenko}, I.~V., \& {Reimer}, O. 2004{\natexlab{a}}, \apj,
  613, 962, \dodoi{10.1086/423193}

\bibitem[{{Strong} {et~al.}(2004{\natexlab{b}}){Strong}, {Moskalenko},
  {Reimer}, {Digel}, \& {Diehl}}]{2004A&A...422L..47S}
{Strong}, A.~W., {Moskalenko}, I.~V., {Reimer}, O., {Digel}, S., \& {Diehl}, R.
  2004{\natexlab{b}}, \aap, 422, L47, \dodoi{10.1051/0004-6361:20040172}

\bibitem[{Tamm {et~al.}(2012)Tamm, Tempel, Tenjes, Tihhonova, \&
  Tuvikene}]{tamm2012stellar}
Tamm, A., Tempel, E., Tenjes, P., Tihhonova, O., \& Tuvikene, T. 2012, A\&A,
  546, A4

\bibitem[{{Tibaldo} {et~al.}(2015){Tibaldo}, {Digel}, {Casandjian},
  {Franckowiak}, {Grenier}, {J{\'o}hannesson}, {Marshall}, {Moskalenko},
  {Negro}, {Orlando}, {Porter}, {Reimer}, \& {Strong}}]{Tibaldo:2015ooa}
{Tibaldo}, L., {Digel}, S.~W., {Casandjian}, J.~M., {et~al.} 2015, \apj, 807,
  161, \dodoi{10.1088/0004-637X/807/2/161}

\bibitem[{Tinivella(2016)}]{tinivella2016review}
Tinivella, M. 2016, arXiv:1610.03672

\bibitem[{van~den Aarssen {et~al.}(2012)van~den Aarssen, Bringmann, \&
  Goedecke}]{van2012thermal}
van~den Aarssen, L.~G., Bringmann, T., \& Goedecke, Y.~C. 2012, PhRvD, 85,
  123512

\bibitem[{Veljanoski {et~al.}(2014)Veljanoski, Mackey, Ferguson, Huxor,
  C{\^o}t{\'e}, Irwin, Tanvir, Pe{\~n}arrubia, Bernard, Fardal,
  {et~al.}}]{veljanoski2014outer}
Veljanoski, J., Mackey, A., Ferguson, A., {et~al.} 2014, MNRAS, 442, 2929

\bibitem[{{Velliscig} {et~al.}(2015){Velliscig}, {Cacciato}, {Schaye}, {Crain},
  {Bower}, {van Daalen}, {Dalla Vecchia}, {Frenk}, {Furlong}, {McCarthy},
  {Schaller}, \& {Theuns}}]{Velliscig:2015ffa}
{Velliscig}, M., {Cacciato}, M., {Schaye}, J., {et~al.} 2015, \mnras, 453, 721,
  \dodoi{10.1093/mnras/stv1690}

\bibitem[{{Vidal} {et~al.}(2015){Vidal}, {Dickinson}, {Davies}, \&
  {Leahy}}]{2015MNRAS.452..656V}
{Vidal}, M., {Dickinson}, C., {Davies}, R.~D., \& {Leahy}, J.~P. 2015, \mnras,
  452, 656, \dodoi{10.1093/mnras/stv1328}

\bibitem[{Vladimirov {et~al.}(2011)Vladimirov, Digel, Johannesson, Michelson,
  Moskalenko, Nolan, Orlando, Porter, \& Strong}]{Vladimirov:2010aq}
Vladimirov, A.~E., Digel, S.~W., Johannesson, G., {et~al.} 2011, Comput. Phys.
  Commun., 182, 1156, \dodoi{10.1016/j.cpc.2011.01.017}

\bibitem[{Welch {et~al.}(1998)Welch, Sage, \& Mitchell}]{welch1998puzzling}
Welch, G.~A., Sage, L.~J., \& Mitchell, G.~F. 1998, ApJ, 499, 209

\bibitem[{Wolleben(2007)}]{wolleben2007new}
Wolleben, M. 2007, ApJ, 664, 349

\bibitem[{Wright(1979)}]{wright1979tail}
Wright, M. 1979, ApJ, 233, 35

\bibitem[{Yang {et~al.}(2016)Yang, Aharonian, \& Evoli}]{yang2016radial}
Yang, R., Aharonian, F., \& Evoli, C. 2016, PhRvD, 93, 123007

\bibitem[{Yun {et~al.}(2001)Yun, Reddy, \& Condon}]{Yun:2001jx}
Yun, M.~S., Reddy, N.~A., \& Condon, J.~J. 2001, ApJ, 554, 803,
  \dodoi{10.1086/323145}

\bibitem[{Yusifov \& K{\"u}{\c{c}}{\"u}k(2004)}]{yusifov2004revisiting}
Yusifov, I., \& K{\"u}{\c{c}}{\"u}k, I. 2004, A\&A, 422, 545

\bibitem[{Zemp {et~al.}(2009)Zemp, Diemand, Kuhlen, Madau, Moore, Potter,
  Stadel, \& Widrow}]{Zemp:2008gw}
Zemp, M., Diemand, J., Kuhlen, M., {et~al.} 2009, MNRAS, 394, 641,
  \dodoi{10.1111/j.1365-2966.2008.14361.x}

\bibitem[{Zhang {et~al.}(2009)Zhang, Yang, Faltenbacher, Springel, Lin, \&
  Wang}]{Zhang:2009gh}
Zhang, Y., Yang, X., Faltenbacher, A., {et~al.} 2009, ApJ, 706, 747,
  \dodoi{10.1088/0004-637X/706/1/747}

\bibitem[{Zhou {et~al.}(2015)Zhou, Liang, Huang, Li, Fan, Feng, \&
  Chang}]{Zhou:2014lva}
Zhou, B., Liang, Y.-F., Huang, X., {et~al.} 2015, PhRvD, 91, 123010,
  \dodoi{10.1103/PhysRevD.91.123010}

\bibitem[{Zucker {et~al.}(2004)Zucker, Kniazev, Bell, Mart{\'\i}nez-Delgado,
  Grebel, Rix, Rockosi, Holtzman, Walterbos, Ivezi{\'c},
  {et~al.}}]{zucker2004new}
Zucker, D.~B., Kniazev, A.~Y., Bell, E.~F., {et~al.} 2004, ApJ Letters, 612,
  L117

\end{thebibliography}


\appendix

\section{Description of the Baseline IEMs} \label{sec:IEM_Summary}

The baseline IEMs are built using GALPROP-based models \citep{Ackermann:2012pya}. The GALPROP code for CR propagation and diffuse emission \citep{Moskalenko:1997gh,Strong:1998pw} has been under development since 1996 and is a {\it de facto} standard in astrophysics of CRs. It solves the CR transport equation with a given source distribution and boundary conditions for all CR species. This includes all relevant processes, such as the Galactic wind (convection), diffusive reacceleration in the interstellar medium, energy losses, nuclear fragmentation, radioactive decay, and production of secondary particles and isotopes. The numerical solution of the transport equation is based on a Crank-Nicholson implicit second-order scheme. Diffusion of CRs in the Galaxy is assumed to be homogeneous and isotropic within a cylindrical volume, defined by the parameters $z$ and $r$, which give the position along the longitudinal and polar axes. The spatial boundary conditions assume free particle escape. For a given halo size the diffusion coefficient, as a function of momentum and propagation parameters, is determined from secondary-to-primary ratios.  

The GALPROP code computes a full network of primary, secondary, and tertiary CR production starting from input source abundances. Starting with the heaviest primary nucleus considered (e.g., $^{64}$Ni, $A=64$), GALPROP uses the dependency tree to compute the source terms for each propagated species, while production and propagation of secondary $e^\pm$ and $\bar{p}$ are calculated at the final steps. Calculations of nuclear fragmentation and production of secondary isotopes are detailed in \citet{PhysRevC.98.034611}. The inelastically scattered $p$ and $\bar{p}$ are treated as separate components (secondary $p$, tertiary $\bar{p}$). GALPROP includes K-capture, electron stripping and pick up, and knock-on electrons.

The \gray{} and synchrotron emissivities are calculated using the propagated CR distributions, including a contribution from secondary $e^\pm$ \citep{2004ApJ...613..962S,2008ApJ...682..400P}. Production of $\pi^0$ and secondary $e^\pm$ is calculated using  parameterizations by \citet{2006ApJ...647..692K}, \citet{2012PhRvD..86d3004K}, and \citet{2014ApJ...789..136K}. The inverse Compton scattering is treated using the formalism for an anisotropic photon field~\citep{Moskalenko:1998gw} with the full spatial and angular distribution of the interstellar radiation field (ISRF) \citep{2005ICRC....4...77P,2008ApJ...682..400P,porter2017high}. The electron Bremsstrahlung calculations are described in~\cite{Strong:1998fr}. Intensity skymaps are then generated using line-of-sight integrations where the gas-related \gray{} intensities ($\pi^0$-decay, Bremsstrahlung) are normalized to the column densities of \hi\ and H$_2$ for Galactocentric annuli based on recent 21-cm and CO survey data. More details can also be found in \citet{2006ApJ...642..902P}, \citet{Strong:2007nh}, \citet{Vladimirov:2010aq}, and \citet{Johannesson:2016rlh} and in the description of the most recent version of GALPROP v.56 \citep[][and references therein]{PoS(ICRC2017)279,porter2017high}.

The interstellar gas distributions and gas-related \gray\ emission are the cornerstones of the analysis presented in this paper. The Galactic gas content is dominated by atomic (\hi) and molecular hydrogen (\htwo{}), which are present in approximately equal quantities ($\sim$$10^{9}\ M_{\odot}$) in the inner Galaxy, but with very different radial distributions. Helium represents $\approx$10\% by number and is usually assumed to be distributed similarly to the neutral gas. There is also a small fraction of low-density ionized hydrogen (\hii). The \htwo{} gas is distributed within $R<10$ kpc, with a peak at $\sim$5 kpc and a scale height of $\sim$70 pc. It is concentrated mainly in dense clouds of typical density $\sim$$10^{3}$ atom cm$^{-3}$ and masses $10^4-10^6 M_\odot$. The \htwo{} gas cannot be detected directly on large scales, but the 115 GHz emission of the molecule $^{12}$CO is a good ``tracer'',  because it forms in the dense \htwo{} clouds \citep[][]{2013ARA&A..51..207B}. The recent result obtained from a complete CO survey and infrared and \hi\ maps gives $X_{\rm CO}\equiv N_{{\rm H}_2}/W_{\rm CO}= (1.8-2.0)\times10^{20}$ cm$^{-2}$ K$^{-1}$ km$^{-1}$ s \citep{2001ApJ...547..792D,2013ARA&A..51..207B}. Observations of the diffuse \gray\ emission from the local medium and the whole Galaxy indicate that the local values are smaller and there are variations even in the local clouds $(0.63-1.0)\times10^{20}$ cm$^{-2}$ K$^{-1}$ km$^{-1}$ s \citep{ackermann2012fermi} while a gradual increase of $X_{\rm CO}$ toward the outer Galaxy is observed on the larger scale \citep{2004A&A...422L..47S,Ackermann:2012pya}.

\hi{} gas is mapped via its 21-cm emission line, which gives both distance (from the Doppler-shifted velocity and Galactic rotation models) and density information. The \hi{} gas extends out to $\sim$35 kpc with a scale height of $\sim$200 pc in the inner Galaxy that increases considerably in the outer Galaxy \citep{2009ARA&A..47...27K}. This results in an increase in the surface density with distance from the Galactic center that peaks at $10-20\  M_\odot$ pc$^{-2}$ at $\sim$12 kpc before it starts falling exponentially with a scale length of 3.75 kpc. The \hi{} disk is asymmetric with warping in the outer disk, and extends up to about 5 kpc above the Galactic plane in the north. The gas density is $\sim$1 atom cm$^{-3}$ in the Galactic plane out to a Galactocentric radius of $\sim$14 kpc, beyond which it decreases quickly. 
Less studied is a cold component of \hi, which does not emit at 21 cm and correlates with the \htwo{} distribution \citep{2002ApJ...578..868K,grenier2005unveiling}. Its presence is detected using absorption spectra measured against bright radio sources or by using the dust reddening maps.

For the purposes of calculation of \gray\ skymaps, all interstellar gas (\hi, \htwo) is assigned to the Galactocentric annuli, providing column den/sity maps for each annulus (so-called gas maps). Since the kinematic resolution vanishes for directions near the Galactic center and Galactic anti-center, the gas maps are interpolated across the regions $|l| < 10^\circ$ and $|180^\circ - l| < 10^\circ$. However, these regions do not overlap with the fields analyzed in this paper. The main uncertainty in deriving the \hi\ column densities $N$(\hi) is the \hi\ spin temperature, which is used to correct for the opacity of the 21 cm line.

Infrared emission from cold interstellar dust is also employed in the determination of the \hi\ and H$_2$ gas maps. Dust reddening maps, $E(B - V)$, are used to correct for uncertainties in $N$(\hi) and $W$(CO), which may not trace all of the neutral gas. Because the quantity of dust traced by $E(B - V)$ cannot be reliably determined in regions with high extinction, two magnitude cuts (2 and 5 mag) are applied to the maps. The extinction is highest along the MW plane, and towards the inner Galaxy, but does not have a significant effect for the M31 field. 

 The ionized hydrogen (\hii) averages only a few percent of the density of the neutral gas. However, because of its extended spatial distribution, it contributes significantly to the \gray\ emission at high latitudes. The modeling of \hii\ is based on pulsar dispersion measurements~\citep{gaensler2008vertical}. 

The ISRF is a major component in calculation of the skymap distribution of the IC emission. The Galactic ISRF (optical, infrared, and cosmic microwave background) is the result of the emission by stars, and scattering, absorption, and re-emission of absorbed starlight by dust in the interstellar medium. Because dust is optically thick to starlight, one has to model the radiation transport to obtain the spatial distribution of the ISRF intensity and spectrum throughout the Galaxy. The major uncertainties in calculations of the ISRF include the distribution of stars and star classes in the Galaxy, and the distribution of dust and its properties. 

\section{Additional Systematic Checks} \label{sec:different_IEMs}

In this section we check some additional systematics pertaining to the observations. Using the M31 IEM we vary the index of the IC components and the \hi-related components. In addition, we repeat the analysis with two different IEMs, namely, the IG IEM and the FSSC IEM. We also take a closer look at the 3FGL point sources in FM31. Lastly, we check the systematics relating to CR contamination in the detector. Note that both of the additional IEMs employ the same underlying gas maps (\hi\ and \htwo), which are also used for the M31 IEM, and so these tests do not address the systematics relating to the 3D gas distribution in the line of sight. In particular, the \hi\ maps for all IEMs use the same cuts in velocity and space for M31. 

\subsection{The M31 IEM} \label{sec:M31_IEM_extra}
\begin{figure}
\centering
\includegraphics[width=0.49\textwidth]{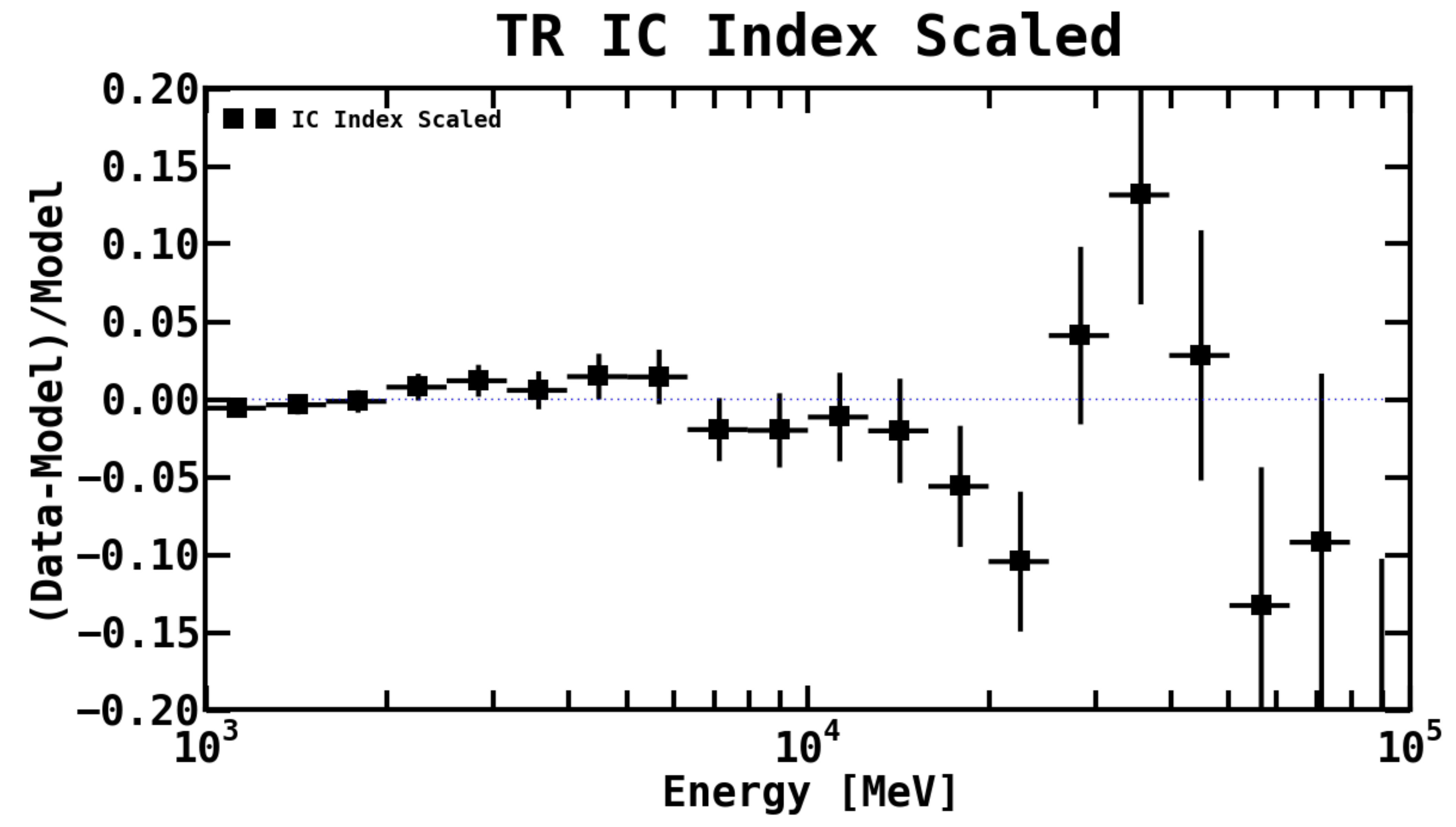}
\includegraphics[width=0.49\textwidth]{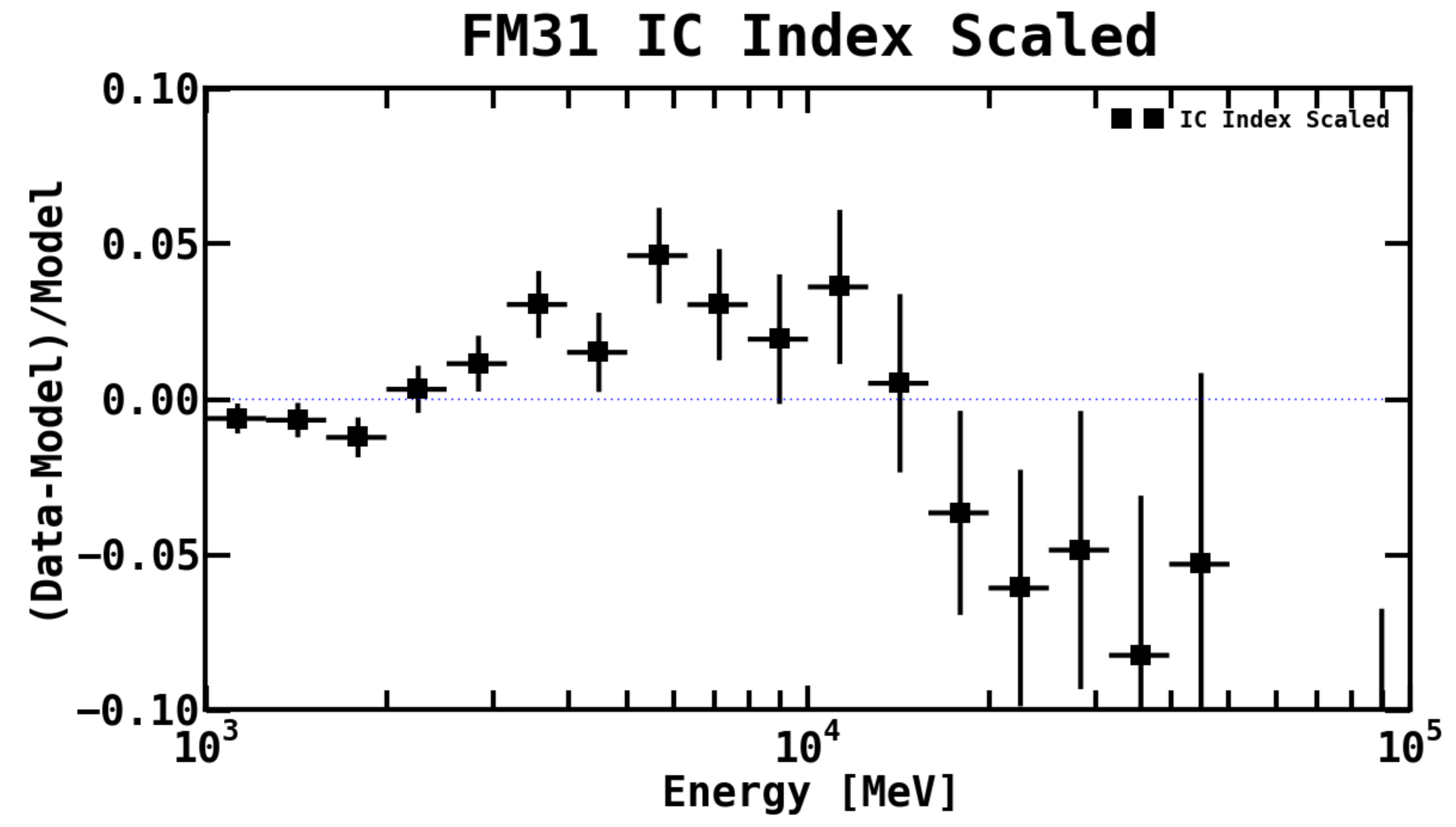}
\caption{Fractional energy residuals resulting from varying the index of the IC components using a PL scaling. Otherwise, the fit is performed in the standard way. The left shows the results for the TR and the right is for FM31.}
\label{fig:IC_index_scaled}
\end{figure}

Using the M31 IEM we perform additional variations of the fit. First, we vary the index of the IC components using a PL scaling. Otherwise, the fit is performed in the standard way, including iterating over all the point sources. Table~\ref{tab:M31_TR_IC_index_table} reports the normalizations and indices of the diffuse components for the fit in the TR, and the resulting fractional count residuals are shown in the left panel of Figure~\ref{fig:IC_index_scaled}. The IC components show a hardening towards the outer Galaxy. The normalization of the isotropic component is higher than the value found in the main analysis, but still consistent within 1$\sigma$. The count residuals are qualitatively consistent with what is found in the main analysis. The improvement in the fit is $-2\Delta$LogL$=6$.

Table~\ref{tab:M31_FM31_IC_index_table} reports the normalizations and indices of the diffuse components for the fit in FM31, and the resulting fractional count residuals are shown in the right panel of Figure~\ref{fig:IC_index_scaled}. The IC components show a hardening towards the outer Galaxy, consistent with the results for the TR. The normalization of \hi\ A5 is 1.22 \p 0.04, compared to its baseline value of 1.04 \p 0.04. The fractional count residuals still show a clear excess between $\sim$3--20 GeV, with the data being over-modeled above and below this range. For this particular fit the isotropic normalization was held fixed to its original value of 1.06. We also tested repeating the fit with the isotropic normalization fixed to 1.12 (from the TR with IC index scaled), but the results where essentially the same. To verify, we also repeated the arc fit with the M31 components, and the results are qualitatively consistent with the results found in the main analysis.  

We also test varying the index of the \hi-related components using a PL scaling. The best-fit normalizations and indices for the fit in the TR are reported in Table~\ref{tab:M31_TR_HI_index_table}. The best-fit index for \hi\ A5 is consistent with zero, thereby returning the original spectral shape. The H I A6 and H I A7 components show significant changes in index and/or normalization, but their contributions at the TR are very minor and not well constrained. Consequently, their normalization can become very high (H I A6) or go to zero (H I A7), but it has a minimal impact on the fit in the TR. 

We next scale the index for the \hi-related components in FM31. Results for this fit are reported in Table~\ref{tab:M31_FM31_HI_index_table}. The \hi-related emission shows a significant hardening towards the outer Galaxy. The \hi\ A5 component obtains a best-fit index $\Delta\alpha$ of --0.13 \p 0.02, in direct contrast to the result from the TR. For the outer Galaxy the \hi\ A7 component obtains a best-fit index of  --0.39 \p 0.11, which corresponds to an effective index of 2.37, compared to its GALPROP prediction of 2.76.

\begin{deluxetable*}{lccccc}
\tablecolumns{3}
\tablewidth{0mm}
\tablecaption{Scaling the Index for the IC Components in the TR\label{tab:M31_TR_IC_index_table}}
\tablehead{
\colhead{Component} &
\colhead{Normalization} &
\colhead{Index, $\Delta\alpha$} 
}
\startdata
\hi\ $\pi^0$, A5 &1.09 \p 0.03& \nodata \\
\hi\ $\pi^0$, A6 &5.2 \p 1.33& \nodata \\
\hi\ $\pi^0$, A7 &0.00 \p 0.06& \nodata \\
\htwo\ $\pi^0$, A5 &2.10 \p 0.12&  \nodata \\
IC, A5 & 2.11 \p 0.15& 0.07 \p 0.04 \\
IC, A6-A7  & 3.7 \p 0.4 & -0.21 \p 0.09\\
Isotropic & 1.12 \p 0.06 & \nodata 
 \enddata
\tablecomments{We vary the index of the IC components using a PL scaling $dN/dE\propto E^{-\Delta\alpha}$. The new effective index is the original index plus the best-fit index, i.e. add the exponents.}
\end{deluxetable*}

\begin{deluxetable*}{lccccc}
\tablecolumns{3}
\tablewidth{0mm}
\tablecaption{Scaling the Index for the IC Components in FM31\label{tab:M31_FM31_IC_index_table}}
\tablehead{
\colhead{Component} &
\colhead{Normalization} &
\colhead{Index, $\Delta\alpha$} 
}
\startdata
\hi\ $\pi^0$, A5 &1.22 \p 0.04& \nodata \\
\hi\ $\pi^0$, A6 &0.35 \p 0.2& \nodata \\
\hi\ $\pi^0$, A7 &2.43 \p 0.4& \nodata \\
\htwo\ $\pi^0$, A5 &2.74 \p 0.3&  \nodata \\
IC, A5 &1.86 \p 0.1 & --0.07 \p 0.03  \\
IC, A6-A7  & 1.35 \p 0.4  & --0.32 \p 0.08 \\
IC, A8  & 43.3 \p 15.8 & --0.6 \p 0.1  
 \enddata
\tablecomments{We vary the index of the IC components using a PL scaling $dN/dE\propto E^{-\Delta\alpha}$. The new effective index is the original index plus the best-fit index, i.e. add the exponents.}
\end{deluxetable*}

\begin{deluxetable*}{lccccc}
\tablecolumns{3}
\tablewidth{0mm}
\tablecaption{Scaling the Index for the \hi-Related Components in the TR\label{tab:M31_TR_HI_index_table}}
\tablehead{
\colhead{Component} &
\colhead{Normalization} &
\colhead{Index, $\Delta\alpha$} 
}
\startdata
\hi\ $\pi^0$, A5 &1.12 \p 0.01 & --0.01 \p 0.01 \\
\hi\ $\pi^0$, A6 &6.05 \p 0.69 & --0.56 \p 0.12 \\
\hi\ $\pi^0$, A7 &0.00 \p 0.1 & 0\p0 \\
\htwo\ $\pi^0$, A5 & 2.1 \p 0.07 & \nodata \\
IC, A5 &2.31 \p 0.02 & \nodata \\
IC, A6-A7  & 3.34 \p 0.15 & \nodata\\
Isotropic & 1.01 \p 0.01 & \nodata
 \enddata
\tablecomments{We vary the index of the \hi-related components using a PL scaling $dN/dE\propto E^{-\Delta\alpha}$. The new effective index is the original index plus the best-fit index, i.e. add the exponents.}
\end{deluxetable*}

\begin{deluxetable*}{lccccc}
\tablecolumns{3}
\tablewidth{0mm}
\tablecaption{Scaling the Index for the \hi-Related Components in FM31\label{tab:M31_FM31_HI_index_table}}
\tablehead{
\colhead{Component} &
\colhead{Normalization} &
\colhead{Index, $\Delta\alpha$} 
}
\startdata
\hi\ $\pi^0$, A5 & 1.39 \p 0.04 & --0.13 \p 0.02 \\
\hi\ $\pi^0$, A6 & 0.61 \p 0.27 & --0.24 \p 0.35 \\
\hi\ $\pi^0$, A7 & 3.01 \p 0.42 & --0.39 \p 0.11 \\
\htwo\ $\pi^0$, A5 & 2.83 \p 0.26 & \nodata \\
IC, A5 & 1.6 \p 0.11& \nodata \\
IC, A6-A7  & 1.13 \p 0.29 & \nodata  \\
IC, A8  & 0.00 \p 0.01  &\nodata 
 \enddata
\tablecomments{We vary the index of the \hi-related components using a PL scaling $dN/dE\propto E^{-\Delta\alpha}$. The new effective index is the original index plus the best-fit index, i.e. add the exponents.}
\end{deluxetable*}

\begin{figure}
\centering
\includegraphics[width=0.45\textwidth]{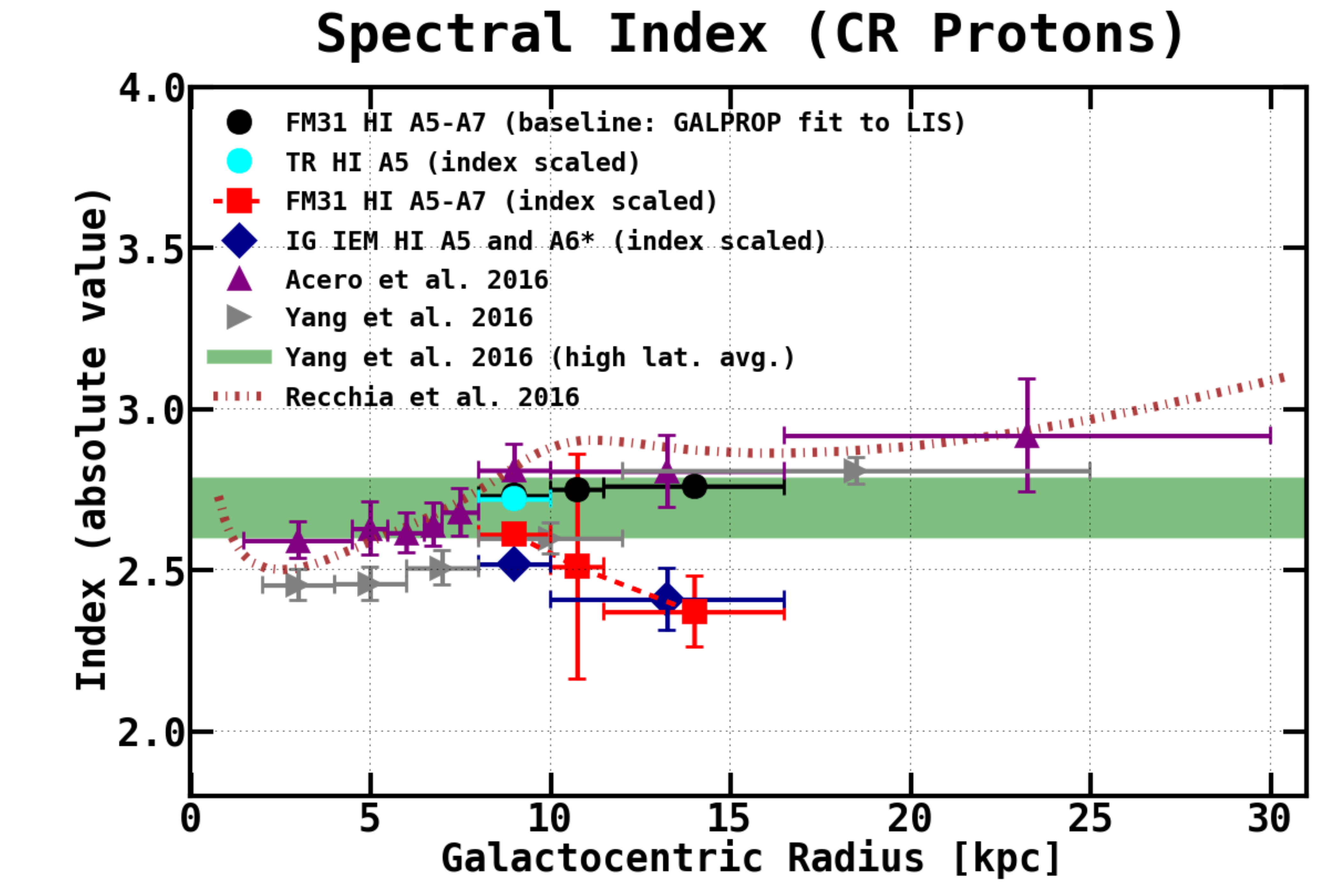}
\includegraphics[width=0.49\textwidth]{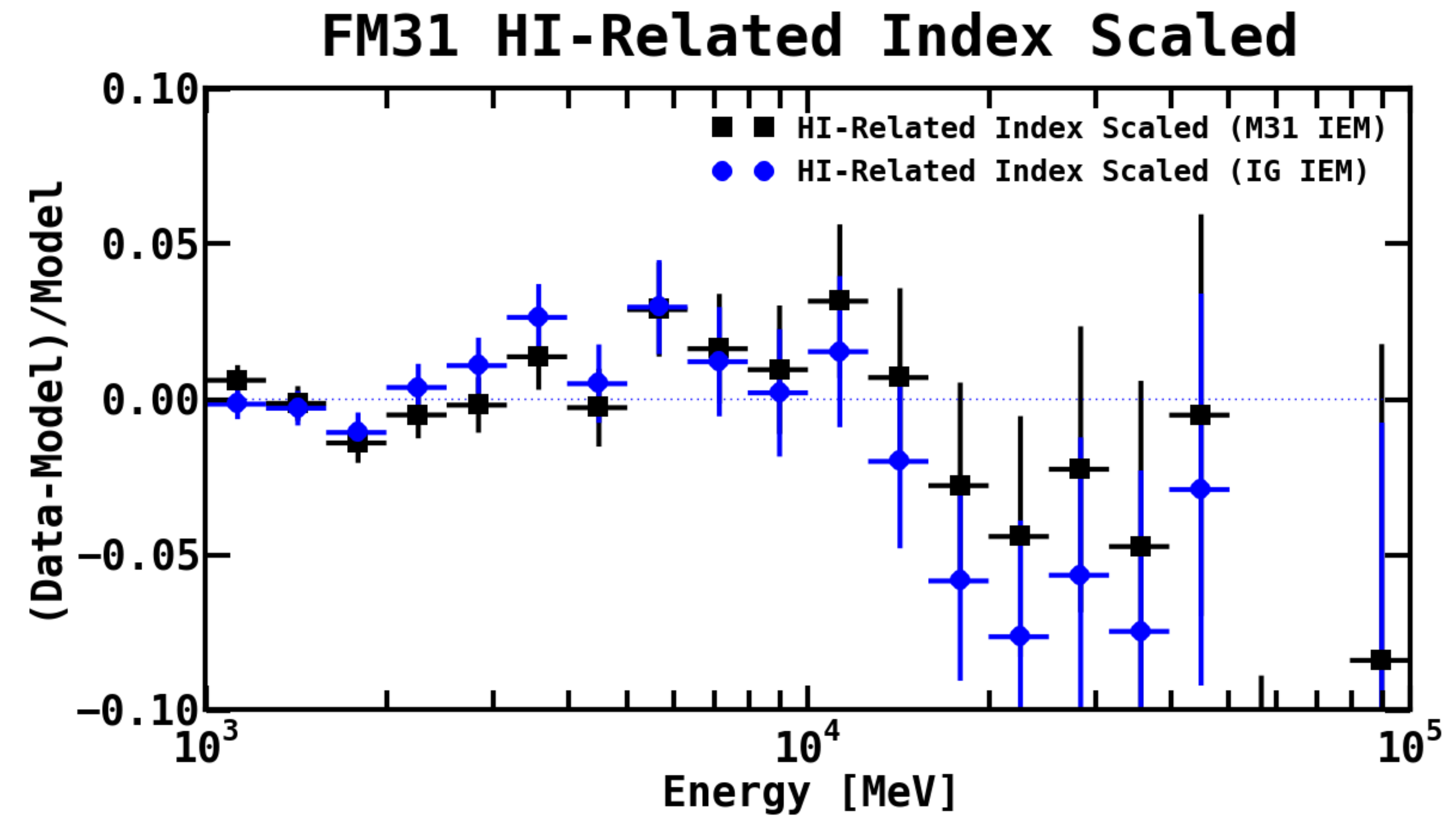}
\caption{\textbf{Left:} The index of the \hi-related emission as a function of Galactocentric radius. The black circles give the baseline index for the M31 IEM corresponding to the GALPROP prediction ($\sim$2.75). The cyan circle is the best-fit index for the local annulus obtained in the TR (using the M31 IEM), which is consistent with the GALPROP prediction. The red squares show the results for scaling the index of the \hi-related components in FM31. The middle ring, A6, has the smallest radial extension, and likewise it has the largest error bars. We also repeat the fit using the IG IEM, which only has one outer ring. The results for the IG IEM are shown with blue diamonds, and they are qualitatively consistent with the M31 IEM. For comparison, we also show other measurements. The purple upward-pointing triangles are from~\citet{Acero:2016qlg}. For the local ring the fit includes all longitudes and $10^\circ<|b|<70^\circ$, and for the outer Galaxy (last two rings) the fit includes all latitudes and $90^\circ<l<270^\circ$. The gray rightward-pointing triangles are from~\citet{yang2016radial}. The fit is performed in the latitude range $|b|<5^\circ$. The green dashed band is also from~\citet{yang2016radial}, and it shows the 1$\sigma$ average photon index (above 2 GeV) in the region $10^\circ<|b|<15^\circ$ and $90^\circ<l<150^\circ$, which corresponds to the M31 direction. Lastly, the brown dashed curve is a model fit from~\citet{recchia2016radial}, which is based on non-linear CR propagation in which transport is due to scattering and advection off self-generated turbulence. These other studies find evidence for a gradual softening towards the outer Galaxy. There is clearly a significant anomaly in FM31. \textbf{Right:} Fractional energy residuals resulting from scaling the index of the \hi-related components, for both the M31 IEM and the IG IEM.}
\label{fig:HI_index_scaled}
\end{figure}

The left panel of Figure~\ref{fig:HI_index_scaled} shows the index as a function of Galactocentric radius. Note that the values reported from this analysis are obtained by fitting a PL to the $\gamma$-ray spectrum for energies above 2 GeV (i.e.~the photon index). Also shown are the results of the template fits from~\citet{Acero:2016qlg} and~\citet{yang2016radial}, as well as a model interpretation from~\citet{recchia2016radial}. The index of the gas-related emission from those fits shows evidence of a gradual softening towards the outer Galaxy, which may also provide a hint to the origin of the flat CR gradient in the outer Galaxy~\citep{strong1996gradient,strong2004distribution,recchia2016radial}. The results obtained in FM31 clearly show an anomaly with respect to these other measurements, as well as an anomaly with respect to the results in the TR and the GALPROP predictions. The anomaly is most clearly evident for the outer Galaxy rings, A6 and A7, and it is also these rings which are found to be partially correlated with the M31 system, as is clearly seen in Figure~\ref{fig:gas_column_densities}. Because of this we also tested the fit with the IG IEM, which only has one outer ring (see Table~\ref{tab:GALPROP_parameters}). The results for the IG IEM are qualitatively consistent with the M31 IEM. These results support our conclusion that the MW IEM may be holding a fraction of gas that actually resides in the M31 system, as already discussed in the main text. 

The fractional count residuals that result from this fit are shown in the right panel of Figure~\ref{fig:HI_index_scaled}. The fit does a better job at flattening the residuals, however, excess emission still remains. To quantify the remaining excess we fit the M31-related components. All other diffuse components are held fixed to the values obtained in the baseline fit, except for the normalizations of the gas components which are rescaled. We also iterate through all the point sources. In this case the spherical halo component is detected at $\sim$ 3$\sigma$ and 4$\sigma$, for the M31 and IG IEMs, respectively. The spectrum remains qualitatively consistent with the results obtained in the main analysis. The far outer halo is not significantly detected.  

The purpose of this test was to allow for a gradual softening towards the outer Galaxy, as has been reported in other studies. We instead found a significant hardening. This result is included here only as a systematic check, but it is not a proper description of the foreground gas-related emission. However, it does further support the conclusion that there is some significant anomaly in FM31.

\subsection{The Inner Galaxy IEM} \label{sec:IG_IEMs}

\begin{figure}
\centering
\includegraphics[width=0.49\textwidth]{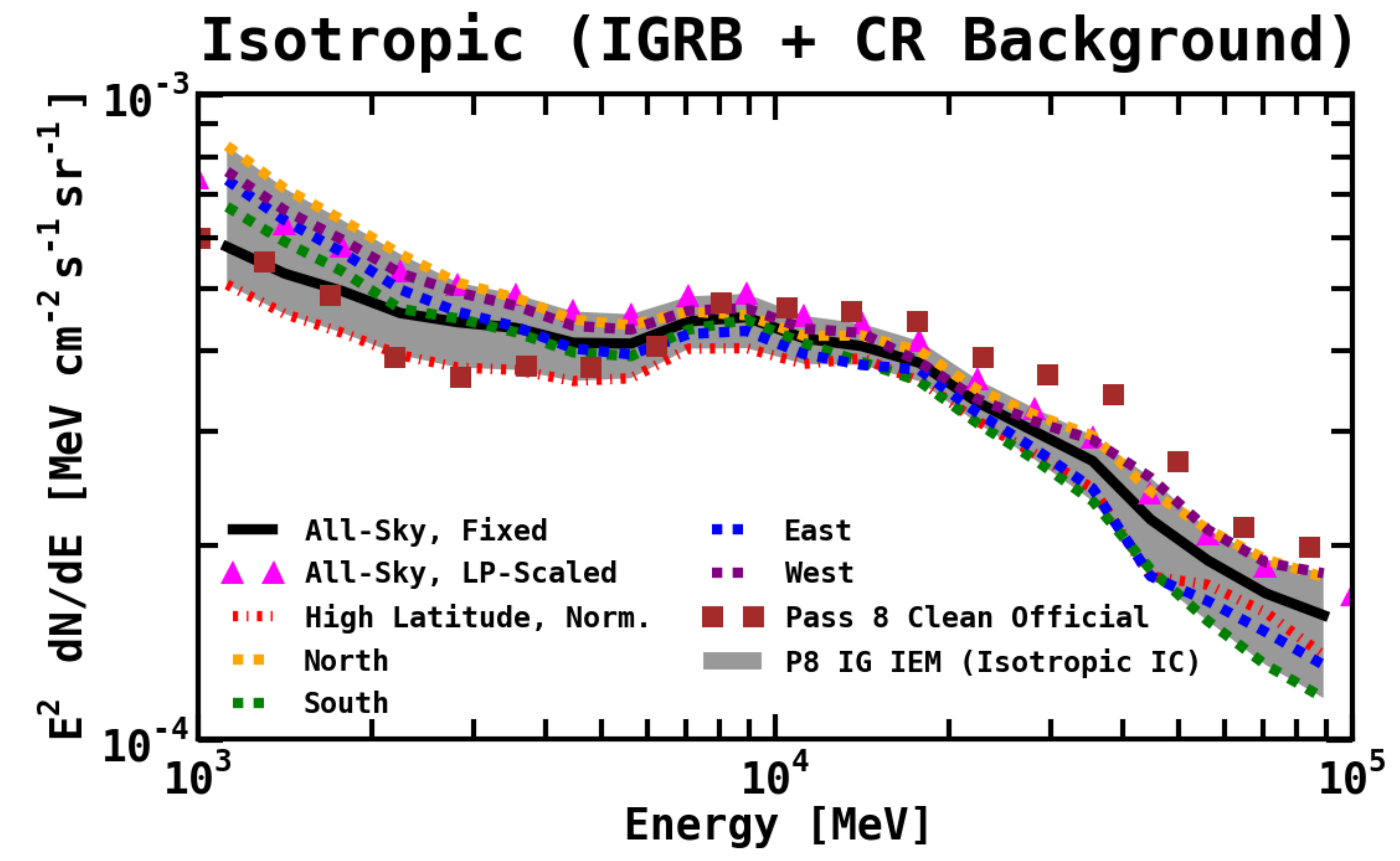}
\caption{The isotropic component includes unresolved diffuse extragalactic emission, residual instrumental background, and possibly contributions from other Galactic components which have a roughly isotropic distribution. The spectrum has a dependence on the IEM and the ROI used for the calculation, as well as the data set. For the IG IEM (which uses the isotropic IC sky maps) we calculate the \textbf{All-Sky} (solid black line) isotropic component in the following region: $|b| \geq 30^\circ, \ 45^\circ \leq  l \leq 315^\circ$. We also calculate the isotropic component in the different sky regions: \textbf{North}: $b\geq 30^\circ, \ 45^\circ \leq  l \leq 315^\circ$ (orange dashed line); \textbf{South}:  $b\leq -30^\circ,\ 45^\circ \leq  l \leq 315^\circ$ (green dashed line); \textbf{East}: $|b|\geq 30^\circ, \ 180^\circ \leq  l \leq 315^\circ$ (blue dashed line); and \textbf{West}: $|b|\geq 30^\circ, \ 45^\circ \leq  l \leq 180^\circ$ (purple dashed line). The calculations are performed using a log parabola (LP) scaling for the diffuse components. In addition, we calculate the isotropic spectrum at high latitudes ($|b|>50^\circ$), scaling just the normalizations of the diffuse components. The brown squares show the official FSSC isotropic spectrum (iso\_P8R2\_CLEAN\_V6\_v06). The gray band is our calculated isotropic component systematic uncertainty for the IG IEM, as shown in Figure~\ref{fig:Isotropic_Sytematics}.}
\label{fig:Isotropic_Sytematics_IG}
\end{figure}

The analysis is repeated with the IG IEM. Like the M31 IEM, the IG IEM is also a GALPROP-based model. There are, however, some important differences between the two, as summarized in Table~\ref{tab:GALPROP_parameters}.  The IG IEM was initially tuned to Pass 7 data, whereas in this analysis we use Pass 8 data. This is fine for the diffuse components, since we retune their normalizations in FM31; however, the isotropic spectrum needs to be recalculated self-consistently with the data set. The isotropic spectrum for the IG IEM recalculated with Pass 8 data is shown in Figure~\ref{fig:Isotropic_Sytematics_IG}, where we show it calculated for the all-sky analysis and separately for specific regions. We use the same data selection as for the main analysis. For the fit in FM31 we use the high latitude spectrum fixed to its nominal value (1.0). This is in contrast to the M31 IEM which uses the all-sky spectrum tuned directly to a TR region south of FM31. Also, the IG IEM uses the isotropic approximation for the IC component, whereas the M31 IEM uses the anisotropic formalism.

\begin{figure}
\centering
\includegraphics[width=0.47\textwidth]{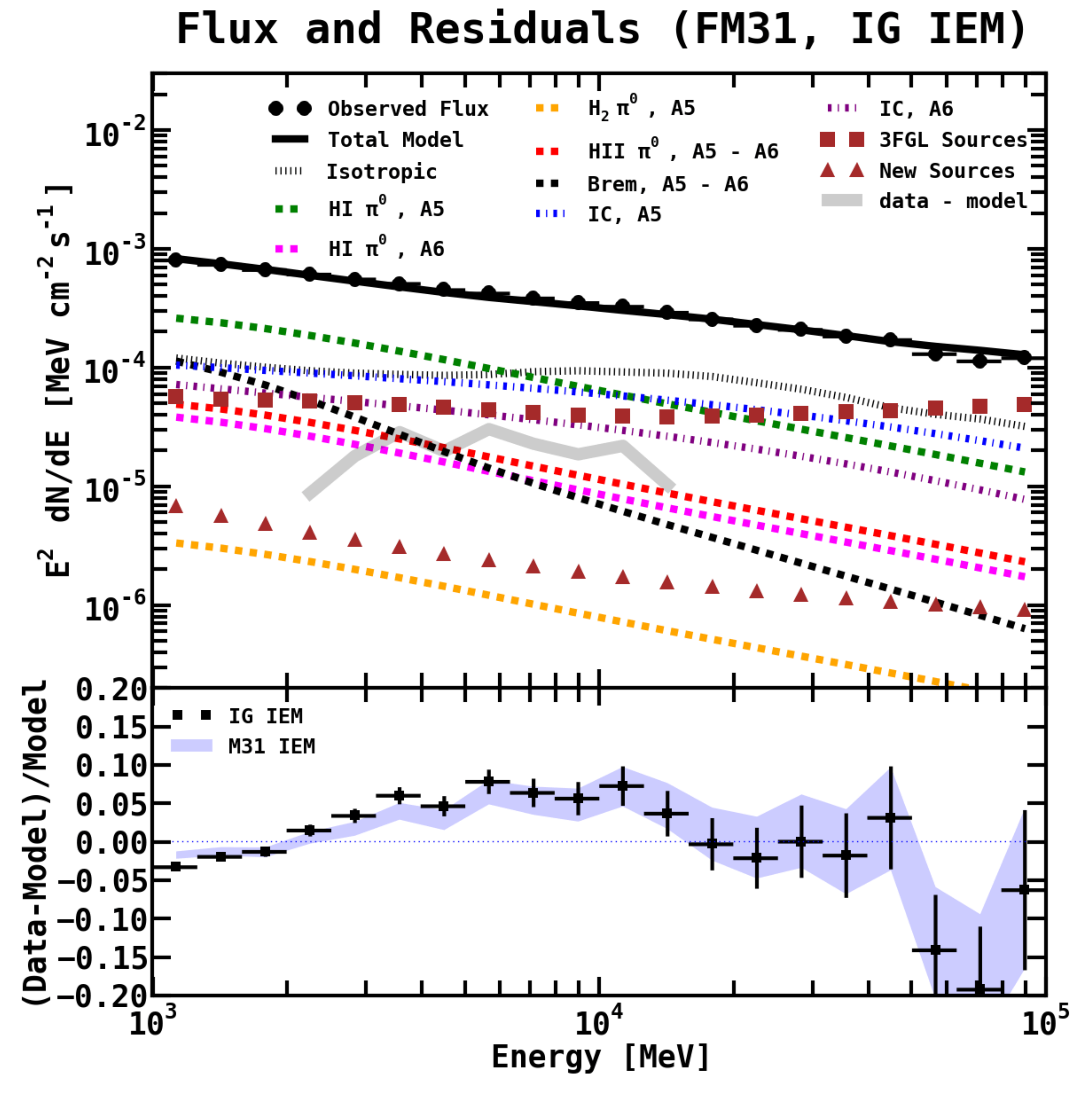}
\caption{Flux (upper panel) and fractional count residuals (lower panel) for the fit in FM31 with the IG IEM. The \hii\ and Bremsstrahlung components are fixed to their GALPROP predictions. The normalizations of the IC, \hi-related, and \htwo-related components are fit to the $\gamma$-ray data in FM31, as well as 3FGL sources within $20^\circ$ of M31, and additional point sources which we find using our point source finding procedure. The fit is performed with the high latitude isotropic component fixed to its nominal value (1.0). The bottom panel shows the fractional residuals, and the blue band shows the corresponding fractional residuals for the baseline fit (with IC scaled) with the M31 IEM. For reference, the residuals (data -- model) are also plotted in the upper panel (faint gray band).}
\label{fig:flux_and_residuals_IG}
\end{figure}

\begin{figure*}
\centering
\includegraphics[width=0.33\textwidth]{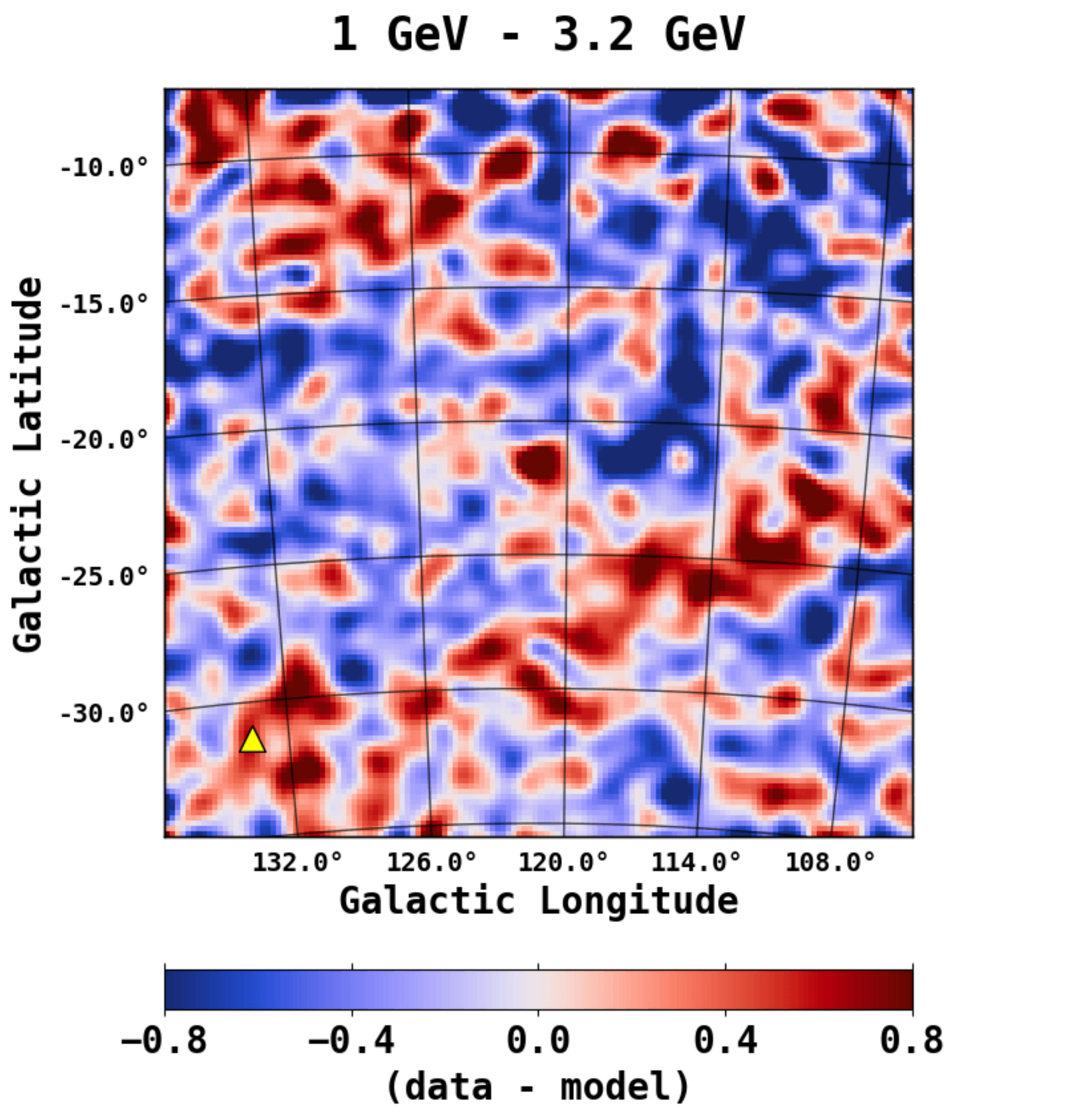}
\includegraphics[width=0.33\textwidth]{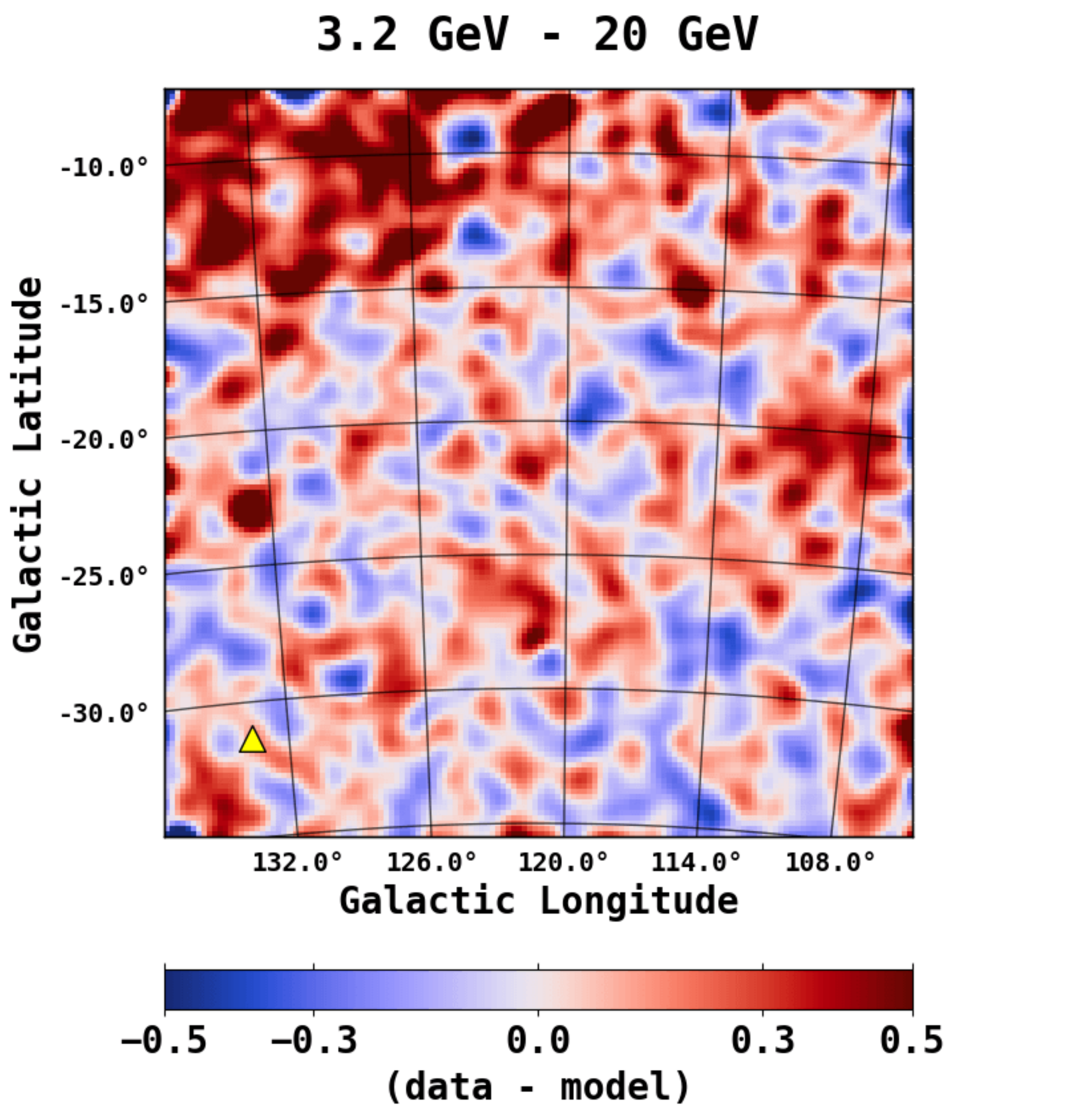}
\includegraphics[width=0.33\textwidth]{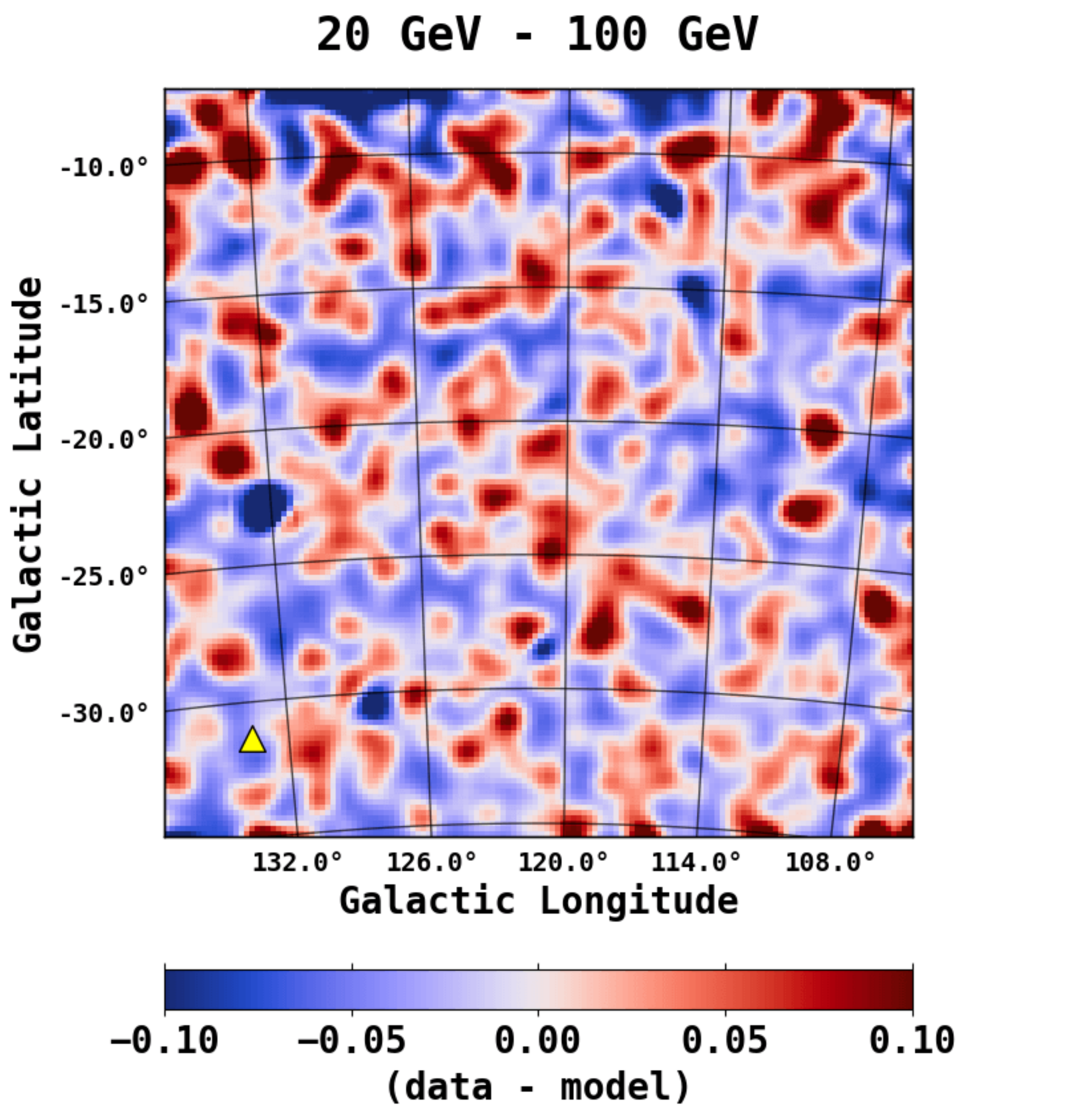}
\caption{Spatial count residuals (data -- model) resulting from the fit in FM31 with the IG IEM for three different energy bands, as indicated above each plot. The energy bins are chosen to coincide with the excess observed in the fractional residuals. The color scale corresponds to counts/pixel, and the pixel size is $0.2^\circ \times 0.2^\circ$. The images are smoothed using a $1^\circ$ Gaussian kernel. This value corresponds to the PSF (68\% containment angle) of \textit{Fermi}-LAT, which at 1 GeV is $\sim$$\mathrm{1^\circ}$. For reference, the position of M33, (l,b) = ($133.61^\circ$,$-31.33^\circ$), is shown with a yellow triangle.}
\label{fig:spatial_residuals_FM31_IG}
\end{figure*}

The baseline fit is performed just as it is in the main analysis. Results for the baseline fit are shown in Figure~\ref{fig:flux_and_residuals_IG}. The top panel in Figure~\ref{fig:flux_and_residuals_IG} shows the best-fit spectra and the bottom panel shows the resulting fractional count residuals. The blue band shows the corresponding residuals for the M31 IEM (baseline with IC scaled). Results for the two IEMs are very similar. Overall, the M31 IEM performs better over the entire energy range, showing marginal improvements compared to the IG IEM. The spatial residuals are shown in Figure~\ref{fig:spatial_residuals_FM31_IG}, and they are also qualitatively similar to those found with the M31 IEM.  The best-fit normalizations and corresponding flux and intensities for the diffuse components are reported in Table~\ref{tab:IG_baseline_normalizations}. The table includes the original best-fit normalizations from~\citet{TheFermi-LAT:2015kwa}, as well as the corrected value, which is obtained by taking the product of the original value with the updated value.

\begin{deluxetable*}{lccccc}
\tablecolumns{6}
\tablewidth{0mm}
\tablecaption{Normalizations of the Diffuse Components and Integrated Flux\label{tab:IG_baseline_normalizations}}
\tablehead{
\colhead{Component} &
\colhead{IG IEM} &
\colhead{Original Value}&
\colhead{Corrected Value}&
\colhead{Flux ($\times 10^{-9})$}&
\colhead{Intensity ($\times 10^{-8})$}\\
&
&
&
&
\colhead{(ph cm$^{-2}$ s$^{-1}$)} &
\colhead{(ph cm$^{-2}$ s$^{-1}$ sr$^{-1}$)} 
}
\startdata
\hi\ $\pi^0$, A5 &0.78 \p 0.02 & 1.21 & 0.94  &179.4 \p 5.8&76.3 \p 2.5\\
\hi\ $\pi^0$, A6 &0.75 \p 0.08 & 1.74 &1.3&25.8 \p 2.7&11.0 \p 1.1\\
\htwo\ $\pi^0$, A5 &1.1 \p 0.2 & 1.4 &1.5&2.3 \p 0.4& 1.0 \p 0.2 \\
IC, A5 &1.8 \p 0.1&1.5&2.7&86.4 \p 5.4&36.7 \p 2.3 \\
IC, A6  &2.0 \p 0.3&1.8 &3.6&54.5 \p 6.9&23.1 \p 2.9 
 \enddata
\tablecomments{Diffuse normalizations and flux for the IG IEM. The original values are from~\citet{TheFermi-LAT:2015kwa}, and they give the initial scaling with respect to the GALPROP predictions. The corrected value is then the product of the original value with the current value (second column). Intensities are calculated by using the total area of FM31, which is 0.2352 sr. The fit uses the high latitude isotropic spectrum fixed to 1.0.}
\end{deluxetable*}

\begin{figure*}
\centering
\includegraphics[width=0.33\textwidth]{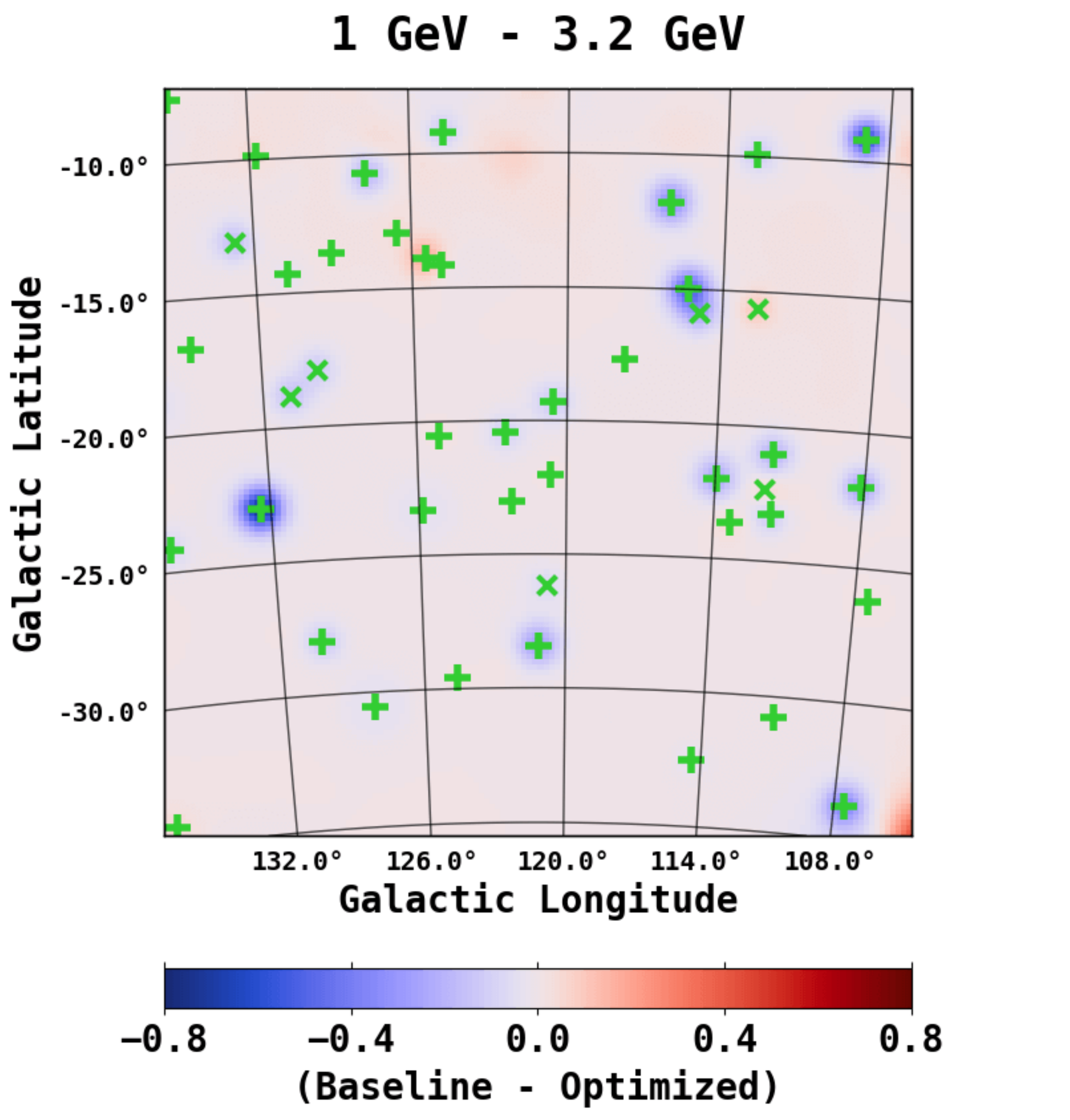}
\includegraphics[width=0.33\textwidth]{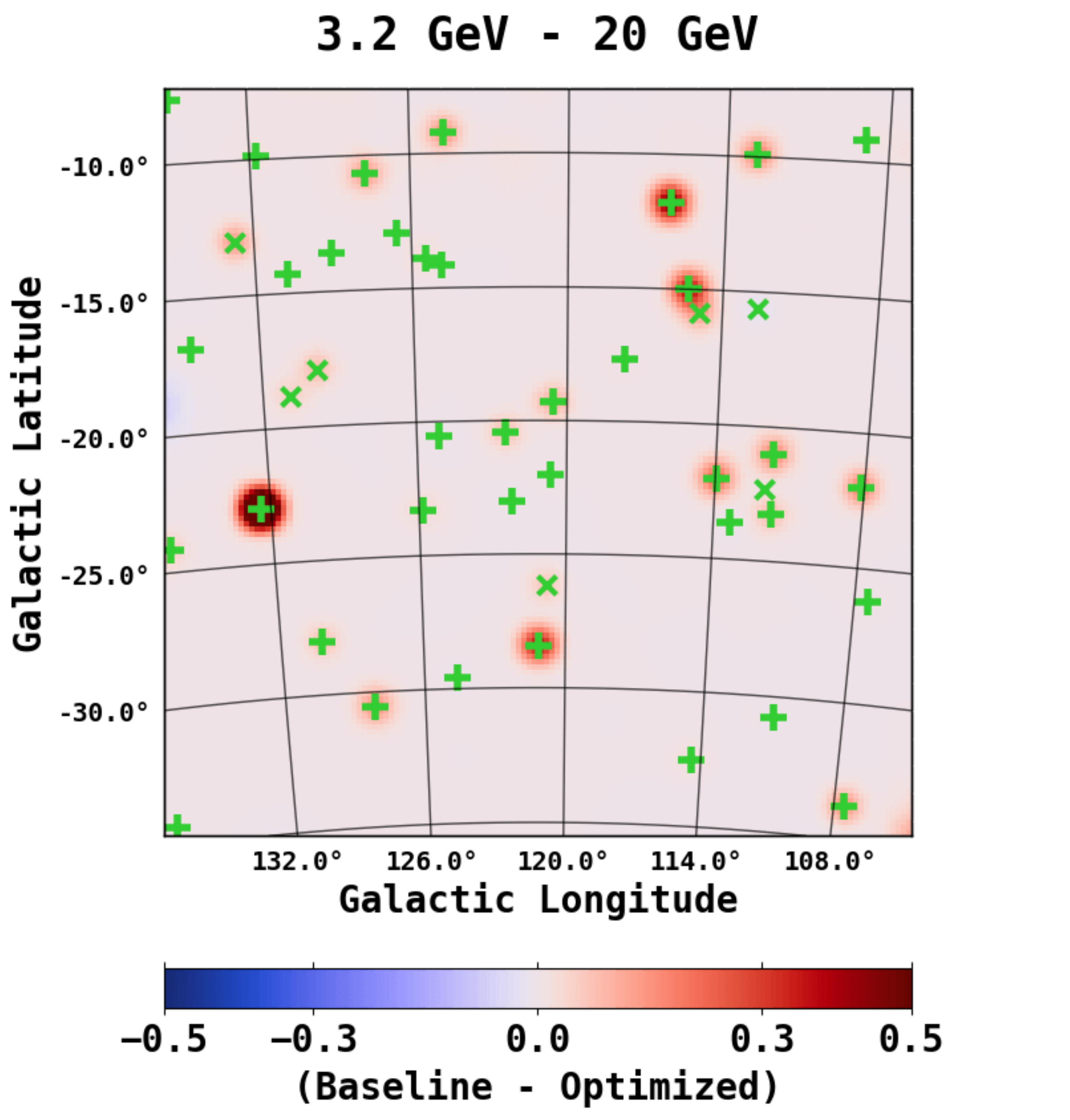}
\includegraphics[width=0.33\textwidth]{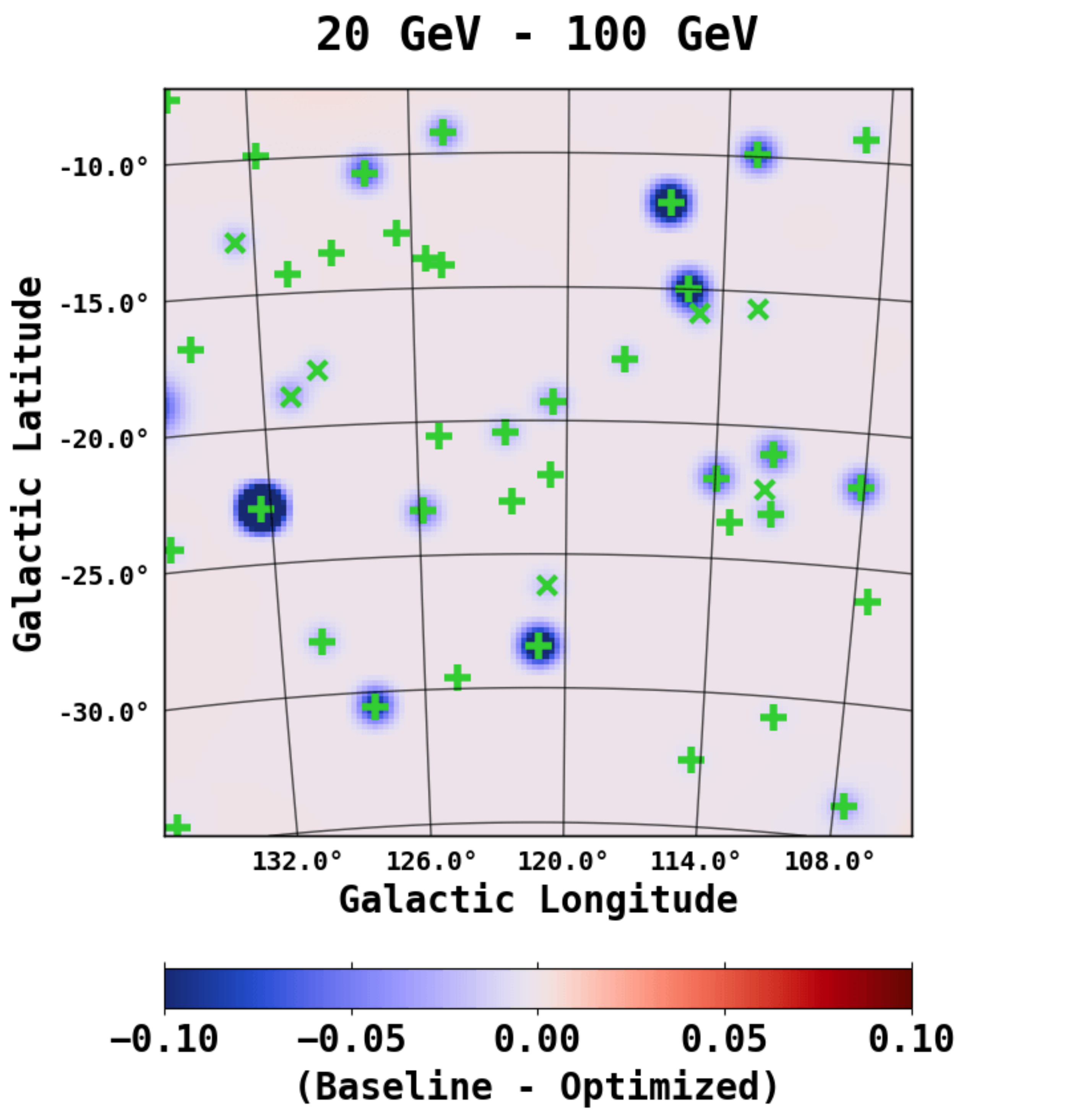}
\caption{The maps show the difference between the spatial residuals resulting from the baseline fit and the spatial residuals resulting from the 3FGL optimized fit. For the optimized fit the PL spectral models are replaced with LogParabola spectral models. Three energy bins are shown, just as in Figure~\ref{fig:spatial_residuals_FM31_tuned}. Green crosses show 3FGL sources with TS$\geq$25 and slanted green crosses show 3FGL sources with 9$\leq$TS$<$25. For the baseline fit, numerous 3FGL sources with PL spectral models were over-modeling in bins 1 and 3, and under-modeling in bin 2, as seen in Figure~\ref{fig:spatial_residuals_gray_FM31_tuned}. And as seen here, in bins 1 and 3 the 3FGL over-modeling is deeper for the baseline fit, resulting in the surrounding blue regions, and in bin 2 the 3FGL under-modeling is more severe for the baseline fit, resulting in the surrounding red regions, i.e.\ numerous 3FGL sources show improvement in the spatial residuals with the optimized fit.}
\label{fig:spatial_residuals_optimized_diff}
\end{figure*}

In addition to the positive residual emission between $\sim$3--20 GeV, the fractional count residuals for both IEMs also show a high energy deficit (HED) in the last few energy bins, reaching as high as $\sim$20\%. Note, however, that the data in these higher bins are limited, and the error bars are fairly large. Likely related to a portion of the HED are the point sources in the field. Most of the 3FGL sources in FM31 are active galactic nuclei, and are modeled with power law spectral models. Many of these sources are over-modeling in the lower and higher energy bins, and under-modeling in the intermediate range, as seen in Figure~\ref{fig:spatial_residuals_gray_FM31_tuned}. This is an indication that some of the sources may be more consistent with a different spectral model, such as a log parabola (LP) $dN/dE=N_0(E/E_b)^{-\alpha-\beta\log(E/E_b)}$. We have tested this by replacing the PL spectral models of all 3FGL sources with LP models. The spectral parameters for each source (norm, $\alpha, \beta, E_b$) are initially set to the corresponding values for the respective PL spectra, with $\beta$ initially set to 0. The fit is otherwise performed in the standard way. Figure~\ref{fig:spatial_residuals_optimized_diff} shows the difference in the spatial residuals for the baseline fit and the optimized fit. In bins 1 and 3 the 3FGL over-modeling is deeper for the baseline fit, resulting in the surrounding blue regions, and in bin 2 the 3FGL under-modeling is more severe for the baseline fit, resulting in the surrounding red regions.

\begin{figure}
\centering
\includegraphics[width=0.4\textwidth]{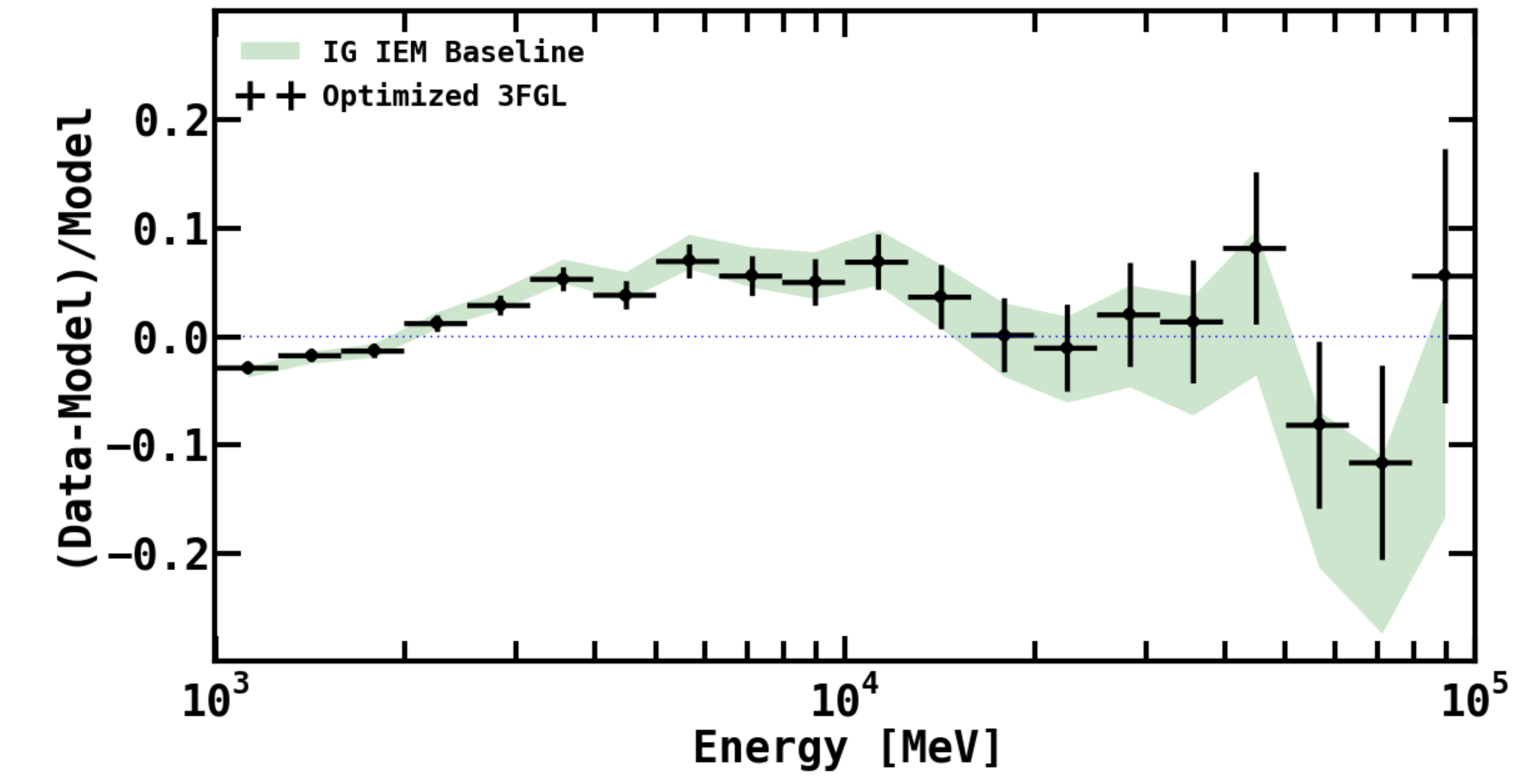}
\caption{All 3FGL sources in FM31 with a PL spectral model are fit with a LogParabola spectral model. The spectral parameters for each source (norm, $\alpha, \beta, E_b$) are initially set to the corresponding values for the respective PL spectra, with $\beta$ initially set to 0. Optimization of the 3FGL sources leads to marginal improvement in the fractional energy residuals, and most notably for the high energy deficit in the last few energy bins. The corresponding differences for each energy bin are reported in Table~\ref{tab:Optimized_3FGL}. For the baseline fit the likelihood value is $-\log L$=143349, and for the optimized 3FGL fit it is $-\log L$=143308.}
\label{fig:optimized}
\end{figure}

The fractional energy residuals resulting from the optimized fit are shown in Figure~\ref{fig:optimized}, and the corresponding difference between the optimized fit and the baseline fit is reported in Table~\ref{tab:Optimized_3FGL}. The optimized 3FGL improve the HED in the last few energy bins by 6--12\%. However, the optimization does not have a significant impact on the positive residual emission between $\sim$3--20 GeV. Properties of all 3FGL sources within $20^\circ$ of FM31, including updated spectral parameters from the baseline fit with the M31 IEM, are summarized in Table~\ref{tab:3FGL_table}. 

\begin{deluxetable}{cc}[p!]
\tablecolumns{2}
\tablewidth{0mm}
\tablecaption{Difference in Fractional Energy Residuals\label{tab:Optimized_3FGL}}

\tablehead{
\colhead{Energy Bin} &
\colhead{Baseline -- Optimized} 
}
\startdata
1& --0.004\\
2& --0.004\\
3& 0.0004\\
4& 0.003\\
5& 0.005\\
6& 0.007\\
7& 0.008\\
8& 0.009\\
9& 0.008\\
10& 0.006\\
11& 0.004\\
12& 0.0003\\
13& --0.004\\
14& --0.01\\
15& --0.02\\
16& --0.03\\
17& --0.05\\
18& --0.06\\
19& --0.08\\
20& --0.12
\enddata
\tablecomments{The corresponding plot is shown in Figure~\ref{fig:optimized}.}
\end{deluxetable}


\begin{deluxetable*}{ccccccccc}
\tablecolumns{7}
\tablewidth{0mm}
\tablecaption{3FGL Parameters for FM31\label{tab:3FGL_table}}
\tablehead{
\colhead{Name} &
\colhead{l} &
\colhead{b}&
\colhead{Significance} &
\colhead{Flux ($\times 10^{-9}$)} &
\colhead{Spectral Type} &
\colhead{Index} &
\colhead{Classification}\\
&\colhead{[deg]} 
&\colhead{[deg]} 
&\colhead{($\sigma = \sqrt{\rm TS}$)}
&
\colhead{[ph cm$^{-2}$ s$^{-1}$]} 
}
\startdata
3FGL J0001.6+3535 & 111.66 & -26.19 & 4.20 & 0.30 & PL & 2.35 & ua \\
3FGL J0003.4+3100 & 110.96 & -30.75 & 6.30 (7.12) & 0.30 (0.37 \p 0.07) & PL & 2.55 (3.02 \p 0.31) & ua \\
3FGL J0003.5+5721 & 116.49 & -4.91 & 5.40 & 0.50 & PL & 2.18 & ua \\
3FGL J0006.4+3825 & 113.33 & -23.61 & 10.50 (7.19) & 0.60 (0.44 \p 0.08) & PL & 2.62 (3.09 \p 0.39) & fsrq \\
3FGL J0006.6+4618 & 114.91 & -15.87 & 4.60 (3.52) & 0.30 (0.15 \p 0.06) & PL & 2.42 (2.19 \p 0.33) & ua \\
3FGL J0007.9+4006 & 113.98 & -22.01 & 4.50 (8.87) & 0.30 (0.36 \p 0.06) & PL & 2.28 (2.18 \p 0.17) & ua \\
3FGL J0008.0+4713 & 115.33 & -15.00 & 24.70 (30.90) & 2.00 (1.88 \p 0.05) & PL & 2.02 (2.06 \p 0.06) & bll \\
3FGL J0009.3+5030 & 116.12 & -11.80 & 36.50 (38.42) & 3.10 (2.51 \p 0.06) & PL & 1.93 (1.97 \p 0.05) & bll \\
3FGL J0014.7+5802 & 118.08 & -4.48 & 5.50 & 0.40 & PL & 1.91 & bll \\
3FGL J0015.7+5552 & 117.91 & -6.65 & 6.40 & 0.40 & PL & 2.11 & bcu \\
3FGL J0018.4+2947 & 114.43 & -32.53 & 5.00 (9.03) & 0.20 (0.25 \p 0.04) & PL & 1.86 (1.75 \p 0.17) & bll \\
3FGL J0022.7+4651 & 117.84 & -15.73 & 4.20 & 0.20 & PL & 2.78 & ua \\
3FGL J0023.5+4454 & 117.75 & -17.68 & 9.00 (6.50) & 0.50 (0.34 \p 0.07) & PL & 2.57 (2.73 \p 0.30) & fsrq \\
3FGL J0032.5+3912 & 118.95 & -23.51 & 5.00 & 0.20 & PL & 2.56 & ua \\
3FGL J0035.9+5949 & 120.99 & -2.98 & 27.50 & 2.70 & PL & 1.90 & bll \\
3FGL J0039.1+4330 & 120.58 & -19.31 & 7.00 (7.27) & 0.30 (0.28 \p 0.06) & PL & 1.96 (2.21 \p 0.22) & bcu \\
3FGL J0040.3+4049 & 120.67 & -22.00 & 6.40 (5.80) & 0.10 (0.16 \p 0.06) & PL & 1.13 (1.80 \p 0.30) & bcu \\
3FGL J0041.9+3639 & 120.82 & -26.17 & 4.60 (4.81) & 0.20 (0.16 \p 0.05) & PL & 1.98 (2.02 \p 0.26) & bll \\
3FGL J0042.0+2318 & 120.14 & -39.52 & 6.10 & 0.40 & PL & 2.35 & fsrq \\
3FGL J0042.5+4117 & 121.13 & -21.55 & 5.80 & 0.40 & PL & 2.56 & gal \\
3FGL J0043.8+3425 & 121.16 & -28.42 & 22.60 (41.16) & 1.60 (2.56 \p 0.06) & PL & 2.04 (1.96 \p 0.05) & fsrq \\
3FGL J0045.3+2126 & 121.06 & -41.40 & 22.30 & 1.40 & PL & 1.90 & bll \\
3FGL J0047.0+5658 & 122.33 & -5.89 & 18.10 & 1.60 & PL & 2.22 & bll \\
3FGL J0047.9+5447 & 122.42 & -8.07 & 5.00 (3.32) & 0.10 (0.10 \p 0.07) & PL & 1.33 (1.95 \p 0.43) & bcu \\
3FGL J0048.0+2236 & 121.91 & -40.26 & 14.20 & 0.90 & PL & 2.33 & fsrq \\
3FGL J0048.0+3950 & 122.22 & -23.02 & 9.20 (17.40) & 0.50 (0.73 \p 0.06) & PL & 1.88 (1.95 \p 0.10) & bll \\
3FGL J0049.0+4224 & 122.45 & -20.46 & 6.90 (6.47) & 0.30 (0.17 \p 0.05) & PL & 1.77 (1.64 \p 0.21) & ua \\
3FGL J0058.3+3315 & 124.59 & -29.58 & 8.30 (6.41) & 0.50 (0.31 \p 0.06) & PL & 2.41 (2.80 \p 0.31) & fsrq \\
3FGL J0102.1+4458 & 124.92 & -17.86 & 4.10 (3.06) & 0.30 (0.11 \p 0.03) & PL & 2.27 (2.08 \p 0.18) & ua \\
3FGL J0102.3+4217 & 125.09 & -20.54 & 12.70 (6.69) & 0.50 (0.32 \p 0.06) & PL & 2.69 (2.84 \p 0.31) & fsrq \\
3FGL J0102.8+4840 & 124.87 & -14.16 & 20.30 (28.19) & 2.20 (2.33 \p 0.10) & PLEXP & 2.20 (1.77 \p 0.24)  & PSR \\
3FGL J0102.8+5825 & 124.43 & -4.41 & 51.50 & 6.20 & LP & 2.25 & fsrq \\
3FGL J0103.4+5336 & 124.74 & -9.21 & 10.40 (11.93) & 0.70 (0.55 \p 0.06) & PL & 2.04 (1.93 \p 0.12) & bll \\
3FGL J0105.3+3928 & 125.85 & -23.32 & 8.10 (11.63) & 0.40 (0.54 \p 0.05) & PL & 2.33 (2.44 \p 0.17) & bll \\
3FGL J0106.5+4855 & 125.48 & -13.87 & 27.50 (36.90) & 3.40 (3.09 \p 0.08) & PLEXP & 2.11 (0.75 \p 0.28) & PSR \\
3FGL J0112.1+2245 & 129.14 & -39.86 & 81.10 & 8.70 & LP & 2.03 & BLL \\
3FGL J0112.8+3207 & 128.19 & -30.53 & 44.50 (36.55) & 3.00 (2.55 \p 0.07) & PL & 2.36 (2.57 \p 0.07) & fsrq \\
3FGL J0113.4+4948 & 126.58 & -12.90 & 12.20 (14.97) & 0.90 (0.86 \p 0.07) & PL & 2.30 (2.26 \p 0.12) & fsrq \\
3FGL J0115.8+2519 & 129.85 & -37.21 & 17.00 & 1.20 & PL & 1.99 & bll \\
3FGL J0121.7+5154 & 127.69 & -10.69 & 6.60 (6.10) & 0.50 (0.37 \p 0.08) & PL & 1.98 (2.30 \p 0.21) & bcu \\
3FGL J0122.8+3423 & 130.28 & -28.03 & 9.60 (8.37) & 0.30 (0.22 \p 0.05) & PL & 1.48 (1.64 \p 0.18) & bll \\
3FGL J0127.4+5433 & 128.19 & -7.95 & 4.60 & 0.40 & PL & 2.53 & ua \\
3FGL J0127.5+5634 & 127.92 & -5.95 & 5.80 & 0.50 & PL & 2.59 & ua \\
3FGL J0127.6+4851 & 129.05 & -13.58 & 5.00 (5.39) & 0.30 (0.34 \p 0.08) & PL & 2.45 (3.19 \p 0.58) & ua \\
3FGL J0127.9+2551 & 133.12 & -36.29 & 7.10 & 0.20 & PL & 3.10 & fsrq \\
3FGL J0128.5+4430 & 129.87 & -17.86 & 7.50 (4.19) & 0.50 (0.19 \p 0.06) & PL & 2.33 (2.47 \p 0.31) & fsrq \\
3FGL J0131.3+5548 & 128.56 & -6.63 & 4.90 & 0.30 & PL & 1.90 & bcu \\
3FGL J0133.3+4324 & 130.96 & -18.82 & 5.00 (3.64) & 0.30 (0.13 \p 0.05) & PL & 2.30 (2.12 \p 0.39) & bcu \\
3FGL J0133.3+5930 & 128.23 & -2.93 & 4.30 & 0.40 & PL & 2.21 & ua \\
3FGL J0134.5+2638 & 134.74 & -35.24 & 12.00 & 0.80 & PL & 1.99 & bcu \\
3FGL J0136.5+3905 & 132.41 & -22.94 & 56.90 (67.43) & 4.90 (4.38 \p 0.07) & PL & 1.70 (1.65 \p 0.03) & bll \\
3FGL J0137.0+4752 & 130.79 & -14.30 & 55.90 (54.10) & 5.10 (4.81 \p 0.11) & LP & 2.27 (1.80 \p 0.21) & fsrq \\
3FGL J0137.8+5813 & 129.02 & -4.09 & 13.50 (4.68) & 1.20 (3823.16 \p 846.66) & PL & 2.06 (0.81 \p 0.10) & bcu \\
3FGL J0144.6+2705 & 137.28 & -34.30 & 50.60 (64.04) & 4.50 (5.72 \p 0.11) & LP & 2.16 (2.22 \p 0.16) & bll \\
3FGL J0148.3+5200 & 131.76 & -9.88 & 9.90 (15.25) & 0.50 (0.55 \p 0.06) & PL & 1.77 (1.61 \p 0.11) & bcu \\
3FGL J0149.6+4846 & 132.72 & -12.99 & 5.10 (3.68) & 0.30 (0.18 \p 0.06) & PL & 2.62 (2.30 \p 0.31) & ua \\
3FGL J0152.2+3707 & 136.19 & -24.17 & 5.60 (5.52) & 0.30 (0.26 \p 0.07) & PL & 2.41 (2.64 \p 0.32) & bcu \\
3FGL J0154.1+4642 & 133.98 & -14.82 & 4.30 & 0.20 & PL & 2.55 & ua \\
3FGL J0154.9+4433 & 134.69 & -16.86 & 7.80 (9.35) & 0.40 (0.47 \p 0.07) & PL & 2.18 (2.28 \p 0.18) & bll 
\enddata
\tablecomments{All 3FGL sources within 20$^\circ$ of M31 are listed, along with their original corresponding parameters from the 3FGL catalog available at \url{http://vizier.cfa.harvard.edu/viz-bin/VizieR}. In total there are 95 sources. Parentheses give updated values from this work (with the M31 IEM). There are 44 sources which are actually in FM31, as can be seen in Figure~\ref{fig:spatial_residuals_gray_FM31_tuned}. Although we scale the parameters for all sources within $20^\circ$ of M31, we only give updated values for the sources with a resulting significance $\geq$3$\sigma$ (which amounts to 48 sources). Note that sources outside of the field (within $30^\circ$ of M31) are included in the model to account for the PSF of the detector, as discussed in the text. The flux is integrated between 1--100 GeV. For the 3FGL catalog the significance is derived from the likelihood test statistic (TS) for 100 MeV -- 300 GeV, and for this work the significance is derived from the energy range 1 GeV -- 100 GeV. The spectral indices also result from the respective energy ranges. The spectral types are either power law (PL), power law with exponential cutoff (PLEXP), or log parabola (LP). The classifications are as follows: ua: unassociated; fsrq: flat spectral radio quasar; bcu: blazar candidate of uncertain type; bll: bl lac type of blazar; psr: pulsar.}
\end{deluxetable*}

\addtocounter{table}{-1}

\begin{deluxetable*}{cccccccccc}
\tablecolumns{8}
\tablewidth{0mm}
\tablecaption{3FGL Parameters for FM31 (Continued)\label{tab:3FGL_table_continued}}
\tablehead{
\colhead{Name} &
\colhead{l} &
\colhead{b}&
\colhead{Significance} &
\colhead{Flux ($\times 10^{-9}$)} &
\colhead{Spectral Type} &
\colhead{Index} &
\colhead{Classification}\\
&\colhead{[deg]} 
&\colhead{[deg]} 
&\colhead{($\sigma = \sqrt{\rm TS}$)}
&
\colhead{[ph cm$^{-2}$ s$^{-1}$]} 
}
\startdata
3FGL J0156.3+3913 & 136.43 & -21.93 & 11.70 & 0.60 & PL & 2.50 & bcu \\
3FGL J0202.5+4206 & 136.78 & -18.84 & 7.90 & 0.50 & PL & 2.28 & bll \\
3FGL J0203.6+3043 & 140.79 & -29.64 & 36.90 & 3.20 & LP & 2.20 & bll \\
3FGL J0204.8+3212 & 140.53 & -28.15 & 9.90 & 0.40 & PL & 2.94 & fsrq \\
3FGL J0208.6+3522 & 140.22 & -24.88 & 6.00 & 0.20 & PL & 1.70 & bll \\
3FGL J0209.5+4449 & 137.20 & -15.88 & 6.70 & 0.40 & PL & 1.97 & bll \\
3FGL J0211.7+5402 & 134.64 & -6.99 & 4.50 & 0.40 & PL & 2.69 & bcu \\
3FGL J0212.1+5320 & 134.93 & -7.65 & 25.10 (29.05) & 2.70 (2.59 \p 0.10) & LP & 2.19 (1.25 \p 0.33) & ua \\
3FGL J0214.4+5143 & 135.77 & -9.07 & 7.20 & 0.40 & PL & 2.04 & bll \\
3FGL J0218.1+4233 & 139.52 & -17.51 & 44.00 & 5.40 & PLEXP & 2.29 & PSR \\
3FGL J0218.9+3642 & 141.87 & -22.92 & 13.90 & 1.00 & PL & 2.58 & bcu \\
3FGL J0221.1+3556 & 142.61 & -23.48 & 48.40 & 4.60 & PL & 2.28 & FSRQ \\
3FGL J0222.6+4301 & 140.15 & -16.77 & 121.30 & 19.30 & LP & 1.94 & BLL \\
3FGL J0223.6+3927 & 141.71 & -20.03 & 7.80 & 0.50 & PL & 2.40 & ua \\
3FGL J2300.0+4053 & 101.24 & -17.24 & 5.20 & 0.20 & PL & 1.51 & ua \\
3FGL J2302.7+4443 & 103.39 & -13.99 & 51.70 & 6.50 & PLEXP & 1.96 & PSR \\
3FGL J2304.6+3704 & 100.34 & -21.06 & 13.00 & 0.70 & PL & 1.82 & bll \\
3FGL J2311.0+3425 & 100.41 & -24.02 & 66.90 & 5.30 & LP & 2.34 & FSRQ \\
3FGL J2313.1+3935 & 103.05 & -19.46 & 5.50 & 0.20 & PL & 2.79 & ua \\
3FGL J2321.3+5113 & 108.87 & -9.18 & 4.30 (6.40) & 0.30 (0.23 \p 0.08) & PL & 2.02 (1.89 \p 0.25) & ua \\
3FGL J2321.6+4438 & 106.57 & -15.37 & 4.60 & 0.30 & PL & 2.64 & fsrq \\
3FGL J2321.9+3204 & 101.69 & -27.08 & 26.40 & 2.10 & LP & 2.25 & fsrq \\
3FGL J2322.5+3436 & 102.87 & -24.78 & 8.90 & 0.30 & PL & 1.44 & bll \\
3FGL J2323.9+4211 & 106.07 & -17.80 & 25.10 & 1.80 & PL & 1.89 & bll \\
3FGL J2325.2+3957 & 105.51 & -19.98 & 23.40 & 2.00 & LP & 2.01 & bll \\
3FGL J2329.2+3754 & 105.54 & -22.17 & 13.20 (3.42) & 0.80 (3.91 \p 1.47) & PL & 1.93 (2.15 \p 0.51) & bll \\
3FGL J2337.5+4108 & 108.26 & -19.62 & 4.40 & 0.30 & PL & 2.23 & ua \\
3FGL J2340.7+3847 & 108.14 & -22.04 & 4.20 (5.94) & 0.30 (0.16 \p 0.05) & PL & 1.92 (1.73 \p 0.22) & ua \\
3FGL J2343.7+3437 & 107.45 & -26.20 & 6.80 (9.91) & 0.30 (0.41 \p 0.05) & PL & 1.75 (1.97 \p 0.15) & bll \\
3FGL J2347.0+5142 & 112.89 & -9.90 & 32.00 (41.02) & 2.40 (2.74 \p 0.07) & PL & 1.78 (1.91 \p 0.05) & bll \\
3FGL J2347.9+5436 & 113.74 & -7.13 & 4.70 & 0.20 & PL & 1.73 & bcu \\
3FGL J2354.0+2722 & 107.57 & -33.78 & 4.20 (5.44) & 0.20 (0.20 \p 0.06) & PL & 2.15 (2.27 \p 0.27) & ua \\
3FGL J2354.1+4605 & 112.67 & -15.64 & 6.30 (3.35) & 0.40 (0.16 \p 0.06) & PL & 2.48 (2.65 \p 0.46) & fsrq \\
3FGL J2356.0+4037 & 111.72 & -21.03 & 9.10 (10.94) & 0.40 (0.44 \p 0.05) & PL & 1.72 (2.00 \p 0.14) & bll \\
3FGL J2358.5+3827 & 111.69 & -23.26 & 5.90 (9.10) & 0.40 (0.37 \p 0.05) & PL & 2.08 (2.11 \p 0.17) & ua \\
3FGL J2358.9+3926 & 112.01 & -22.31 & 4.70 (4.87) & 0.30 (0.19 \p 0.07) & PL & 2.21 (2.14 \p 0.33) & fsrq 

\enddata
\tablecomments{All 3FGL sources within 20$^\circ$ of M31 are listed, along with their original corresponding parameters from the 3FGL catalog available at \url{http://vizier.cfa.harvard.edu/viz-bin/VizieR}. In total there are 95 sources. Parentheses give updated values from this work (with the M31 IEM). There are 44 sources which are actually in FM31, as can be seen in Figure~\ref{fig:spatial_residuals_gray_FM31_tuned}. Although we scale the parameters for all sources within $20^\circ$ of M31, we only give updated values for the sources with a resulting significance $\geq$3$\sigma$ (which amounts to 48 sources). Note that sources outside of the field (within $30^\circ$ of M31) are included in the model to account for the PSF of the detector, as discussed in the text. The reported flux is integrated between 1--100 GeV. For the 3FGL catalog the significance is derived from the likelihood test statistic (TS) for 100 MeV -- 300 GeV, and for this work the significance is derived from the energy range 1 GeV -- 100 GeV. The spectral indices also result from the respective energy ranges. The spectral types are either power law (PL), power law with exponential cutoff (PLEXP), or log parabola (LP). The classifications are as follows: ua: unassociated; fsrq: flat spectral radio quasar; bcu: blazar candidate of uncertain type; bll: bl lac type of blazar; psr: pulsar.}
\end{deluxetable*}

\subsection{The FSSC IEM} \label{sec:FSSC_IEM}

We also repeat the analysis using the official IEM provided by the Fermi Science Support Center (FSSC IEM) for point source analysis. We note, however, that in general the FSSC IEM \emph{is not} intended for extended source analysis\textsuperscript{\ref{caveats}} \citep{Acero:2016qlg}. Construction of the FSSC IEM is based on a template fitting approach. In this approach, the intensities of the model components are not calculated based on CR data and propagation models, as they are for the GALPROP-based IEMs; rather, a linear combination of gas and IC components is fit to the $\gamma$-ray data, based on corresponding spatial correlations~\citep{Acero:2016qlg}. The different gas column density maps offer spatial templates for $\gamma$-ray photons originating mainly from $\pi^0$-decay and Bremsstrahlung emission. For the IC component, there is no direct observational template, and so it must be calculated. The FSSC IEM employs an IC template from the GALPROP code. We note that the FSSC IEM contains patches to account for extended excess emission (EEE) of unknown origin, also referred to as the ``rescaled IC component''. The region towards the north of FM31 (primarily in the MW plane) contains such a patch. 

\begin{figure}
\centering
\includegraphics[width=0.47\textwidth]{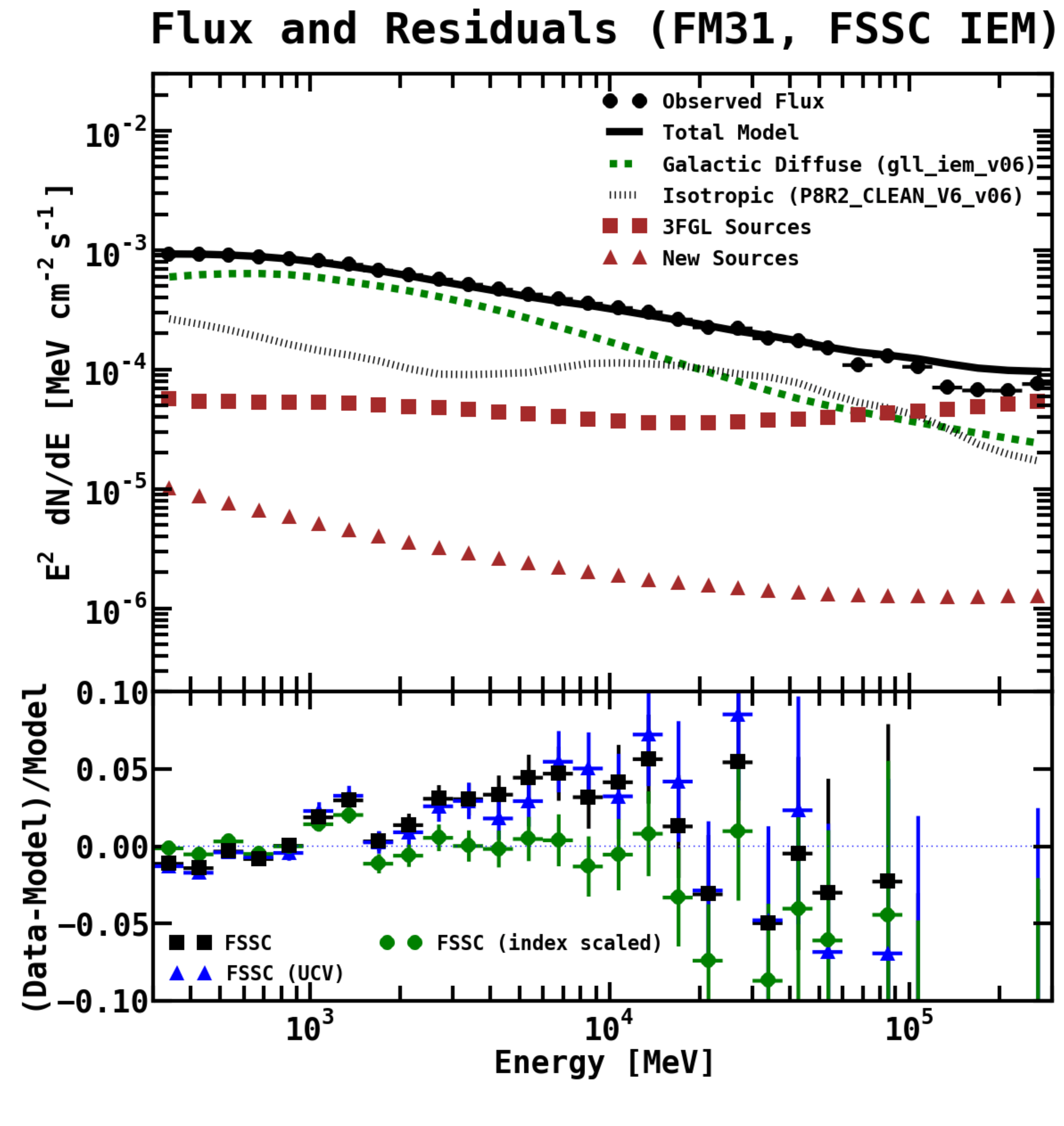}
\caption{The top panel shows the best-fit spectra resulting from the FSSC IEM using the Clean data class (with the Galactic diffuse index fixed). The bottom panel shows the resulting fractional count residuals. Black squares are for the Clean class (with the Galactic diffuse index fixed), green circles show the same fit but with the index of the Galactic diffuse component freed, and blue triangles are for the UltraCleanVeto (UCV) class (with Galactic diffuse index fixed). All components are fit in FM31, including the isotropic.}
\label{fig:flux_and_residuals_FSSC}
\end{figure}

\begin{figure*}
\centering
\includegraphics[width=0.33\textwidth]{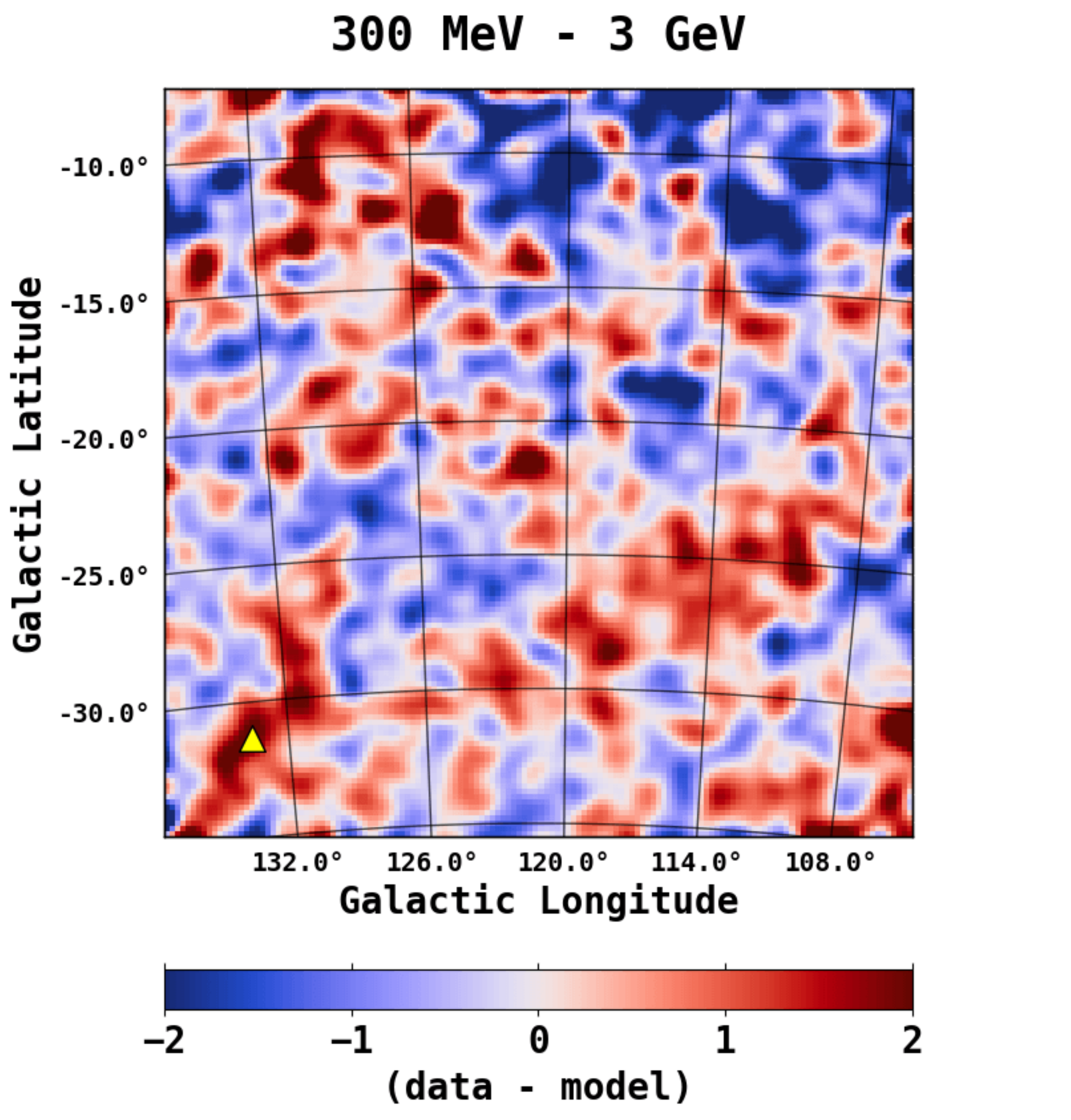}
\includegraphics[width=0.33\textwidth]{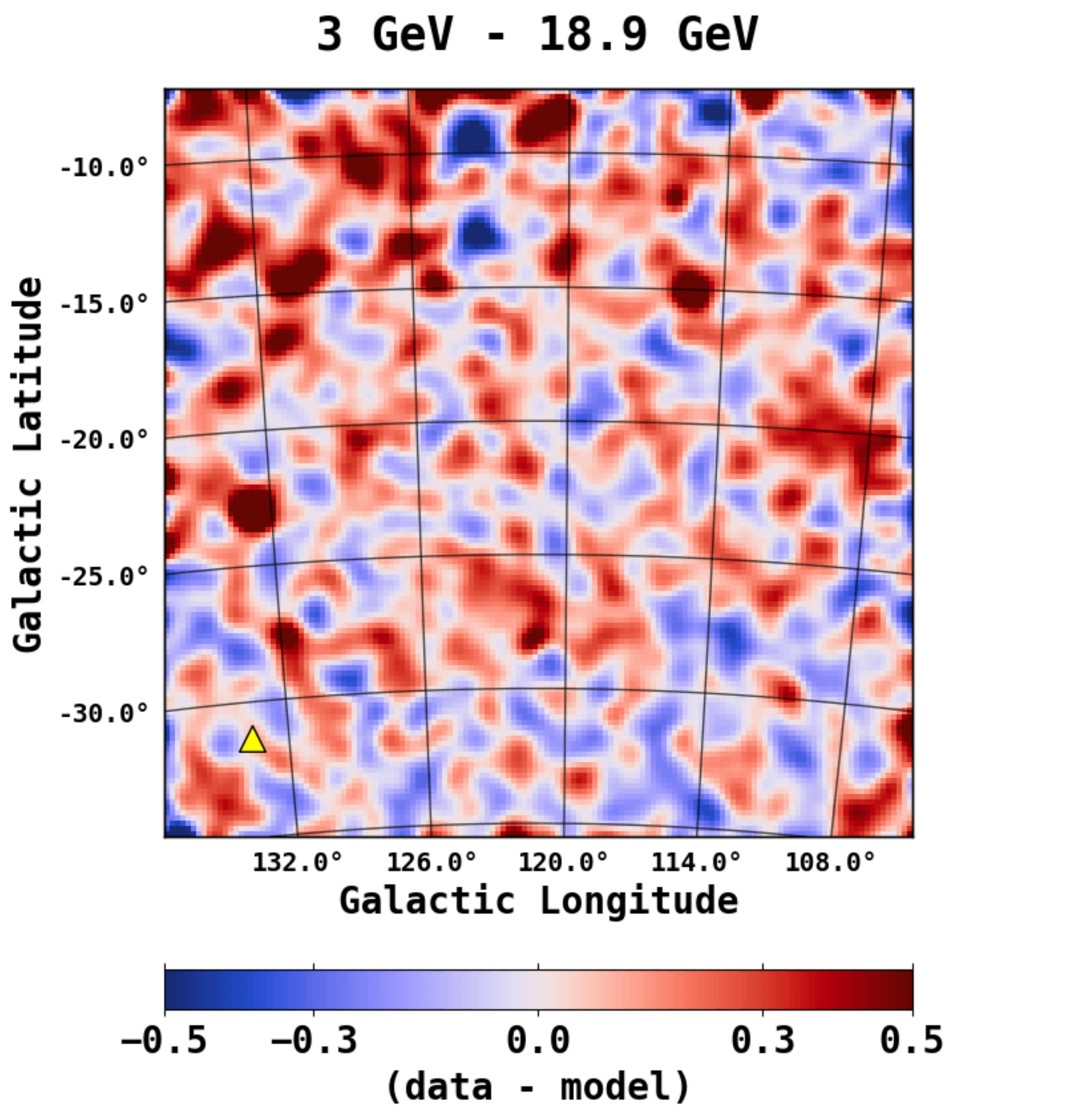}
\includegraphics[width=0.33\textwidth]{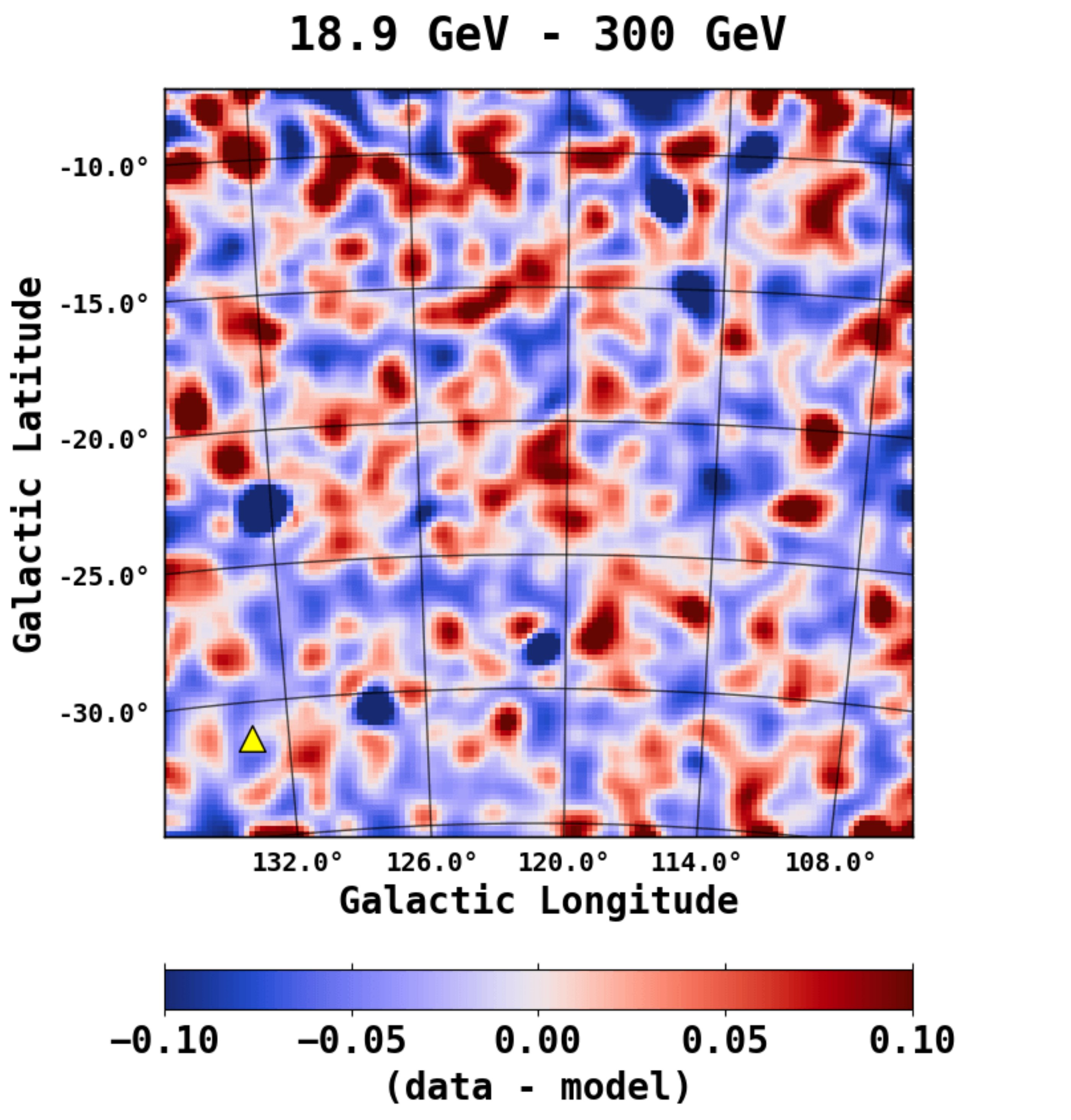}
\caption{Spatial count residuals (data -- model) resulting from the fit in FM31 with the FSSC IEM for three different energy bands, as indicated above each plot. The energy bins are chosen to coincide with the excess observed in the fractional residuals. The color scale corresponds to counts/pixel, and the pixel size is $0.2^\circ \times 0.2^\circ$. The images are smoothed using a $1^\circ$ Gaussian kernel. This value corresponds to the PSF (68\% containment angle) of \textit{Fermi}-LAT, which at 1 GeV is $\sim$${1^\circ}$. For reference, the position of M33, (l,b) = ($133.61^\circ$,$-31.33^\circ$), is shown with a yellow triangle.}
\label{fig:spatial_residuals_FM31_FSSC}
\end{figure*}

Figure~\ref{fig:flux_and_residuals_FSSC} shows results for the baseline fit with the FSSC IEM. The fit is performed over the energy range 300 MeV -- 300 GeV, using the same ROI as for the main analysis. The normalizations of the diffuse components, Galactic and isotropic, are freely scaled in FM31, as well as the 3FGL point sources, and additional sources that we find with our point source finding procedure. Note that the index of the Galactic diffuse component is held fixed. The top panel in Figure~\ref{fig:flux_and_residuals_FSSC} shows the best-fit spectra using the Clean data class. The bottom panel shows the fractional count residuals. Black squares are for the Clean selection. The best-fit normalizations and flux for the isotropic and Galactic diffuse components are reported in Table~\ref{tab:FSSC_baseline_normalizations}. The spatial residuals resulting from the baseline fit are shown in Figure~\ref{fig:spatial_residuals_FM31_FSSC}. They are qualitatively consistent with the results for the GALPROP IEMs.

We also repeat the fit with freeing the index of the Galactic diffuse component.The best-fit index and normalizations are reported in Table~\ref{tab:FSSC_index_scaled} and the fractional count residuals are shown with green circles in the bottom panel of Figure~\ref{fig:flux_and_residuals_FSSC}. As can be seen, this variation is able to flatten the excess between $\sim$3--20 GeV. We again stress that the FSSC IEM is not intended for extended source analysis, especially for weak sources, and this result illustrates how an application of an improper IEM can alter the physical results.

Using the FSSC IEM we also repeat the observations with the UltraCleanVeto data selection. The isotropic background was found to be enhanced by a factor of $\sim$2 at 1--3 GeV within $20^\circ$ of the Ecliptic/Equator compared to the poles, ascribed to primary CRs misclassified as photons\footnote{For a discussion regarding the enhanced isotropic emission within $20^\circ$ of the Ecliptic/Equator see \url{https://fermi.gsfc.nasa.gov/ssc/data/analysis/LAT_caveats.html}}. The UltraCleanVeto data selection removes this anisotropy. More generally, the UltraCleanVeto selection is the cleanest of all data classes, with respect to CR contamination in the detector. Shown in the bottom panel of Figure~\ref{fig:flux_and_residuals_FSSC} are the fractional residuals resulting from the UltraCleanVeto selection (blue triangles). The brackets in Table~\ref{tab:FSSC_baseline_normalizations} give the best-fit normalizations for the fit. Note that the index of the Galactic diffuse component is held fixed. The results are qualitatively consistent with the clean data selection.

\begin{deluxetable*}{lccc}
\tablecolumns{4}
\tablewidth{0mm}
\tablecaption{Normalizations of the Diffuse Components and Integrated Flux for the FSSC IEM\label{tab:FSSC_baseline_normalizations}}
\tablehead{
\colhead{Component} &
\colhead{FSSC IEM} &
\colhead{Flux ($\times 10^{-9})$}&
\colhead{Intensity ($\times 10^{-8})$}\\
&
&
\colhead{(ph cm$^{-2}$ s$^{-1}$)} &
\colhead{(ph cm$^{-2}$ s$^{-1}$ sr$^{-1}$)} 
}
\startdata
Galactic Diffuse &0.98 \p 0.0002 [1.002 \p 0.004]   &1861.1 \p 0.4& 791.1\p 0.2 \\
Isotropic &1.04 \p 0.005 [1.04 \p 0.02] &635.0 \p 3.3 & 270.0 \p 1.4 

 \enddata
\tablecomments{Diffuse normalizations and flux for the FSSC IEM with the Clean data selection. For reference, the normalizations in brackets are for the UltraCleanVeto data selection. Intensities are calculated by using the total area of FM31, which is 0.2352 sr.}
\end{deluxetable*}

\begin{deluxetable*}{lccc}
\tablecolumns{4}
\tablewidth{0mm}
\tablecaption{Scaling the Index of the Galactic Diffuse Component for the FSSC IEM\label{tab:FSSC_index_scaled}}
\tablehead{
\colhead{Component} &
\colhead{Normalization} &
\colhead{Index, $\Delta\alpha$}&
 }
\startdata
Galactic Diffuse & 0.900 \p 0.007& --0.033 \p 0.002 \\
Isotropic &1.07 \p 0.01 & \nodata 

 \enddata
\tablecomments{For this fit we free the index of the Galactic diffuse component $dN/dE\propto E^{-\Delta\alpha}$. The fit is otherwise performed in the standard way.}
\end{deluxetable*}

\begin{figure*}
\centering
\includegraphics[width=0.33\textwidth]{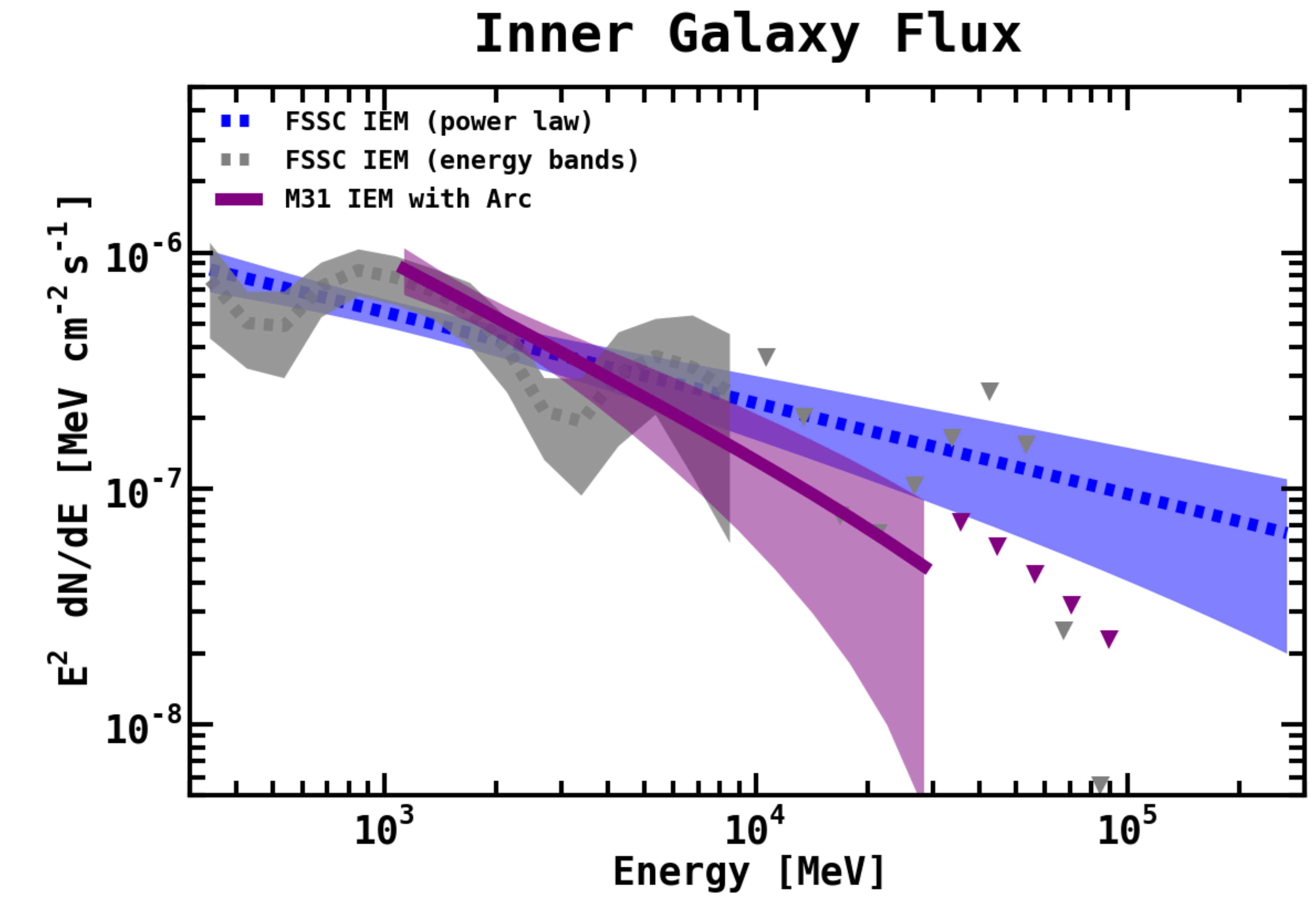}
\includegraphics[width=0.33\textwidth]{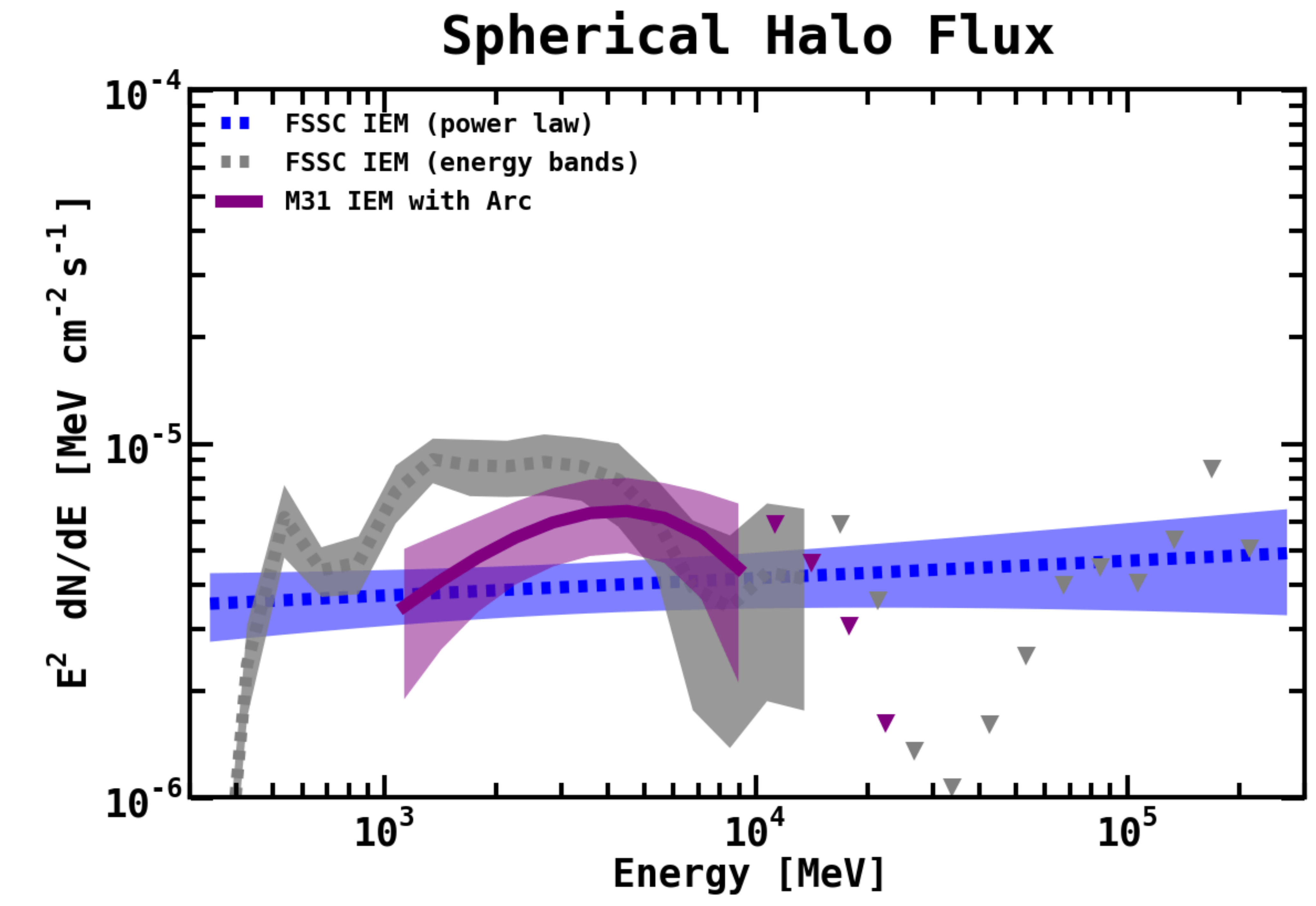}
\includegraphics[width=0.33\textwidth]{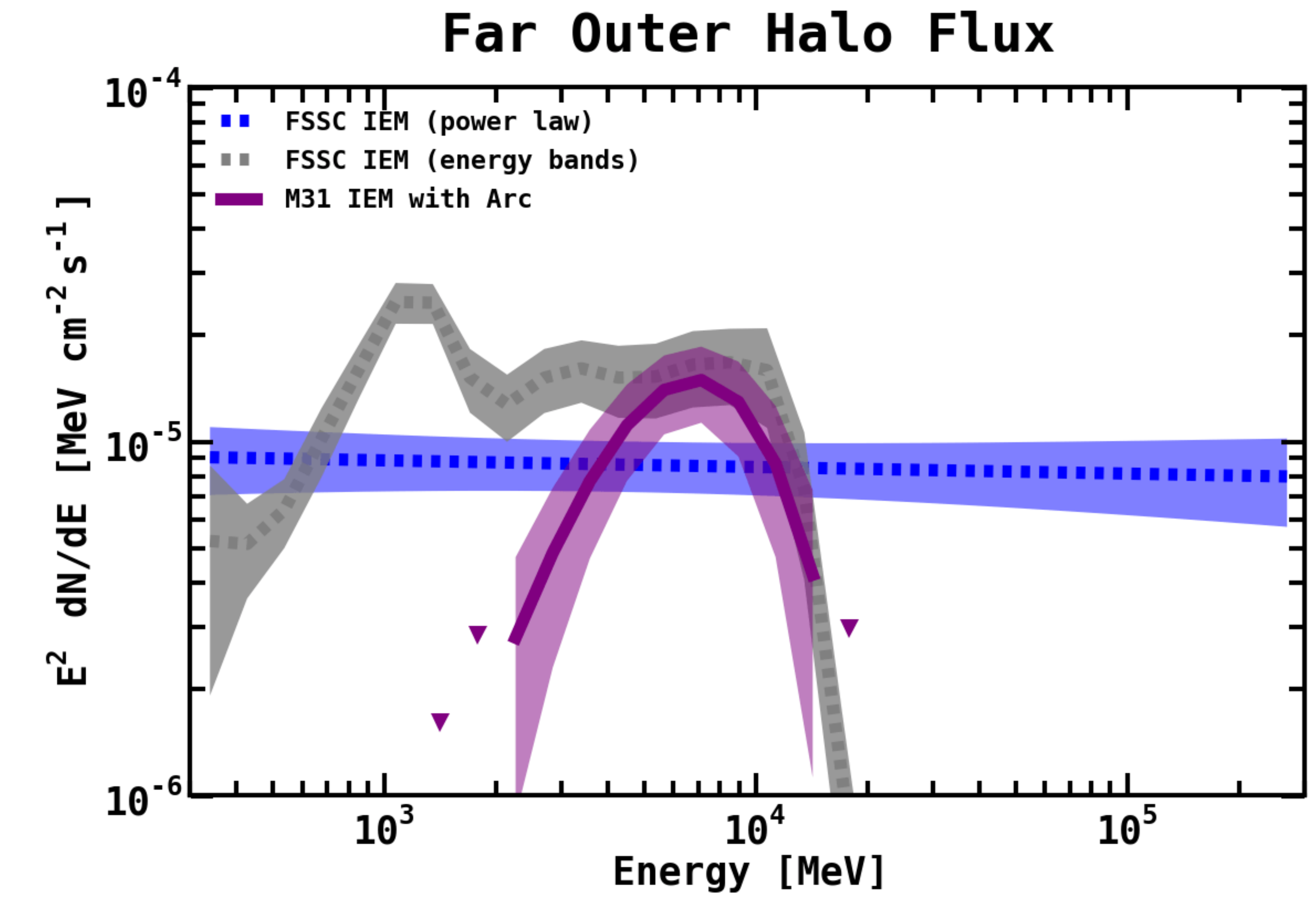}
\caption{The M31-related components  (not including the arc template) are added to the model and fit with the FSSC IEM. For this fit the normalization of the isotropic component is held fixed to its best-fit value obtained in the baseline fit (1.04). All other components are fit simultaneously in the standard way. Two variations of the fit are performed. In one variation the M31-related components are given PL spectral models (dashed blue curves).  In the second variation the M31-related components are fit with a power law per every other energy band over the range 0.3--300 GeV (dashed gray curves). The free parameters include an overall normalization, as well as the index in each respective energy bin. Corresponding results for the M31 IEM are shown with solid purple curves.}
\label{fig:M31_components_flux_and_residuals_FSSC}
\end{figure*}

We fit the M31-related components (not including the arc template) using the FSSC IEM, with the index of the Galactic diffuse component fixed. The best-fit spectra are shown in Figure~\ref{fig:M31_components_flux_and_residuals_FSSC}. For this fit the normalization of the isotropic component is held fixed to its best-fit value obtained in the baseline fit (1.04). The fit can also be performed by freeing the normalization of the isotropic component, but in this case the M31-related components are assigned more counts, and the isotropic normalization is decreased. All other components are fit simultaneously in the standard way. Two variations of the fit are performed. In one variation the M31-related components are given PL spectral models. In the second variation the M31-related components are fit with a power law per every other energy band (EB) over the range 0.3--300 GeV. The free parameters include an overall normalization, as well as the index in each respective energy bin. Note that for the extended energy range a PLEXP spectral model does not provide a good fit. Defining the null model to be everything but the spherical halo and far outer halo components, and the alternative model to include the halo components, the significance of the fit is $-2\Delta$logL$=$176.

The inner galaxy component shows a harder spectrum when fitting over the energy range 300 MeV -- 300 GeV, compared to 1 GeV --  100 GeV. Interestingly, the EB spectrum for the inner galaxy component appears to show two distinct features, one bump near 7 GeV and a second bump near 1 GeV. However, the data becomes limited as the energy approaches 10 GeV, and for higher energies only upper limits are obtained. 

The spectra for the spherical halo component are pretty consistent for the two IEMs. For the far outer halo component, the FSSC IEM shows a bump near 1 GeV. For this fit we do not include the arc template, and the bump at $\sim$1 GeV may be related to inaccuracies in the foreground model, akin to that which is accounted for using the arc template. Otherwise, the spectra for the far outer halo are consistent for the different IEMs. Both the spherical halo and far outer halo components show a sharp spectral cutoff at lower energies.

\subsection{The Inner Galaxy}

The inner galaxy of M31 has previously been detected, and as a consistency check we compare our results with the results from these other studies. To be consistent with the other studies, we use the FSSC IEM and a $14^\circ \times 14^\circ$ ROI. We use an energy range of 0.3--300 GeV, with the P8R2\_SOURCE event class, and the same time range as for the main analysis. Our baseline model consists of the FSSC isotropic (iso\_P8R2\_SOURCE\_V6\_v06), Galactic diffuse (gll\_iem\_v06), and 3FGL point sources. The model includes all 3FGL sources within $20^\circ$ of M31, but only sources within $10^\circ$ are freely scaled in the fit (normalization and index).

\begin{figure*}
\centering
\includegraphics[width=0.49\textwidth]{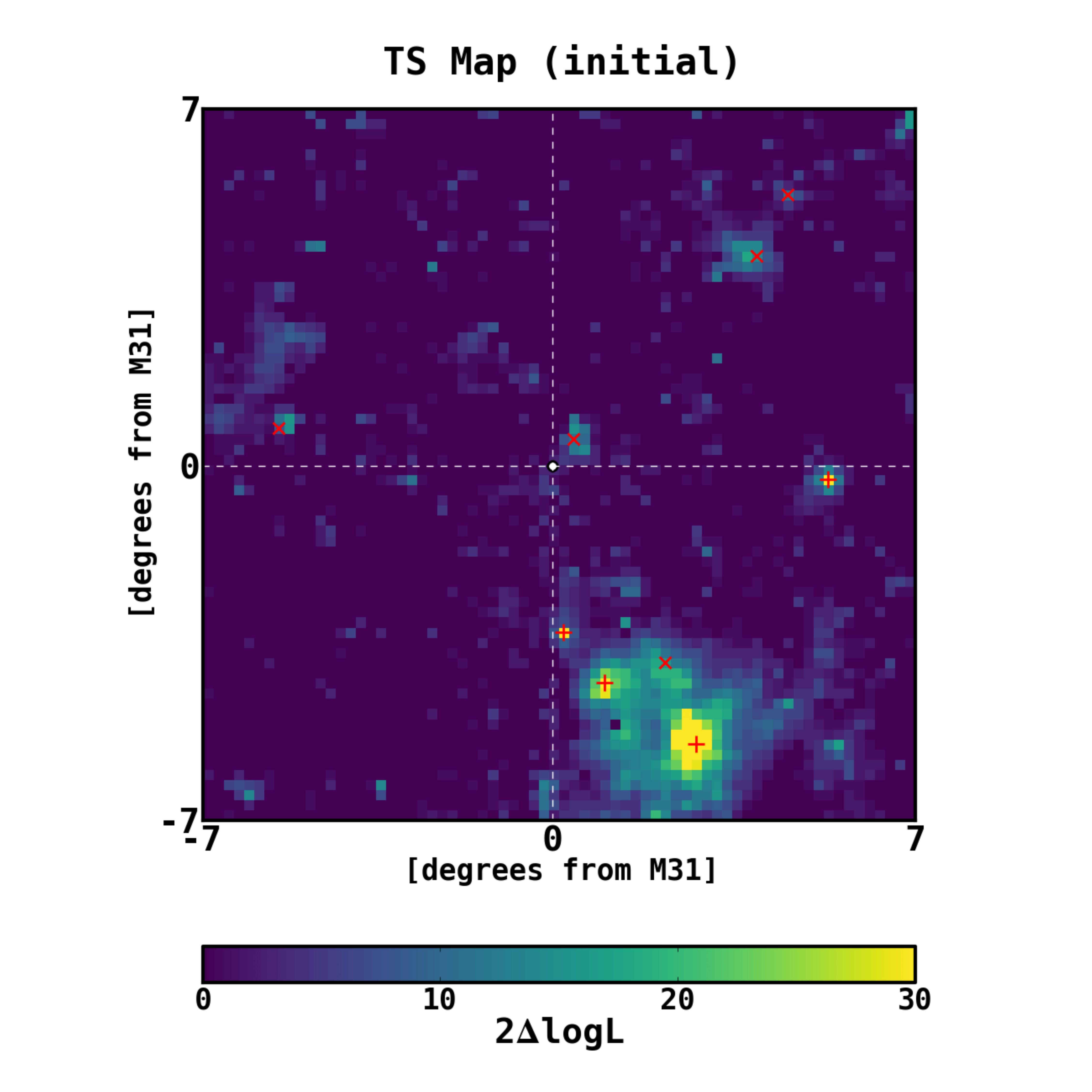}
\includegraphics[width=0.49\textwidth]{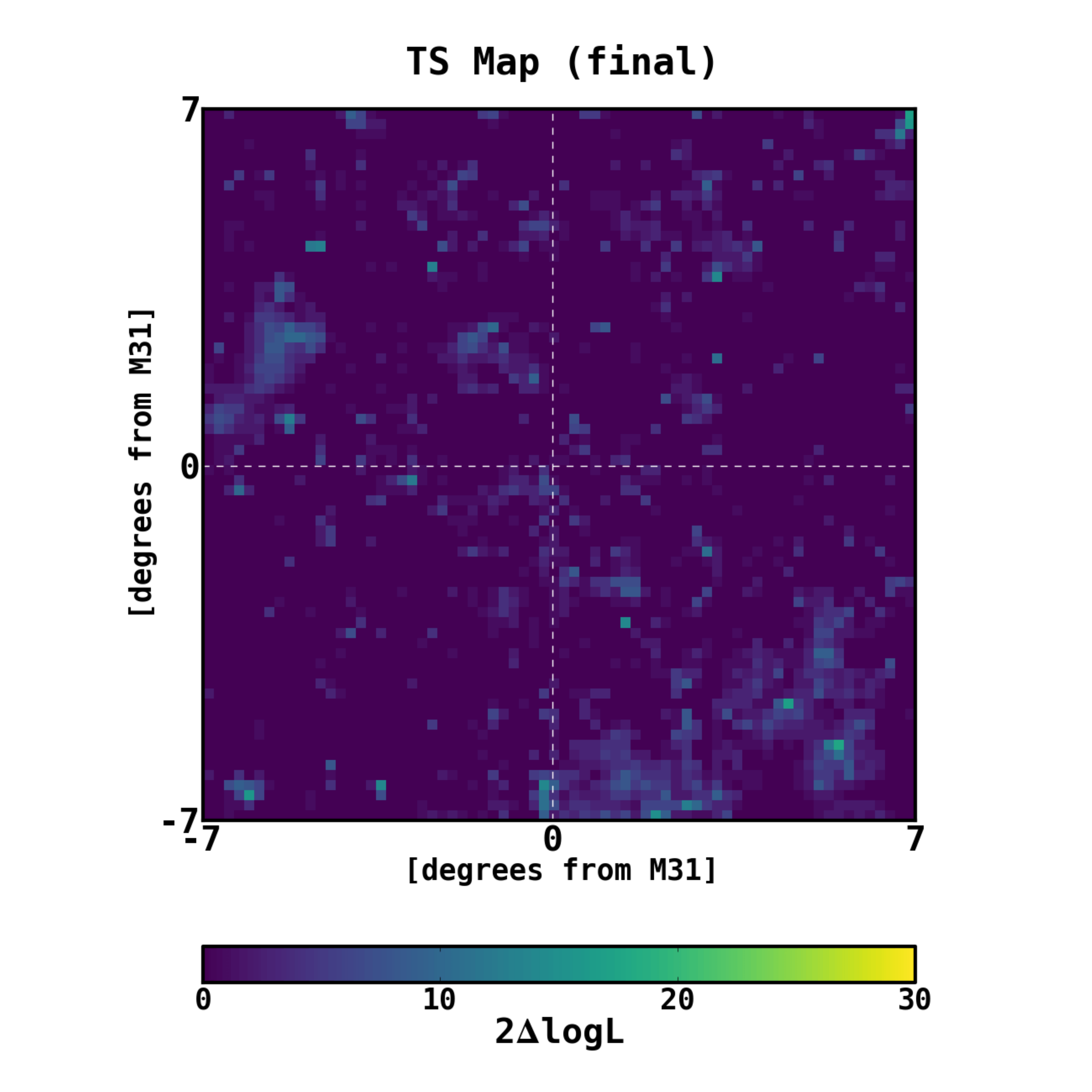}
\caption{TS map before and after the additional point sources are included in the model. Note that M31 is modeled with an elliptical template based on the IRIS 100 $\mu$m map of the galaxy The region shown is a $14^\circ \times 14^\circ$ square, centered at M31 (white circle). The color scale corresponds to the TS value ($2\Delta \log L$), as calculated by {\it gttsmap}. Overlaid on the initial TS map are the positions of the additional point sources that we find with our point source finding procedure. Point sources with TS $\geq$ 25 are shown as red crosses, and sources with 9 $\leq$ TS $<$ 25 are shown as angled red crosses.}
\label{fig:TS_FSSC}
\end{figure*}

We also find new sources using our point source finding procedure. The procedure is employed self-consistently with the FSSC IEM, and we include an M31 template based on the IRIS 100 $\mu$m map of the galaxy. TS maps for the region with and without the additional sources are computed  using the {\it gttsmap} package (included in the \fermilat\ ScienceTools) and  are shown in Figure~\ref{fig:TS_FSSC}. The additional point sources are overlaid. Point sources with TS $\geq$ 25 are shown with  red crosses,  while  sources with  9 $\leq$ TS $<$ 25 are shown with red angled crosses. We find that the agreement between the data and the model significantly improves with the additional point sources, i.e.\ the initial TS map shows a bright extended region in the lower right-hand corner. The point source finding procedure models this region as three point sources (all with very soft spectra), and as can be seen in the final TS map, these sources do a fairly good job in absorbing the excess. However, this structure is more likely part of a larger extended component, as discussed in the main analysis. We also find that the peaks in the initial TS map are in good agreement with the positions of the additional point sources. 

One of the new point sources is located close to M31, as seen in Figure~\ref{fig:TS_FSSC}. Other studies have observed a source at a similar location~\citep{Pshirkov:2016qhu,Ackermann:2017nya}. In~\citet{Pshirkov:2016qhu} the source is detected with TS $=$ 11.2, and is attributed to a nearby AGN (FSRQ B3 0045+013). In~\citet{Ackermann:2017nya} the source was found to have a TS = 12, where the TS map is calculated for energies between 1--100 GeV, in a $3.5^\circ \times 3.5^\circ$ region around M31, with M31 modeled as a point source. From our point source finding procedure we find this source to have a TS = 14. We point out here, for the first time, that this source is spatially coincident with NGC 205, having an angular separation of $\sim$$0.16^\circ$. NGC 205 is an irregular dwarf elliptical galaxy of M31, and it has a number of open issues that are associated with its star formation history, total gas content, and its orbital history \citep{welch1998puzzling,demers2003carbon,marleau2006mapping,howley2008darwin,monaco2009young}. Theoretical arguments support a history of violent supernova explosions in NGC 205; however, no supernova remnants have been detected~\citep{lucero2007radio,de2012herschel}. We leave further discussion to a forthcoming analysis.

\begin{deluxetable*}{lcccccccc}
\tablecolumns{7}
\tablewidth{0mm}
\tablecaption{Comparison of measured values for the inner galaxy of M31\label{tab:comparison_1}}
\tablehead{
\colhead{Template}&
\colhead{$-\log(L)$}&
\colhead{TS}&
\colhead{Index}&
\colhead{Flux ($\times 10^{-9})$}&
\colhead{Isotropic}&
\colhead{Galactic}&
\colhead{Energy}&
\colhead{Reference}\\
&
&
&
&
\colhead{(ph cm$^{-2}$ s$^{-1}$)} &
\colhead{Normalization}&
\colhead{Normalization}&
\colhead{Range}&
}
\startdata
M31 IEM &\nodata&54 & 2.8 \p 0.3& 0.5 \p 0.1 &\nodata &\nodata & 1--100 GeV & this study\smallskip\\

Baseline FSSC	&--4745 &\nodata &\nodata &\nodata &0.970 \p 0.002 &1.0167 \p 0.0009 & 0.3--100 GeV & this study\\
Point	 FSSC	&--4771 &52 & 2.75 \p 0.15 &2.16 \p 0.35 &0.96 \p 0.03 &1.02 \p 0.01 & 0.3--100 GeV & this study\\
IRIS	FSSC		&--4776	&62 		& 2.46 \p 0.07		&3.29 \p 0.24		&0.96 \p 0.01		&1.017 \p 0.003 & 0.3--100 GeV & this study\smallskip\\

Fermi Point	&\nodata	&25.5		& 2.5 \p 0.3		&\nodata			&\nodata			&\nodata & 0.2\ --\ 20 GeV & \citet{Fermi-LAT:2010kib}\\
Fermi IRIS	&\nodata	&29			& 2.1 \p 0.3		&11.0 \p 6.7		&\nodata			&\nodata& 0.2\ --\ 20 GeV & \citet{Fermi-LAT:2010kib}\\
Fermi 0.4$^\circ$ Disk	&\nodata	&97	& 2.4 \p 0.1		&10.0 \p 2.0		&\nodata			&\nodata &0.1--100 GeV&\citet{Ackermann:2017nya}\smallskip\\

Pshirkov Point	&\nodata	&62			& 2.64 \p 0.15		&1.9 \p 0.3		&\nodata			&\nodata&0.3--100 GeV&\citet{Pshirkov:2015hda,Pshirkov:2016qhu}\\
Pshirkov IRIS	&\nodata	&79			& 2.4 \p 0.12		&2.6 \p 0.4		&\nodata			&\nodata&0.3--100 GeV&\citet{Pshirkov:2015hda,Pshirkov:2016qhu}
\enddata
\tablecomments{Comparison of the $\gamma$-ray emission when modeling M31 both as a point source and an extended source, where the extended source is based on estimates from the IRIS 100 $\mu$m map of the galaxy. Our baseline model consist of the FSSC isotropic (iso\_P8R2\_SOURCE\_V6\_v06), Galactic diffuse (gll\_iem\_v06), and 3FGL point sources. For this calculation we use a $14^\circ \times 14^\circ$ ROI, in order to be consistent with the other studies. The reported flux is integrated over the entire energy range. The TS is defined as $-2\Delta\log L$.}
\end{deluxetable*}

\begin{deluxetable*}{lccccccc}
\tablecolumns{7}
\tablewidth{0mm}
\tablecaption{Comparison for the gas and bubble templates \label{tab:comparison_2}}
\tablehead{
\colhead{Template}&
\colhead{$-\log(L)$}&
\colhead{TS}&
\colhead{Index}&
\colhead{Flux ($\times 10^{-9})$}&
\colhead{Isotropic}&
\colhead{Galactic}&
\colhead{Reference}\\
&
&
&
&
\colhead{(ph cm$^{-2}$ s$^{-1}$)} &
\colhead{Normalization}&
\colhead{Normalization}&
}
\startdata
Baseline FSSC			&--4875			&\nodata			&\nodata 				&\nodata				&0.91 \p 0.01				&1.017 \p 0.004 & this study\\
Gas 	FSSC			&--4880			&10			& 2.30 \p 0.05			&2.8 \p 0.3				&0.909 \p 0.005			&1.016 \p 0.002 & this study\\
Bubbles FSSC			&--4880			&10		& 2.33 \p 0.07 			&1.96 \p 0.20			&0.910 \p 0.006			&1.017 \p 0.003 & this study\smallskip\\
Pshirkov Gas		&\nodata		&22			& 2.3 \p 0.1			&3.2 \p 1.0			&\nodata					&\nodata &\citet{Pshirkov:2015hda,Pshirkov:2016qhu}\\
Pshirkov Bubbles	&\nodata		&28			& 2.3 \p 0.1			&2.6 \p 0.6			&\nodata				&\nodata&\citet{Pshirkov:2015hda,Pshirkov:2016qhu}
\enddata
\tablecomments{
Our baseline model consists of the FSSC isotropic (iso\_P8R2\_SOURCE\_V6\_v06), Galactic diffuse (gll\_iem\_v06), 3FGL point sources, new sources determined from our point source finding procedure, and an M31 template based on the IRIS 100 $\mu$m map of the galaxy. For this calculation we use a $14^\circ \times 14^\circ$ ROI. We use an energy range of 300 MeV -- 300 GeV. The gas template is a uniform $0.9^\circ$ disk centered at M31, and the bubble template consists of two uniform $0.45^\circ$ disks perpendicular to the M31 disk. For each respective fit, the respective template is added in addition to the IRIS template. Note that~\citet{Pshirkov:2016qhu} find an additional source (TS $=$ 11.2) near M31, and when including this source in the fit the TS for the gas template reduces to 15, and the change in TS for the bubble template was not reported. We found a source in a similar location with TS $=$ 14. The TS is defined as $-2\Delta\log L$.}
\end{deluxetable*}

 The initial analysis by the \emph{Fermi}-LAT collaboration modeled the emission in M31 both as a point source and using an elliptical template. All of the relevant measurements from that analysis are given in Table~\ref{tab:comparison_1}. Also shown in Table~\ref{tab:comparison_1} are the updated measurements from~\citet{Ackermann:2017nya}, the measurements from~\citet{Pshirkov:2015hda,Pshirkov:2016qhu}, and our current measurements (using the FSSC IEM). When M31 is modeled as a point source using a power law spectrum, the best fit index has a value of 2.75 \p 0.15, the total flux integrated over the energy range 0.3--300 GeV is (2.16 \p 0.35) $\times 10^{-9}$ ph cm$^{-2}$ s$^{-1}$, and the significance is TS $=$ 52. When modeled using the IRIS elliptical template the spectrum is harder, having an index of 2.46 \p 0.07, the total flux is (3.29  \p 0.24) $\times 10^{-9}$ ph cm$^{-2}$ s$^{-1}$, and the statistical significance is TS $=$ 62. 

The seeming discrepancies could be attributed to the different energy ranges and exposure times. With our GALPROP-based IEM, and using the energy range 1--100 GeV, the index for the inner galaxy template (0.4$^\circ$ uniform disk) is 2.8 \p 0.3. This is a somewhat softer spectrum than that measured in~\citet{Ackermann:2017nya} (2.4 \p 0.1), although the two values are still consistent within 1$\sigma$. This is likely attributed to the energy range used for the fit. The spectral analysis in~\citet{Ackermann:2017nya} shows a flattening of the flux below 1 GeV, which would result in a hardening of the spectrum. When repeating the fit with the FSSC IEM between 300 MeV -- 300 GeV, we obtain a best-fit index of 2.4 \p 0.1 for the inner galaxy component. 

Evidence of a spherical \gray\ halo around M31 with a $0.9^\circ$ extension is reported in~\citet{Pshirkov:2015hda}. In~\citet{Pshirkov:2016qhu} the morphology of the extended \gray\ emission is reported to consists of two bubbles symmetrically located perpendicular to the M31 disk, akin to the MW Fermi bubbles. We have tested both these templates. We found that there is evidence of an extended \gray\ halo around M31. However, we found no statistical preference between the disk template and the bubble template. The corresponding measured values are given in Table~\ref{tab:comparison_2}. In their analysis they used a disk (bubble) template $+$ the IRIS template. Performing the fit in this way we found that both templates, disk and bubbles, have a TS $=$ 10. Moreover, the characterization of  the \hi\ along the line of sight is a significant systematic uncertainty when it comes to determining the actual morphology of the extended emission from M31.

\section{Details for the Dark Matter Radial Profiles} \label{sec:DM}

Observational evidence for DM in M31 comes from measurements of its rotational velocity curve. Some of the earliest of these measurements were published by~\citet{babcock1939rotation}, \citet{Rubin:1970zza}, and \citet{roberts1975rotation}. These observations provide coarse-grained properties of the dark matter distribution near the central regions of the halo where the galaxy resides. With the existing data, the fine-grained structure of DM and its distribution outside of the galaxy is primarily inferred from simulated halos.

In 1997 a variety of studies culminated in the realization that the spherically averaged mass distribution of DM halos can be accurately described by an approximately universal profile, determined by the halo mass and halo characteristic density, as introduced by Navarro, Frenk, and White (NFW)~\citep{Navarro:1996gj,Hayashi:2006es}. However, individual DM halos are not necessarily expected to be smooth, nor spherically symmetric, especially on galactic scales~\citep{Kamionkowski:1997xg,Braun:1998ik,blitz1999high,deHeij:2002ne,Helmi:2003pp,Braun:2003ey,Bailin:2004wu,Allgood:2005eu,Hayashi:2006es,Bett:2006zy,Diemand:2006ik,Kuhlen:2007ku,Springel:2008cc,Banerjee:2008kt,Law:2009yq,Zemp:2008gw,Saha:2009dt,Banerjee:2011rr,Velliscig:2015ffa,Bernal:2016guq,Moline:2016pbm,pawlowski2017lopsidedness}.

DM halos can form irregular shapes depending on their environment and formation history. In general, the geometry may either be spherical, ellipsoidal with an allowed minor to major axis ratio $c/a$ as low as $\sim$0.4, or even lopsided. In addition, the local filament structure of the cosmic web may also affect the halo geometry~\citep{Zhang:2009gh,Carlesi:2016qqp,pawlowski2017lopsidedness}. Moreover, M31 and the MW cannot necessarily be considered as two isolated halos, as, in fact, it is possible that they are interacting and may be connected by a DM filament~\citep{Carlesi:2016qqp,pawlowski2017lopsidedness}, and such a feature would predict additional DM substructures in the M31 field of view.

Simulations of cold DM, extended to smaller scales with semi-analytic models of hierarchical structure formation, indicate that halo substructure amounts to as much as $\sim$10--40\% of the total (MW-size) halo mass; however, simulation resolution remains a limiting factor in these studies~\citep{Diemand:2006ik,Kuhlen:2007ku,Springel:2008cc,Zemp:2008gw,Moline:2016pbm}. The presence of substructure is especially important for indirect detection, as it provides a significant boost to the annihilation intensity, such that the substructures dominate over the NFW profile, except near the halo center. Note that the flux enhancement is most important for more massive halos, as they enclose more hierarchical levels of  structure formation~\citep{Ng:2013xha,sanchez2014flattening}. Thus, in the case of dwarf galaxies, a boost factor is not expected to be as important for indirect searches. The main uncertainty pertaining to the boost factor is the low mass behavior of the halo substructures, which includes the number of low-mass halos and their individual density profiles. In addition, the presence of the galactic disk is predicted to have an effect on the substructure content. Tidal forces near the disk may act to break apart the substructure, resulting in a smaller substructure fraction within a radial distance of $\sim$50 kpc~\citep{garrison2017not}. 

The prompt \gray\ flux for DM annihilation is given by
\begin{equation} \label{eq1}
\frac{dN_\gamma}{dE} =\bigg[\sum_{f}\frac{ \langle \sigma_f v \rangle}{4 \pi \eta  m_\chi ^2} \frac{dN_\gamma^f}{dE}\bigg]J,
\end{equation}
where the summation is over annihilation final states \textit{f} (i.e.\ up-type quarks, down-type quarks, leptons, etc.), $dN_\gamma^f/dE$ is the number of photons produced for a single annihilation into final state \textit{f}, $\langle \sigma_f v \rangle$ is the thermally averaged cross section for final state \textit{f}, $m_\chi$ is the DM particle mass, and $\eta = 2(4)$ for self-conjugate (non-self-conjugate) DM. The total cross section $\langle \sigma v \rangle$ is the sum of the cross sections for all final states $\langle \sigma_f v \rangle$. The quantity in large brackets depends on the particle nature of the DM and is referred to as the \emph{DM attribute quantity.} Note that we take all annihilation channels to be $s$-wave dominated (for which $\sigma v$ does not depend on $v$) over the kinetic energies found in the M31 halo. Other velocity-dependent scenarios can be considered, but in these cases the velocity distribution of M31 must be modeled and the average $\langle \sigma v\rangle$ must be included in the line-of-sight integration of $J$~\citep{Campbell:2010xc}.  

The $J$-factor ($J$) characterizes the spatial distribution of the DM, and is given by the integral of the mass density squared, over the line of sight. When describing the dark matter distribution as an ensemble of disjoint dark matter halos, the $J$-factor is:
\begin{equation} \label{eq2}
J=\sum_i\int_{\Delta\Omega}d\Omega\int_{\text{LoS}}ds\rho_i^2(\mathbf{r}_i(s,\mathbf{n})),
\end{equation}
summed over all halos in the line of sight (LoS), where $\rho_i(\mathbf{r})$ is the density distribution of halo $i$, and $\mathbf{r}_i(s,\mathbf{n})$ is the position within that halo at LoS direction $\mathbf{n}$ and LoS distance $s$. The spherically-averaged DM halo density profile $\rho(r)$ for each halo is often taken to be a generalized NFW profile (however, other profiles are also possible):
\begin{equation} \label{eq3}
\rho (r) = \rho_s \left ( \frac{r}{R_s}\right)^{-\gamma}\left(1 + \frac{r}{R_s}\right)^{\gamma - 3}.
\end{equation}
Here, $\gamma$ specifies the inner spectral slope of the profile, $R_s$ is the scale radius, and $\rho_s$ is the scale density, often determined for the MW halo from the local DM density. 

$J$-factors determined from these spherically-averaged profiles (denoted $J$) are an underestimate of the total $J$-factor (denoted $J^\prime$) because of the effect of the non-spherical structure. This underestimate is typically encoded with a boost factor ($B$) such that
\begin{equation}
J^\prime = B  J,
\end{equation}
with $B$ determined from the model of halo substructure.

To give a sense of the DM properties typically implicated by the GC excess, for the \emph{DM attribute quantity} we use the results from~\citet{Karwin:2016tsw} (the pseudoscalar interaction model), which is based on the IG IEM. The spectral characteristics for this model favor a DM particle with a mass in the range $\sim$50--190 GeV and annihilation cross section $\langle \sigma v \rangle$ in the range $\sim$$3 \times 10^{-26} - 8 \times 10^{-25}$ cm$^{3}$ s$^{-1}$ (for Dirac DM). For masses above $\sim$175 GeV the annihilation final state is mostly up-type quarks, and below 175 GeV the annihilation final state is mostly down-type quarks. For each case, a small fraction ($\lesssim$5\%) of the annihilation also goes to leptonic final states. Note that other DM scenarios are also possible, but the corresponding DM mass and annihilation cross section typically fall in the ranges given above. Also note that an excess in the AMS-02 anti-proton spectrum has recently been reported which implies a DM interpretation consistent with the GC excess~\citep{Cuoco:2016eej,Cholis:2019ejx} (see~\citet{Reinert:2017aga} for a less optimistic view).

\begin{figure*}
\centering
\includegraphics[width=0.4\textwidth,center]{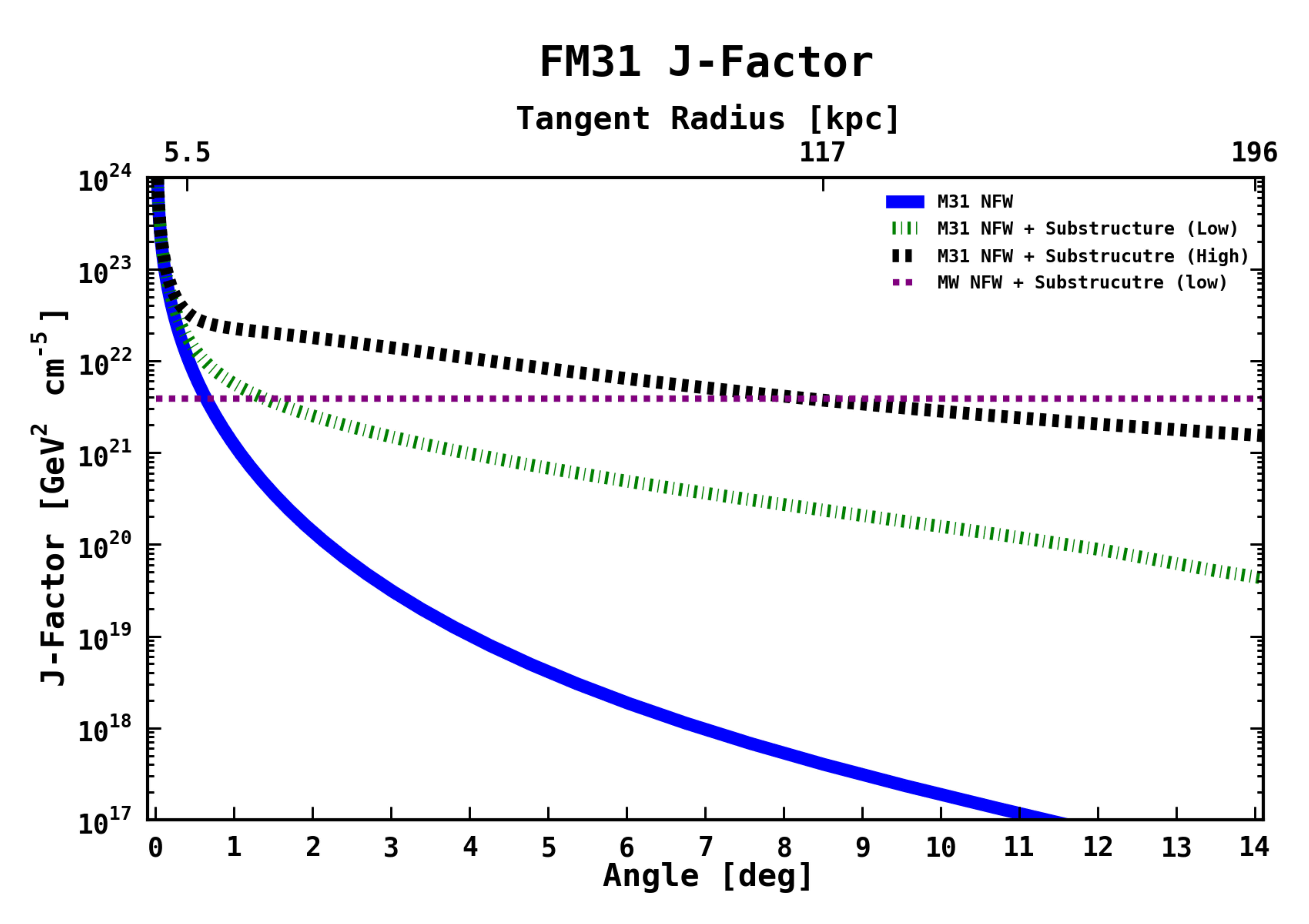}
\caption{The blue solid line shows a smooth NFW halo appropriate for warm DM models that do not produce significant structure below the dwarf galaxy scale. The parameters of the NFW profile are as follows: mass $= 10^{12} M_\odot$, concentration = 11.2, $R_{\rm virial} = 210$ kpc, $R_{\rm scale} = 18.9$ kpc, $\gamma= 1.0$. The green (lower) dashed line, labeled NFW $+$ Substructure (Low), shows the expected DM for a typical $\Lambda$CDM cosmology with thermal WIMP DM. For the NFW $+$ Substructure (Low): overall boost factor $=$ 4.2, substructure fraction $=$ 13\%, minimum halo mass $=$ $10^{-6} M_\odot$. The corresponding MW contribution along the line of sight is shown with a thin dashed purple line. The black (upper) dashed line, labeled NFW $+$ Substructure (high), shows a scenario in which DM is produced very cold such that the minimum mass structures form with very high concentrations. These smallest structures would dominant the annihilation signal. For the NFW $+$ Substructure (High) we use results from \citep{Gao:2011rf}.}
\label{fig:_radial_J_Factor}
\end{figure*}

The hatched regions in Figure~\ref{fig:M31_radial_profile} show predicted intensity profiles for DM annihilation in M31 corresponding to three different DM substructure scenarios. The blue (slanted hatch) region shows a smooth NFW halo appropriate for warm DM models that do not produce significant structure below the dwarf galaxy scale~\citep{pagels1982supersymmetry,peebles1982large,olive1982cosmological,Colombi:1995ze,maccio2012cores,lovell2014properties,bose2016substructure,Bose2016irl}. Other alternatives to warm DM are the scenarios of late kinetic decoupling~\citep{bringmann2016suppressing} and dark acoustic oscillations through self-interactions, through interactions with the standard model~\citep{boehm2001constraining,boehm2005constraints,hooper2007possible}, or interactions with a dark sector thermal bath~\citep{feng2009hidden,cyr2013cosmology,van2012thermal,cyr2014constraints,buckley2014scattering,Cherry:2014xra,cyr2016ethos,Binder:2016pnr,huo2018signatures}. The parameters of the NFW profile are as follows: mass = $10^{12} M_\odot$, concentration = 11.2, $R_{\rm virial} = 210$ kpc, $R_{\rm scale} = 18.9$ kpc, $\gamma = 1.0$. 

The presence of a (DM) \gray\ halo around M31 would likely indicate that the line of sight extends through a similar halo surrounding the MW. To obtain a simple estimate for this scenario, we model M31 and the MW as two isolated spherical halos separated by 785 kpc, both described by the same NFW mass profiles (as given above), and both with the same substructure content. We calculate $J$-factors for an observer inside the MW halo at a distance of 8.5 kpc from the Galactic center. Our standard cold DM halo substructure is modeled with a radial dependent subhalo mass function with tidally truncated density profiles, as described in \citet{ludlow2016mass} and \citet{han2016unified}. The relevant parameters are as follows: overall (M31) boost factor = 4.16, substructure mass fraction = 13\%, minimum halo mass = $10^{-6} M_\odot$. This scenario corresponds to the expected DM signal for a typical $\Lambda$CDM cosmology with thermal WIMP DM. The corresponding intensity profiles are plotted in Figure~\ref{fig:M31_radial_profile}, and labeled as NFW + Substructure (Low). The green (slanted cross-hatch) region shows the M31 component, and the black band shows the corresponding MW component in the line of sight. 

The black (vertical cross-hatch) region, labeled NFW + Substructure (High), shows a scenario in which DM is produced very cold such that the smallest substructures around $10^{-6} M_\odot$ have large concentrations which dominate the $J$-factor, as described in~\citet{Gao:2011rf}. Corresponding $J$-factors are given in Figure~\ref{fig:_radial_J_Factor}.

\end{document}